\documentclass[referee]{mn} 
\usepackage{epsfig}
\usepackage{amsfonts}
\usepackage{subfigure}
\usepackage{enumerate}

\title{Quasar candidates selection in the Virtual Observatory era}
\author[D'Abrusco et al 2007]
 {D'Abrusco$^{1,2}$, R., Longo$^{1,3,4}$, G., Walton$^{2}$, N. A.\\ \\
 1 - Department of Physical Sciences, University of Napoli Federico II, via Cinthia 9, 80126 Napoli, ITALY\\
 2 - Institute of Astronomy, Cambridge, Madingley Road, UK\\
 3 - INAF - Osservatorio Astronomico di Capodimonte, via Moiariello 16, 80131, Napoli ITALY\\
 4 - INFN - Napoli Unit, Dept. of Physical Sciences, via Cinthia 9, 80126, Napoli, ITALY}
\date{Accepted ;
      Received ;
      in original form }
\pagerange{\pageref{}--\pageref{}} \pubyear{}

\begin{document}

\label{firstpage}
\maketitle

\begin{abstract}
We present a method for the photometric selection of candidate quasars in multiband surveys. The method makes use of a priori knowledge derived from a subsample of spectroscopic confirmed QSOs to map the parameter space. The disentanglement of QSOs candidates and stars is performed in the colour space through the combined use of two algorithms, the Probabilistic Principal Surfaces and the Negative Entropy clustering, which are for the first time used in an astronomical context. Both methods have been implemented in the VONeural package on the Astrogrid VO platform. Even though they belong to the class of the unsupervised clustering tools, the performances of the method are optimized by using the available sample of confirmed quasars and it is therefore possible to learn from any improvement in the available ''base of knowledge''. The method has been applied and tested on both optical and optical plus near infrared data extracted from the visible SDSS and infrared UKIDSS-LAS public databases. In all cases, the experiments lead to high values of both efficiency and completeness, comparable if not better than the methods already known in the literature. A catalogue of optical candidate QSOs extracted from the SDSS DR7 Legacy photometric dataset has been produced and is publicly available at the URL {\it http://voneural.na.infn.it/qso.html}. 
\end{abstract}


\section{Introduction}

Over the years, serendipitous discoveries and systematic searches have caused the number 
of confirmed quasars to grow dramatically, but we are still far from having discovered 
even a significant fraction of the $\sim$ 1.6  million QSOs which are expected to populate 
the universe out to $z \simeq 3$ (e.g. \cite{richards_2004}). Thanks to their high intrinsic 
brightness, such lack of coverage is not due to the limited depth of the observations aimed at 
discovering QSOs, but mainly to the difficulties encountered, first in disentangling normal 
stars from QSO candidates, and then in confirming their nature through additional data 
(such as spectroscopy, radio or X-ray fluxes) which are usually either difficult or very 
time-consuming to provide for statistically significant samples of objects. This is very 
unfortunate since, as it has been stressed by many authors \cite{richards_2002,richards_2004}, 
sizeable samples of quasars covering a broad range of redshifts and selected with uniform 
and well controlled criteria, are greatly needed to address many relevant issues such as the 
evolution of the quasar luminosity function or the spatial clustering of quasars as a function of the redshift. 

\noindent The most notable recent efforts at building extensive samples have been the 
quasars searches in the Sloan Digital Sky Survey (hereafter SDSS, \cite{york_2000}) and 
the main competing project, the 2dF QSO redshift survey \cite{croom_2001} which will 
soon be joined by ongoing or planned multiband photometric survey projects such as, for 
instance, the Palomar Quest \cite{PQ}, or the VST \cite{VST} and VISTA \cite{VISTA} 
surveys. From the photometric standpoint, all quasar candidates selection algorithms are 
based on a few simple facts: 

\begin{enumerate}
\item the spectral energy distribution of stars is roughly blackbody in shape, while quasars have 
SEDs that are characterized by featureless blue continua and strong emission lines. In particular, at low redshifts ($z \leq 2.2$), the lack of a Balmer jump in quasars separates them from hot stars; at higher redshifts the presence of a strong $Ly_{\alpha}$ emission line and of absorption by the $Ly_{\alpha}$ forest results in the broadband colours of quasars to become increasingly redder with redshift. These, in general, is the reason why quasars can be distinguished by stars when looking at their position in a colour space generated by observations in a broadband photometric system: stars and QSOs may have the same colour in some wavelength region, but the colours in the other regions will be different; 
\item as stressed by \cite{richards_2001}, the overall shape of the continuum of quasars is well 
approximated by a power law. The combined effect of the different shapes of the continuum emission and the presence of spectral features not observed in stars is that QSOs show colours which vary significantly as a function of redshifts, but locally (in redshift) the distribution of colours is narrow and not-degenerate;
\item at high redshifts, the density of absorption lines in the spectral energy distribution of QSOs increases, so that, when the  region of the quasar rest frame spectrum shortwards of 912 $\AA$ enters the observed band, absorption due to the intervening optically thick clouds further reddens the spectrum, causing the quasar to move away from the region of the colours space occupied by Galactic stars. 
\end{enumerate}

\noindent In the particular case of infrared wavelengths, the availability of large-field detectors on 
large field of view telescopes has provided the opportunity to undertake surveys capable 
of establishing the importance of the main mechanism of reddening, i.e. the extinction by 
dust, on the observed population of quasars. Since the spectral energy distributions of 
quasars vary significantly as a function of the wavelength, flux measurements at widely 
separated wavelengths are used to characterize fully the spectral properties of the quasar 
population. More precisely, two methods exploiting the differences between the power-law 
nature of quasar spectra and the convex spectra of stars have been proposed to 
select candidate quasars, based on the fact that quasars are significantly brighter than stars 
at both short wavelengths - the UVX method \cite{sandage_1965,hewitt_1993} - and long wavelengths - 
the KX method \cite{warren_2000}. In order to build large samples of quasars a major goal 
is to improve the reliability and efficiency of the algorithms used to extract from multiband 
survey data the list of quasar candidates. Most current algorithms are typically more than 
60\% efficient for UV-excess (UVX) quasars at relatively bright magnitudes, but the selection 
efficiency drops at fainter magnitudes where the photometric errors are largest and most of the observable 
objects resides. As shown by \cite{richards_2004}, however, it is possible to build algorithms 
achieving levels of accuracy and completeness (for a definition of these terms see Section 
(\ref{Sec:methods})) which can mitigate the need for confirming spectra. One additional 
fact that needs to be noticed is that the efficiency of all quasar candidate selection algorithms 
depends on some degree on the fine tuning of the algorithms on what we could call the \emph{a priori} 
knowledge (hereafter ''Base of Knowledge'' or BoK) of what quasars are and on what 
their main characteristics are. This BoK may be either built out of synthetic spectra or from 
available quasar samples. The latter approach, being based on fewer \emph{a priori} 
assumptions, seems preferable but, on the other end, it keeps all biases introduced by 
the selection criteria (to be more explicit: if a specific subclass of objects is not present 
in the BoK, the algorithm will not be able to recognize them). In a near future, however, the large 
amount of data which will be made available to the community through the Virtual 
Observatory \cite{walton_2002}, will provide an ever growing (both in size and accuracy) 
BoK which will allow to overcome at least some of the existing limitations. In this paper 
we present a method based on unsupervised clustering capable to map the photometric parameter 
space using the information contained in the BoK and to disentangle stars from candidate quasars. 
Finding candidate QSOs is only the first step in the process of the characterization of the distribution of such sources in the Universe, the second being the production of a three dimensional map of their position through the estimate of the redshifts for each of them. A method for the calculation of accurate photometric redshifts based on a machine learning approach similar in its philosophy to the one described in \cite{dabrusco_2007} and intended to complement the present QSOs candidate selection algorithm is in preparation, and will be discussed in a forthcoming paper by the same authors.
Even though it is applied to the SDSS and/or UKIDSS public data, the method is of general validity 
and can be easily adapted to any other data set given that a proper BoK is available.
It needs to be explicitly stressed that the effectiveness and accuracy of the method presented
in this paper strongly depend on the completeness and accuracy of the assumed "a priori" knowledge
contained in the BoK, and, in this sense, this method does not overcome the biases contained in the BoK. 
However, with the advent of VO technology the quality and accuracy of the BoK's 
is bound to increase dramatically, since the growing availability of federated data sets
observed at different wavelengths will remove most of the systematic selection effects encountered 
in the current data and will improve the completeness of the selection of members of the BoK. 
In this respect, one of the main advantages of this particular method is that it can easily adapts to 
any improvement of the BoK. In section (\ref{Sec:thedata}) we present the main characteristics of 
the data used for the experiments and in Section (\ref{Sec:methods}) we shortly summarize the 
main methods used so far for the selection of candidate QSOs from the SDSS catalogues. In Section 
(\ref{Sec:method}) we introduce the clustering algorithm and the agglomerative method and in 
Section (\ref{Sec:experiments}) we discuss the results of the experiments performed. The catalogue 
of optical QSO candidates extracted from ther SDSS photometric dataset and the base of knowledge 
used for the candidates selection are described in Section (\ref{Sec:catalogues}), while the conclusions 
are drawn in Section (\ref{Sec:conclusions}). Besides the experiment described in section (\ref{Sec:experiments}), in order to guide the reader through the subtleties of the method presented here, in section (\ref{Sec:methods}), together with the theoretical aspects of the algorithms employed, an 
experiment performed on a simplified version of one of the data samples is described. A restricted
BoK and a different set of parameters for the algorithms will be used as well for the sake of clarity.
In two forthcoming papers \cite{dabrusco_2008,cavuoti_2008} we shall discuss the application of 
the method to the selection of heavily obscured quasars and to the physical classification of AGNs, 
respectively.

\section{The data}
\label{Sec:thedata}

\subsection{SDSS data}
\label{sdssdata}

The Sloan Digital Sky Survey is a digital survey aimed at covering $\sim 10,000$ sq. deg. 
mainly in the Northern emisphere \cite{stoughton_2002} in five specifically designed bands 
$(u,g,r,i,z)$ \cite{fukugita_1996} and is complemented by an extensive redshift survey for 
about $10^6$ objects (mainly galaxies and QSOs). The SDSS data are made available to the 
community through a public archive which at the moment is distributing its Seventh
Data Release Legacy Survey (hereafter DR7) \cite{abazajian_2008}. So far, SDSS provides the best data 
set to date for those interested in mining for photometrically selected 
subsamples of objects. As such, it has been extensively studied in almost all its 
aspects and an impressive amount of literature has been produced providing an 
accurate knowledge of completeness, selection effects etc. \cite{adelman_2007}.
As to quasar selection from the SDSS data, it is worth to recall a few facts.
The SDSS photometric system does not allow the detection of quasars with 
$z >6$ and; with the additional constraint of having the objects detected in 
at least two bands, this limit reduces to $z \sim 5.8$ \cite{fan_2001b,richards_2002}.
At the low- end, the design of the $u$ filter and the location of the gap 
between the $u$ and $g$ filters were chosen to emphasize the difference between 
objects with power-law spectral energy distributions (SEDs), such as quasars at 
$z < 2.2$, and objects that are strongly affected by the Balmer decrement, 
in particular A stars, which are recognized as the prime contaminants in multicolor 
optical searches for low-redshift quasars.  

\subsection{UKIDSS data}
\label{ukidssdata}

The United Kingdom Infrared Deep Sky Survey (hereafter UKIDSS) \cite{lawrence_2007},  
is a near-infrared sky survey that will cover $7500$ square degrees of the Northern sky, 
extending over both high and low Galactic latitudes, in $(Y, J, H, K)$ bands down to $K\simeq 18.3$, thus reaching three 
magnitudes deeper than 2MASS \cite{jarrett_2000}. UKIDSS has been designed and operated to be the 
near-infrared counterpart of the SDSS survey; it is made up of five separate 
surveys and includes two deep extra-Galactic elements, one covering 35 square degrees 
down to $K=21$, and the other reaching $K=23$ over 0.77 square degrees of the sky. 
In this work we make use of the UKIDSS Large Area Survey (hereafter LAS) which aims at 
covering an area of 4,000 deg$^{2}$ overlapping with the SDSS. The LAS is expected to be 
completed in 2012, after an observing period of seven years. LAS is surveying the sky in four 
photometric bands $(Y, J, H, K)$ with typical limiting magnitudes $[20.5, 20.0, 18.8, 18.4]$ and 
astrometric accuracy typically $< 0^{''}.1$. The UKIDSS DR1 \cite{dye_2006} release overlaps 
a subset of the SDSS northern and southern areas with photometric and astrometric 
performances similar to the SDSS.

\subsection{The bases of knowledge}

\noindent In this work, three different samples of objects were used as BoKs in order to evaluate 
the performances of our candidate selection algorithm with respect to the original SDSS quasars 
candidate selection technique (see paragraph (\ref{Subsec:SDSSselecalgorithm})) and the algorithm 
proposed by Richards and his collaborators \cite{richards_2004} (see paragraph (\ref{Subsec:kdeselecalgorithm})) 
respectively. The first and third datasets (S-A and S-S) employ only optical photometric and spectroscopic 
data from the SDSS survey, while the second dataset (S-UK), despite based on the same spectroscopic 
classification provided by Sloan database, uses both optical and near infrared photometry, the latter 
extracted from the UKIDSS database (for details, see paragraph (\ref{ukidssdata}). This last dataset 
has been used to perform an experiment aimed at comparing the efficiencies and completenesses 
provided by our method when the photometric parameter space where the selection takes place, 
is extended by adding infrared colours to the optical colours. 
\noindent The first sample (hereafter S-A) is formed by candidate quasars selected 
from the SDSS-DR5 database, classified as unresolved (i.e. belonging to the table 
"Star"), and for which the spectroscopic classification index "specClass" was available 
together with a spectroscopic redshift for each object.  Such index classifies objects in 6 
different classes:  $SP=0$ unknown sources, $SP=1$ stars, $SP=2$ galaxies, $SP=3$ nearby AGN, $SP=4$ quasars; 
$SP=5$ sky, $SP=6$ late type stars. Since most SDSS QSOs fall into the star-like category, 
the possible values of the "specClass" index for the objects in our data sets are 0, 1, 3, 4 \& 6. 
The objects considered in the S-A sample were selected inside a roughly rectangular 
patch of the sky, situated in the equatorial region matching with the area covered by the 
data release 1 of UKIDSS LAS observations (see next paragraph). For 2519 sources 
matching the selection criteria, the point spread function magnitudes "psfMag" in the five 
SDSS bands $(u, g, r, i, z)$ were retrieved.  

\noindent The second sample (hereafter S-UK) is formed by all objects belonging to 
the SDSS-DR5 "Star" (containing all star-like photometric sources) table (with 
spectroscopic classification available) positionally matching with UKIDSS-DR1 LAS 
objects which have been also classified as stars according to the "mergedClass" 
classification index (requiring that "mergedClass" = -1). The matching was performed 
selecting all unresolved sources in LAS "lasSource" database table placed within a maximum
distance of 10 arc-seconds from the SDSS source and selecting the closest to the optical source. 
Only in few cases ($\sim 5\%$) the separation between the members of matched couples exceeds 
4 arc-seconds, as a result of the good relative astrometry of SDSS and UKIDSS surveys, and these 
couples have been checked visually in order to reduce the number of false matches. On the total of 
120 matchings of this kind, only $\sim 20$ have been rejected. For this sample optical PSF magnitudes 
from SDSS and near infrared PSF magnitudes in the four LAS UKIDSS $(J, Y, H, K)$ bands have been retrieved. 
A total of 2192 candidate quasars were successfully selected according to these 
prescriptions. 

\begin{figure}
\centering
\includegraphics[width=14cm]{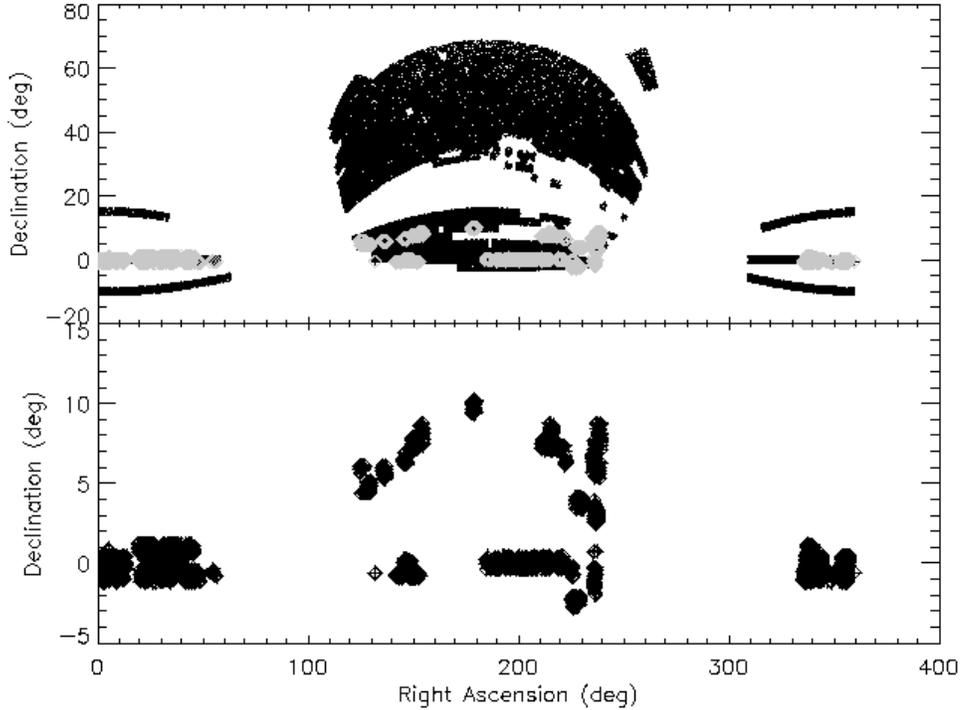}
\caption{Positions of objects belonging to the three samples (see text). The upper figure represents the 
S-S sample while the lower shows the positions of the S-A sample members as black crosses and 
those of the S-UK sample members as grey diamonds.}
\label{samples_pos}
\end{figure}

\noindent The third sample (hereafter S-S) is composed by star-like sources 
belonging to the SDSS-DR5 "Target" table and selected as candidate quasars according 
to the algorithm described in \cite{richards_2002}. For all these objects spectroscopic 
classification ("specClass") and redshifts are available. The only additional constraint 
applied to these objects is that their "psf" magnitudes are required to be measured correctly
in all photometric bands (using the photometric attribute 'flags' contained in the PhotoObjAll 
table). The number of objects selected according to these requirements 
is 94196. 

\begin{table*}
\caption{Total number of members and "specClass" distribution of the three samples.}             
\label{table:2}      
\centering          
\begin{tabular}{l c c c c c c}    
\hline\hline       
Sample & Total & SP = 0 & SP = 1 & SP = 3 & SP = 4 & SP = 6\\ 
\hline
S-A   & 2719 & 43 (1.6\%)&1176 (43.3\%) &  827 (30.4\%) & 73 (2.6\%) & 600 (22.0\%) \\  
S-UK&  2159  & 23 (1.0\%)&954   (43.5\%) &  773 (35.3\%) &  69 (3.1\%) & 373 (17.0\%) \\
S-S   &  94197   & 2609 (2.4\%) & 22636 (20.4\%) & 53554 (48.4\%) & 4661 (4.2\%)  & 10737 (9.67\%) \\ 
S-SCat & 158345 & 1782 (1.1\%) & 58801 (37.1\%)& 59052 (37.3\%) & 6332 (4\%) & 32360 (20.4\%) \\
\hline                  
\end{tabular}
\end{table*}

\begin{figure}
\centering
\includegraphics[width=8cm]{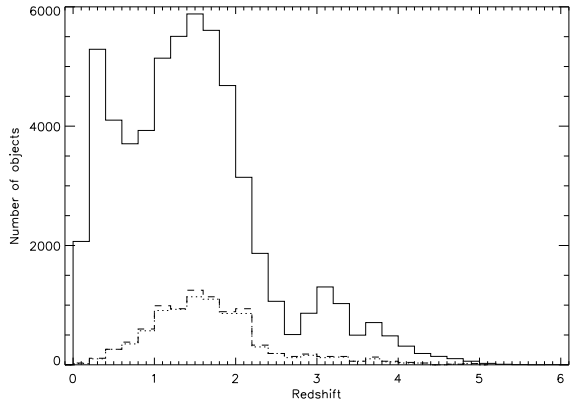}
\caption{Redshift distribution for the three samples used in this work. S-A (dashed line) and 
S-UK (dotted line) samples counts have been multiplied by ten to increase their visibility.}
\label{samples_red}
\end{figure}

In figures (\ref{samples_pos}) and (\ref{samples_red}), we plot the positions and the redshift distribution for the members of the samples. The S-A and S-UK samples differ only for a few objects for which one or more UKIDSS magnitudes have not been measured and therefore their distributions in redshift space are almost identical. The distribution in redshift of S-S objects is characterized by a peak at $z\sim 1.7$. The composition of the different datasets described above in terms of the spectroscopic classification index "specClass", in addition to the composition of the dataset called "S-SCat" used to extract the catalogue of photometric candidate QSOs described in paragraph (ref{Sec:catalogues}), are given in table (\ref{table:2}).

\section{Photometric selection of quasars}
\label{Sec:methods}

As a result of their distinct colours, the general idea behind all quasar candidate selection algorithms working in photometric parameter spaces is that quasars tend to lay far away from the regions 
occupied by normal stars (i.e. what is usually called \textsf{stellar locus}). Therefore, from the mathematical point of view, the problem of quasars candidate selection can be regarded as that of properly partitioning the parameter space in order to isolate the regions populated by quasars, minimizing the level of contamination from stars and the number of missed quasars. Spectroscopic observations of candidate quasars are crucial for quasars confirmation and to test the relative performances of the algorithm used to identify candidates. Such performances are usually expressed by two parameters, called respectively ''completeness $c$'' and ''efficiency $e$'', and defined as follows:

\begin{equation}
c = \frac{\textsf{N candidates }}{\textsf{N \emph{a priori} known QSOs}}; \ \ \ \ e = \frac{\textsf{N confirmed}}{\textsf{N candidates}}
\end{equation}

It is apparent from the above definitions that $c$ provides a measure of how good is the method at retrieving all quasars in the sample, while $e$ provides a measure of the contamination in the list of candidates selected by the algorithm. In what follows we shall identify as ''base of knowledge'' 
(BoK) the spectroscopically studied objects which can be used to build the samples of \emph{a priori} known and ''confirmed'' quasars. The optimal balance (if anything like that exists at all...) between completeness and efficiency is a delicate one since stars outnumber quasars by several orders of magnitude, and improving the efficiency by rejecting objects in regions of color space in which both stars and quasars are located, necessarily affects the completeness. Due to the unavoidable incompleteness in spectroscopic surveys (and as a matter of fact, of any other type of selection criterion), the BoK is always affected by biases which reflect into the value of $c$. In other words, in a given photometric catalogue, in order to have an exhaustive list of \emph{a priori} known quasars, all objects should be observed spectroscopically and the same holds true for the list of candidates. 
The BoK is also needed to evaluate the errors. As an example, the knowledge of the intrinsic spread in quasar spectral indices translates into a lack of knowledge of the intrinsic spread in quasar colors \cite{richards_2002}. The first attempt to produce a list of candidate quasars from multicolor survey data was by \cite{sandage_1965}. This pioneeristic attempt was soon followed by many others 
\cite{koo_1982,schmidt_1983,warren_1990,warren_1991}, and more recently by \cite{hewett_1995,hall_1996,croom_2001,richards_2002,richards_2004}. In what follows we shall shortly summarize two of them, focusing on those which have been tested on the SDSS data set and with whom our method will be compared.

\subsection{Original SDSS selection algorithm}
\label{Subsec:SDSSselecalgorithm}

The official SDSS quasars candidate selection algorithm \cite{richards_2002}  (R02) is sensitive to quasars at all redshifts lower than $z \leq 5.8$ (i.e. very close to the theoretical limit predicted for the SDSS), and to atypical AGNs such as broad absorption line quasars and heavily reddened quasars. Performances of this algorithm, as stated in the paper, are a completeness $c \sim 90 \%$ and an efficiency $e \sim 65\% $. The R02 algorithm under-samples certain regions of the colours space where degeneracy between colours of quasars in the redshift range $[2.2, 3.0]$ arises. Since this redshift range is crucial for the cosmological applications which were the primary target of SDSS \cite{stoughton_2002}, objects falling in these regions were nonetheless selected paying the price of a worse overall efficiency. \\
Star-like objects are selected in a four dimensional colour space defined by the $(u, g, r, i, z)$ SDSS bands. Non stellar candidates are selected via their colours and by matching unresolved sources to the FIRST radio catalogues. \noindent The R02 algorithm can be summarised as follows: i) objects with spurious and/or problematic fluxes in the imaging data are rejected; ii) matches to unresolved FIRST radio sources are preferentially targeted without reference to their colors;  iii) the sources remaining after the first step are compared to the distribution of normal stars and galaxies in two distinct three-dimensional colour spaces, one for low-redshift quasar candidates (based on the $ugri$ colors) and one for high-redshift quasar candidates (based on the $griz$ colours). The two groups are selected down to the limiting magnitudes $i^* \sim 19.1$ and $\sim 20.2$ respectively. Colour selection is performed accordingly to the their distance from a modelled, fixed hypersurface containing the stellar locus which, for a given photometric system, has been shown to be rather stable with respect to changes in stellar populations (e.g. \cite{richards_2002}). No specific line is drawn between quasars and other types of active galactic nuclei. The completeness of the SDSS algorithm for QSOs candidate selection was also empirically measured by \cite{vandenberk_2005}, using spectra of nearly 20,000 unresolved SDSS quasars to a reddening corrected magnitude of $i = 19.1$. On the 278 $deg^{2}$ of the total area of the completeness survey, the combined colour and radio selection algorithms yield a completeness of $\sim 94.9^{+2.6}_{-3.8}\%$ at the $90\%$ confidence level. The completeness for the separate colour and radio selection algorithms  are $\sim 93.8^{+2.6}_{-3.8}\%$ and $\sim 6.8^{+0.9}_{-0.9}\%$ respectively at the $90\%$ confidence level, leading to a higher estimate than the value reported by the R02 paper on a significantly larger sample of sources which have been spectroscopically observed.

\subsection{Kernel Density Estimation-based QSO selection algorithm}
\label{Subsec:kdeselecalgorithm}

The method for candidate QSOs selection discussed in \cite{richards_2004} is based on the application of a nonparametric bayesian classification algorithm to evaluate the probability distribution
function of different classes of sources inside the four dimensional SDSS photometric parameter space defined by the $(u, g, r, i, z)$ bands. The so-called kernel density estimation (hereafter KDE) algorithm is employed for the evaluation of the probabiliy density functions of stars and quasars distributions respectively, on a training set of labelled sources observed during the spectroscopic SDSS survey. After the training and test of the method, each photometric source is associated to the class of sources (stars or quasars) for which the estimated likelihood is higher. The performance of this method is expressed by an efficiency of 95.0\% and a completeness of $\sim 90\%$ evaluated on the catalogue of unresolved UVX-excess candidates extracted from the DR1 SDSS photometric database. This method has been applied to the whole SDSS DR6 sample yielding a catalogue of about 1 million candidate QSOs \cite{richards_2008}, with a training set for QSOs comprising SDSS DR5 quasars dataset as listed by \cite{schneider_2007}, quasars discovered during the first observing season of the OAAOmega-UKIDSS-SDSS QSO survey, the whole high-redshift ($z > 5.7$) sample of quasars discovered in SDSS data to date \cite{fan_2006} and the 920 point sources  selected as highly likely quasars from cross-comparison of SDSS and Spitzer data, according to two different mid-IR colour selections criteria \cite{lacy_2004,stern_2005}. While the completeness is slightly smaller than the value measured using the first catalogue ($\sim 93.4\%$), the efficiency is almost unchanged, reaching an overall value of $\sim 95\%$.

\section{Unsupervised quasar selection}
\label{Sec:method}

Quasars candidates detection can be achieved using unsupervised clustering 
algorithms inside photometric colours space distribution of objects for which a
reliable base of knowledge is available. The method presented in this paper 
follows a hierarchical approach which, starting from a preliminary clustering 
performed on the objects inside the parameter space, is followed by a second 
phase of agglomeration which reduces the initial number of clusters produced 
in the first step to an \emph{a priori} unknown numbers of final clusters. 
We than have a phase which we shall call "labelling", based on the existing 
BoK, i.e. on the objects for which independent spectroscopic confirmation is 
available. This labelling is used to refine the partition of the parameter space 
in order to define the stellar and quasar loci. The characterization of the final 
clusters is then used to select \emph{ex novo} the candidate QSOs.

\noindent The unsupervised\footnote{PPS as most other unsupervised 
algorithms require the number of clusters to be provided by the user. In our
approach this limitation can be circumvented by assuming a number of clusters 
much higher than what could be realistically be present in the data.} clustering
is accomplished using the Probabilistic Principal Surfaces algorithm which, 
strictly speaking is not a clustering algorithm but rather a nonlinear 
generalization of principal components particularly suited for dimensionality 
reduction purposes. As it will be shown, PPS project the input data onto a 
lower dimensionality space defined by what we shall call 'latent variables' 
which act as attractors of input vectors and, therefore, can be interpreted as 
cluster centroids. The algorithm used for the second step is the so called 
Negative Entropy Clustering algorithm (hereafter NEC), which has been 
selected after comparative testing against other similar algorithms among 
the wide class of unsupervised hierarchical agglomerative clustering algorithms 
according to its high efficiency and reliability \cite{ciaramella_2005}.
One advantage, which is as well a limitation, of this technique needs to be stressed:
the distribution in the parameter space of the objects belonging to each 
cluster selected by the NEC is approximated by multivariate gaussians. 
Consequently, the projection of cluster members positions along each axis 
of the parameter space can be modelled as a one-dimensional gaussian, and 
common statistics quantities such as the average or the standard deviation 
can be used to describe the distribution of the members of each cluster over  
the entire parameter space. On the other hand, the assumption of gaussian 
shape for the clusters requires further discussion (see section (\ref{Sec:NEC})).

\subsection{Latent variables and the PPS algorithm}
\label{Sec:PPS}

The Probabilistic Principal Surfaces model \cite{chang_2000,chang_2001,staiano_phd} 
belongs to the family of the so called {\it latent variables} methods 
\cite{bishop_1999} and can be regarded as an extension of the Generative Topographic 
Mapping \cite{bishop_1998}. Since the one described here is the first application of this 
method to astrophysical issues, in what follows we shortly summarize the mathematical
background, referring to the above quoted papers of a more detailed discussion. 
Uninterested readers may skip this paragraph and go directly to paragraph (\ref{Subsec:PPS_clustering}).
\noindent The goal of any latent variable model is to express the distribution 
$p(\mathbf{t})$ of the variable $\mathbf{t}=(t_{1},\ldots,t_{D}) \in \mathbb{R}^{D}$ in 
terms of a smaller number of latent variables $\mathbf{x}=(x_{1},\ldots,x_{Q}) \in 
\mathbb{R}^{Q}$ where $Q<D$. 
In order to achieve it, the joint distribution $p(\mathbf{t},\mathbf{x})$ is decomposed 
into the product of the marginal distribution $p(\mathbf{x})$ of the latent variables and 
the conditional distribution $p(\mathbf{t}|\mathbf{x})$ of the data variables given the latent
variables. It is convenient to express the conditional distribution as a factorization over the 
data variables, so that the joint distribution becomes:

\begin{equation}
\label{joint_p}
p(\mathbf{t},\mathbf{x})=p(\mathbf{x})p(\mathbf{t}|\mathbf{x})=p(\mathbf{x})\prod_{d=1}^{D}
p(t_{d}|\mathbf{x})
\end{equation}

\noindent The conditional distribution $p(\mathbf{t}|\mathbf{x})$ is then expressed in terms of a 
mapping from latent variables to data variables, so that

\begin{equation}
\label{latent_mapping}
\mathbf{t}=\mathbf{y}(\mathbf{x};\mathbf{w})+\mathbf{u}
\end{equation}

\noindent where $\mathbf{y}(\mathbf{x};\mathbf{w})$ is a function of the latent variable $\mathbf{x}$ 
with parameters $\mathbf{w}$, and $\mathbf{u}$ is an $\mathbf{x}$-independent noise process. 
If the components of $\mathbf{u}$ are uncorrelated, the conditional distribution for $\mathbf{t}$ 
will factorize as in (\ref{joint_p}). From the geometrical point of view, the function $\mathbf{y}(\mathbf{x};\mathbf{w})$ 
defines a manifold in the data space given by the image of the latent space. The definition of the 
latent variable model needs to be completed by specifying the distribution $p(\mathbf{u})$, the 
mapping $\mathbf{y}(\mathbf{x};\mathbf{w})$, and the marginal distribution $p(\mathbf{x})$. The 
type of mapping $\mathbf{y}(\mathbf{x};\mathbf{w})$ determines the specific latent variable model. 
The desired model for the distribution $p(\mathbf{t})$ of the data is then obtained by marginalizing 
over the latent variables:

\begin{equation}
\label{p_inte}
p(\mathbf{t})=\int p(\mathbf{t}|\mathbf{x})p(\mathbf{x})d\mathbf{x}.
\end{equation}

\begin{figure}
\centering
\includegraphics[width=14cm]{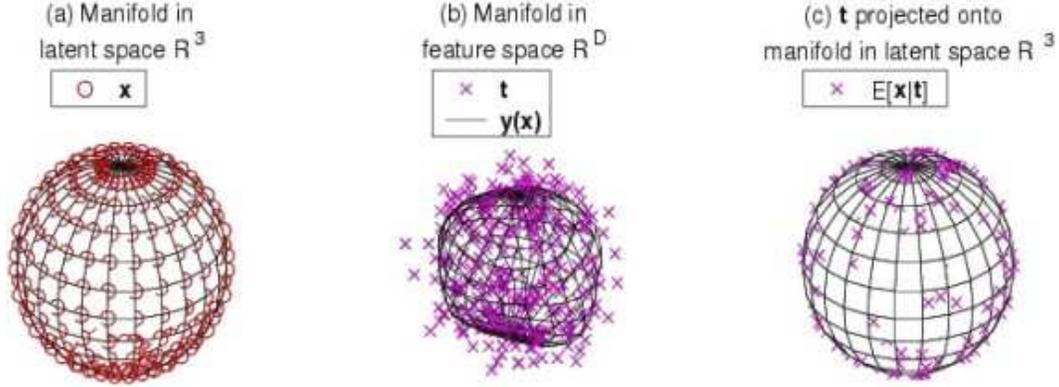}
\caption{Schematic representation of the spherical manifold in the three dimensional 
latent space $R^{3}$ (a), the same manifold distorted in the feature space $R^{D}$ 
together with points associated to data (b), and the projection of the points distribution 
onto the surface of the spherical manifold embedded in $R^{3}$ latent space.}
\label{spherical_pps}
\end{figure}

\noindent This integration will, in general, be analytically intractable except for specific forms of the 
distributions $p(\mathbf{t}|\mathbf{x})$ and $p(\mathbf{x})$. PPS define a non-linear, parametric 
mapping $\mathbf{y}(\mathbf{x};\mathbf{W})$, where $\mathbf{y}$  is defined continuous and 
differentiable, which projects every point in the latent space to a point into the data space. Since 
the latent space is $Q$-dimensional, these points will be confined to a $Q$-dimensional manifold 
non-linearly embedded into the $D$-dimensional data space. This implies that data points projecting 
near a principal surface node (i.e., a Gaussian center of the mixture) have higher influences on that 
node than points projecting far away from it (cf. figure (\ref{spherical_pps})). Each of these nodes 
$\mathbf{y}(\mathbf{x};\mathbf{w})$, $\mathbf{x}\in \{\mathbf{x}_m\}_{m=1}^M$ has covariance 
expressed by:

\begin{equation}
\label{pps_covariance}
    \mathbf{\Sigma}(\mathbf{x})=\frac{\alpha}{\beta}\sum_{q=1}^Q
    \mathbf{e}_{q}(\mathbf{x})\mathbf{e}_{q}^{T}(\mathbf{x})+
    \frac{(D-\alpha Q)}{\beta (D-Q)}\sum_{d=Q+1}^{D}\mathbf{e}_{d}(\mathbf{x})\mathbf{e}_{d}^{T}(\mathbf{x}),
    \end{equation}
\begin{displaymath}
0<\alpha<\frac{D}{Q} \end{displaymath} where
\begin{itemize}
    \item $\{\mathbf{e}_{q}(\mathbf{x})\}_{q=1}^{Q}$ is the set of orthonormal
    vectors tangential to the manifold at $\mathbf{y}(\mathbf{x};\mathbf{w})$,
    \item $\{\mathbf{e}_{d}(\mathbf{x})\}_{d=Q+1}^{D}$ is the set of orthonormal
    vectors orthogonal to the manifold in $\mathbf{y}(\mathbf{x};\mathbf{w})$.
\end{itemize}

\noindent The complete set of orthonormal vectors $\{\mathbf{e}_{d}(\mathbf{x})\}_{d=1}^{D}$ spans $\mathbb{R}^D$ 
and the parameter $\alpha$ is a clamping factor and determines the orientation of the covariance matrix. The unified 
\emph{PPS} model reduces to GTM for $\alpha=1$ and to the manifold-aligned GTM for $\alpha>1$:

\begin{displaymath}
        \mathbf{\Sigma}(\mathbf{x})  = \left\{
     \begin{array}{ccc}
       0< \alpha <1 &  & \perp \mbox{to the manifold } \\
      \alpha = 1 &  & \mbox{$I_D$ or spherical } \\
      1<\alpha<D/Q &  & \parallel \mbox{to the manifold. } \\
    \end{array}
    \right.
\end{displaymath}
   
\noindent In order to estimate the parameters $\mathbf{W}$ and $\beta$ we used 
the Expectation--Maximization (EM) algorithm \cite{dempster_1977},  while the 
clamping factor is fixed by the user and is assumed to be constant during the 
EM iterations. In a $3D$ latent space, then, a spherical manifold can be constructed 
using a PPS with nodes $\{\mathbf{x}_m\}_{m=1}^M$ arranged regularly on the 
surface of a sphere in $\mathbb{R}^3$ latent space, with the latent basis functions 
evenly distributed on the sphere at a lower density. The motivation behind such 
a spherical manifold is that spherical PPS are particularly well suited to capture 
the sparsity and periphery of data in large input spaces \cite{bishop_1995}. 
In order to better explain this issue let us consider the following low-D analogy first 
proposed by \cite{chang_2000}: \emph{... imagine fitting a rubber band 
($2-D$ spherical manifold) to data distributed uniformly on the surface of a 
sphere in $\mathbb{R}^3$. Any fit bisecting the sphere into two equal halves 
will be optimal. On the other hand, consider using a piece of string to fit the same 
data. The string has a significantly lower probability of finding the optimal fit as it is 
open-ended...} After a spherical PPS model is fitted to the data, the data themselves 
are projected into the latent space as points onto a sphere (figure (\ref{spherical_pps})).

\noindent The latent manifold coordinates ${\mathbf{\hat{x}}}_n$ of each data point 
${\mathbf{t}}_n$ are computed as:

\begin{displaymath}
{\mathbf{\hat{x}}}_n\equiv\langle {\mathbf{x}}|{\mathbf{t}}_n
\rangle=\int {\mathbf{x}}
p({\mathbf{x}}|{\mathbf{t}})d{\mathbf{x}}=\sum_{m=1}^M
r_{mn}{\mathbf{x}}_m
\end{displaymath}

\noindent where $r_{mn}$ are the latent variable responsibilities defined as:

\begin{eqnarray}{ll}
r_{mn}=p(\mathbf{x}_m|\mathbf{t}_n)=& \frac{p(\mathbf{t}_n| \mathbf{x}_m)P(\mathbf{x}_m)}{\sum_{m\prime=1}^{M}p(\mathbf{t}_n| \mathbf{x}_{m\prime})P(\mathbf{x}_{m\prime})}\\
& =\frac{p(\mathbf{t}_n| \mathbf{x}_m)}{\sum_{m\prime=1}^{M}p(\mathbf{t}_n|\mathbf{x}_{m\prime})}
\end{eqnarray}

\noindent Since $\|\mathbf{x}_m\|=1$ and $\sum_{m} r_{mn} = 1$, for $n=1,\ldots,N$, these coordinates 
lay within a unit sphere, i.e. $\|{\mathbf{\hat{x}}}_n\|\leq 1$. An interesting issue is the assessment of 
the incidence of each input data feature on the latent variables which helps to understand the relation 
between the features and the clusters found. The feature incidences are computed by evaluating the 
probability density of the input vector components with respect to each latent variable. More specifically, 
let $\{\mathbf{t}_n\}_{n=1}^N$ be the set of the D-dimensional input data, i.e 
$\mathbf{t}_n=(t_{n1},\ldots,t_{nD}) \in \mathbb{R}^D$, and $\{\mathbf{x}_m\}_{m=1}^M$ be the set of 
latent variables with $\mathbf{x}_m \in \mathbb{R}^3$. For each data point $\mathbf{t}_n=(t_{n1},\ldots,t_{nD})$ 
we want to evaluate $p(t_{ni}/t_{n1},\ldots,t_{ni-1},t_{ni+1},\ldots,t_{nD},\mathbf{x}_m)$, for $m=1,\ldots,M$ and $i=1,\ldots,D$. 
In detail:

\begin{eqnarray}
\label{ygivenyx}
p(t_{ni}/t_{n1},\ldots,t_{ni-1},t_{ni+1},\ldots,t_{nD},\mathbf{x}_m)=\\
=\frac{p(t_{n1},t_{n2},\ldots,t_{nD},\mathbf{x}_m)}{p(t_{n1},\ldots,t_{ni-1},t_{ni+1},\ldots,t_{nD},\mathbf{x}_m)}=\\
=\frac{p(t_{n1},\ldots,t_{nD}/\mathbf{x}_m)P(\mathbf{x}_m)}{p(t_{n1},\ldots,t_{ni-1},t_{ni+1},\ldots,t_{nD}/\mathbf{x}_m)P(\mathbf{x}_m)}=\\
\frac{p(t_{n1},\ldots,t_{nD}/\mathbf{x}_m)}{p(t_{n1},\ldots,t_{ni-1},y_{ni+1},\ldots,t_{nD}/\mathbf{x}_m)}.
\end{eqnarray}

\noindent The last term is easily obtained since the numerator is simply the $m$-th Gaussian 
component of the mixture computed by the PPS model with mean 
$y(\mathbf{x}_m;\mathbf{W})$ and oriented variance $\Sigma_m$, while the 
denominator is the same Gaussian component in which the $i$-th component is 
missing. Finally the mean of  expression (\ref{ygivenyx}) over the $N$ input data points, 
for each $\mathbf{x}_m$, is computed. This explains why spherical PPS can be used as 
a "reference manifold" for classifying high-D data.  A reference spherical manifold is 
computed for each class during the training phase. In the test phase, a data previously 
unseen by the network is classified to the class of its nearest spherical manifold. 
Obviously, the concept of "nearest" implies a distance computation between a data point 
$\mathbf{t}$ and the nodes of the manifold. Before doing this computation, the data point 
$\mathbf{t}$ must be linearly projected onto the manifold. Since a spherical manifold 
consists of square and triangular patches, each one defined by three or four manifold 
nodes, what is computed is an approximation of the distance. The PPS framework provides 
three approximation methods:

\begin{itemize}
    \item Nearest Neighbour: finds the minimal square distance to
    all manifold nodes;
    \item Grid Projections: finds the shortest projection distance
    to a manifold grid;
    \item Nearest Triangulation: finds the nearest projection
    distance to the possible triangulation;
\end{itemize}

\noindent In what follows we used the Nearest Neighbour approximation method 
because it allows to evaluate distances of each data point in the feature space 
to all nodes embedded in the spherical manifold; even if computationally heavier 
than the other two methods, the Nearest Neighbour approximation provides 
the most trustworthy choice of the node (or nodes, in case of multiple nodes at the same
distance from a given point) that each data point has to be assigned to.
\noindent Another way to use PPS as classifiers consists in choosing the class $C$ with 
the maximum posterior class probability for a given new input $\mathbf{t}$. 
Formally speaking, let us suppose to have $N$ labelled data points $\{\mathbf{t}_1, 
\ldots, \mathbf{t}_N\}$, with $\mathbf{t}_i\in \mathcal{R}^D$ and labels $\emph{class}$ 
in the set $\{1, \ldots, C\}$, then the posterior probabilities may be derived from 
the class-conditional density $p(\mathbf{t}|\emph{class})$ via the Bayes' theorem:

\begin{displaymath}
    P(\emph{class}|\mathbf{t})=\frac{p(\mathbf{t}|\emph{class})P(\emph{class})}{p(\mathbf{t})}\propto
    p(\mathbf{t}|\emph{class})P(\emph{class}).
\end{displaymath}

\noindent In order to approximate the posterior probabilities $P(\emph{class}|\mathbf{t})$ we estimate
$p(\mathbf{t}|\emph{class})$ and $P(\emph{class})$ from the training data. 
Finally, an input $\mathbf{t}$ is assigned to the class with maximum $P(class|\mathbf{t})$. 
In \cite{staiano_phd} and \cite{chang_2000} the effectiveness of PPS classifier is reported. 
A more detailed exposition of PPS as data mining framework can be found in 
\cite{staiano_phd,staiano_2004}. We want to emphasise that the non-linear relation between the features and the latent variables 
evaluated by PPS cannot be expressed mathematically in a closed form since it is completely empirical 
in nature. A representation of this relation can be recovered observing the positions of the same groups 
of objects in both the original parameter space and in the latent space, after the projection onto the 2-dimensional 
surface embedded in the latent space (in this case, a spherical surface was chosen for the sake of clarity). 
In this way, it is possible to determine qualitatively the composition of latent variables, each associated to 
one and only one of the nodes of an equally-spaced grid overlaying the spherical surface, in terms of 
original data features. As an illustrative example of how the PPS algorithm works, a simplified experiment 
exploiting the S-A sample (see paragraph (\ref{Sec:thedata})) has been performed. 
The PPS algorithm has been applied to this sample after setting the number of latent variables to 14, 
so that 14 different pre-clusters were produced at the end of the process. Since this experiment is only 
aimed at showing the properties of the application of our algorithm to real data and does not pretend 
to identify any real clustering of the QSOs photometric parameter space, the initial resolution of the 
experiment was set to a smaller value than the one used for the real experiments 
(see paragraph (\ref{Sec:experiments})).

\begin{figure}
\centering
\includegraphics[width=12cm]{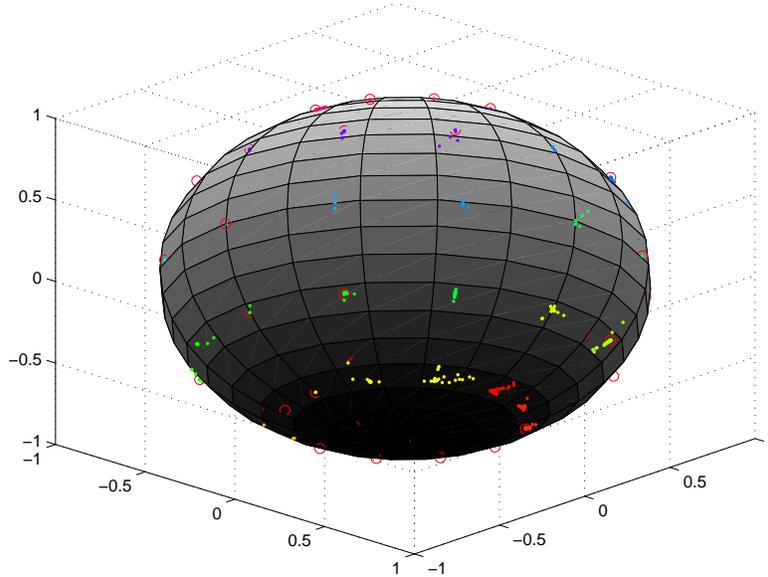}
\caption{The distribution of points of S-A sample projected onto a spherical surface embedded into the 
latent space after PPS application. Each distinct pre-cluster produced by PPS algorithm from the original distribution of points into the 
parameter space is assigned a different colour and only points with confidence $p_{i} > 0.8$ are shown. 
Arbitrary dimensions normalised to the radius of the sphere are shown on the three axes.}
\label{sphere_proj_example}
\end{figure}
 
\begin{figure}
\centering
\includegraphics[width=12cm]{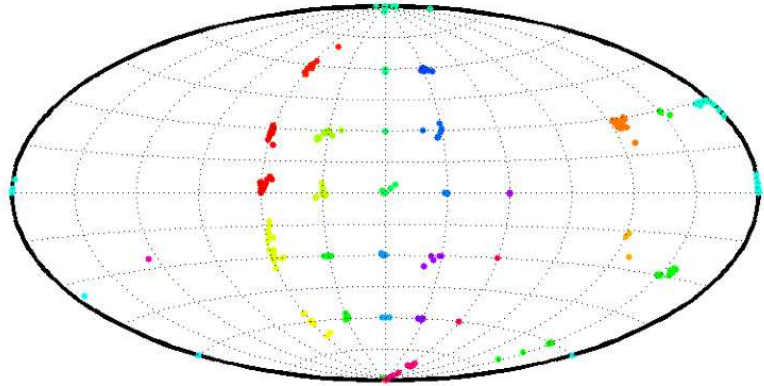}
\caption{Aitoff projection of the distribution of points on the spherical surface embedded in the latent space as 
determined by PPS for the S-A sample of objects. In this figure, each distinct pre-cluster is assigned the 
same colour used in the previous plot, and only the points with confidence $p_{i} >  0.8$ are drawn.}
\label{aitoff_proj_example}
\end{figure} 
 
\begin{figure}
\centering
\subfigure[($u - g$) vs ($g - r$)]{\includegraphics[width=8cm]{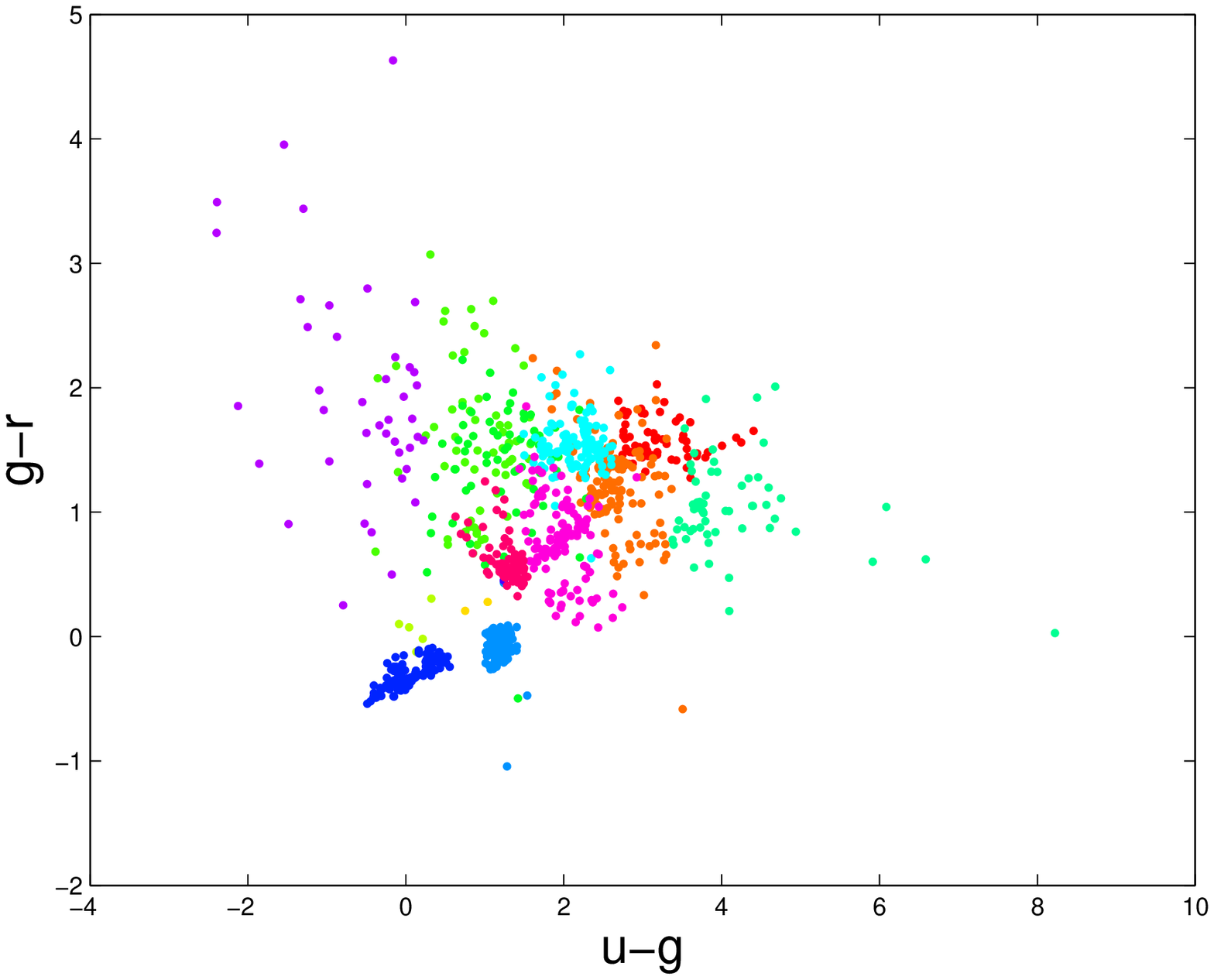}}
\subfigure[($g - r$) vs ($r - i$)]{\includegraphics[width=8cm]{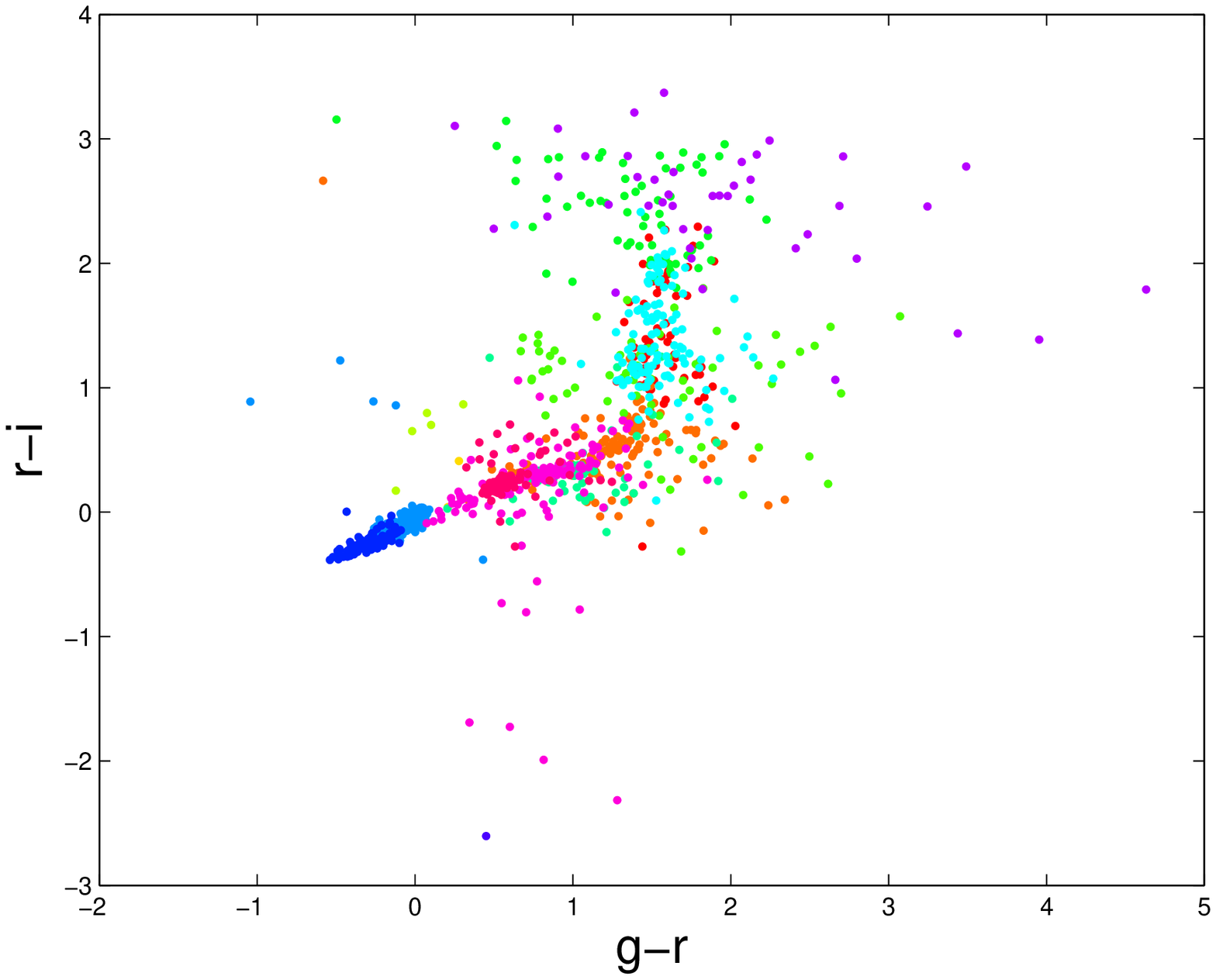}}\\
\subfigure[($r - i$) vs ($i - z$)]{\includegraphics[width=8cm]{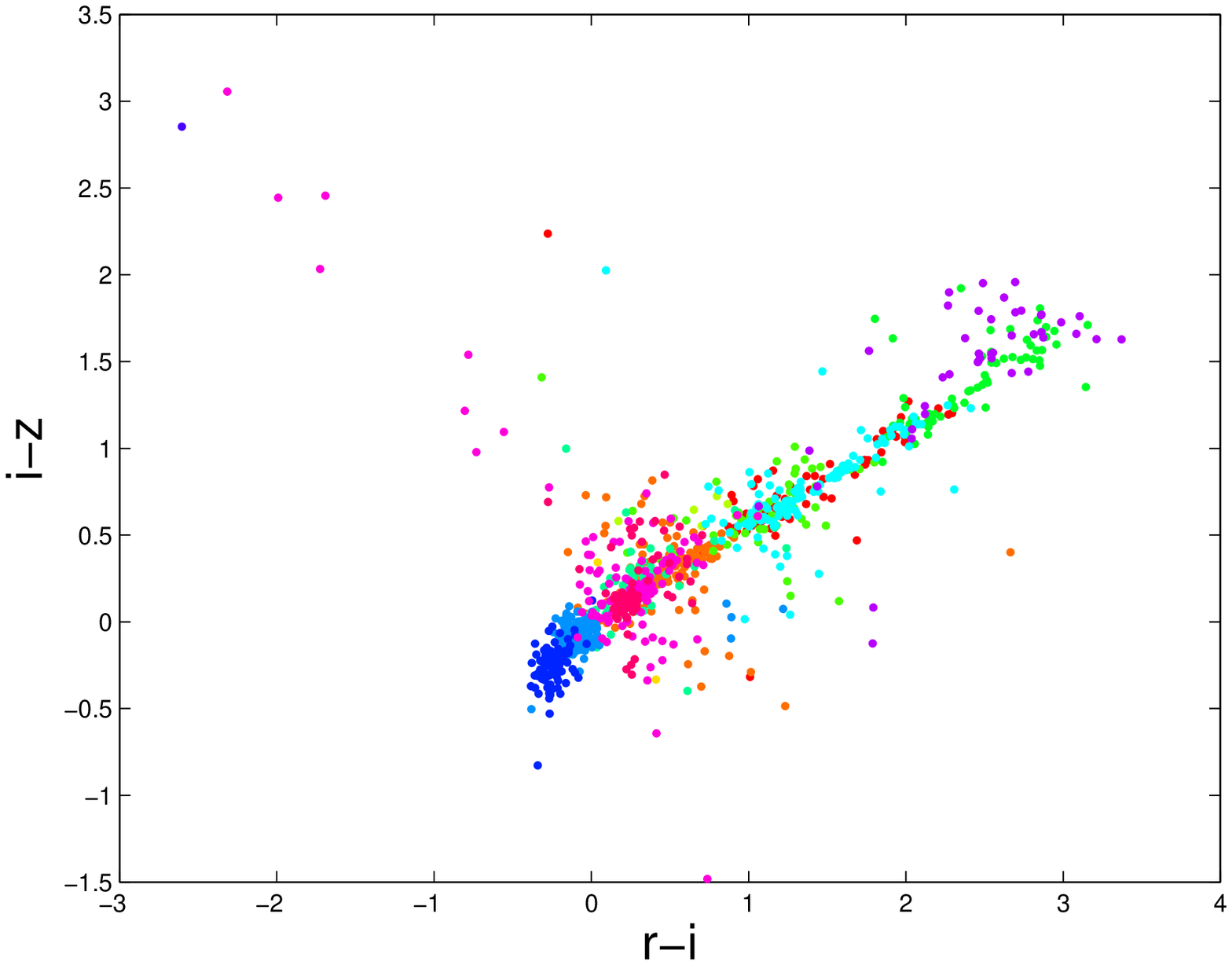}}
\caption{Original parameter space distribution projected on three planes of the members of sample S-A after the application on the PPS 
algorithm. Different pre-clusters are drawn with different colours following the same colour-scheme used in figures
(\ref{sphere_proj_example}) and (\ref{aitoff_proj_example}). In all these figures, only the points with confidence $p_{i} >  0.8$ are drawn.}
\label{planes_proj_example}
\end{figure}

\noindent Figure (\ref{sphere_proj_example}) shows the pre-clusters produced by PPS inside the S-A sample and 
the projection of the members of this sample on the spherical surface embedded into the latent space. As already stated previously, each point in the initial parameter space is assigned to a unique pre-cluster according to the nearest latent variable; this proximity criterion is nonetheless applied only inside the latent variable space and does not necessarily translate on closeness of the corresponding projections of the points on the surface of the sphere. The final product of the PPS algorithm is a set of $N$ pre-clusters, whose $i$-th member $PC_{i}$ is 
formed by all closest points to the latent variable $LV_{i}$,  which is associated to a node $N_{i}$ on the surface of 
the spherical surface. Only points having probability (also called confidence) $p_{i}$  to be assigned to the closest 
latent variable $LV_{i}$ greater than 0.8, have been plotted to make clearly visible the different regions of the spherical 
surface, and as a consequence, of the latent space occupied by different pre-clusters. Each pre-cluster has been 
plotted with a different colour, which has been used throughout the plots to label the same pre-cluster. Figure  
(\ref{aitoff_proj_example}) shows the Aitoff projection of the same pre-clusters produced by PPS. In order to prevent 
from misleading interpretations of figures (\ref{sphere_proj_example}) and (\ref{aitoff_proj_example}), it is worth 
noticing that the number of nodes appearing on both the spherical surface and its Aitoff projection is greater than 
the actual number of latent variables used for the experiment. The grid is used only to ease the comprehension of 
the distribution of pre-clusters. The composition of the latent variables responsible for each pre-cluster in terms of the original features of 
the parameter space is shown in figure (\ref{planes_proj_example}), where points belonging to the same pre-clusters 
are drawn in the planes obtained by projecting the original parameter space distribution onto some of the possible 
couples of parameters, using the same colour code used in both plots (\ref{sphere_proj_example}) and 
(\ref{aitoff_proj_example}). A qualitative guess of the composition of each latent variable found by PPS in terms of 
the variables of the original parameter space is possible comparing the positions of the members of each 
pre-cluster in the original parameter space and in the latent space. As it can be seen in figure 
(\ref{planes_proj_example}), the method is much more effective than the classical clustering algorithms in 
disentangling groups of objects which appear indistinguishable in the simple colour-colour diagrams. Except 
for few clusters placed at the extremes of the nearly one-dimensional locus visible in ($g - r$) vs ($r - i$) and 
($r - i$) vs ($i - z$) plots, all the other clusters appear to be almost inextricably mixed, so that discriminating them 
even on the bases of an unusual partition of these planes appears to be impossible. 

\subsection{PPS as a clustering tool}
\label{Subsec:PPS_clustering}

It needs to be explicitly noted that, as already mentioned, even though strictly speaking 
PPS are not a clustering algorithm, they can be effectively used for clustering purposes. 
Each latent variable, in fact, defines an attractor for points which are projected near to 
it and therefore the input space is partitioned in a number of clusters coinciding with the 
number of latent variables. The number of latent variables can therefore be regarded as 
the initial 'resolution' of clustering process but, provided that this number is not too low or 
too high (to avoid respectively a rough and unspecific or sparse clustering), it is found 
empirically that every reasonable choice leads to consistent results. In fact it suffices to 
set it to a higher value than the number of clusters realistically expected to appear 
in the data and then to use an agglomerative algorithm capable of recombining clusters 
of points artificially split into two or more chunks of smaller size. In our case we used the 
Negative Entropy Clustering described in the following paragraph.

\subsection{The hierarchical clustering algorithm}
\label{Sec:NEC}

\begin{figure}
\centering
\includegraphics[width=10cm]{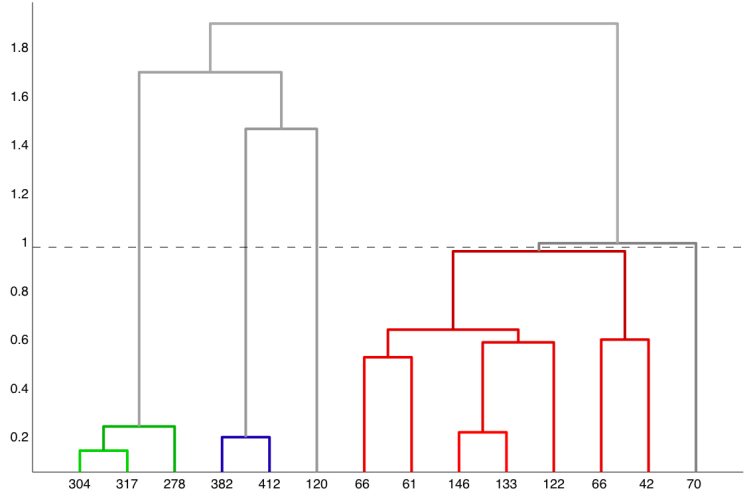}
\caption{An example of dendrogram (see text for details) used as a representation of an agglomerative clustering process
performed by the NEC algorithm. On the x and y axes are respectively reported the numeric labels of the initial
clusters and the values of the dissimilarity threshold. The dashed line intersects the lines of the tree-like graph associated
with the clusters produced by the clustering process when the value of the dissimilarity threshold has been 
fixed to $T_{cr} = 1$.}
\label{dendrogram}
\end{figure}

\begin{figure}
\centering
\subfigure[(u - g) vs (g - r)]{\includegraphics[width=8cm]{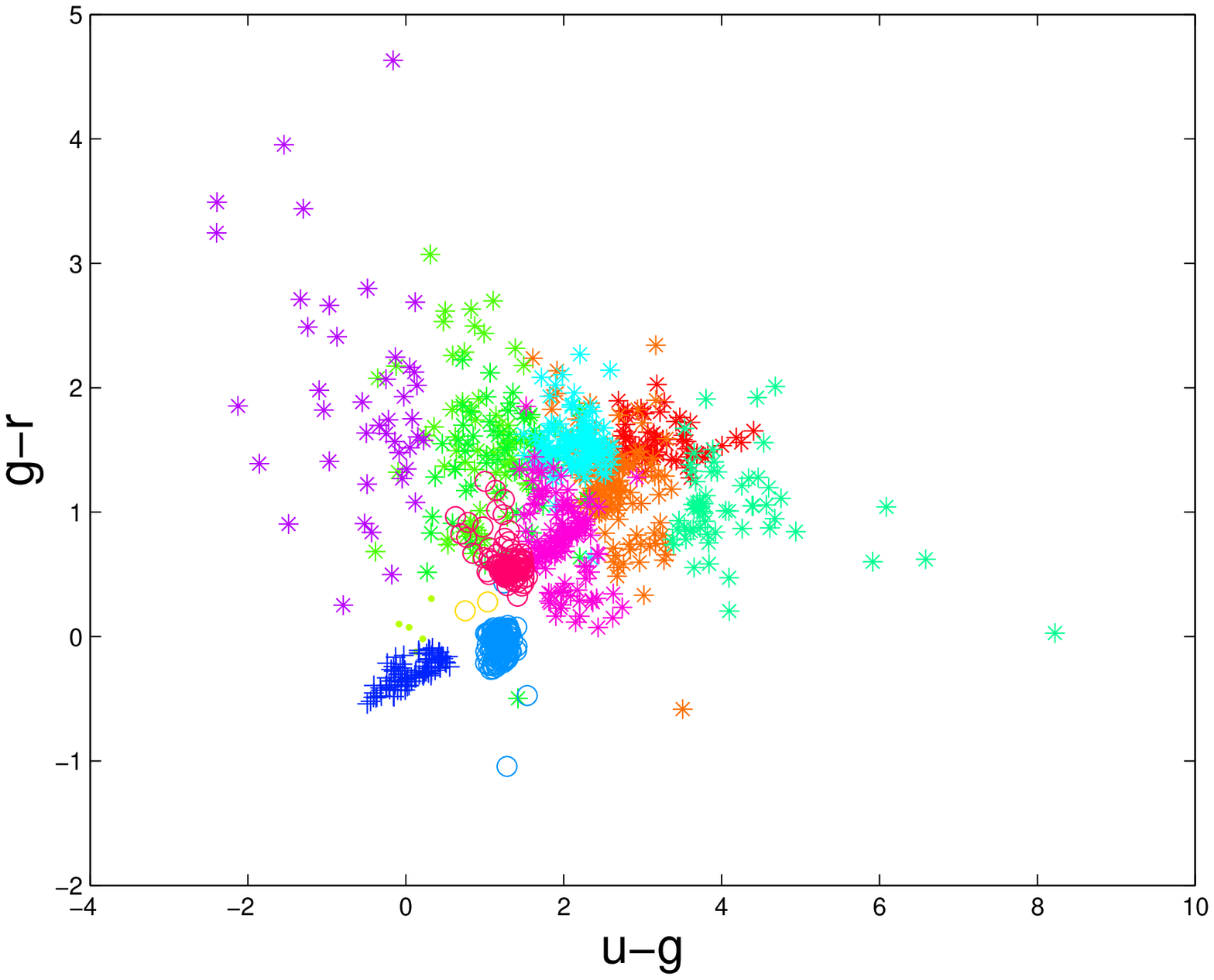}}
\subfigure[(g - r) vs (r - i)]{\includegraphics[width=8cm]{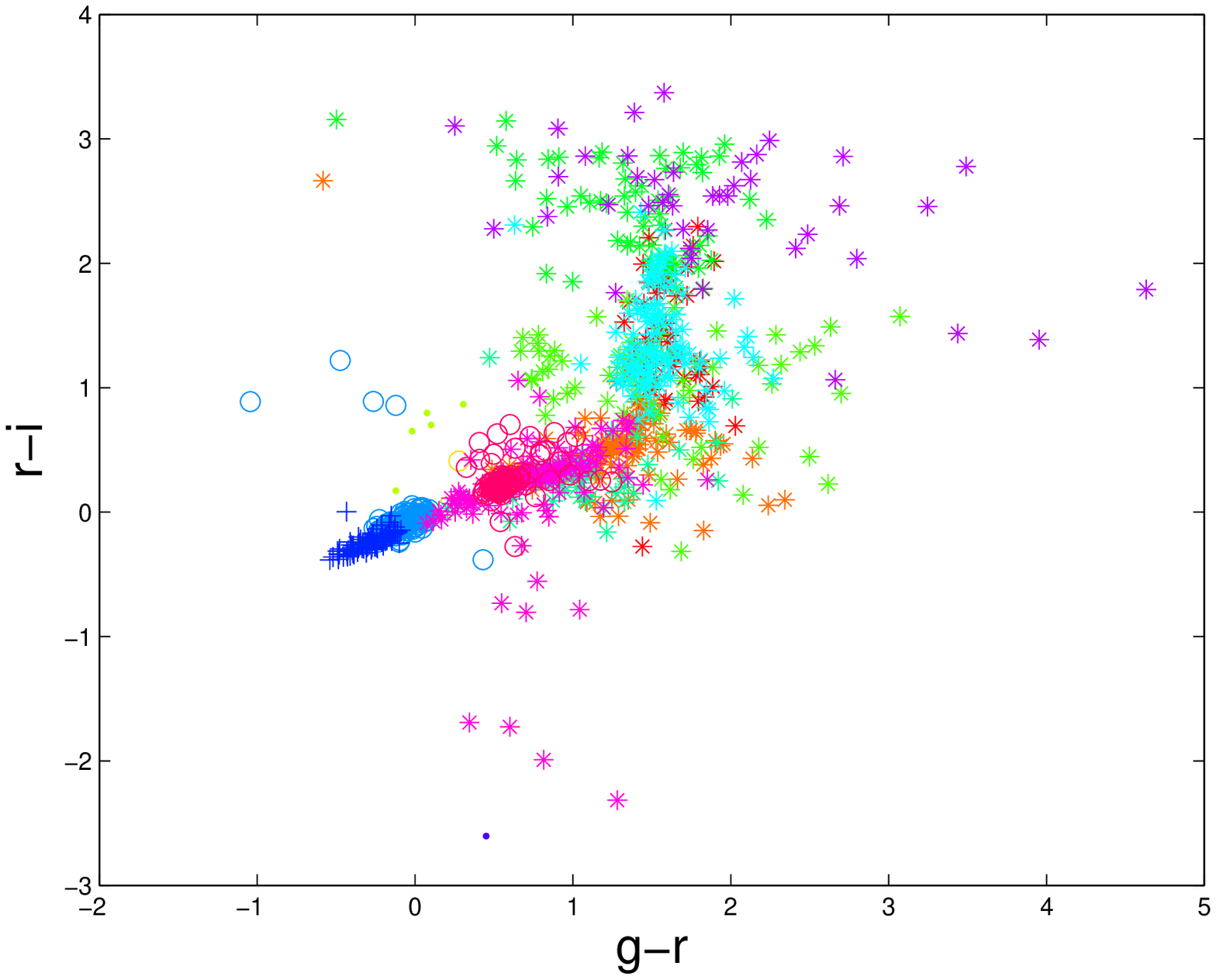}}\\
\subfigure[(r - i) vs (i - z)]{\includegraphics[width=8cm]{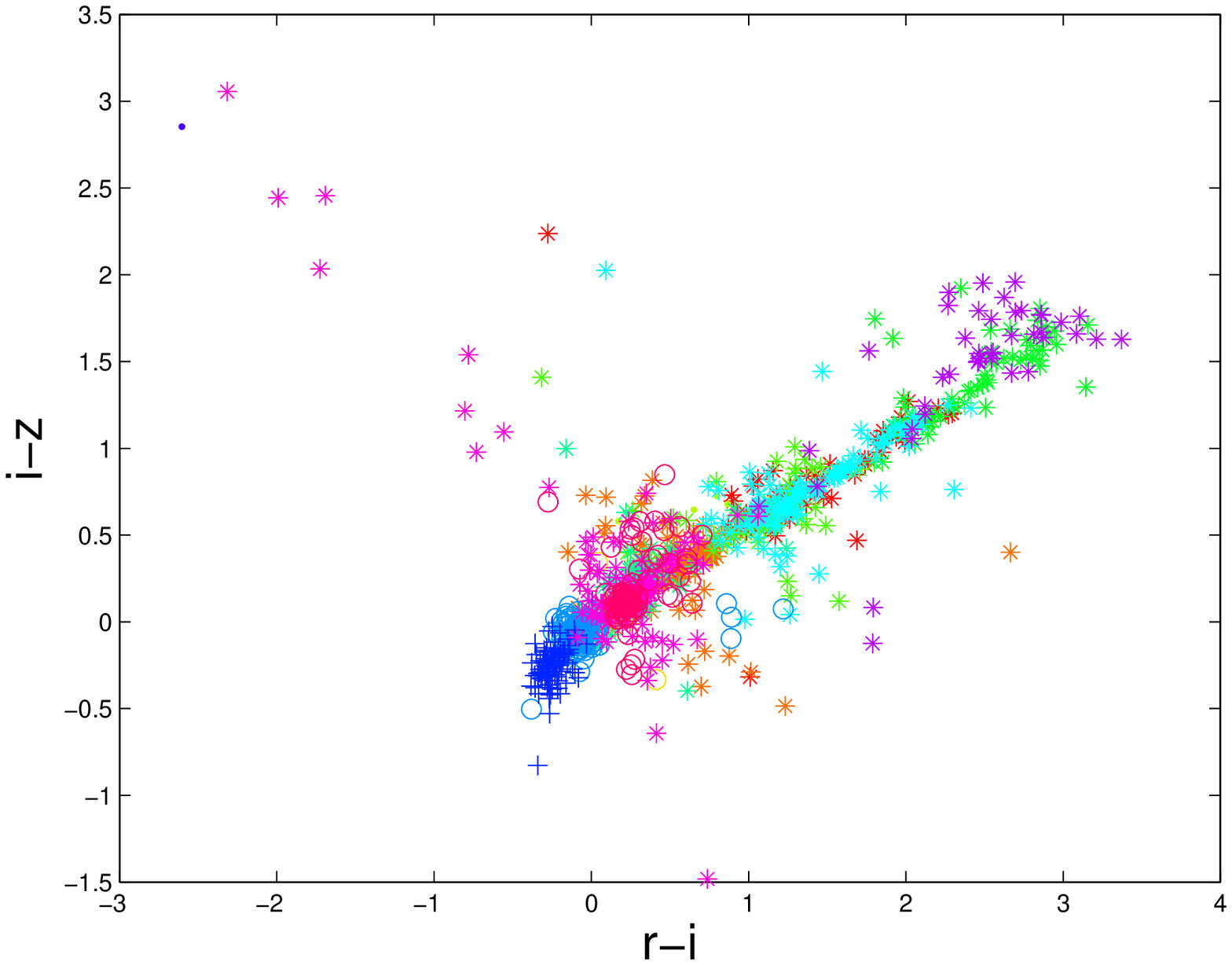}}
\caption{Original parameter space distribution projected on three colour-colour planes of the members of sample S-A 
after the application of the Negative Entropy Clustering algorithm to the pre-clusters produced by PPS. Former pre-clusters 
are drawn with different colours following the same colour-scheme used in figures (\ref{sphere_proj_example}) and (\ref{aitoff_proj_example}). 
Each final cluster is separated and singled out over the pre-clusters distribution through different marker symbols.}
\label{planes_proj_example_NEC}
\end{figure}

Most unsupervised methods require the number of clusters to be provided \emph{a priori}. 
This circumstance represents a serious problem when exploring large complex data sets 
where the number of clusters can be very high or, in any case, largely unpredictable. 
A simple threshold criterion is not satisfactory in most astronomical applications due to the 
high degeneracy and noisiness of the data which can often lead to an erroneous agglomeration. 
A classical agglomerative clustering algorithm is completely specified by assigning a definition 
of distance between clusters and a linkage strategy, i.e. a rule according to which clusters 
separated up to a given value of the distance are merged and others are not. Several definitions 
of distances can be found in the literature (distance between centroids of the clusters, maximum 
distance between members of the clusters, minimum distance, etc.) and many linkage strategies 
are used for common tasks (for example simple linkage, average, complete, Ward's, etc.). 
Independently from the choice of the distance definition, successive generations of merging are carried 
out using updated distances between clusters: the resulting structure of clusters can be represented by a 
tree-like graph, called dendrogram (see below for more details) until some convergence or threshold 
criterion is satisfied (so that, at the end of the process, there are no clusters to be merged).
A strictly geometrical interpretation of distances between clusters in the parameter space can be 
relaxed in order to generalise this class of algorithm. In this framework, the distance between 
clusters is a generic function of the composition of the clusters in terms of parameter space coordinates. 
We chose an approach to the hierarchical clustering based on the combination of a similarity criterion 
founded on the notion of 'Negative Entropy' and the use of a dendrogram to investigate the structure 
of clusters produced during the agglomerative process. We made use of the Fisher's linear discriminant 
which is a classification method that first projects high-dimensional data onto a line, and then performs a 
classification in the projected one-dimensional space \cite{bishop_1995}. The projection is performed 
in such a way to maximize the distance between the means of the two classes while minimizing the 
variance within each class. At the same time, the differential entropy H of a random vector is defined as:
$${\mathbf y} = (y_1, \ldots, y_n)^T$$ with density $f({\mathbf y})$:
$$ H({\mathbf y}) = \int f({\mathbf y}) \log f({\mathbf y}) d{\mathbf y}$$ so that negentropy $J$ can be defined as: 
$$ J(\mathbf{y}) = J({\mathbf y}_{Gauss}) - H(\mathbf{y})$$ where $\mathbf{y}_{Gauss}$ is a Gaussian 
random vector of the same covariance matrix as $\mathbf{y}$. 
\noindent The Negentropy can be interpreted as a measure of non-Gaussianity and, since 
it is invariant for invertible linear transformations, it is obvious that finding an invertible 
transformation that minimizes the mutual information is roughly equivalent at finding 
directions in which the Negentropy is maximised. Our implementation of the method 
uses an approximation of Negentropy that provides a good compromise between the 
properties of the two classic non-Gaussianity measures given by Kurtosis and Negentropy. 
Negentropy clustering algorithm can be used to perform unsupervised agglomeration 
of pre-clusters found by the PPS algorithm during the first step of our method. 
The only \emph{a priori} information needed by NEC is a particular scalar value of the Negentropy 
called dissimilarity threshold $T$. We suppose to have $n$ $D$-dimensional preclusters $X_{i}$ 
with $i=1,\ldots,n$ that have been determined by the PPS;  these clusters are passed to the Negentropy 
Clustering algorithm which, in practice, ascertains whether each couple of contiguous 
clusters (according to the Fisher's linear discriminant) can or cannot be more efficiently 
modelled by one single multivariate gaussian distribution. In other words, NEC algorithm 
determines if two clusters belonging to a given couple can be considered to be substantially 
distinct or parts of a greater more general data set (i.e. cluster). This method can be easily 
generalized to other model distributions; we preferred to use a multivariate 
gaussian model only because the normal distribution can be considered a good approximation of 
any reasonably shaped peaked distribution, since the colours of objects belonging to the same 
observational family of QSOs are widespread around a central value due to several physical 
mechanism (differential scattering, absorption, etc.) and observational effects. Similarly to what was done for the PPS 
algorithm, in order to elucidate the interpretation of the results from this unsupervised 
agglomerative method, we used NEC to generate a set of final clusters 
from the pre-clusters produced by the PPS algorithm inside the S-A sample already used
in the illustrative example described in paragraph (\ref{Sec:PPS}). The pre-clusters underwent the 
clustering process displayed in the figure (\ref{dendrogram}) in the form of a dendrogram, i. e. a 
tree-diagram frequently used to illustrate the arrangement of the clusters produced by agglomerative 
or divisive algorithms. This kind of representation of a clustering process consists of many U-shaped 
lines connecting objects in a hierarchical tree: the height of each U represents the value of a distance 
function evaluated between the two objects being connected; the distance definition depends on the 
specific kind of clustering algorithm considered. In our case, the quantity reported on the vertical 
axes is the negative entropy, a measure of the non-gaussianity of the multivariate distribution 
obtained merging the couple of clusters connected by the branches of the U-shaped lines.  
The optimal set of clusters (in a sense that will be stated clearly in paragraph (\ref{Sec:label})) is obtained for 
a certain value of negentropy, called the critical value of the dissimilarity threshold $T = T_{cr}$. In 
the illustrative experiment considered in paragraph (\ref{Sec:PPS}), only points with confidence greater 
than 0.8 were retained for clustering, so that the final number of clusters is equal to 3, corresponding 
to the critical dissimilarity threshold $T_{cr} = 1.46$. The composition of these clusters in terms of individual 
points of the S-A sample and of the pre-clusters which have been passed to the NEC algorithm, is 
showed in figure (\ref{planes_proj_example_NEC}), where the positions of points in the 4 dimensional original 
parameter space are projected onto the same colour-colour planes used in figure (\ref{planes_proj_example}).

\noindent Colours and marker shapes of each point are respectively associated to the pre-cluster and final 
cluster membership of the objects, so that becomes feasible the analysis of the relationship between the input and output of NEC 
algorithm in terms of the original distribution of points inside the parameter space and the pre-clusters used as input 
of the clustering algorithm. The three figures in (\ref{planes_proj_example_NEC}) show that the cluster whose members
are marked as crosses appears to be isolated in the first plot and clearly distinguishable from the other two in the remaining
colour-colour plots, while the clusters represented by circles and asterisks are entwined in all the plots and their members
can be barely discriminated.
  
\subsection{The labelling phase}
\label{Sec:label}

\noindent The results of the agglomeration performed by the NEC algorithm 
depend crucially on the value of the dissimilarity threshold $T$. Since different 
clustering processes of the data set correspond to different values of this 
constant, it is necessary to apply an objective criterion for the determination of 
the best (hereafter critical) value of the dissimilarity threshold $T_{cr}$, i.e the 
value leading the best performance in terms of the selection algorithm completeness 
and efficiency. To this aim, we use the BoK to label as  "goal-successful" those clusters $C_{j}$ 
(with $j \in \{1,...,N_{cl}\}$ (where $N_{cl}$ is the total number of clusters for 
a given value of $T$), for which the following relation is satisfied:

\begin{equation}
sr_{j}^{(g)} = \frac{\textsf{Nm confirmed "goal" members in } C_{j}}
{\textsf{Nm members in } C_{j}} \geq \widetilde{sr}^{(g)}
\end{equation}

\noindent i.e., the fraction $sr^{(g)}$ of "goal" objects contained in the cluster 
$C_{j}$ must be higher then a given value $\widetilde{sr}^{(g)}$. 
At the same time "notgoal-successful" clusters are defined as clusters for which 
the following relation holds:

\begin{equation}
sr_{j}^{(ng)} = \frac{\textsf{Nm confirmed "notgoal" members in } C_{j}}
{\textsf{Nm members in } C_{j}} \geq \widetilde{sr}^{(ng)}
\end{equation}

\noindent with a similar meaning of the symbols. In other words, "goal-successful" 
("notgoal-successful") clusters are defined as the clusters containing "goal" 
("notgoal") objects fractions above a given threshold. A third type of clusters which 
we shall simply call "not successful" are those which do not fulfill any of the two 
above definitions as they are formed by comparable fractions of "goal"  and "notgoal" 
objects. In the specific case addressed here, they will be composed by a mixture of 
confirmed quasar and other types of objects (mainly stars). The critical value $T_{cr}$ 
of the dissimilarity threshold is therefore defined as the one which, given a set of initial 
clusters provided by the PPS algorithm, produces the maximum number of 
"goal-successful" clusters, or, in a more quantitative fashion, as the value which 
maximizes the normalised success ratio NSR:

\begin{equation}
NSR(T) = \frac{\textsf{Nm successful clusters}}{\textsf{Nm clusters}} 
\end{equation}  
 
\noindent A further requirement is imposed to select $T_{cr}$ is that the number of clusters 
produced has to range between 25 $\%$ and 75 $\%$ of the number of initial clusters. This 
last constraint excludes from the selection values of  $T_{cr}$ producing unreasonable numbers 
of clusters, namely an excessive number of poor clusters or few very reach clusters (for a 
detailed discussion see section (\ref{Sec:NECparameter})), even if this requirement has been
loosened during each experiment in order to check if any feasible clustering combination had 
been dropped. The process described hitherto is recursive: once $\widetilde{sr}^{(g)}$ and 
$\widetilde{sr}^{(ng)}$ have been fixed, the suitable value of the dissimilarity 
threshold is identified and a first clustering is performed using $T_{cr}$ as an imput to 
the NEC algorithm. All successful clusters produced in this first generation of clustering 
are 'frozen' and the efficiency $e_{1}$ is estimated; unsuccessful clusters are merged, 
forming the input data set for the following iteration of the selection process. After this second iteration, 
the new "goal-successful" and "notgoal-successful" clusters, if any, are retrieved and stored. The procedure is 
repeated until no other successful clusters are determined. Critical values of the dissimilarity threshold 
for each generation are fixed according to the same criteria explained above. The total efficiency 
$e_{tot}$ of the selection is defined as the sum of the efficiencies of each of $M$ 
generations weighted according to the total number of objects belonging to the "goal-successful" 
clusters of that generation:

\begin{equation}
e_{tot} = \frac{\sum_{i=1}^{M} n_{i}e_{i}}{\sum_{i=1}^{M}n_{i}} = \frac{\sum_{i=1}^{M} n_{i}^{(goal)}}{N_{tot}^{(goal)}}   \ \ \ \ \ 
\end{equation}
where $n_{i} $  is the total number of objects belonging to "goal-successful" 
clusters of the $i_{th}$ generation such that $\sum_{i=0}^{M}n_{i} = N_{tot}^{(goal)}$ 
and $n_{i}^{(goal)}$ is the number of confirmed goal objects contained in 
all "goal-successful" clusters selected in the $i$-th generation. 

\begin{figure}
\centering
\subfigure[Efficiency in $z$ vs $u-g$ plane ("goal-successful")]{\includegraphics[width=8cm]{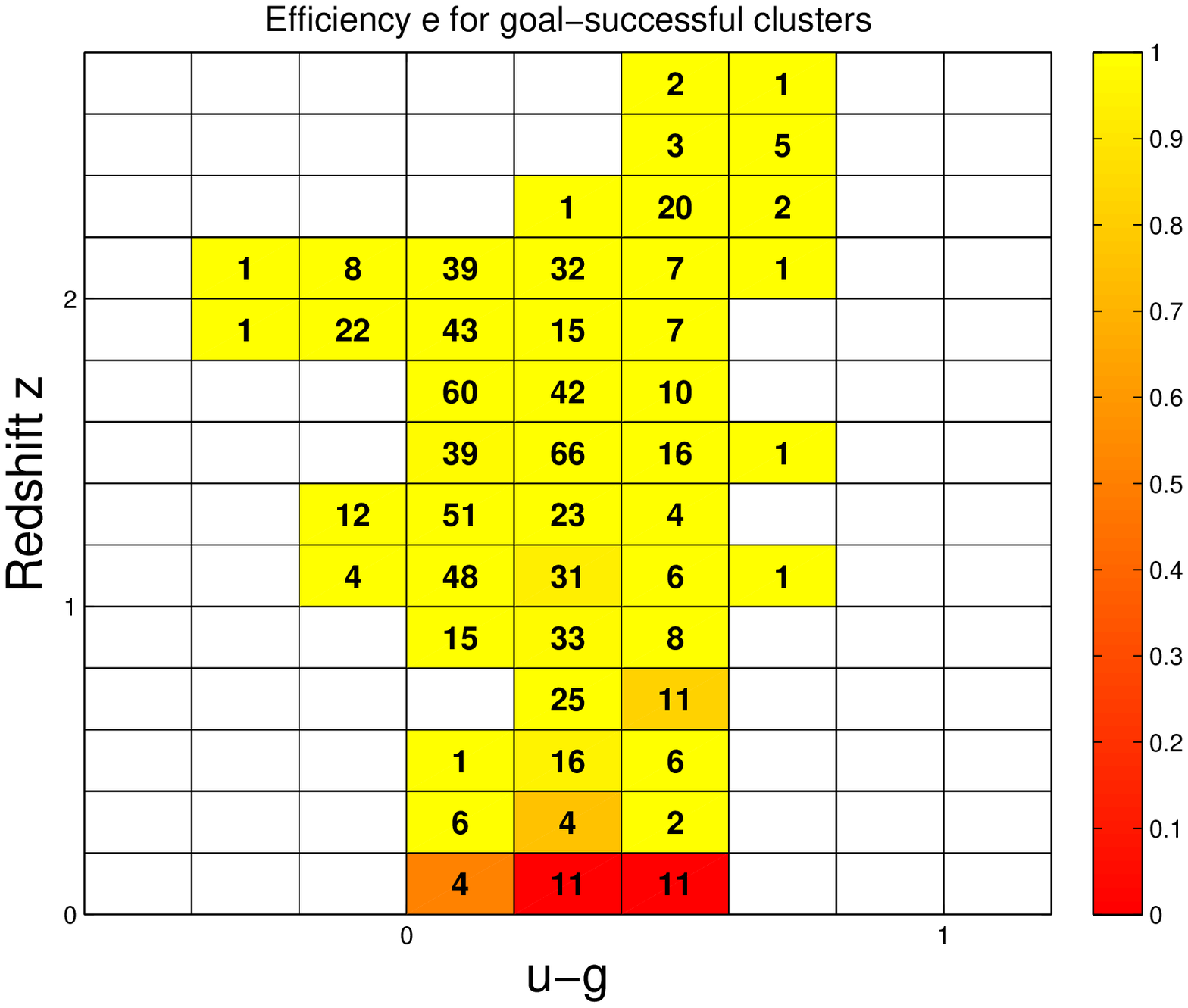}}
\subfigure[Completeness in $z$ vs $u-g$ plane ("goal-successful")]{\includegraphics[width=8cm]{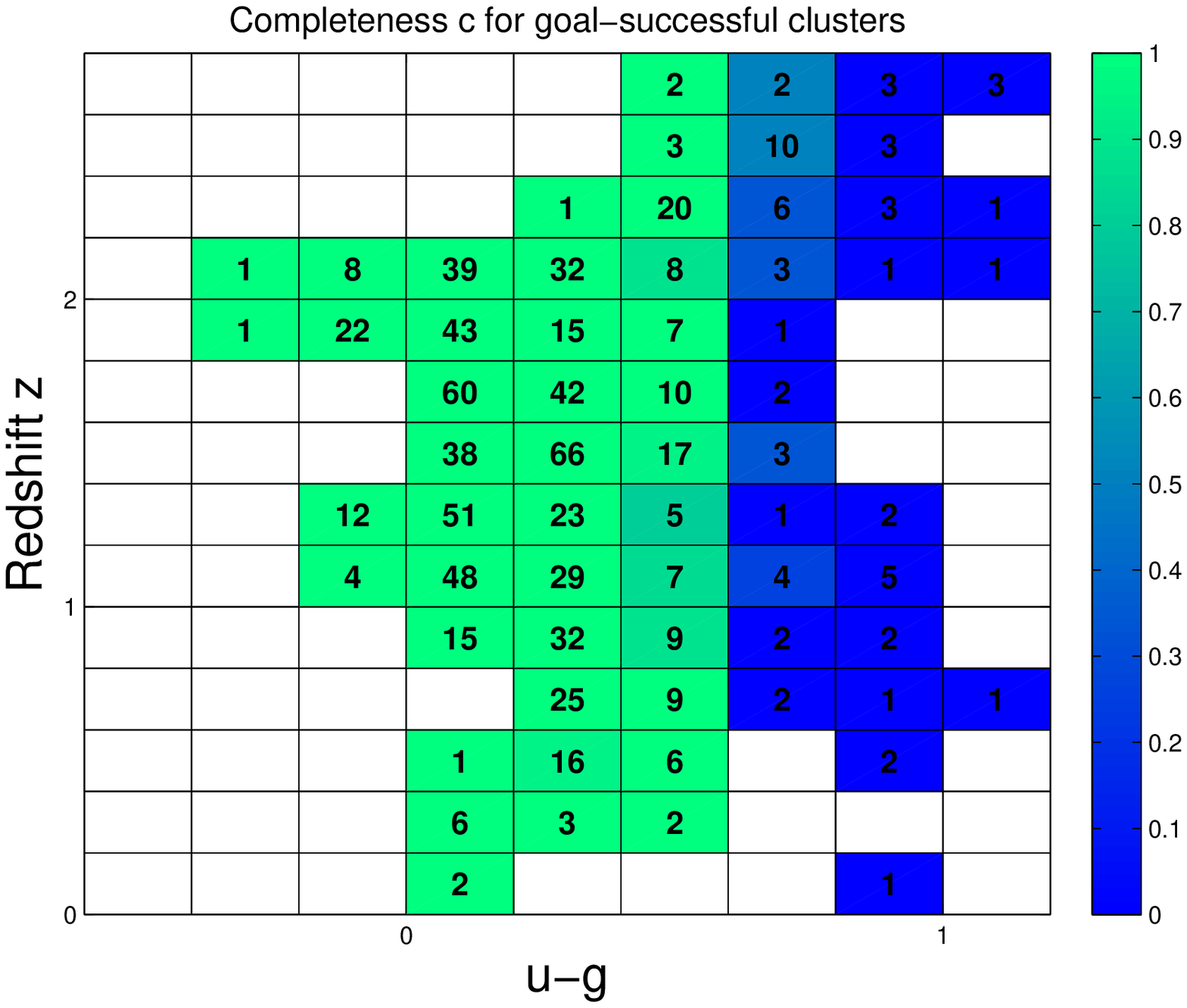}}
\caption{Efficiency (left-hand panel) and completeness (right-hand panel) in the redshift vs ($u-g$) plane for the labelling
phase of the illustrative experiment. Colour of the cell is associated to the efficiency and completeness of the selection process
of the "goal-successful" clusters, while the numbers contained in each cell represent the total number of candidates in that region of the
$z$ vs $u-g$ plane.}
\label{fig:paramparam_training_illustrative}
\end{figure}

\noindent The total completeness $c_{tot}$ of the process is similarly defined:

\begin{equation}
c_{tot} = \frac{\sum_{i=1}^{M}n_{i}^{(goal)}}{N_{tot}^{(all)}}
\end{equation}

\noindent where $N_{tot}^{(all)}$ is total number of goal objects contained in the data set
used for the experiment. Extensive testing showed that, within the range $\left[0.65, 0.90\right]$ 
the values of the constant thresholds $\widetilde{sr}^{(g)}$ and $\widetilde{sr}^{(ng)}$ for "goal-successful" and 
"notgoal-successful" clusters respectively, do not affect the final efficiency and completeness of 
the method but only the number of generations of the process needed to 
achieve the final result. The selection performances of the labelling phases are estimated
for all the experiments described in this paper as functions of one of the parameters used for the clustering 
and the redshift of the members of the data set considered, using the available BoK to measure the local 
values of efficiency and completeness in the resulting redshift vs parameter plane. As an example, 
in the upper left-hand panel and right-hand panel of figure (\ref{fig:paramparam_training_illustrative}) 
are shown respectively the efficiency $e$ and the completeness $c$ in the redshift vs ($u-g$) plane  
of the illustrative experiment discussed in the previous sections. The colours of the cells are associated 
to the normalized values of efficiency and completeness expressed by the colourbars visible on the right 
of the plots. In the plots of the efficiency, each cell contains the total number of candidates selected by the 
algorithm in that region of the plane (the number of confirmed candidate being the product of this number 
by the efficiency in the cell), while in the plots showing the local values of completeness the numbers 
represent the amount of confirmed QSOs according to the BoK (the number of confirmed objects selected 
being the product of this number by the completeness in the cell). Blank cells represent regions of the 
plane where no candidate belonging to the BoK is found. The lower panels of figure 
(\ref{fig:paramparam_training_illustrative}) contain respectively a graphical representation of the partition 
of the redshift vs ($u-g$) plane in terms not only of the "goal-successful" clusters used for candidate selection, but also in terms of the other two types
of clusters produced by our algorithm, namely the "notgoal-successful" and "unsuccessful" ones.
Each cell containing at least one object belonging to the reference catalogue is marked with a different symbol
according to the type of the cluster (circles for "goal-successful", crosses for "notgoal-successful" and 
asterisks for "unsuccessful" clusters) contributing the higher efficiency to the selection of QSOs candidates
in that region of the plane. Also in this case, the colours of the cells represent the maximum values of 
efficiency and completeness respectively.

\subsection{Selection of candidate quasars from photometric samples.}
\label{subsec:photoselection}

After the labelling phase, which provides the most suitable partition of 
the parameter space in terms of selection of "goal-successful" and "notgoal-successful"
clusters, candidate quasars extraction from a purely photometric data set (i.e., for which 
no spectroscopic BoK is available) can be carried out using one of the three different 
approaches explained in the following paragraphs.

\subsubsection{Method I: "re-labelling"}
\label{method1} 

The first method, hereby called "re-labelling", is based on the assumption 
that confirmed QSOs in the BoK are tracers of successful clusters containing 
mainly goal objects even when other objects are added to the sample. The 
data-set used for the labelling and the photometric sample are merged and the 
whole process described in the previous paragraphs is repeated using this 
extended group of objects. The selection of candidate quasars is then carried 
out considering as candidates all non-BoK objects belonging to clusters where 
spectroscopic confirmed quasars (membersof the sample used for the 
labelling) are dominant. This simple and, at least in theory, straightforward 
method unfortunately is applicable only when the non-BoK sample is composed by 
few objects, namely a small fraction of the number of BoK objects. The reason 
is that PPS algorithm determines the best projection from the parameter space 
to the latent space by modelling a probability distribution which is a function of the initial distribution of BoK sample 
objects inside the initial space. New objects added to the labelling sample modify 
the shape of the probability density function and the final result of the unsupervised 
clustering, so that the efficiency and completeness estimated during the labelling 
phase are not appropriate. It is worthwhile to emphasize that this method has to be 
preferred when limited amount of data are added to an existing data set but that it 
cannot be used when the amount of new data to be processed is large.

\subsubsection{Method II: colours cuts}
\label{method2}
 
The second approach at the extraction of candidate QSOs from the photometric dataset
is based on the geometrical characterization of the distribution of the sources belonging 
to the "goal-successful" clusters in the parameter space. In general, for any "goal-successful" 
cluster in a $n$-dimensional parameter space, 2$n$ constraints are identified as some 
values of statistical meaning associated to the distribution of points of the cluster along 
each axis of the parameter space. The candidate objects are selected applying to the photometric data set the cuts 
derived by the 2$n$ generic constraints as the objects which satisfy these requirements.
This method is more flexible than the previous one since the selection of candidate 
QSOs can be fine-tuned by modifying the cuts applied to the parameters in order to 
achieve different performance goals. Obviously, this choice yields a trade-off between 
efficiency and completeness of the selection. For instance, loose constraints allow to select 
a larger number of candidates in the outskirts of the clusters, where the contamination from 
"notgoal" sources is higher, thus resulting in a increased completeness but in a lower efficiency 
of the overall selection process. On the other hand, tight constraints increase the efficiency 
of the algorithm at the cost of a lower completeness, by selecting only the central regions of 
the parameter space occupied by successful clusters. In the present work two different 
prescriptions have been used to determine the cuts in the parameter space. According 
to the first prescription, $n$-haedrons whose vertices are set by the extremal values of 
cluster members distribution for each parameter for both "goal-successful" and 
"notgoal-successful" clusters, were chosen to establish the regions of the parameter 
space containing candidate quasars. More precisely, all photometric objects found inside 
the $n$-haedrons containing "goal-successful" clusters and not selected by any of the 
$n$-haedrons derived by "notgoal-successful" clusters are selected as candidate quasars.
Errors on parameters have been used to estimate the distance of each object from the 
surfaces of the $n$-haedrons generated by "goal-successful" clusters, in order to avoid 
the possible contamination from spurious objects placed near to the borders of the 
"goal-successful" regions. More precisely, only inner objects with distance from the 
surfaces of the "goal-successful" $n$-haedrons larger than 3$\sigma$ have been 
retained as candidates. The second prescription is more conservative (in the sense of ensuring higher efficiency 
and lower completeness) than the first one: in order to reduce the fraction of contaminants, 
the vertices of the $n$-haedrons describing the positions of "notgoal-successful" clusters 
are set to the positions $\bar{x}_{i} \pm \sigma_{i}$ along each axis of the parameter space, 
where $\bar{x}_{i}$ is the average value and $\sigma_{i}$ the standard deviation of the 
distribution of points along the $i$-th axis. All objects placed inside these $n$-haedrons 
are discarded and the remaining are selected as candidate according to the first prescription.
 
\subsubsection{Method III: Mahalanobis' distance}
\label{subsubsec:method3}

The third method for the extraction of photometric candidate objects is based on the notion 
of Mahalanobis' distance. This particular type of distance differs from the euclidean one since it 
takes the dispersion of the each variable and correlations among multiple variables into account 
when determining the distance between two points. In general, given two generic points 
$\vec{x} = (x_{1},...,x_{p})^{t}$ and $\vec{y} = (y_{1},...,y_{p})^{t}$ in the $t$-dimensional space $R^{t}$, 
the Mahalanobis' distance between them is defined as:

\begin{equation}
d_{S}=\sqrt((\vec{x}-\vec{y})^{t}S^{-1}(\vec{x}-\vec{y}))
\end{equation}

\noindent where $S^{-1}$ is the covariance matrix. Mahalanobis' distance is widely employed in cluster 
analysis and other classification techniques, since it is possible to use it to classify a test point 
as belonging to one of a set of more clusters defined in the same space where the test point resides. 
The first step is the estimation of the covariance matrix of the clusters, usually achieved through a set of
realizations of the populations of each clusters represented by samples of points drawn from the same 
statistical distribution. Then, given a test point, Mahalanobis' distances $d_{Mal}$ of this point from all 
clusters are calculated and the point is assigned at that cluster for which $d_{Mal}$ is the smallest. In 
the probabilistic interpretation given above, this is equivalent to selecting the class to which is more probable 
that the given point belongs. In this work, given a set of clusters produced by the combined PPS and NEC 
algorithms, Mahalanobis' distance has been used to assign each member of the photometric data 
set to the its nearest cluster. Only objects associated to "goal-successful" clusters have been considered 
QSOs candidates. An interesting consideration about this method can be made by recalling that the 
usage of Mahalanobis' distance to evaluate the membership of objects to a given set of "goal-successful" 
clusters with a threshold value (a maximum distance beyond which an object is not assigned to any 
clusters) can be exploited to select outliers or rare objects were not included in the BoK because of their scarcity. 

\subsection{Comparison of the photometric candidate selection methods}
\label{subsec:comparisonmethods}

\begin{figure}
\centering
\subfigure[Efficiency $e$ for the second method in $z$ vs $u-g$ plane]{\includegraphics[width=8cm]{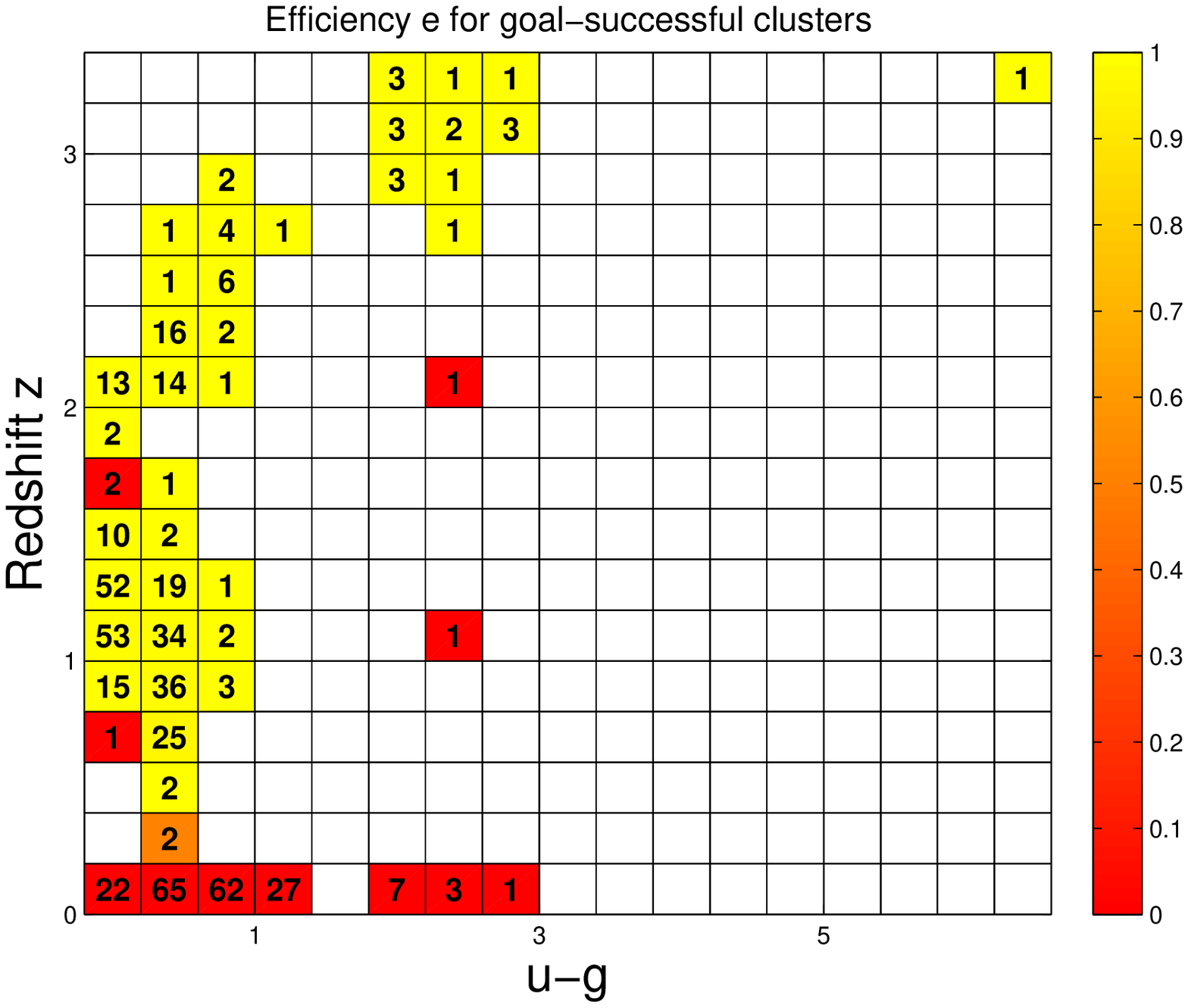}}
\subfigure[Completeness $c$ for the second method in $z$ vs $u-g$ plane]{\includegraphics[width=8cm]{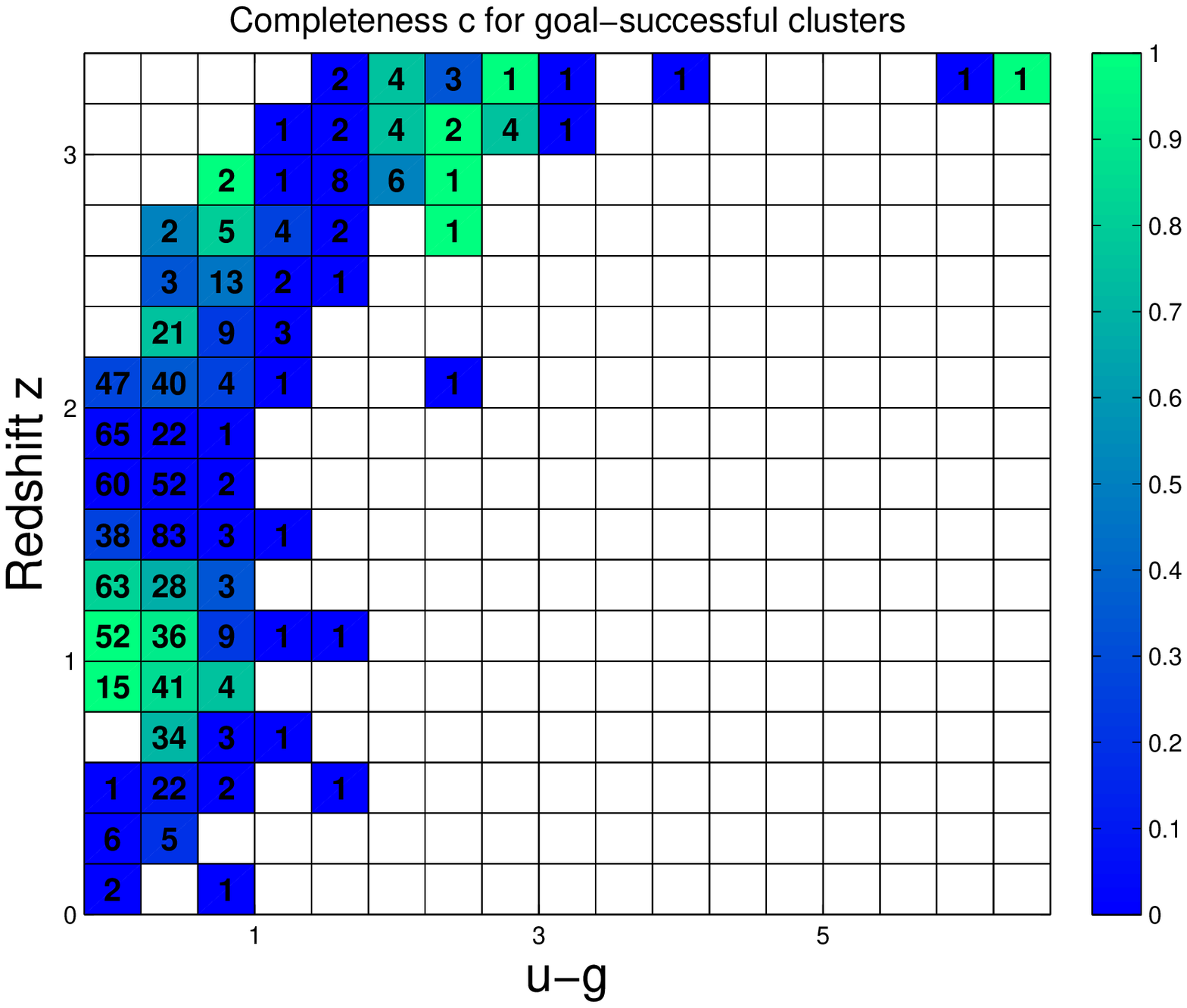}}\\
\subfigure[Efficiency $e$ for the third method in $z$ vs $u-g$ plane]{\includegraphics[width=8cm]{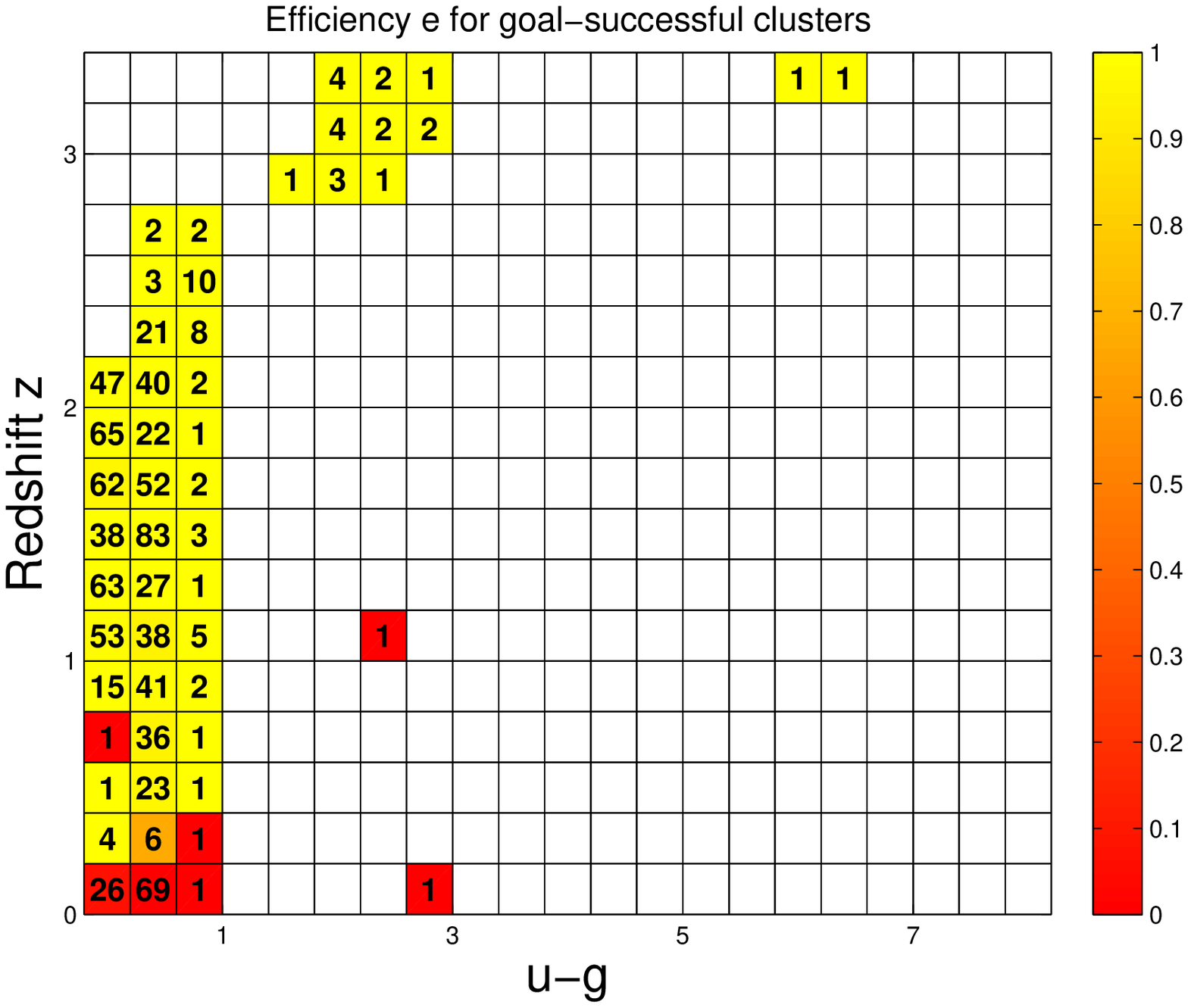}}
\subfigure[Completeness $c$ for the third method in $z$ vs $u-g$ plane]{\includegraphics[width=8cm]{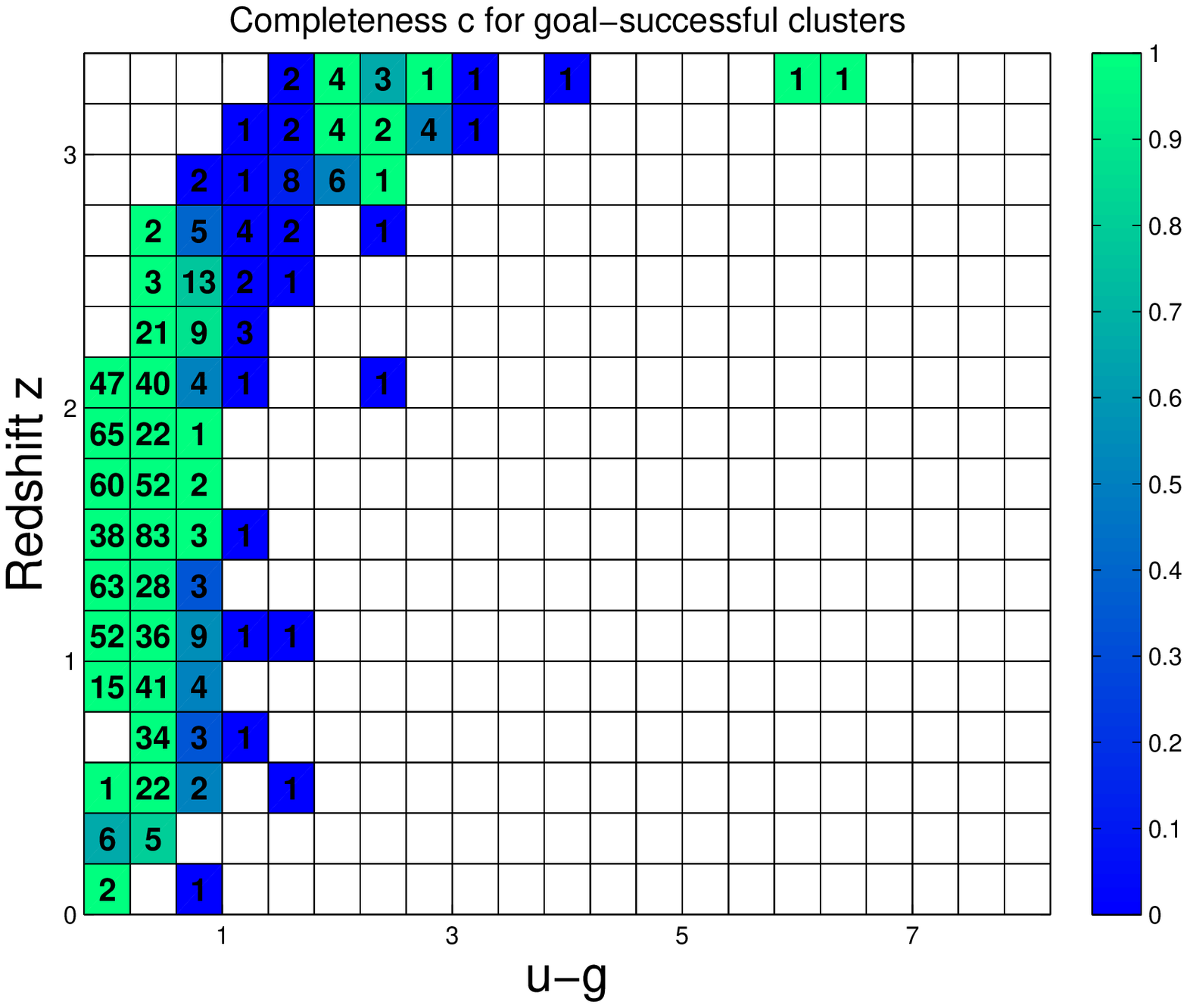}}
\caption{Efficiency (left-hand panel) and completeness (right-hand panel) in the redshift vs $u-g$ plane for the colours cuts (upper panels) and 
Mahalanobis' distance (lower panels) methods of photometric candidates selection. Colours of the cells express the local values of efficiency 
and completeness while integers represent the numbers of candidates selected by the algorithm and confirmed QSOs according to the BoK for efficiency and completeness plots respectively.}
\label{fig:paramparam_training}
\end{figure}

In order to compare the performances of the methods described 
in the previous paragraphs, we used instead of the set of clusters produced by the illustrative experiment
described in paragraphs (\ref{Sec:PPS}) and (\ref{Sec:NEC}), the set of clusters produced in the first experiment 
(see paragraph (\ref{subsec:first})) since our goal, in this case, is to evaluate the best method of photometric 
candidate selection in terms of the clusters labelling produced in a real scientific application of the method. 
First of all, it is worthwhile to remind the reader that the "re-labelling" method can be effectively used only in 
case the number of photometric objects is much smaller than the number of objects composing the base of knowledge 
(i.e., the objects whose clustering in the parameter space has been determined, c.f. paragraph (\ref{method1})). 
In addition, this approach is very time-consuming since a whole new application of PPS and NEC algorithms is 
needed every time that a new sample of photometric sources is considered, even if the BoK remains unchanged. 
For these reasons, this method though theoretically interesting is not well suited for large scale data sets and will 
not be further considered in the following comparison. The other two methods have been compared by applying 
both of them to the same sample of objects randomly drawn from the S-A sample used for the first phase of the 
experiment. Only the original parameters associated to these objects, namely the four colours derived from 
the optical magnitudes measured in SDSS $(u, g, r, i, z)$ photometric system, were used for the selection. 
The total performances, in terms of efficiency $e$ and completeness $c$, evaluated for the "re-labelling", 
Mahalanobis' distance and the "colours cuts" methods (with the two different prescriptions described in (\ref{method2})) 
on the "goal-successful" clusters produced by the first experiment are shown in the table (\ref{table_comparison}), 
exploiting the BoK associated to the S-A sample. A more detailed comparison of the performances of the two 
methods considered for the selection of photometric candidates is shown in figures (\ref{fig:paramparam_training}), 
where the two indicators are plotted as functions of the redshift and of one of the photometric parameters, namely 
the $u-g$ colour of the source (the choice of the particular colour is not important in this case since the 
methods of selection show similar relative performances when other colours are considered). Each cell 
in the redshift vs colour plane is coloured according to the local value of efficiency and completeness as 
specified by the colourbar, while the integers written inside the cells represent the numbers of candidates 
selected by the algorithm and confirmed QSOs according to the BoK for efficiency and completeness plots, respectively
(for details, see section (\ref{Sec:label})).

\begin{table}
\centering                 
\begin{tabular}{c c c c}        
\hline\hline                
Method   &   $e$   &   $c$ \\    
\hline 
   Re-labelling (few photo objects)  &   81.3\%   & 89.4\% \\                  
   Colours cuts (1$^{st}$ prescription)  &    79.8\%        &    87.6\% \\
   Colours cuts (2$^{nd}$ prescription) &   81.1\%        &     79.4\% \\ 
   Mahalonobis' distance  &  81.0\%     &    88.8\%  \\ 
\hline                                   
\end{tabular}
\caption{Efficiency and completeness calculated for three methods of selection of 
photometric candidates using the clusters labelling obtained in the first experiment.}
\label{table_comparison}
\end{table}

\noindent The comparison of the upper and lower panels in figure (\ref{fig:paramparam_training}) clearly 
shows that the Mahalanobis' distance method offers higher efficiency and completeness in large swaths
of the redshift vs ($u-g$) plane than the colours cuts method, improving in particular the completeness of the
selection in the redshift range $[1, 2]$. Since the best method in terms of both local and total efficiency 
and completeness is the Mahalanobis' distance\footnote{Neither threshold nor maximum value are fixed. 
The only necessary condition is that all "notgoal-successful" clusters are more distant than 
the "goal-successful" cluster the point is assigned to.}, it has been used for the extraction of the 
lists of QSOs photometric candidates (for details see paragraph (\ref{Sec:conclusions})) determined on the 
bases of the experiments described in this paper. 

\section{The experiments}
\label{Sec:experiments}

Four different sets of experiments based the samples labelled from 1 to 4 and described above have been performed in order
to evaluate the ability of the method to select candidate QSOs using different BoKs (although based on the same "specClass"
index from the SDSS) in different photometric parameter spaces. The following are the set-up of the four experiments:

\begin{itemize}

\item The first experiment is based on the S-A sample, composed by all SDSS star-like objects 
for which a spectroscopic classification is available and falling into the region of overlap with the UKIDSS-DR1 
LAS. Colours have been calculated using optical "psf" magnitudes to define the parameter space 
where the clustering is carried out. 
\item The second and third sets of experiments exploit  the S-UK sample, \emph{i. e.} all 
stellar objects in the SDSS having spectroscopic classification and matching with a stellar NIR 
counterpart belonging to the UKIDSS-DR1 LAS catalogue. We made use of optical plus infrared 
colours only (second experiment) and only optical colours (third experiment).
\item The fourth set of experiments explored the distribution of S-S data set (formed by all candidate 
QSOs in SDSS-DR5 according to the official Sloan selection algorithm for which all 5 "psf" magnitudes 
were correctly measured) inside the photometric parameter space determined by their optical colours. 
\end{itemize}

\begin{table*}
\label{table_results}      
\caption{Total efficiency and completeness of the experiments.}             
\centering          
\begin{tabular}{llrrr}    
\hline\hline       
Experiment & Sample  & $e_{tot}$ & $c_{tot}$ & $n_{gen}$\\ 
\hline                    
   1 & S-A & 81.5\% &  89.3\% & 2\\  
   2 & S-UK & 92.3\% &  91.4\% & 1\\
   3 & S-UK & 97.3\% &  94.3\% & 1\\
   4 & S-S & 95.4\% & 94.7\% & 3\\
\hline                  
\end{tabular}
\end{table*}

The colours used for all these experiments were calculated using adjacent bands: $u-g$, $g-r$, $r-i$, $i-z$ for the optical bands, and $Y-J$, $J-H$, $H-K$ for 
the near infrared ones. The dependance of the results on the choice of the colours will be discussed in section (\ref{dependance_parameters}).
The final results of these experiments in terms of total efficiency and completeness of the candidate quasars selection are summarised in table 
(\ref{table_results}) while the number of successful clusters and the fraction of confirmed quasars and stars inside each "goal-successful" 
cluster for each experiment are reported in table (\ref{table_detailed}). A colour-colour plot is shown for each experiment for the comparison of the 
results of the candidate selection method exposed in this paper with the classical colour-colour selection techniques commonly employed in the 
astronomical literature. At the same time, plots of the local values of the efficiency and completeness in the redshift vs ($u-g$) plane are displayed 
for each experiment and will be discussed in section (\ref{Sec:conclusions}). The same plots for all the others colours for each experiment are shown in 
the appendix (\ref{appendix}).

\subsection{First experiment}
\label{subsec:first}

\begin{figure}
\centering
\subfigure[NEC diagnostics]{\includegraphics[width=8cm]{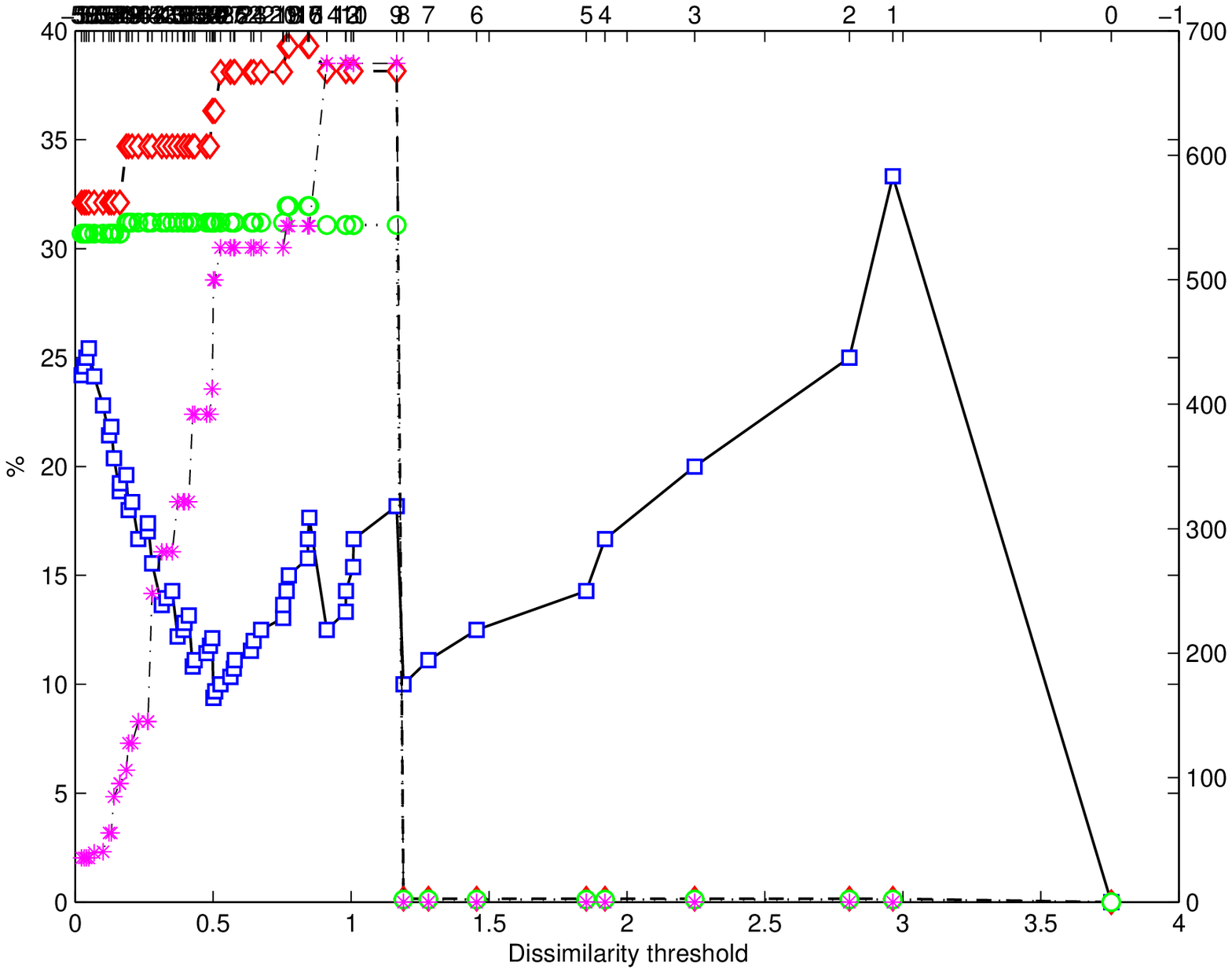}} 
\subfigure[Efficiency and completeness]{\includegraphics[width=8cm]{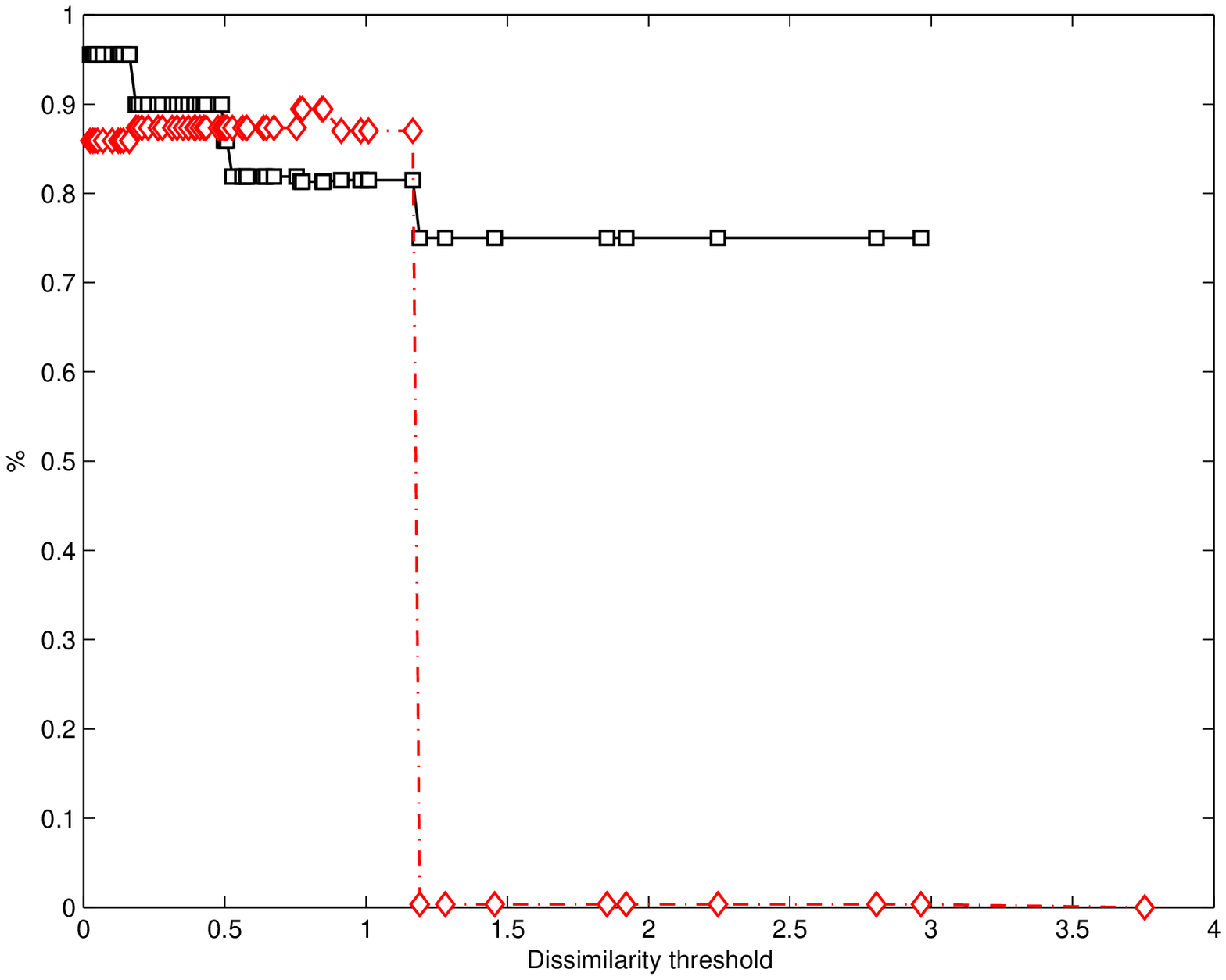}}\\ 
\subfigure[specClass vs redshift plane distribution of clusters]{\includegraphics[width=8cm]{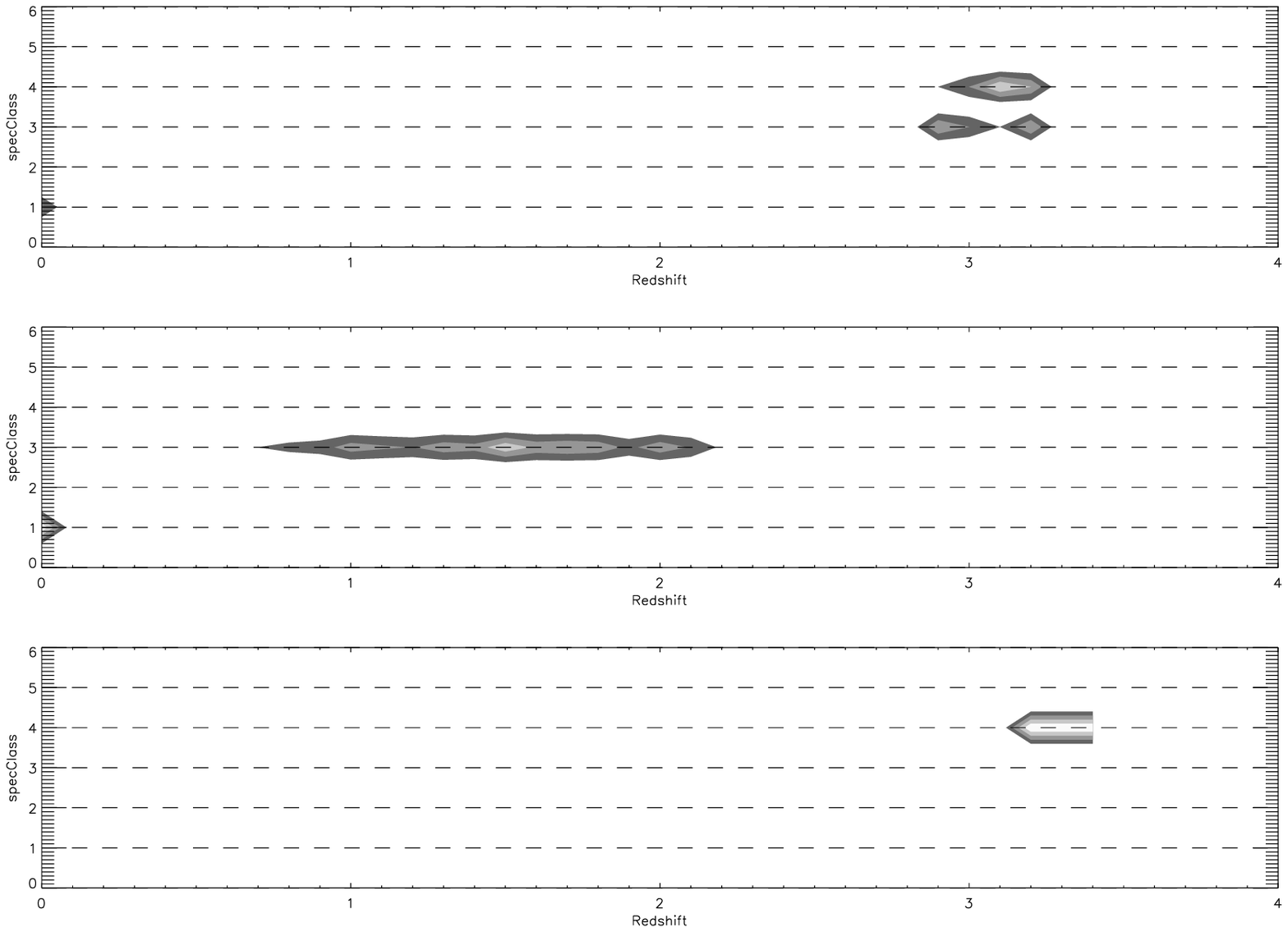}} 
\subfigure["Box and wisker" plot of clusters]{\includegraphics[width=8cm,angle = -90]{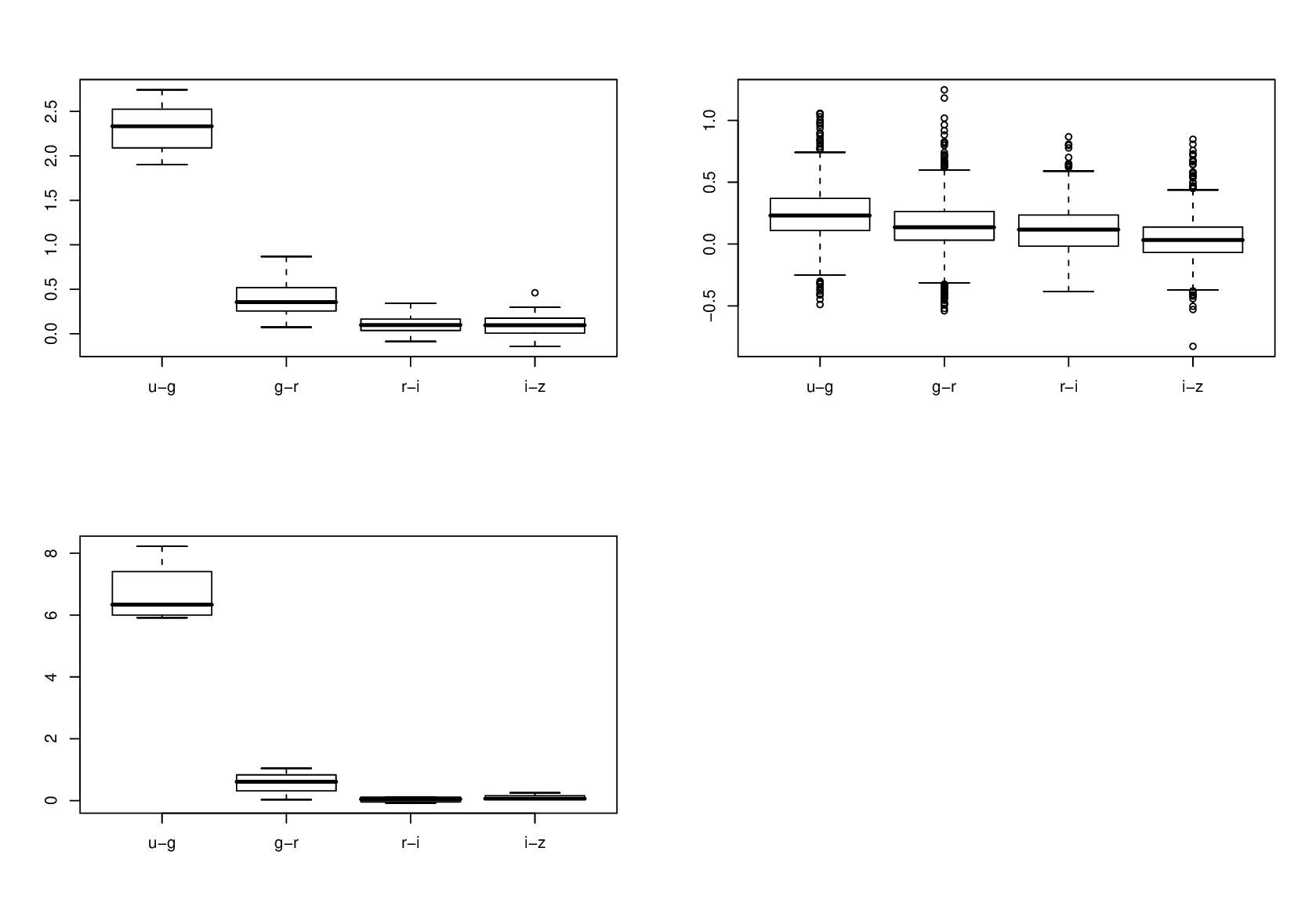}}
\caption{Upper left-hand panel: In this figure, the fraction of 'goal-successful' clusters 
(squares), the total percentage of all sources objects belonging to 'goal-successful' 
clusters (circles), the total percentage of objects belonging to 'goal-successful' 
clusters irrespective of their spectroscopic classification (diamonds) and variance 
of 'goal-successful' clusters (asterisks) are plotted as a function of the dissimilarity 
threshold for the first experiment. Upper right-hand panel: total estimated 
efficiency $e_{tot}$ (square symbols) and completeness $c_{tot}$ (diamond symbols) 
of the candidate quasars selection process as functions of the dissimilarity threshold 
for the first experiment. Lower left-hand panel: distribution of members of 
"goal-successful" clusters in the first experiment in the specClass-redshift plane. 
Lower right-hand panel: "box and whisker" plot of the distribution of different clusters 
produced in the first experiment. The black rectangle indicates the 95\% confidence 
interval for the median while the greater rectangle is contained between the first 
and third interquartile. The notches extend to $\pm 1.58$ of the interquartile range 
times the inverse of the square root of the number of objects. Outliers are represented 
as dots.}
\label{exp_1}
\end{figure}

This experiment is aimed at comparing the SDSS native selection algorithm to 
our method. The number of latent variables (i.e. initial clusters) used for the PPS 
pre-clustering was 62, and the critical value of the dissimilarity threshold was 
chosen according to the criteria explained above. The normalised success ratio 
and other statistical indicators of the clustering process are plotted as functions of 
the dissimilarity threshold in the upper left-hand panel of figure (\ref{exp_1}). The estimated 
efficiency and completeness of the selection process are shown in the upper right-hand 
panel of figure (\ref{exp_1}) as functions of the dissimilarity threshold as well.
The distribution of sources as a function of redshift and spectroscopic 
classification "specClass" in the "goal-successful" clusters selected in this 
experiment is shown in the lower left-hand panel of figure (\ref{exp_1}) while, 
a "box and whisker" plot of the distribution of candidate QSOs for each 
"goal-successful" cluster selected in this experiment is shown in remaining panel of 
figure (\ref{exp_1}).

\noindent The three "goal-successful" clusters selected in this experiment are evenly 
distributed in terms of colours:

\begin{itemize}
\item The first cluster is formed by objects having $u-g$ ranging form 2.0 to 2.5 and 
the other colours averaging between 0.5 and 0 with little dispersions. 
These objects have redshift around 3 and they are mainly classified as high-redshift 
QSOs in terms of spectroscopic "specClass" classification, with a little contamination 
by misidentified stars ("specClass" = 1).
\item The second cluster contains the vast majority of objects and is mainly formed by 
normal QSOs ("specClass" = 3), with redshift ranging from 0.8 to 2.0, except for a little 
fraction of stars. In terms of colours distribution, this cluster shows all colours having average values 
near 0.0 with a higher and asymmetric dispersion, caused by outliers mainly having higher 
colours values than the average.  
\item The third "goal-successful" cluster members are characterized by a very high $u-g$ 
values, compared to the average values of the others colours which are similar to those 
found for the first cluster. It is formed by only high redshift ($z > 3$) QSOs with spectroscopic 
classification "specClass" = 4. 
\end{itemize}

\begin{figure}
\centering
\subfigure[Efficiency in $z$ vs $u-g$ plane ("goal-successful")]{\includegraphics[width=8cm]{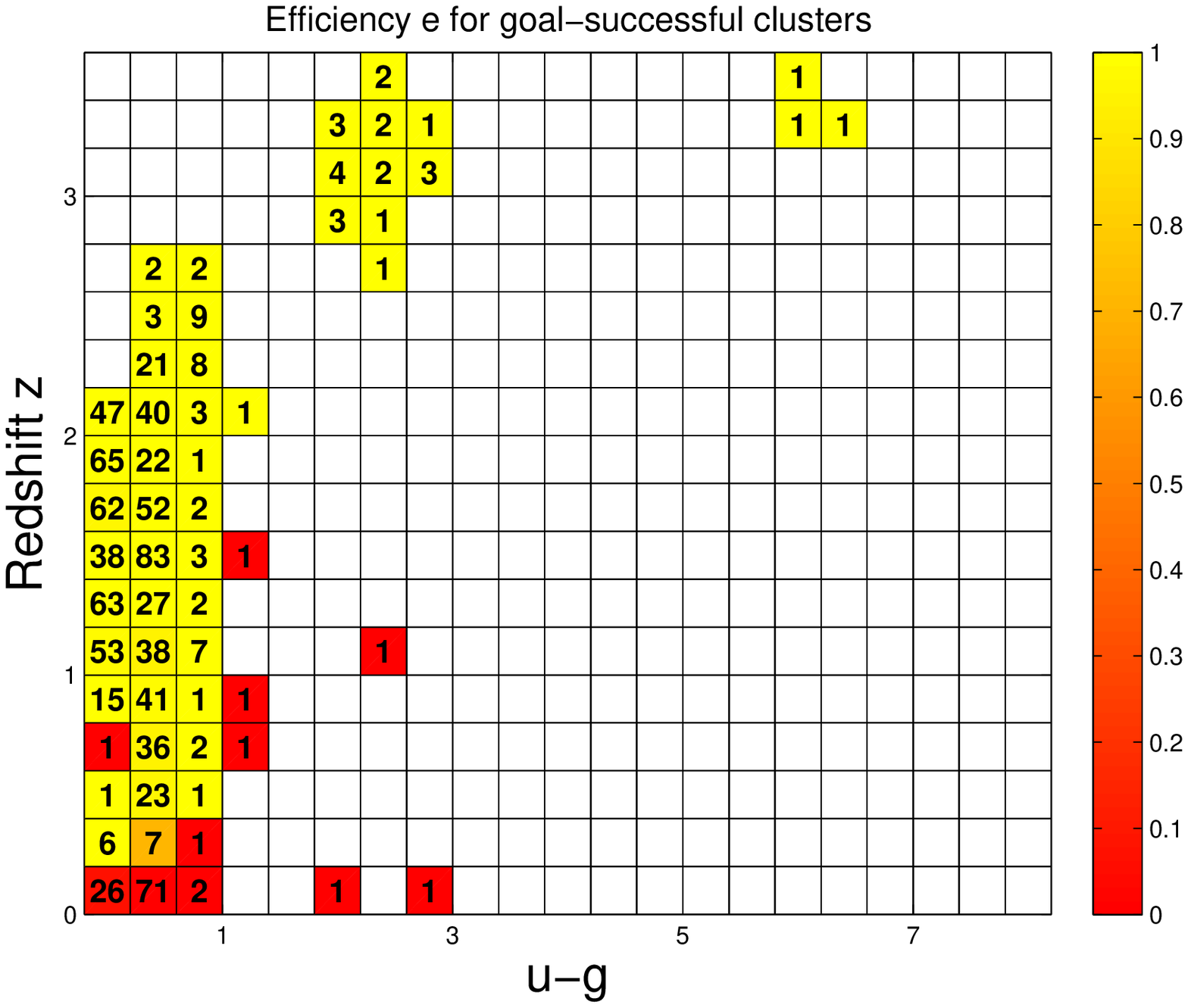}}
\subfigure[Completeness in $z$ vs $u-g$ plane ("goal-successful")]{\includegraphics[width=8cm]{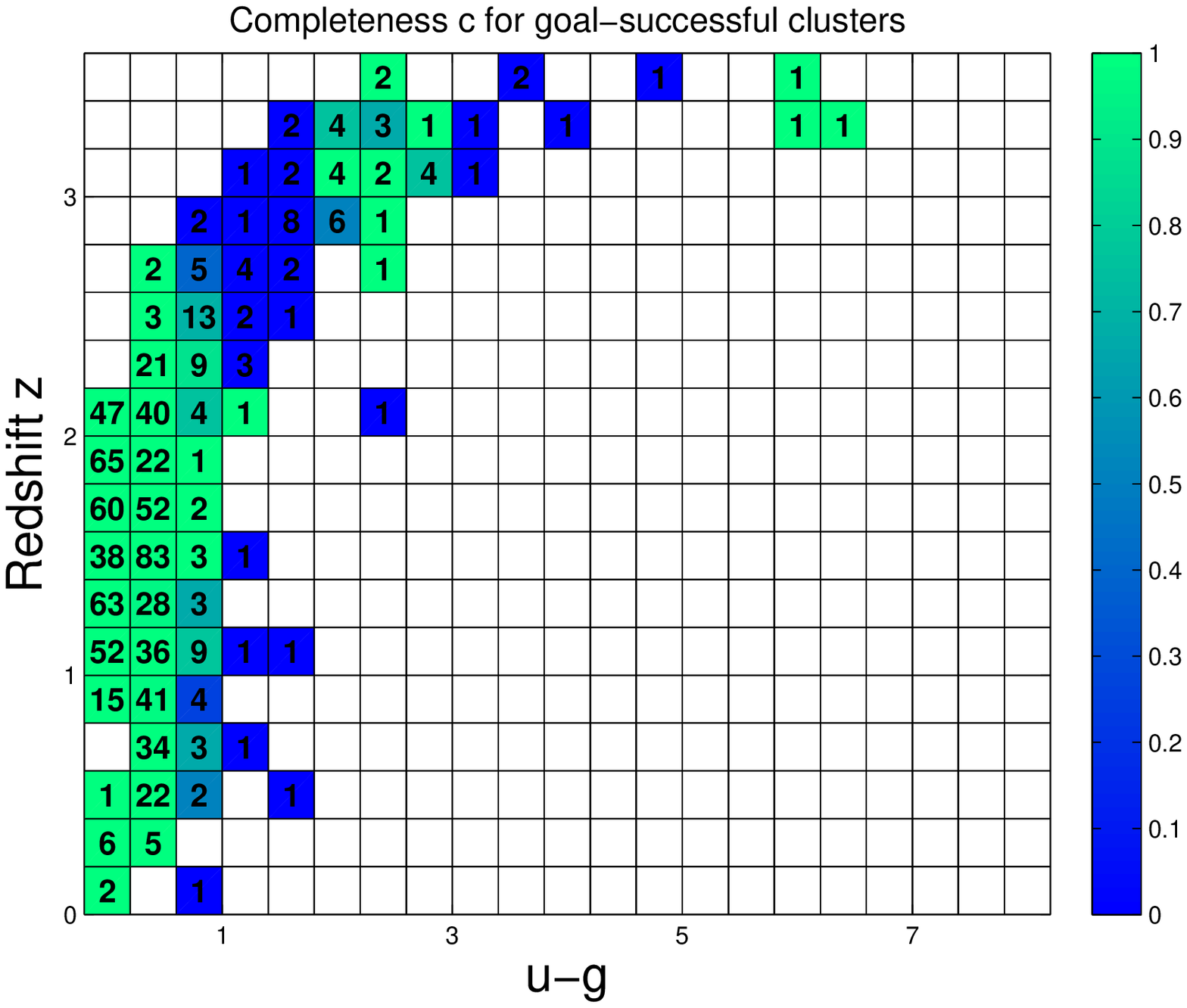}}\\
\subfigure[Efficiency in $z$ vs $u-g$ plane (all clusters)]{\includegraphics[width=8cm]{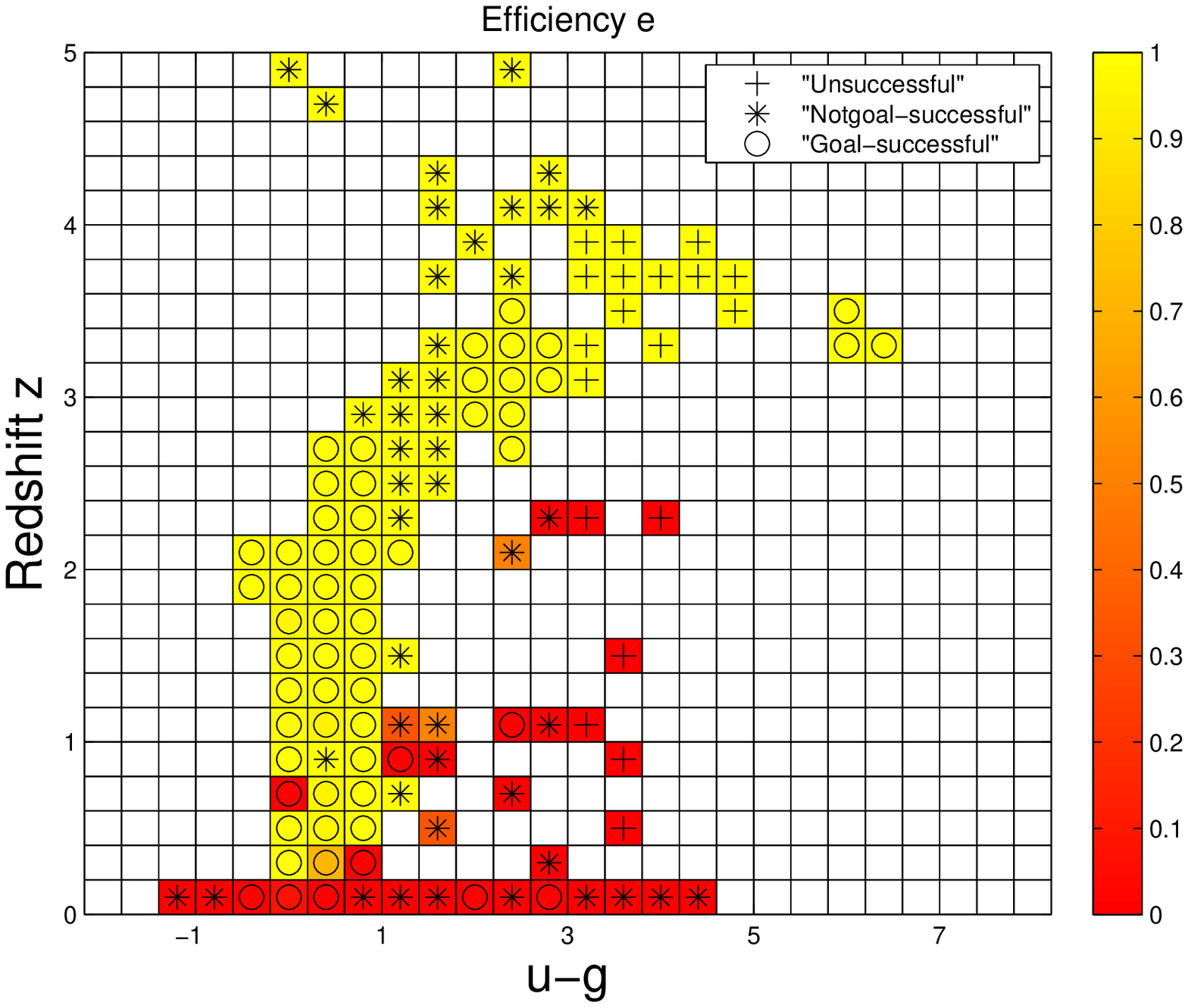}}
\subfigure[Completeness in $z$ vs $u-g$ plane (all clusters)]{\includegraphics[width=8cm]{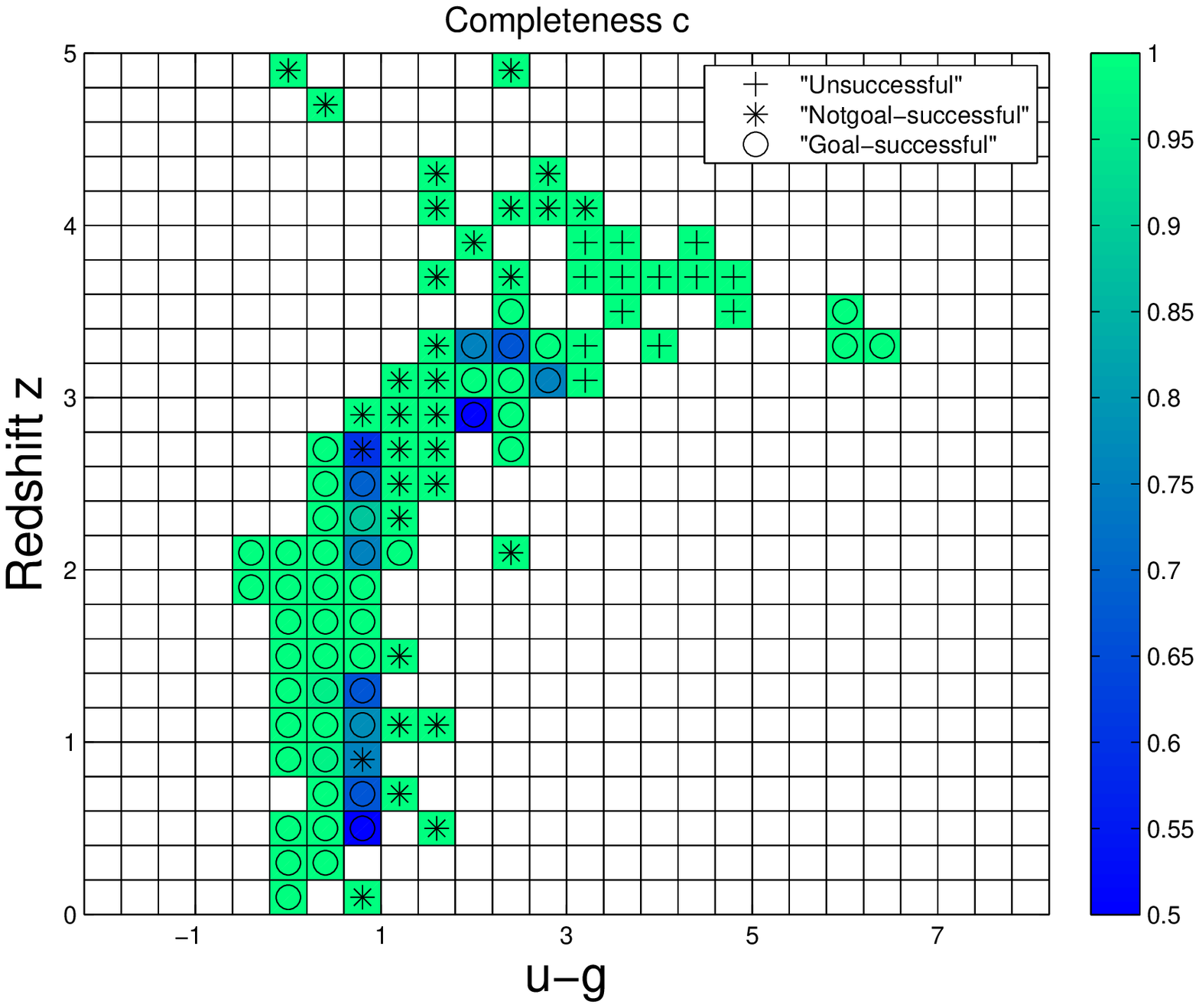}}\\
\caption{Efficiency (upper left-hand panel) and completeness (upper right-hand panel) in the redshift vs ($u-g$) plane for the labelling
phase of the first experiment. Colour of the cell is associated to the efficiency and completeness of the selection process
of the "goal-successful" clusters, while the numbers contained in each cell represent the total number of candidates in that region of the
$z$ vs $u-g$ plane. Lower left-hand and right hand panels contain the maximum efficiency and completeness for the selection 
process of the three types of clusters produced by the algorithm. The class of clusters contributing the maximum fraction to the efficiency
and completeness represented by the symbol contained in each cell of the plane.}
\label{fig:paramparam_first_experiment_ug}
\end{figure}

\begin{figure}
\centering
\includegraphics[width=12cm]{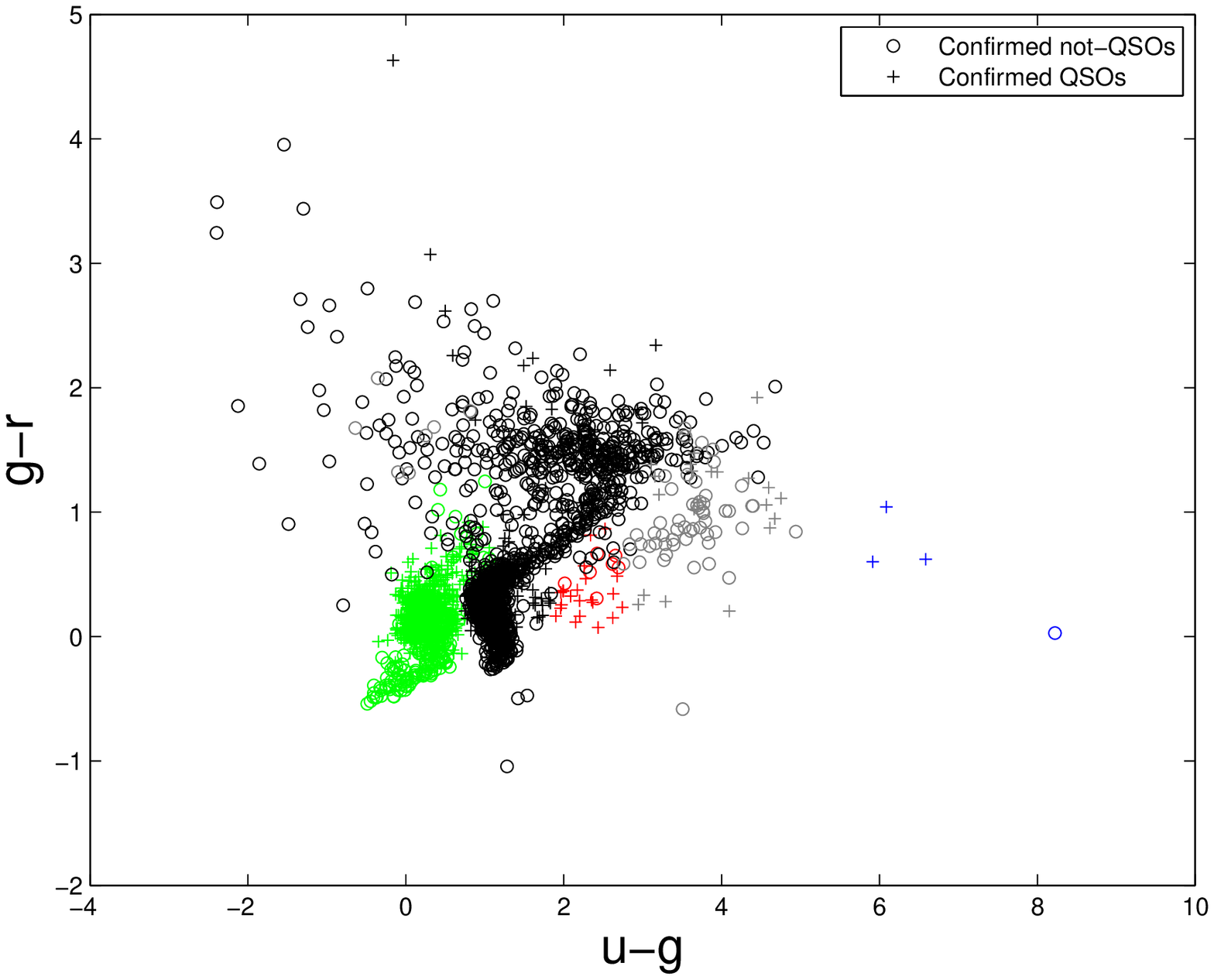}
\caption{Distribution of S-A sample points in the ($u-g$) vs ($g-r$) plane after the labelling 
phase of the first experiment. Black and grey symbols are associated respectively to members of
the "notgoal-successful" and "unsuccessful" clusters, while each single "goal-successful" cluster
is drawn using a different colour. All confirmed QSOs and not QSOs, regardless of their 
membership, are represented by crosses and circles respectively. Three final "goal-successful" 
clusters, green, red and blue respectively, are selected.}
\label{ugvsgr_first}
\end{figure}

The distribution of objects in the ($u-g$) vs ($g-r$) plane and belonging to the 
"goal-successful" clusters selected during the first experiment is shown in the figure 
(\ref{ugvsgr_first}). A  combined colour and symbol code has been used: 
confirmed QSOs according to the BoK are drawn using a cross symbol, while objects 
which the BoK identifies as confirmed not QSOs are drawn as circles. On the other hand, all the 
points belonging to "notgoal-successful" and "unsuccessful" clusters are painted using 
black and grey respectively, while different "goal-successful" clusters involve different colours.
The local values of the efficiency and completeness of the selection are shown in figure 
(\ref{fig:paramparam_first_experiment_ug}). The upper panels show the $e$ and $c$ distributions 
in the redshift vs ($u-g$) plane of the "goal-successful" clusters only, together with the number of 
confirmed QSOs falling in each cell, while the lower panels show the distribution of all the clusters 
identified during the labelling phase of the first experiment ("goal-successful", "notgoal-successful" and 
"unsuccessful") and associate each cell to the class of clusters yielding the higher fraction of
confirmed QSOs selection. 

\subsection{Second experiment}
\label{subsec:second}

\begin{figure}
\centering
\subfigure[NEC diagnostics]{\includegraphics[width=8cm]{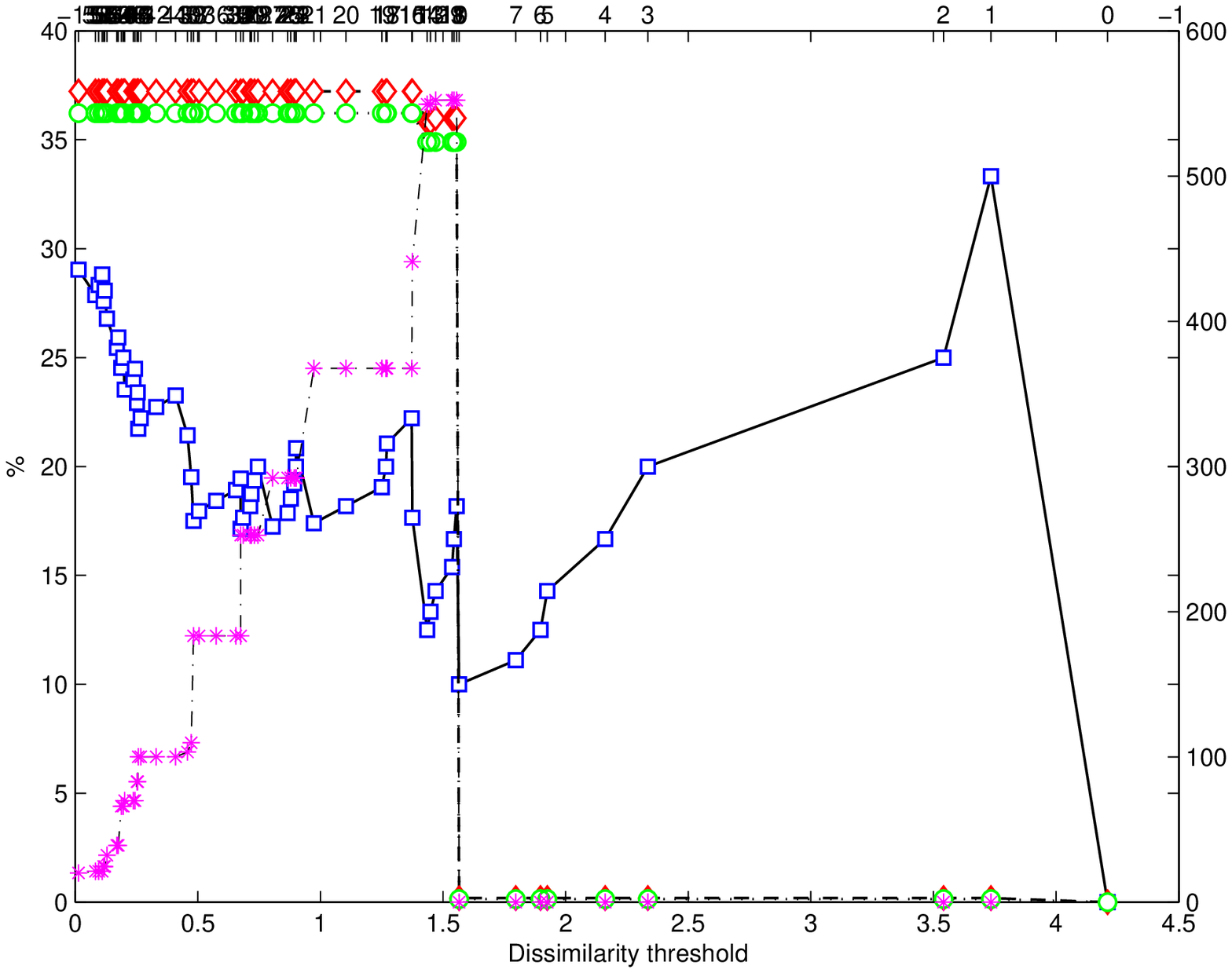}}
\subfigure[Efficiency and completeness]{\includegraphics[width=8cm]{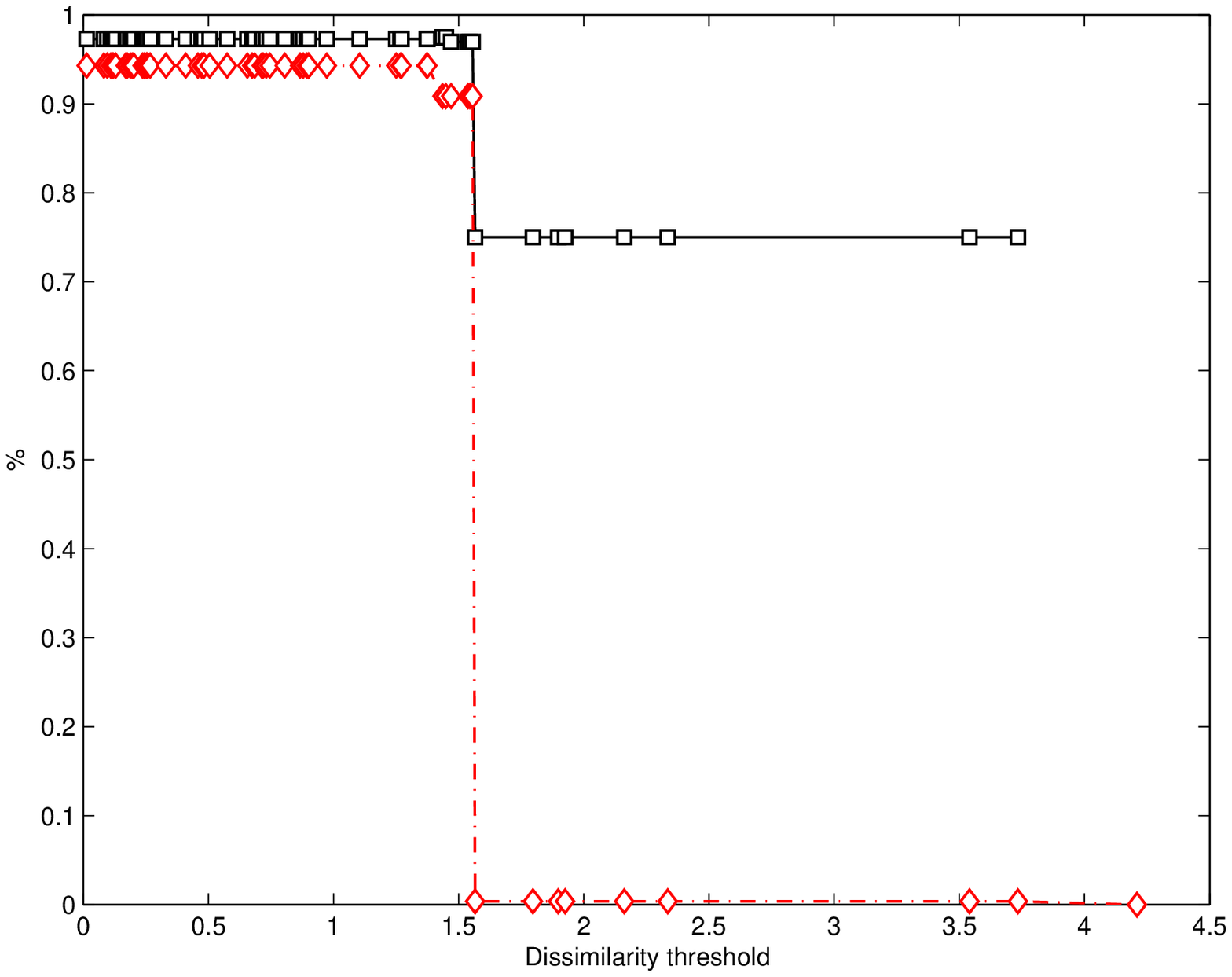}}\\
\subfigure[specClass vs redshift plane distribution of clusters]{\includegraphics[width=8cm]{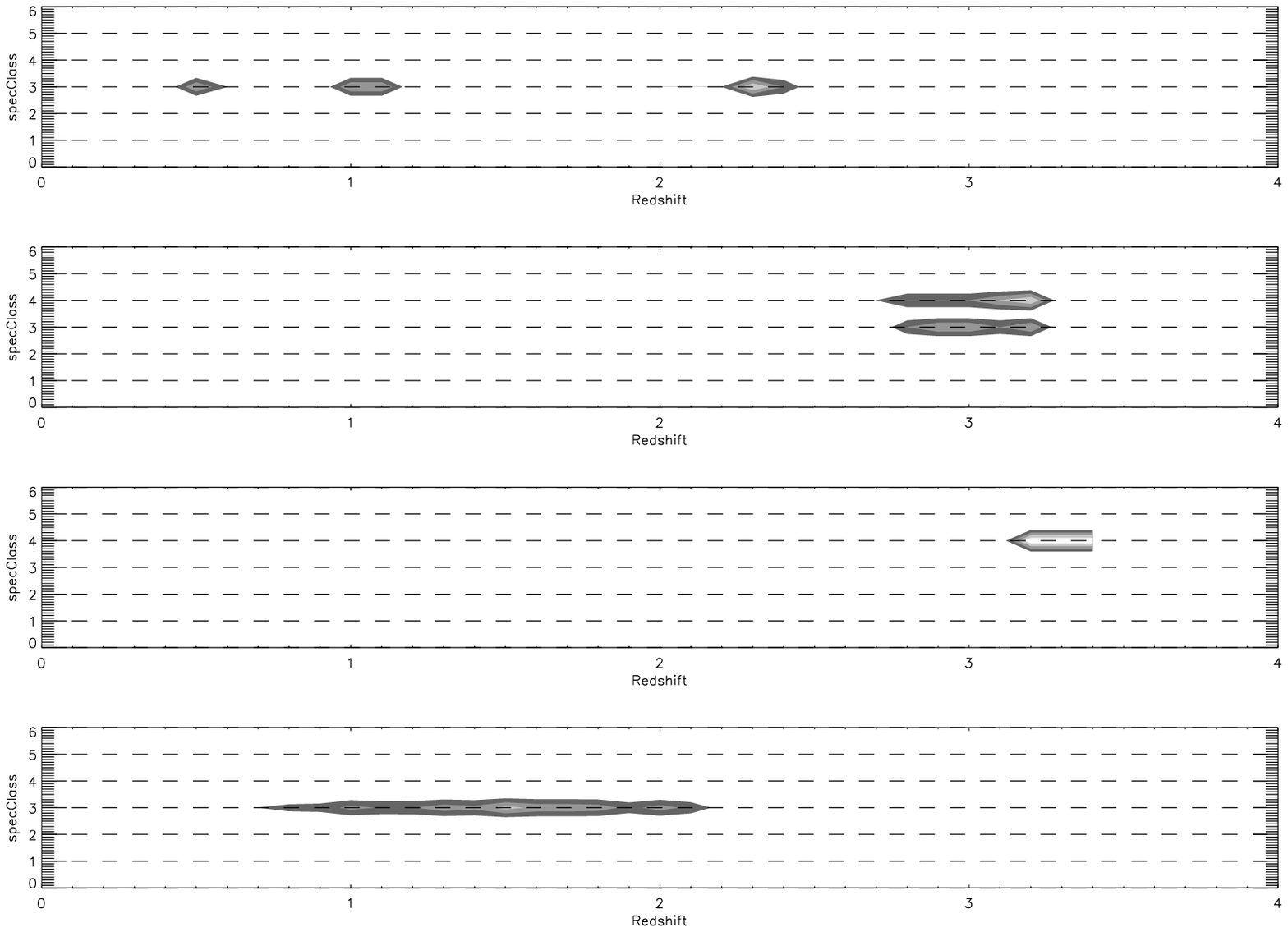}}\\
\subfigure["Box and wisker" plot of clusters]{\includegraphics[width=8cm,angle=-90]{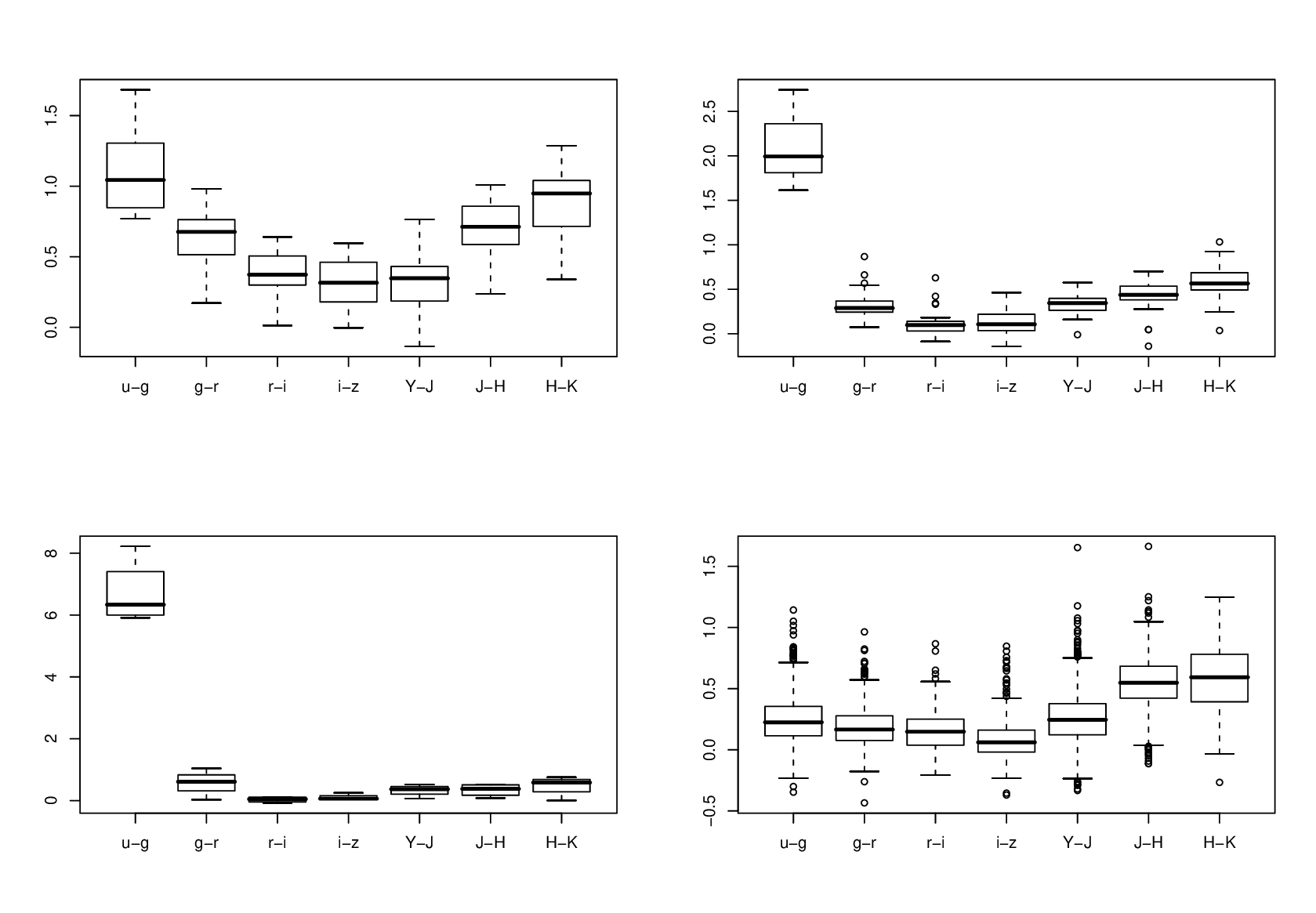}}
\caption{Same as in the previous figure for the second experiment.}
\label{Exp_2}
\end{figure}

The goal of the second experiment was the selection of candidate quasars 
inside the optical colour space, using a sample of star-like objects selected 
in both optical and near infrared catalogues. The number of latent variables used 
for PPS pre-clustering was again fixed to 62 in order to ease the comparison with the first experiment. 
The labelling process has been repeated as described in the previous section; 
the fraction of "goal-successful" clusters and other parameters of the clustering 
are shown in the upper left panel of figure (\ref{Exp_2}) as functions of the 
dissimilarity threshold. The estimated total efficiency and completeness are 
also plotted as functions of the dissimilarity threshold in the upper right panel 
of figure (\ref{Exp_2}). The distribution of sources as a function of redshift and 
spectroscopic classification "specClass" in the "goal-successful" clusters 
selected in this experiment is shown in the middle panel of figure (\ref{Exp_2}) 
and the "box and whisker" plot of the distribution of candidate quasars for each 
of "goal-successful" clusters is given in the bottom panel of figure (\ref{Exp_2}).
The results of this experiment show that our method can reach a significantly higher 
efficiency and completeness level respect to the optical-infrared candidate selection 
algorithms found in the literature \cite{richards_2002,richards_2004,warren_1991}, 
using a base of knowledge formed by both "specClass" for SDSS sources with 
spectroscopic classification but not selected as quasars candidates and 
spectroscopic classification of quasars external to SDSS. 

\begin{figure}
\centering
\subfigure[Efficiency in $z$ vs $u-g$ plane ("goal-successful")]{\includegraphics[width=8cm]{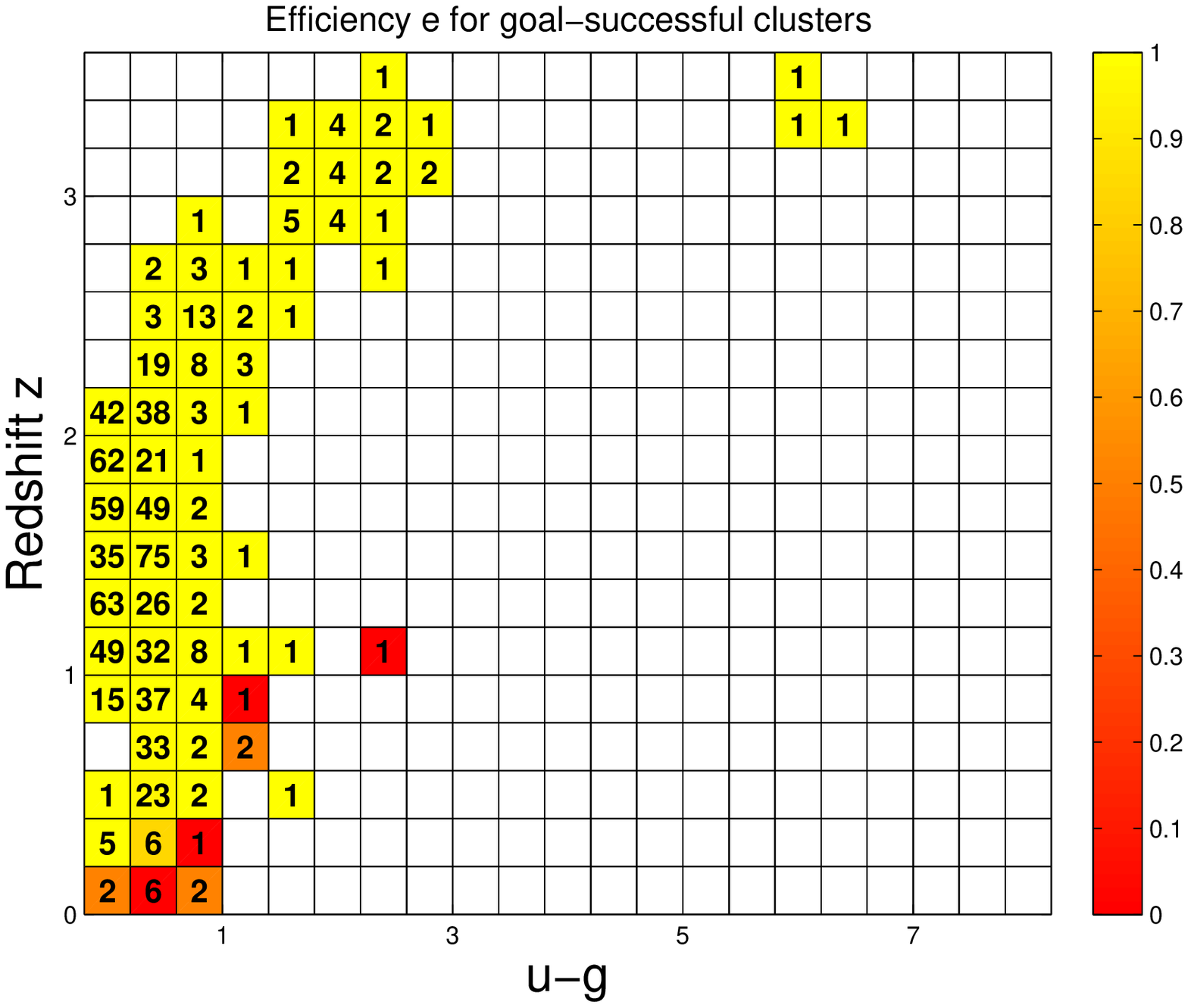}}
\subfigure[Completeness in $z$ vs $u-g$ plane ("goal-successful")]{\includegraphics[width=8cm]{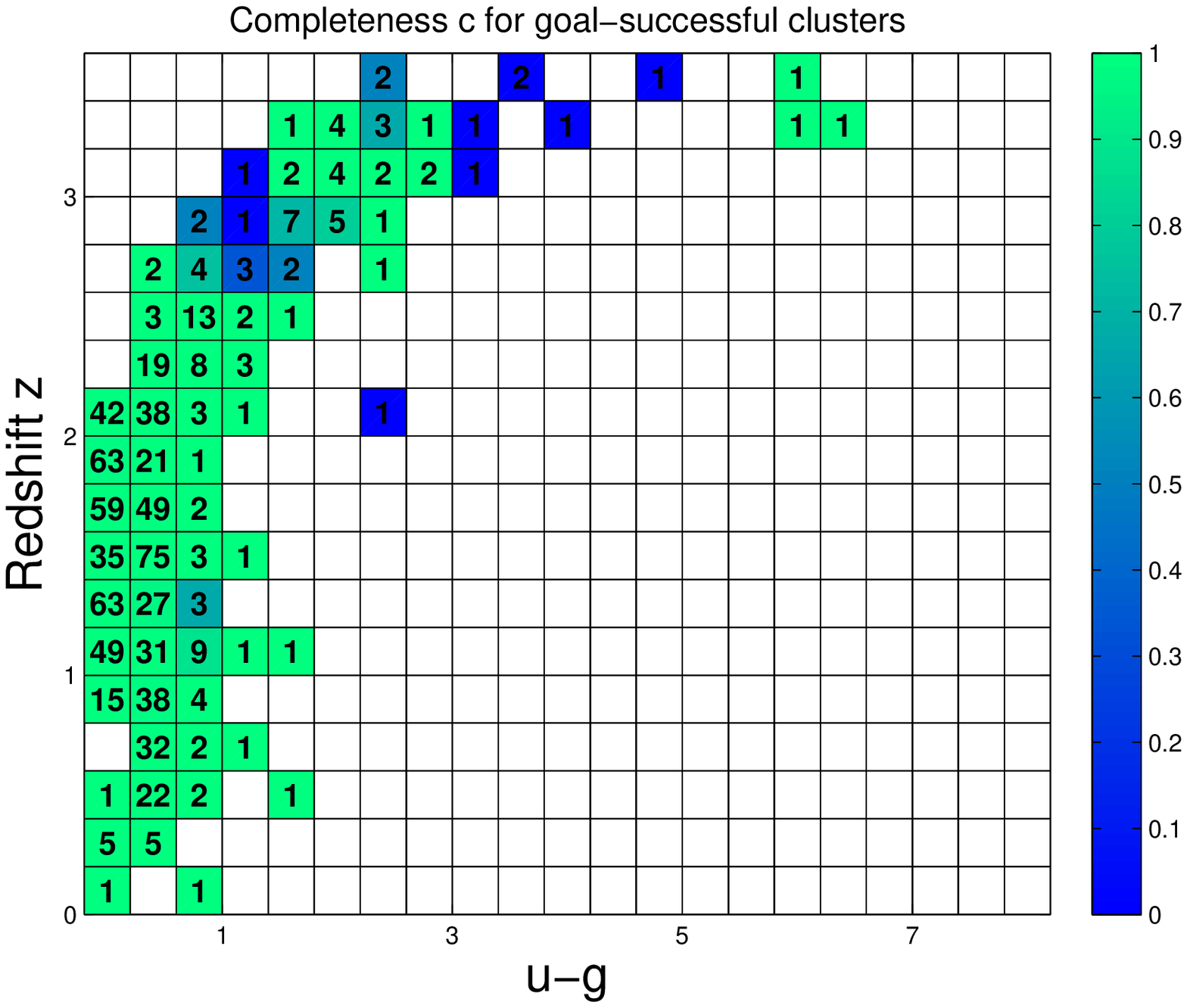}}\\
\subfigure[Efficiency in $z$ vs $u-g$ plane (all clusters)]{\includegraphics[width=8cm]{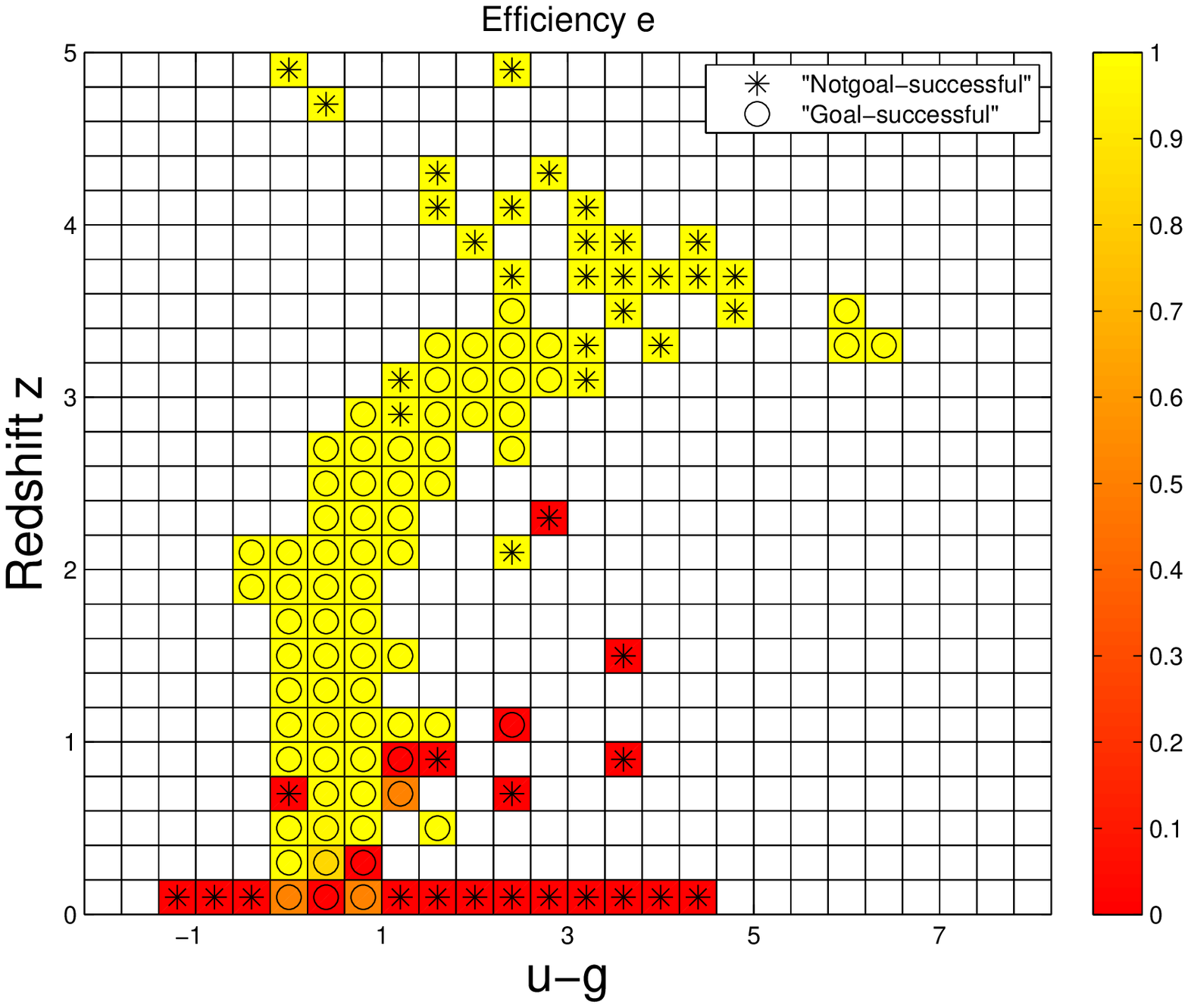}}
\subfigure[Completeness in $z$ vs $u-g$ plane (all clusters)]{\includegraphics[width=8cm]{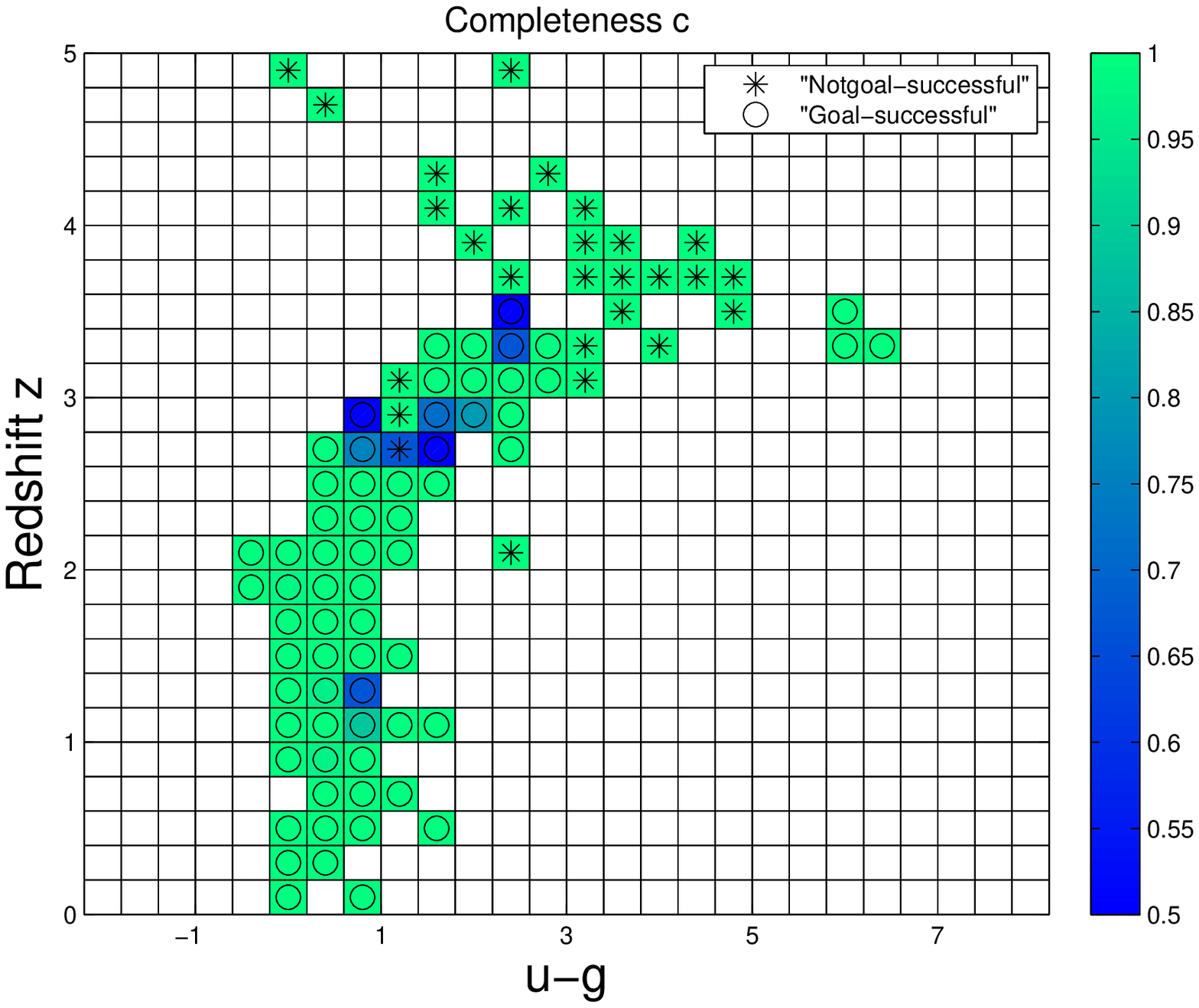}}\\
\caption{Efficiency (upper left-hand panel) and completeness (upper right-hand panel) in the redshift vs ($u-g$) plane for the labelling
phase of the second experiment based on the S-UK data set. Colour of the cell is associated to the efficiency and completeness of the selection process of the "goal-successful" clusters, while the numbers contained in each cell represent the total number of candidates in that region of the
$z$ vs $u-g$ plane. Lower left-hand and right hand panels contain the maximum efficiency and completeness for the selection 
process of the three types of clusters produced by the algorithm. The class of clusters contributing the maximum fraction to the efficiency
and completeness represented by the symbol contained in each cell of the plane.}
\label{fig:paramparam_sec_experiment_ug}
\end{figure} 

\begin{figure}
\centering
\subfigure[Efficiency in $z$ vs $Y-J$ plane ("goal-successful")]{\includegraphics[width=8cm]{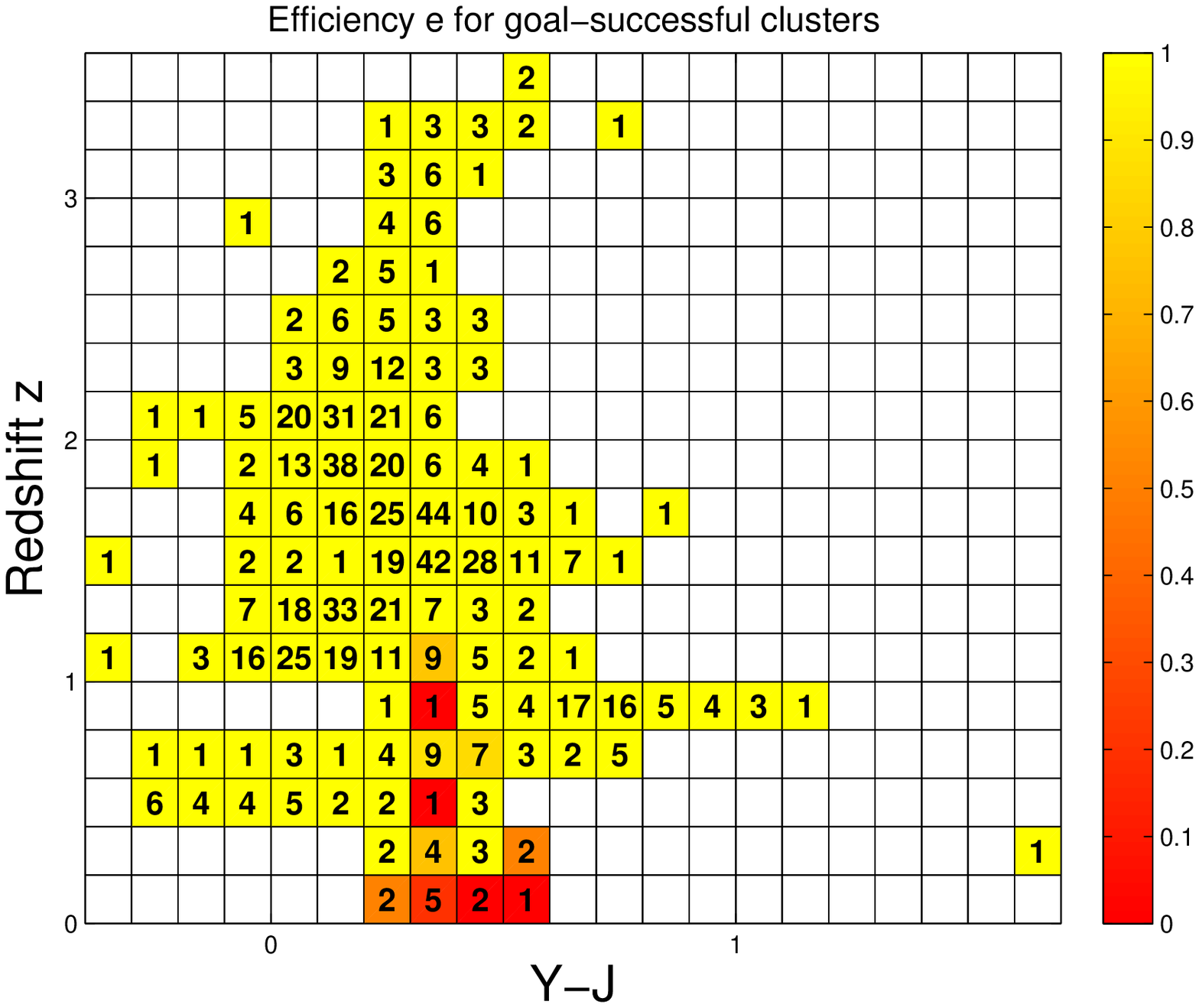}}
\subfigure[Completeness in $z$ vs $Y-J$ plane ("goal-successful")]{\includegraphics[width=8cm]{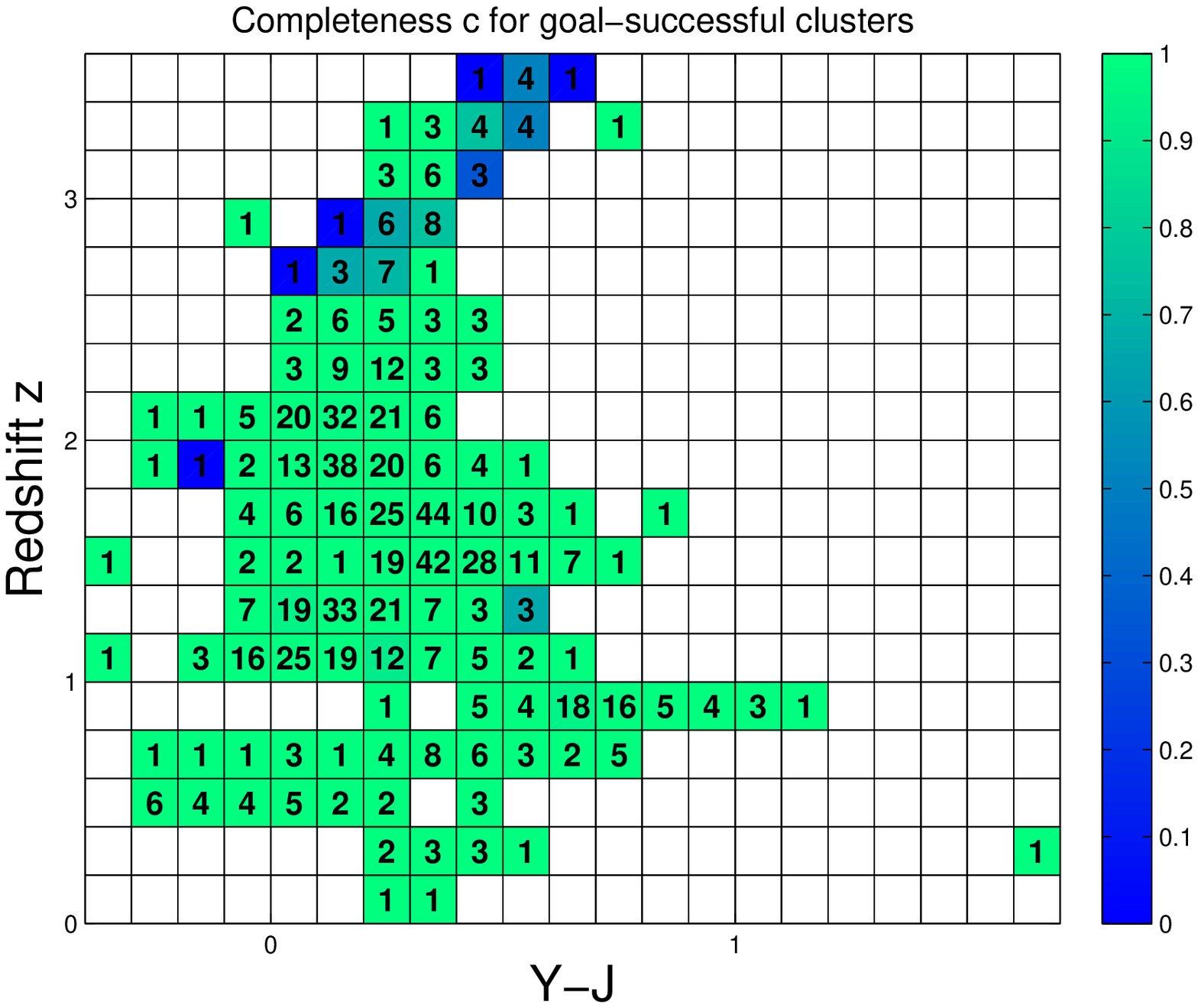}}\\
\subfigure[Efficiency in $z$ vs $Y-J$ plane (all clusters)]{\includegraphics[width=8cm]{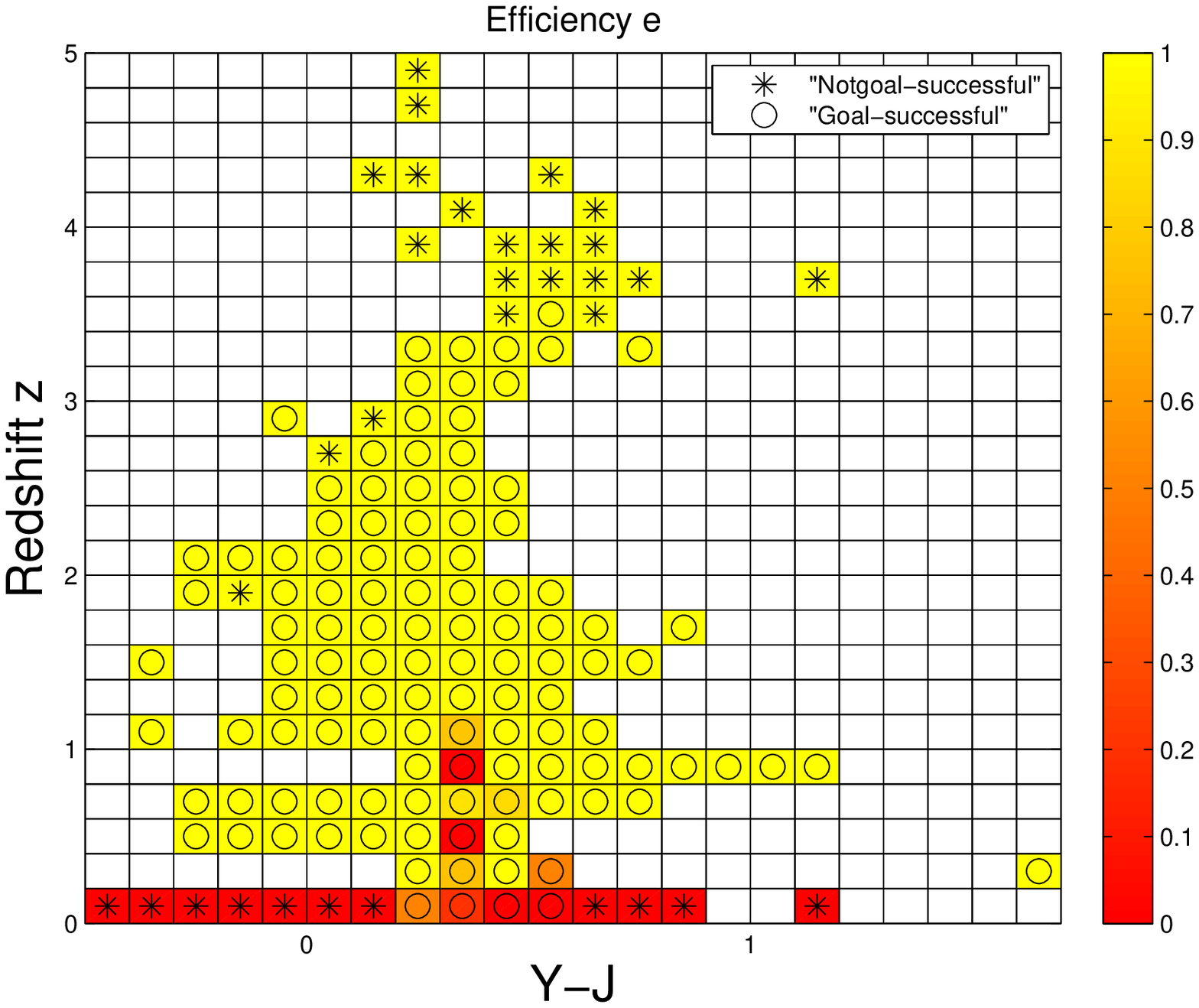}}
\subfigure[Completeness in $z$ vs $Y-J$ plane (all clusters)]{\includegraphics[width=8cm]{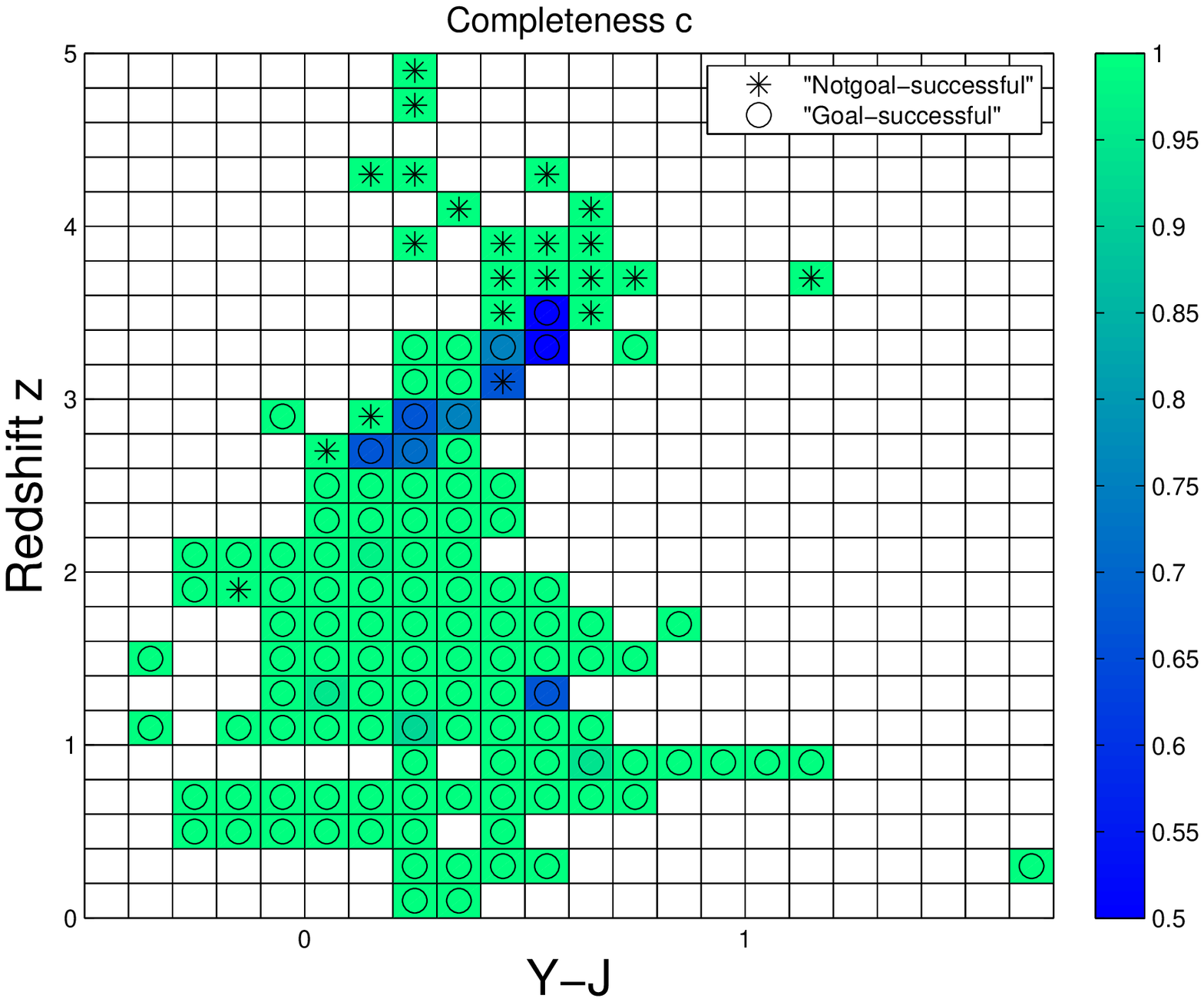}}\\
\caption{Efficiency (upper left-hand panel) and completeness (upper right-hand panel) in the redshift vs ($Y-J$) plane for the labelling
phase of the second experiment based on the S-UK data set. Colour of the cell is associated to the efficiency and completeness of the selection process of the "goal-successful" clusters, while the numbers contained in each cell represent the total number of candidates in that region of the
$z$ vs $Y-J$ plane. Lower left-hand and right hand panels contain the maximum efficiency and completeness for the selection 
process of the three types of clusters produced by the algorithm. The class of clusters contributing the maximum fraction to the efficiency
and completeness represented by the symbol contained in each cell of the plane.}
\label{fig:paramparam_sec_experiment_yj}
\end{figure} 

\noindent For what the "goal-successful" clusters are concerned, we can notice that:

\begin{itemize}
\item The first cluster is composed of sources with optical $u-g$ colour 
concentrated around 1.0, while the other optical colours $g-r$, $r-i$ and $i-z$ 
have average values falling from around 0.7 to 0.5 respectively. 
The infrared colour $Y-J$ averages at about 0.5, while $J-H$ and $H-K$ 
increase to mean values of about 0.8 and 1.0. 
The redshift distribution of this cluster members, all ranked as normal 
QSOs according to the SDSS spectroscopic classification ("specClass" = 3), 
shows three different groups of objects situated approximately at $z \sim 0.5, 
1.1$ and 2.2.  
\item The second cluster is characterized by a colour distribution similar 
to the one of the first cluster, except for the optical $u-g$ colour whose 
mean value increases to 2.0. 
This cluster is composed of equal fractions of normal and far QSOs according 
to the "specClass" index, with a distribution spanning about 0.5 in redshift 
around 3.  
\item The third cluster is formed by far QSOs only, with a higher redshift 
($z \sim 3.3$) mean value and a colour distribution similar to the second 
cluster except for a much higher value of $u-g$, with well defined mean at 
approximately 6.1.
\item The last cluster is composed mainly by "specClass" = 2 sources inside a 
large redshift interval spanning from $\sim 0.7$ to 2.2. All colours have mean 
values included between 0.0 and 0.5. 
\end{itemize}

\begin{figure}
\centering
\subfigure[]{\includegraphics[width=8cm]{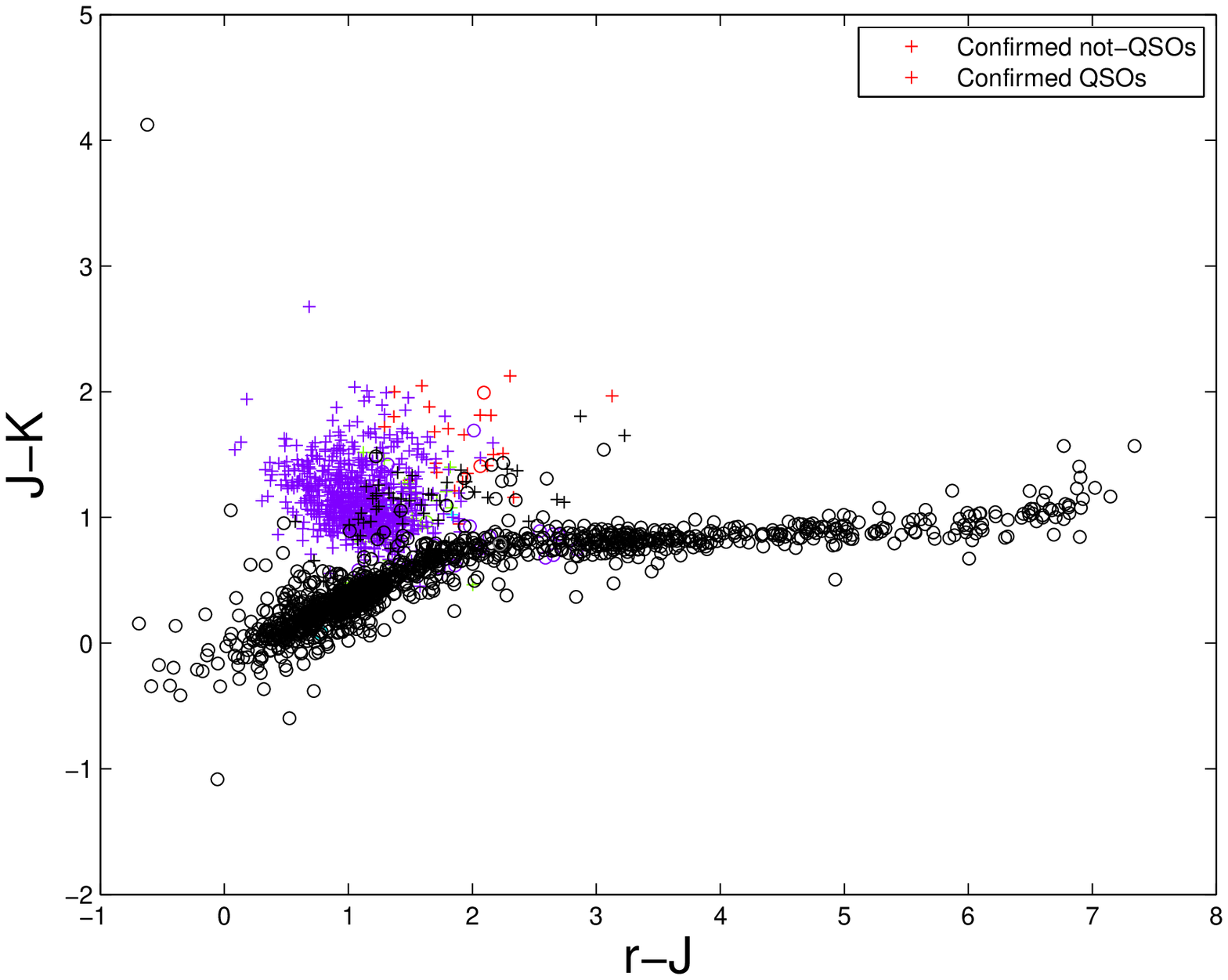}}
\subfigure[]{\includegraphics[width=8cm]{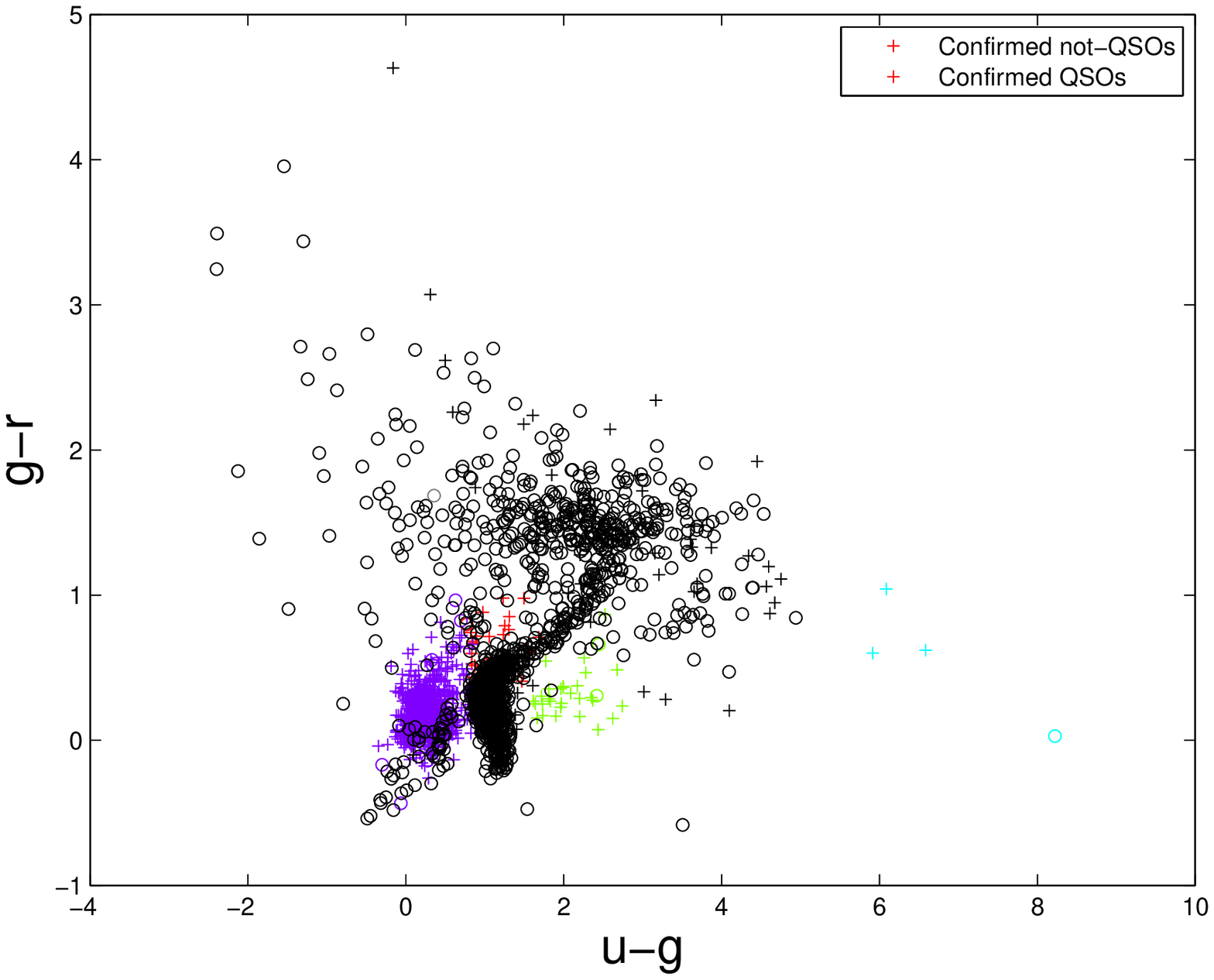}}
\caption{Distribution of S-UK sample points in the ($r-J$) vs ($J-K$) plane (left-hand panel) 
and ($u-g$) vs ($g-r$) plane (right-hand panel) after the labelling 
phase of the second experiment. Black and grey symbols are associated respectively to members of
the "notgoal-successful" and "unsuccessful" clusters, while each single "goal-successful" cluster
is drawn using a different colour. All confirmed QSOs and not QSOs, regardless of their 
membership, are represented by crosses and circles respectively.}
\label{colourcolour_second}
\end{figure}

The clusters from second to fourth appear to be very similar to the "goal-successful" clusters selected in the first experiment both in terms of colours and redshift distribution and spectroscopic type composition. As for the first experiment, the distribution of objects associated to "goal-successful" clusters during this experiment in two interesting colour-colour planes, namely the mixed optical-infrared colours plane ($r-J$) vs ($J-K$) and the purely optical ($u-g$) vs ($g-r$) plane, is shown in the left-hand and right-hand panel of figure (\ref{colourcolour_second}) respectively. In the figures (\ref{fig:paramparam_sec_experiment_ug}) and (\ref{fig:paramparam_sec_experiment_yj}) the local values of efficiency and completeness are shown in the redshift vs ($u-g$) and redshift vs ($Y-J$) planes, respectively, with meaning and features of the single plots already explained in the previous paragraph. 

\subsection{Third experiment}
\label{subsec:third}

\begin{figure}
\centering
\subfigure[NEC diagnostics]{\includegraphics[width=8cm]{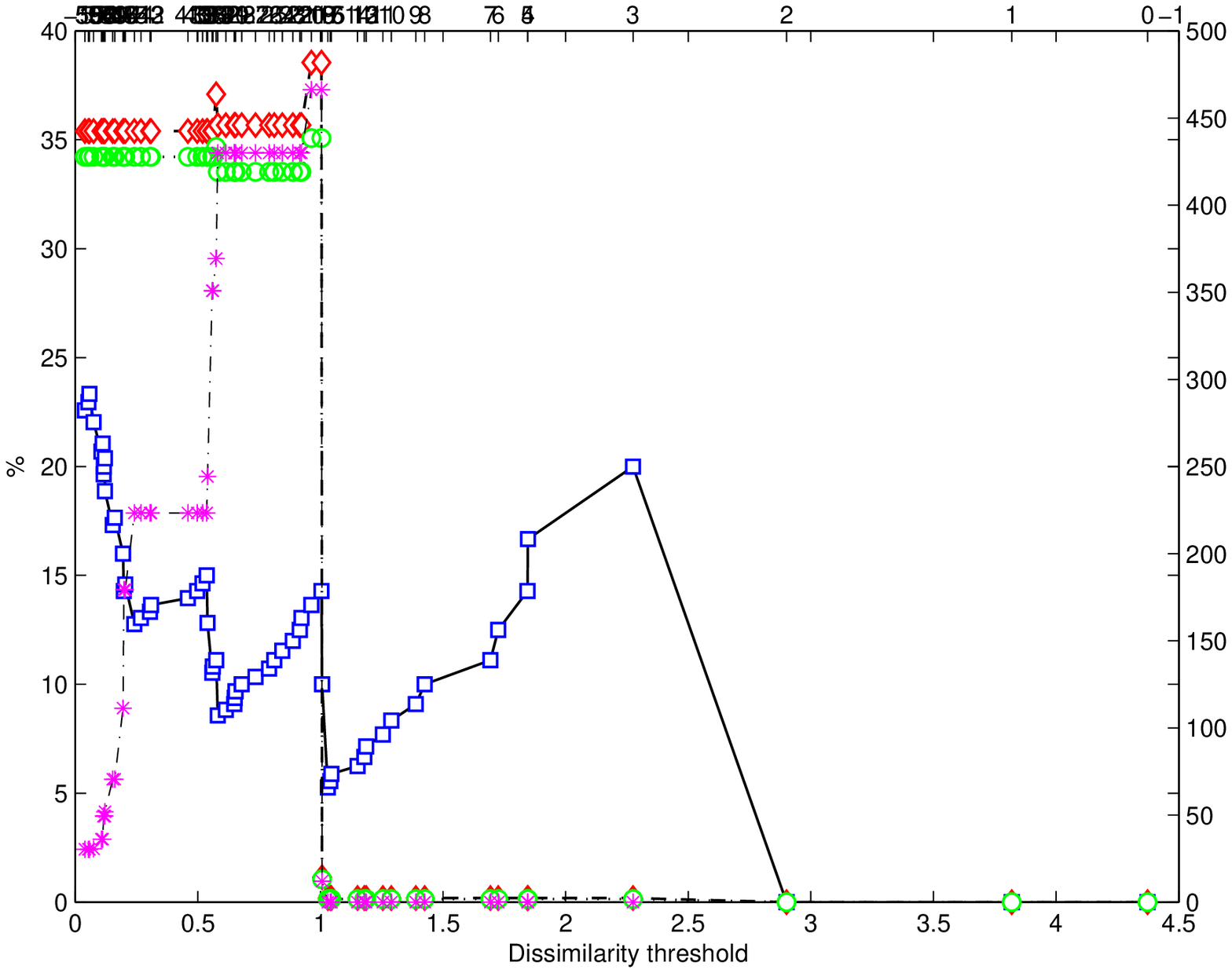}}
\subfigure[Efficiency and completeness]{\includegraphics[width=8cm]{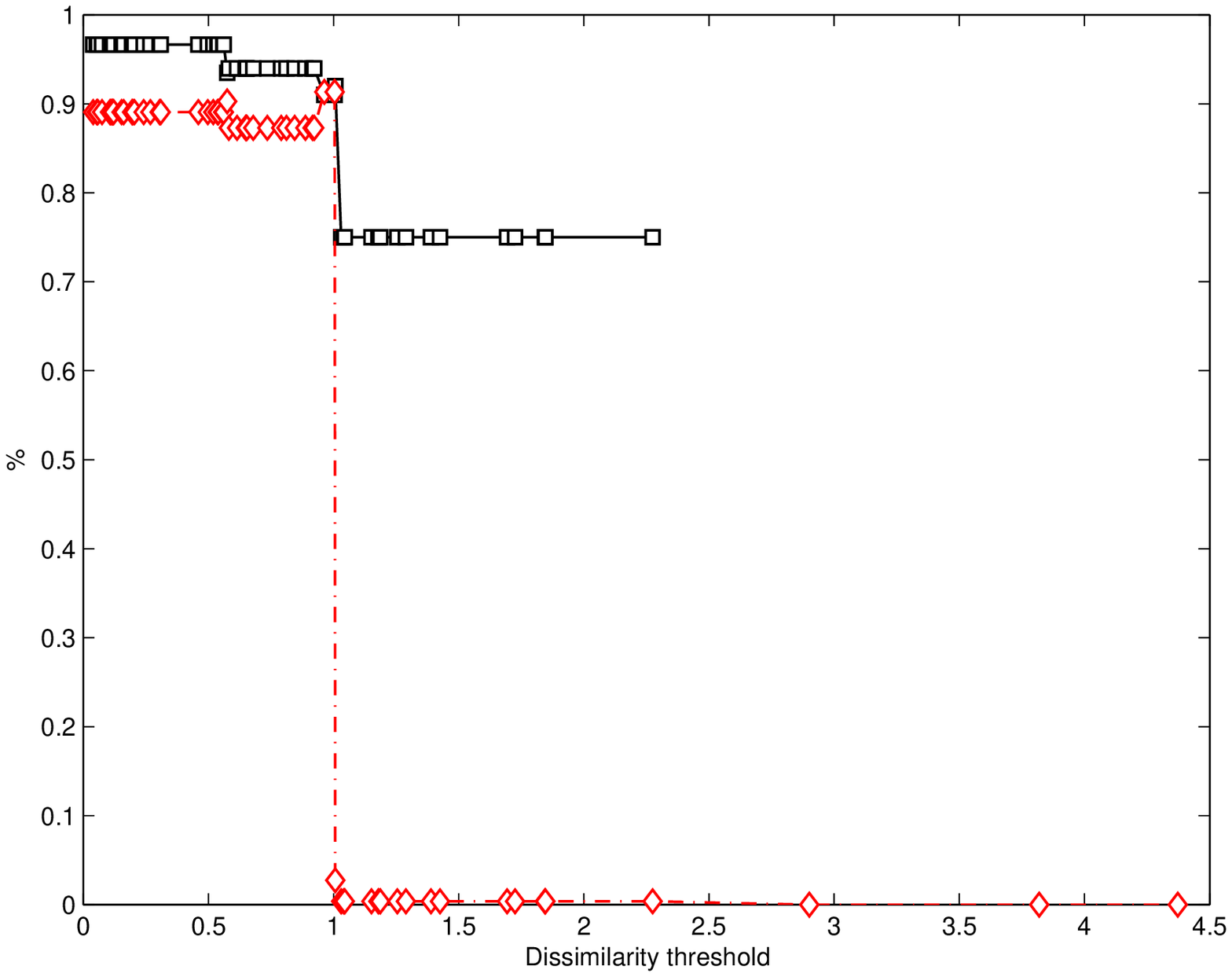}}\\
\subfigure[specClass vs redshift plane distribution of clusters]{\includegraphics[width=8cm]{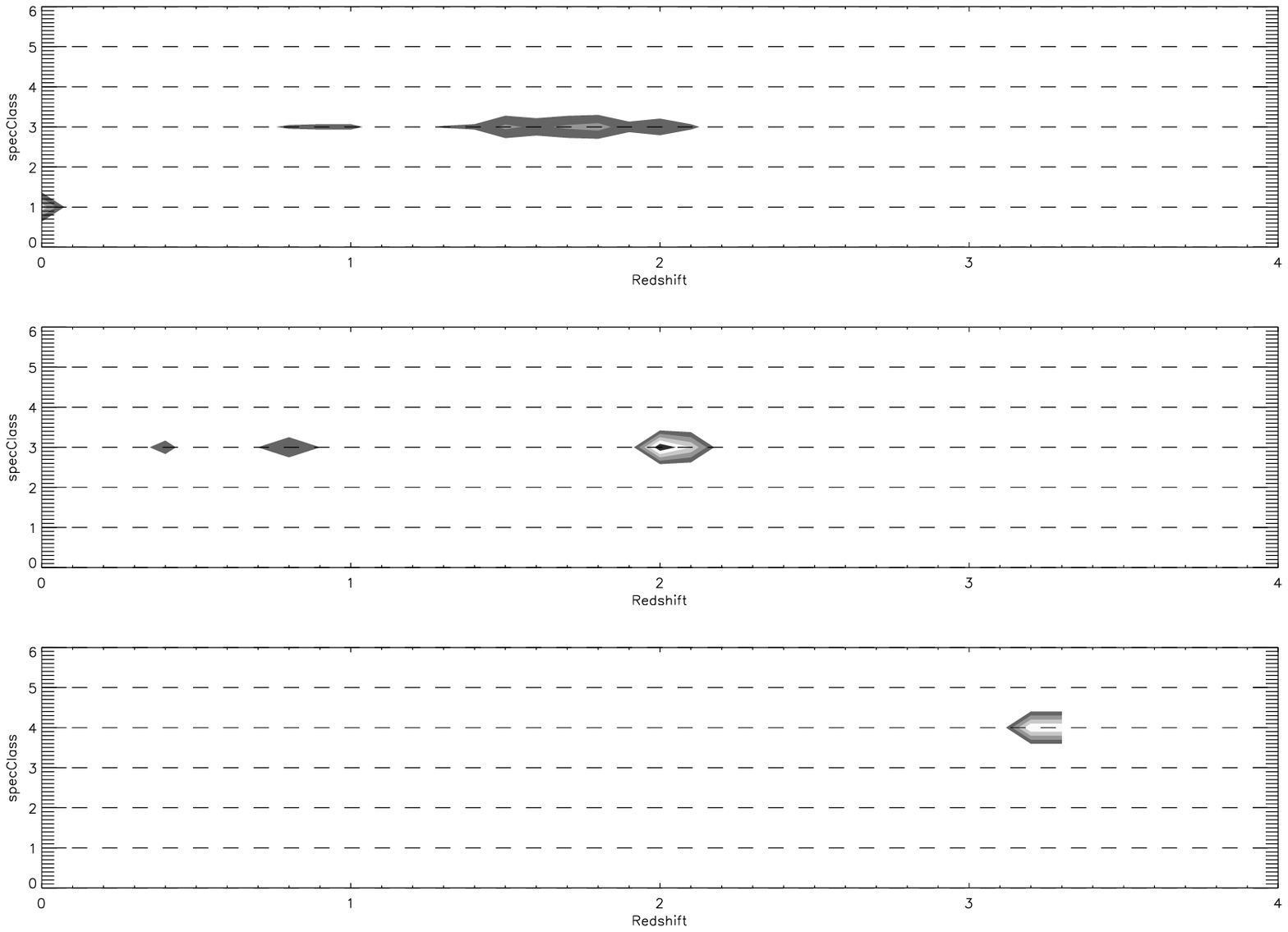}}\\
\subfigure["Box and wisker" plot of clusters]{\includegraphics[width=8cm,angle=-90]{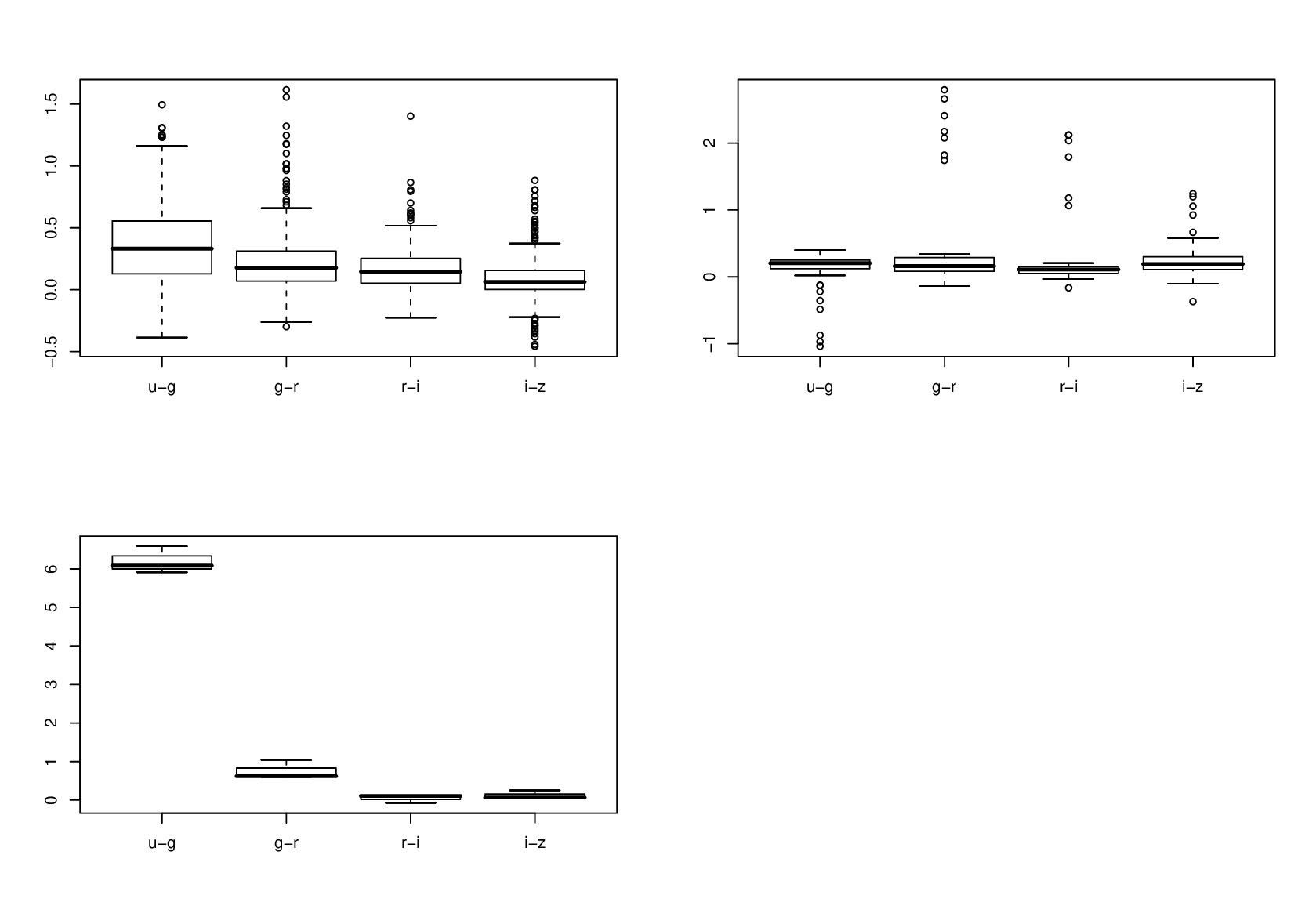}}
\caption{Same as in the previous figure for the third experiment.}
\label{exp_3}
\end{figure}

The third experiment was carried out in order to test whether the addition of the 
near infrared colours to the optical colours already used as parameters for the 
first two experiments improves or not the total efficiency and completeness of 
candidate quasars selection. In conformity to the previous experiments, the 
number of latent variables was fixed to 62, resulting in an equal number of 
pre-clusters produced. As in the previous experiments, all relevant information 
are reported in figure (\ref{exp_3}).

\begin{table*}
\caption{Description of the contents of "goal-successful" clusters selected in the 
experiments carried out. For each cluster, the total number of members $n$ and 
the fraction of confirmed quasars are reported.}             
\label{table_detailed}      
\centering          
\begin{tabular}{ccrrr}    
\hline\hline       
Exp.  & Cluster &  \% quasars &  $n$ \\ 
\hline                    
    1 & 1  & 75.9\% & 29 \\  
       & 2  & 81.5\%  & 957 \\
       & 3 &  75.1 \% & 4\\
    2 & 1 &  76.1\%  & 652 \\  
       & 2  &  83.3\%  & 48 \\
       & 3 &  100.0\%  & 4\\
    3 & 1  & 92.3\%  & 26  \\  
       & 2  & 93.5\%  & 31  \\
       & 3 &  75.0\%  & 4\\
       & 4  &  97.7\%  & 755\\
    4 & 1  &   86.2\% & 2121\\  
       & 2  &   93.1\% & 52190\\
       & 3 &   77.9\% & 2433\\
       & 4 &  75.8\% & 198 \\
       & 5 &   78.6\% & 126\\
       & 6 &   90.6\% & 171\\
       & 7 &  91.5.\% & 298\\
       & 8 &   78.9\% & 90\\
       & 9 &  79.0\% & 76 \\
       & 10 &  86.1\% & 92 \\
\hline                  
\end{tabular}
\end{table*}

The results of this experiment are summarised in the following description of the "goal-successful" clusters selected.

\begin{itemize}
\item The mean values of colours distribution of the members of the first 
cluster range from 0.0 to 0.5 with outliers mainly situated at higher values, 
while the distribution of normal QSOs in redshift spans from 0.7 to 2.2 and 
a little contamination from "specClass = 1" stars is present. 
\item The second cluster, entirely composed by "specClass" = 2 sources, shows 
a colours distribution almost identical to the previous cluster, while the 
distribution in redshift peaks at $z\sim 0.4, 0.7 $ and 2.1.
\item The third clusters contains sources with a value of $u-g \sim 6$ and 
$g-r \sim 1$, while the other two colours have distributions consistent with zero. 
This cluster is formed by far QSOs with redshift higher than 3.0. 
\end{itemize}

\begin{figure}
\centering
\subfigure[Efficiency in $z$ vs $u-g$ plane ("goal-successful")]{\includegraphics[width=8cm]{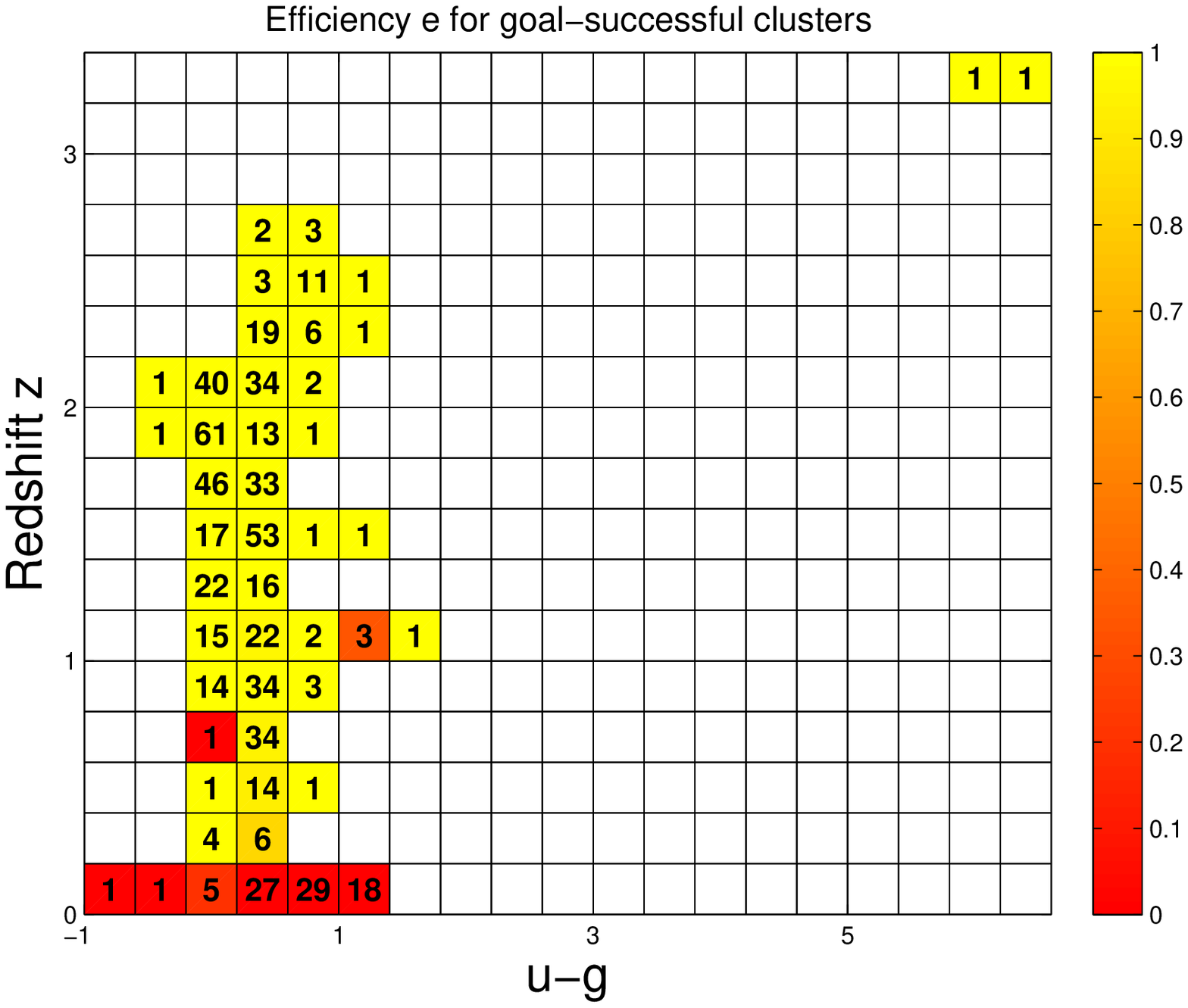}}
\subfigure[Completeness in $z$ vs $u-g$ plane ("goal-successful")]{\includegraphics[width=8cm]{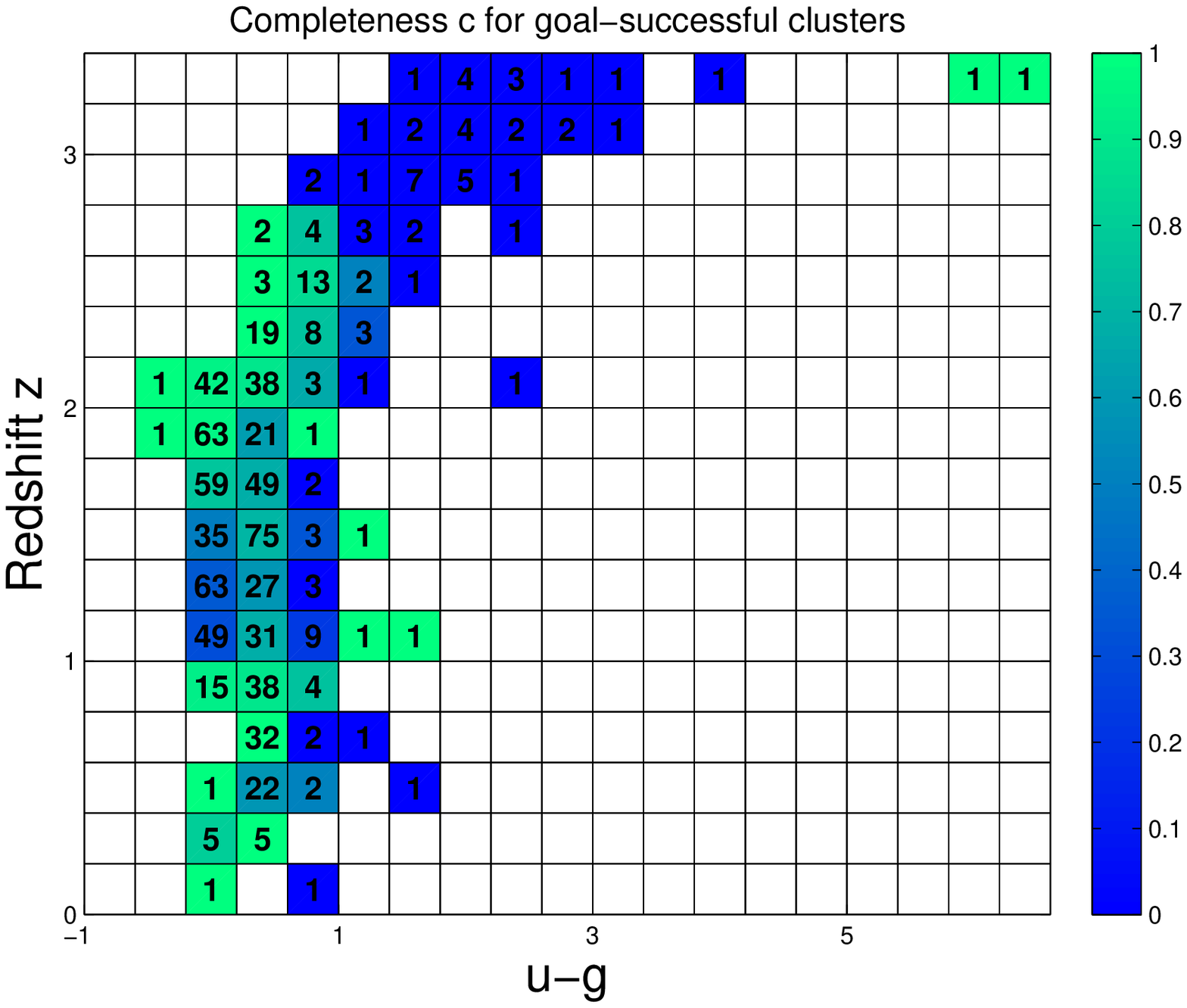}}\\
\subfigure[Efficiency in $z$ vs $u-g$ plane (all clusters)]{\includegraphics[width=8cm]{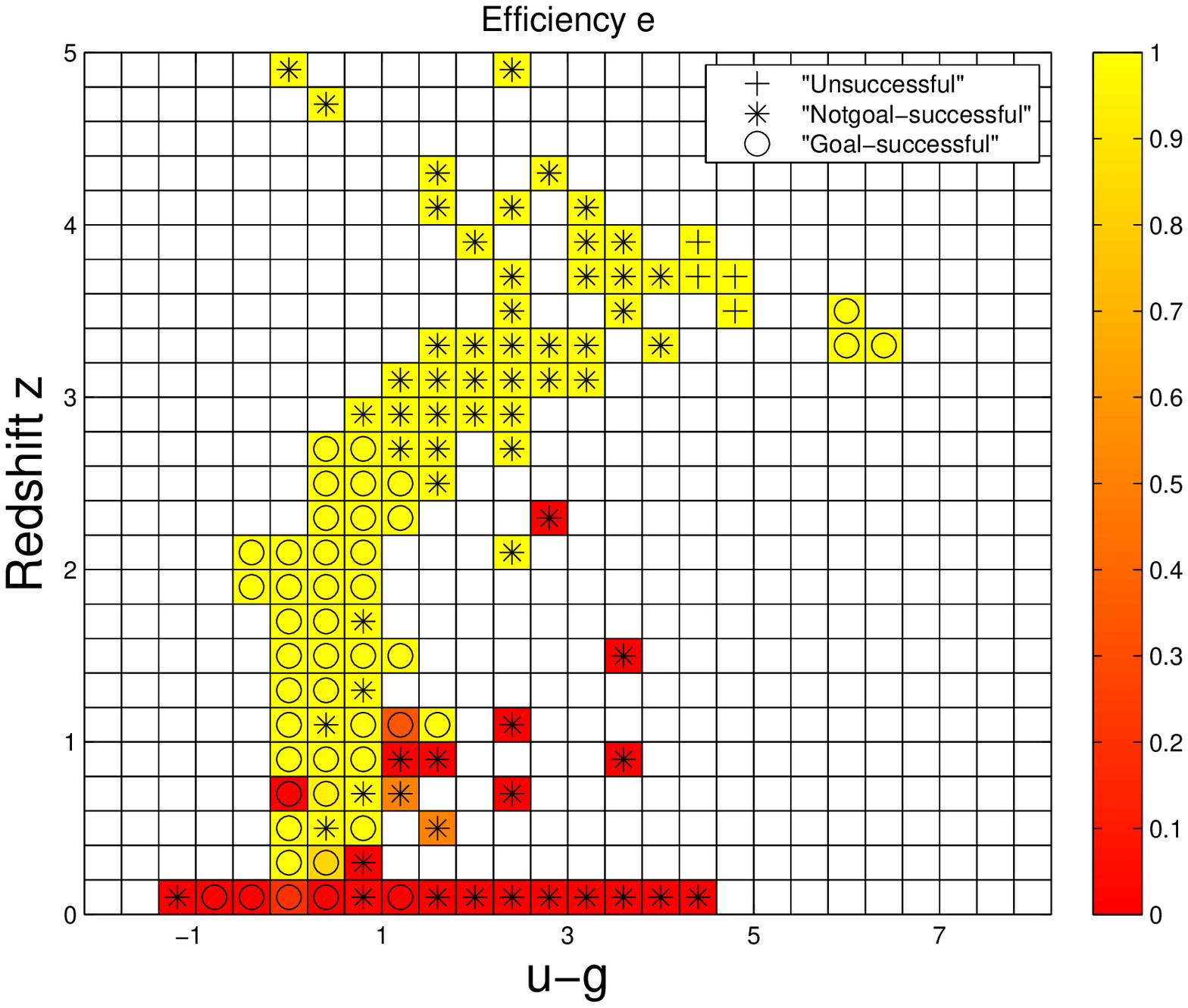}}
\subfigure[Completeness in $z$ vs $u-g$ plane (all clusters)]{\includegraphics[width=8cm]{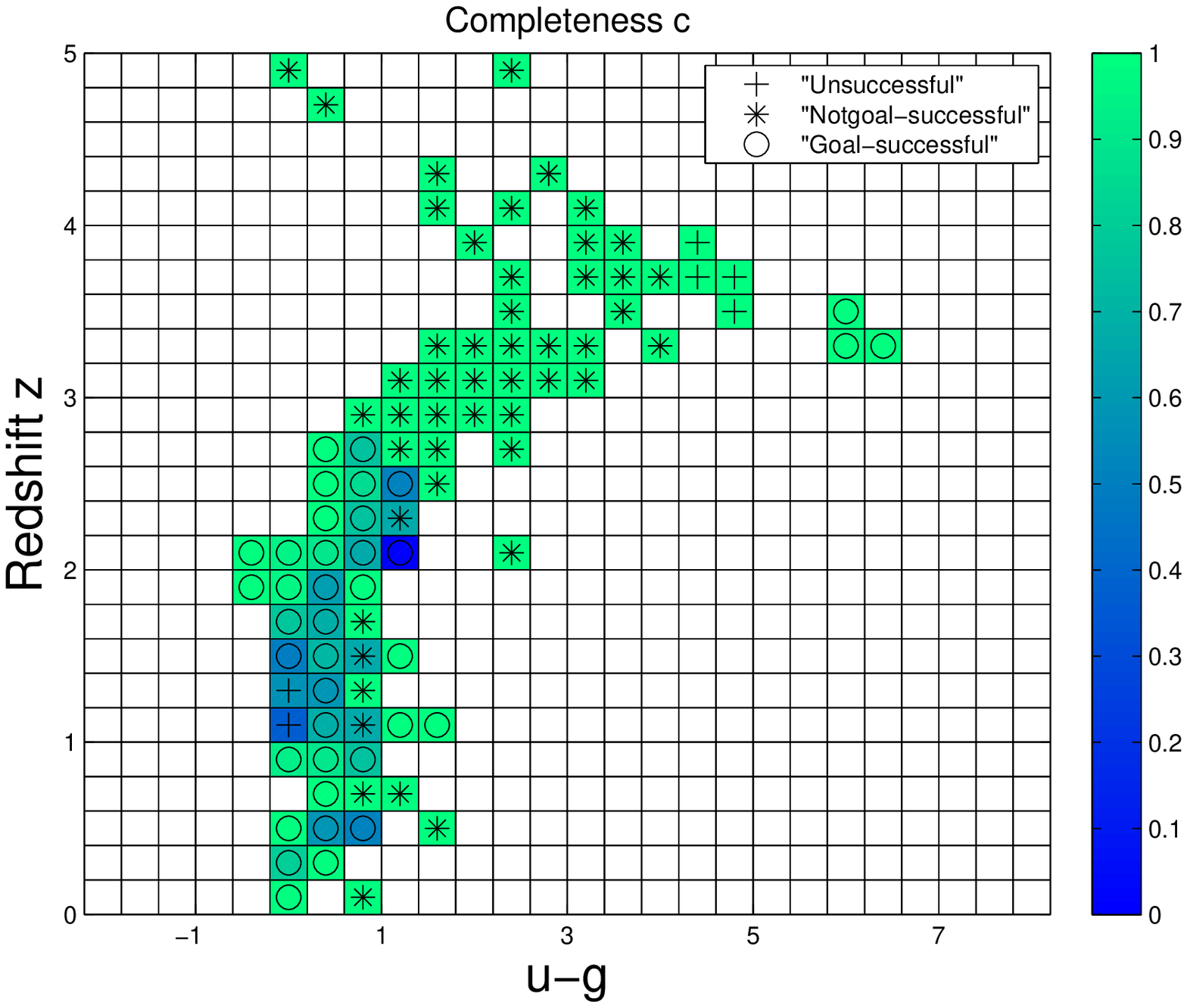}}\\
\caption{Efficiency (upper left-hand panel) and completeness (upper right-hand panel) in the redshift vs ($u-g$) plane for the labelling phase of the third experiment based on the S-UK data set though using only optical photometry. Colour of the cell is associated to the efficiency and completeness of the selection process of the "goal-successful" clusters, while the numbers contained in each cell represent the total number of candidates in that region of the $z$ vs $u-g$ plane. Lower left-hand and right hand panels contain the maximum efficiency and completeness for the selection process of the three types of clusters produced by the algorithm. The class of clusters contributing the maximum fraction to the efficiency and completeness represented by the symbol contained in each cell of the plane.}
\label{fig:paramparam_third_experiment}
\end{figure}

\begin{figure}
\centering
\includegraphics[width=12cm]{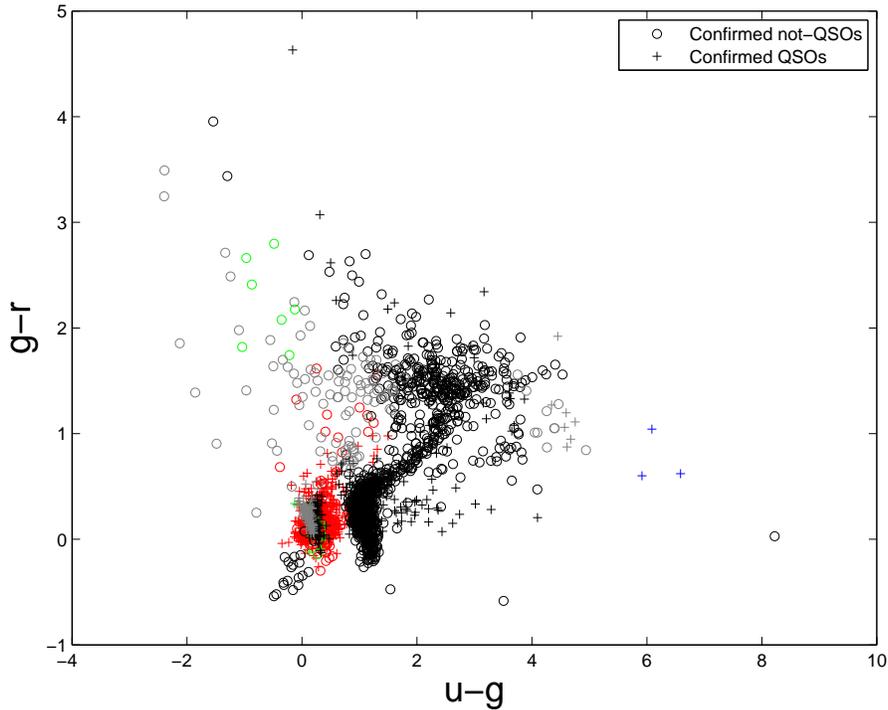}
\caption{Distribution of S-UK sample points in the ($u-g$) vs ($g-r$) plane after the labelling 
phase of the third experiment. Black and grey symbols are associated respectively to members of
the "notgoal-successful" and "unsuccessful" clusters, while each single "goal-successful" cluster
is drawn using a different colour. All confirmed QSOs and not QSOs, regardless of their 
membership, are represented by crosses and circles respectively.}
\label{ugvsgr_third}
\end{figure}

Also in this case, the clusters selected are very similar in terms of colours and 
redshift distributions to the "goal-successful" clusters selected in the previous, 
two experiments.  The distribution of "goal-successful" candidates in the optical
colour-colour ($u-g$) vs ($g-r$) plane is shown in figure (\ref{ugvsgr_third}). The figure
(\ref{fig:paramparam_third_experiment}) contains the local values of the efficiency and
completeness in the redshift vs ($u-g$) plane for the third experiment for the "goal-successful" only 
and all the three kind of clusters determined by the labelling in the upper and lower panels, respectively.  

\subsection{Fourth experiment}
\label{subsec:fourth}

\begin{figure}
\centering
\subfigure[NEC diagnostics]{\includegraphics[width=8cm]{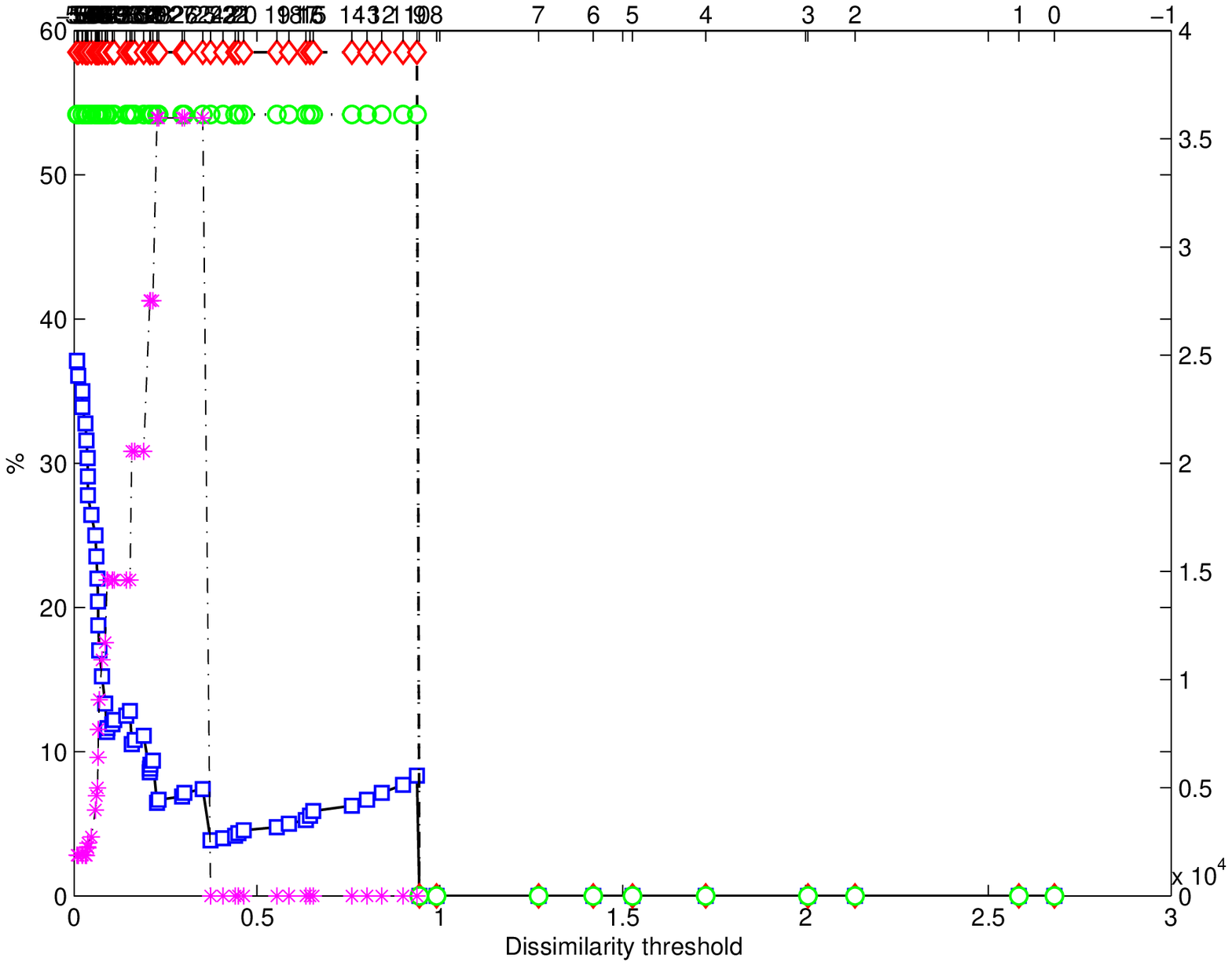}}
\subfigure[Efficiency and completeness]{\includegraphics[width=8cm]{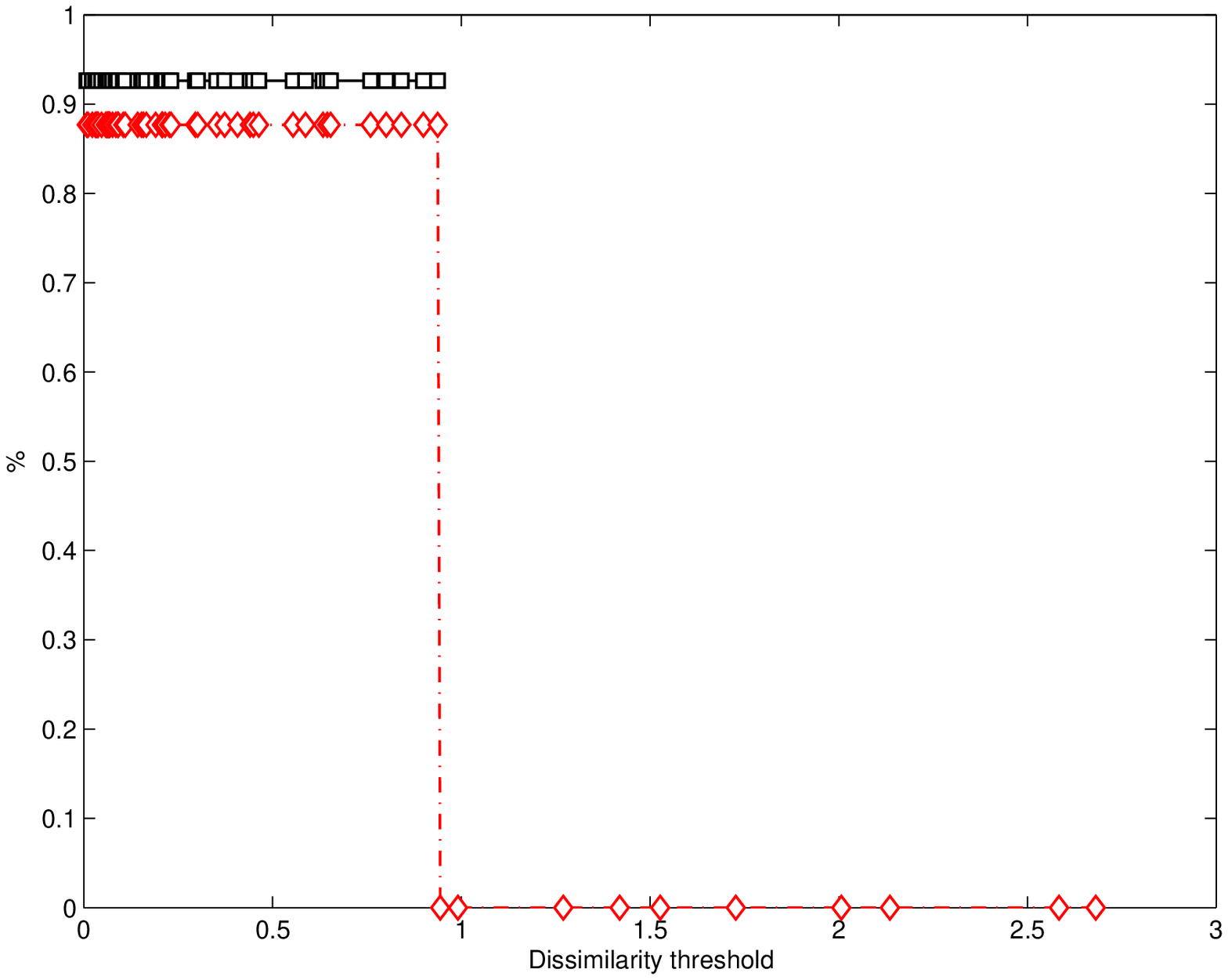}}\\
\subfigure[specClass vs redshift plane distribution of clusters]{\includegraphics[width=7cm,angle=-90]{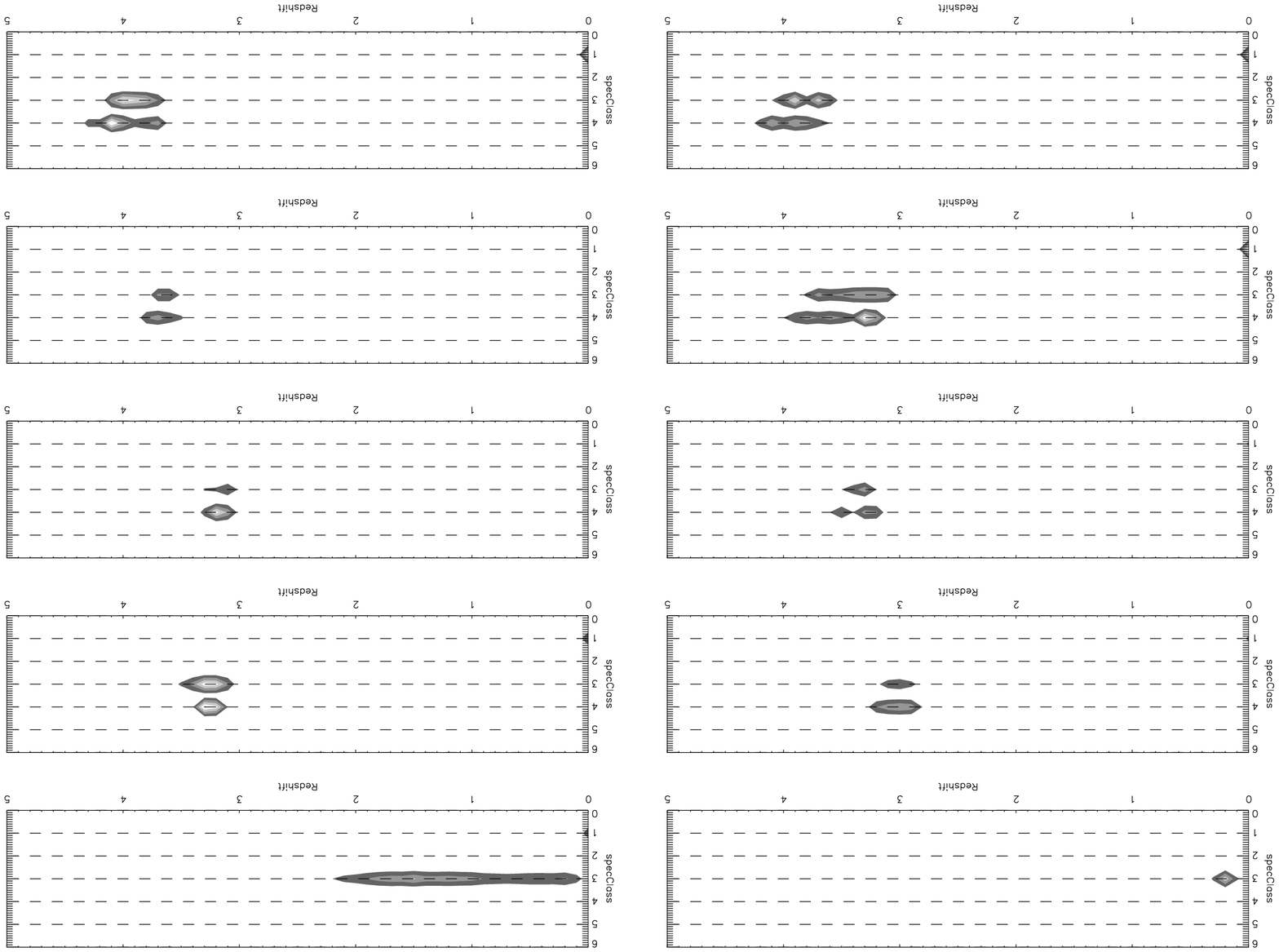}}\\
\subfigure["Box and wisker" plot of clusters]{\includegraphics[width=7cm,angle=-90]{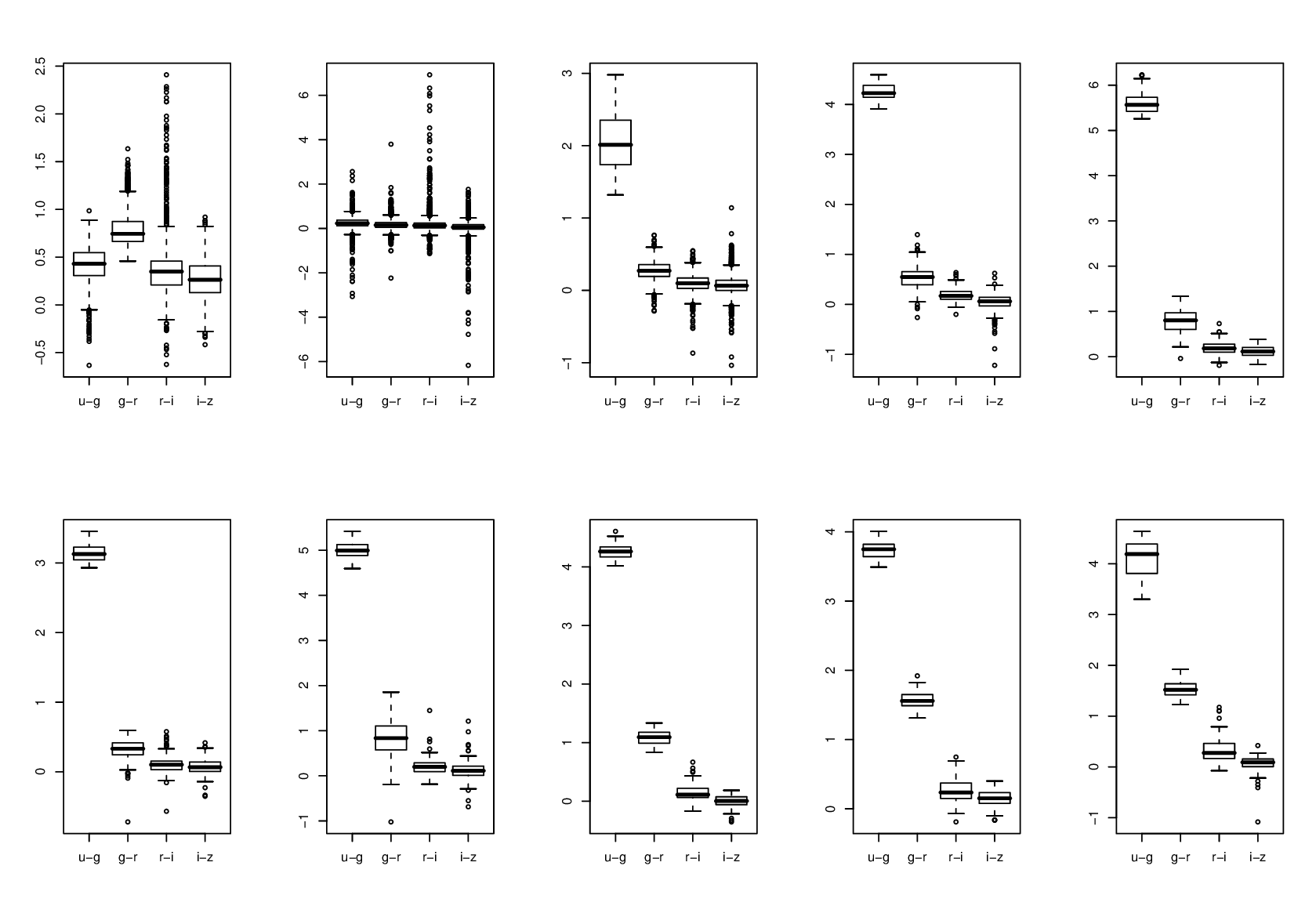}}
\caption{Same as in the previous figure for the fourth experiment.}
\label{exp_4}
\end{figure}

The fourth experiment was carried out as an application to the SDSS 
candidate quasars data set of the algorithm described in this paper. 
Also in this case the number of latent variables for the PPS algorithm 
was fixed to 62. Results are shown in figure (\ref{exp_4}). 

\begin{itemize}
\item The first cluster is composed mainly by normal QSOs situated 
at low redshift, having $u-g$, $r-i$ and $i-z$ colours with means 
ranging form 0.0 to 0.5, and the $g-r$ colour with an average value 
slightly higher and centred on $\sim 1.0$.
\item The distributions of colours of the members of the second cluster 
all peak at about 0.0 with outliers reaching higher values especially 
in $r-i$ and $i-z$ colours. 
The distribution in redshift of the sources of this cluster, formed 
exclusively by "specClass" = 3 normal QSOs, spans the whole interval 
from $\sim 0.0$ to 2.2.     
\item The remaining "goal-successful" clusters selected in this 
experiment show as common feature distributions of $u-g$ with higher 
means than previous clusters, ranging from $\sim 2$ to $\sim 5.5$. 
The $g-r$ colours are instead distributed between 1.8 and $\sim 0.2$ 
and the others colours have similar mean values around 0. 
All these clusters are formed by a mixture of normal and far QSOs 
according to the SDSS spectroscopic classification, with redshifts 
going from 3 to 4.2. 
\end{itemize}

\begin{figure}
\centering
\subfigure[Efficiency in $z$ vs $u-g$ plane ("goal-successful")]{\includegraphics[width=8cm]{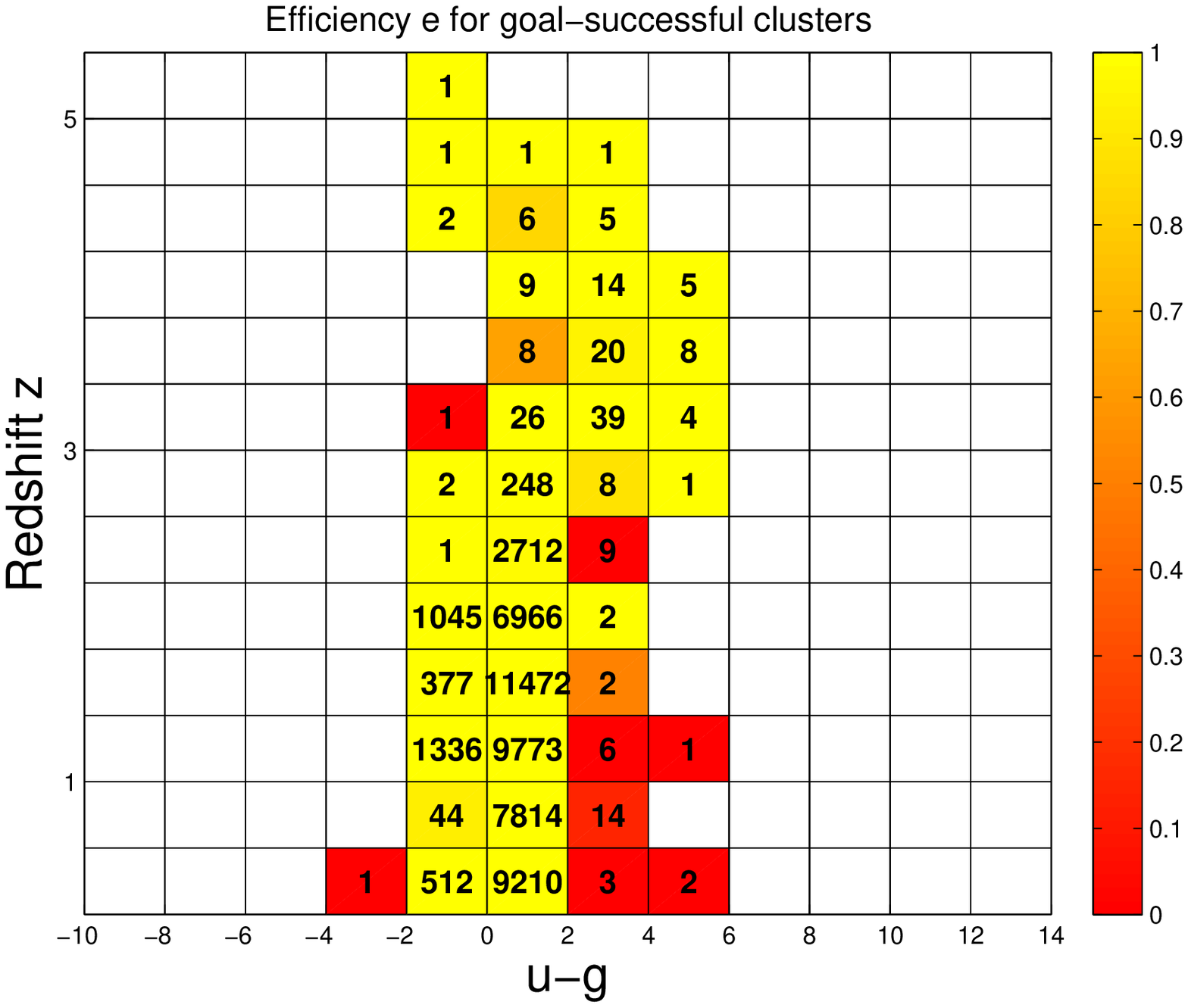}}
\subfigure[Completeness in $z$ vs $u-g$ plane ("goal-successful")]{\includegraphics[width=8cm]{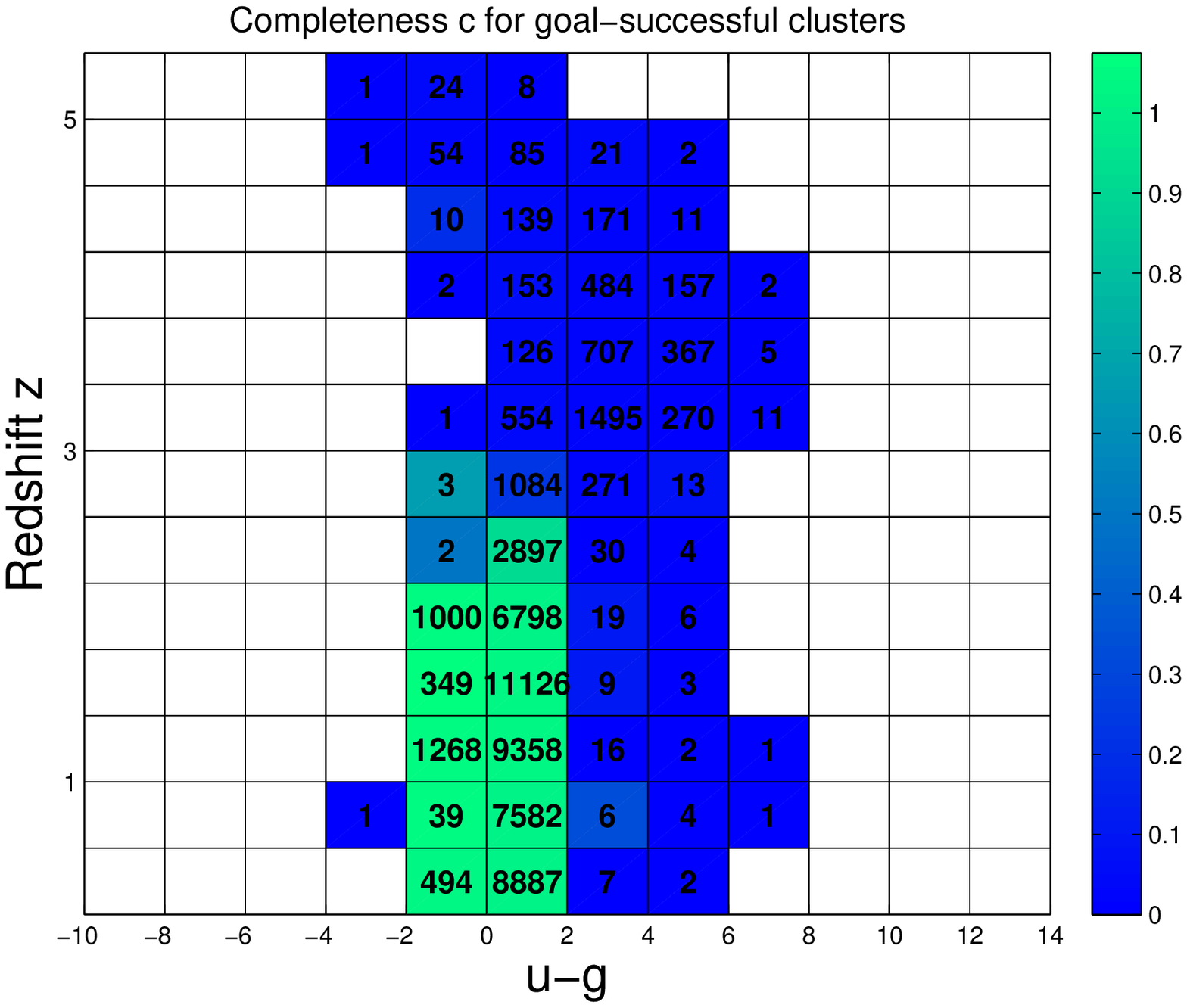}}\\
\subfigure[Efficiency in $z$ vs $u-g$ plane (all clusters)]{\includegraphics[width=8cm]{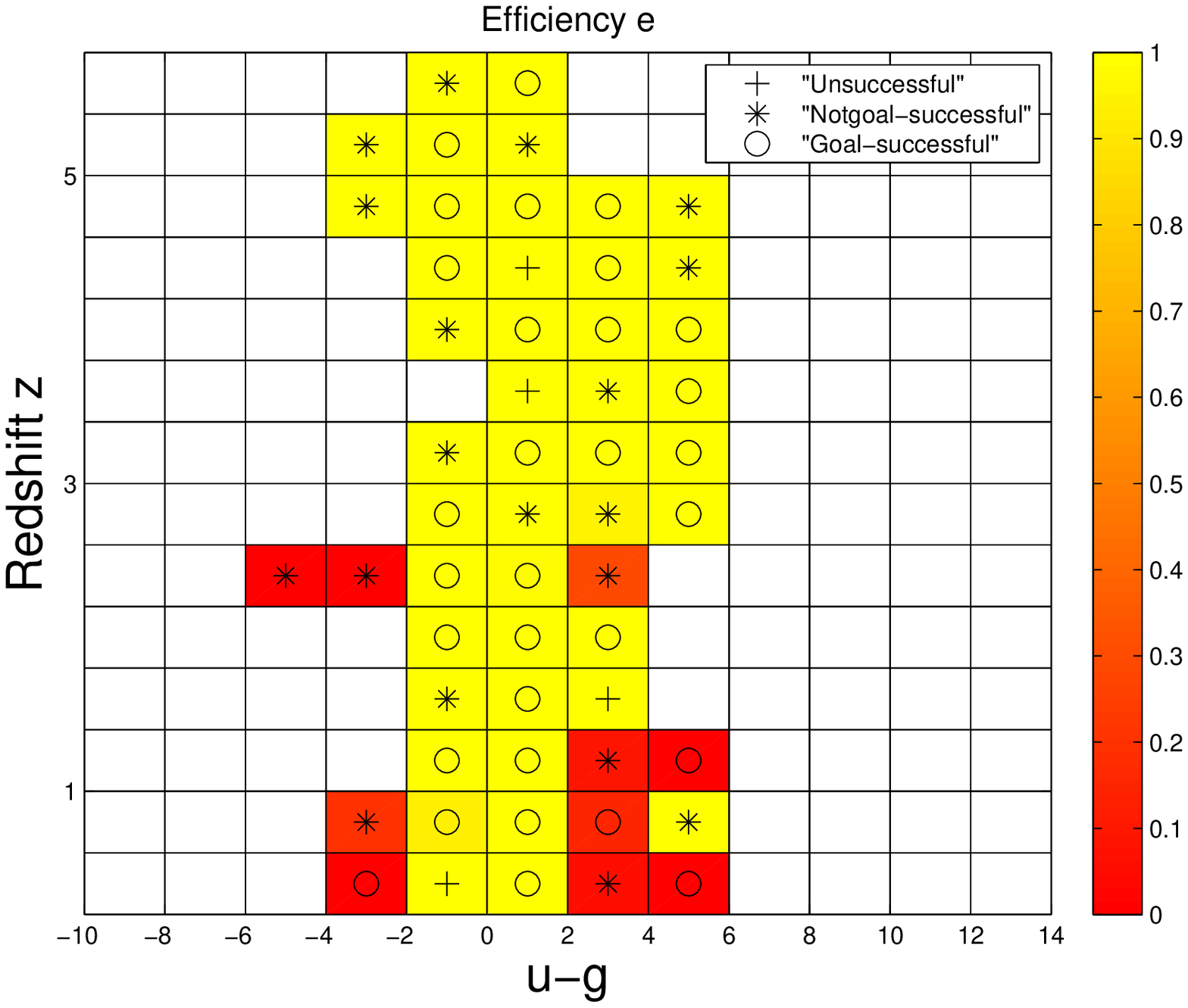}}
\subfigure[Completeness in $z$ vs $u-g$ plane (all clusters)]{\includegraphics[width=8cm]{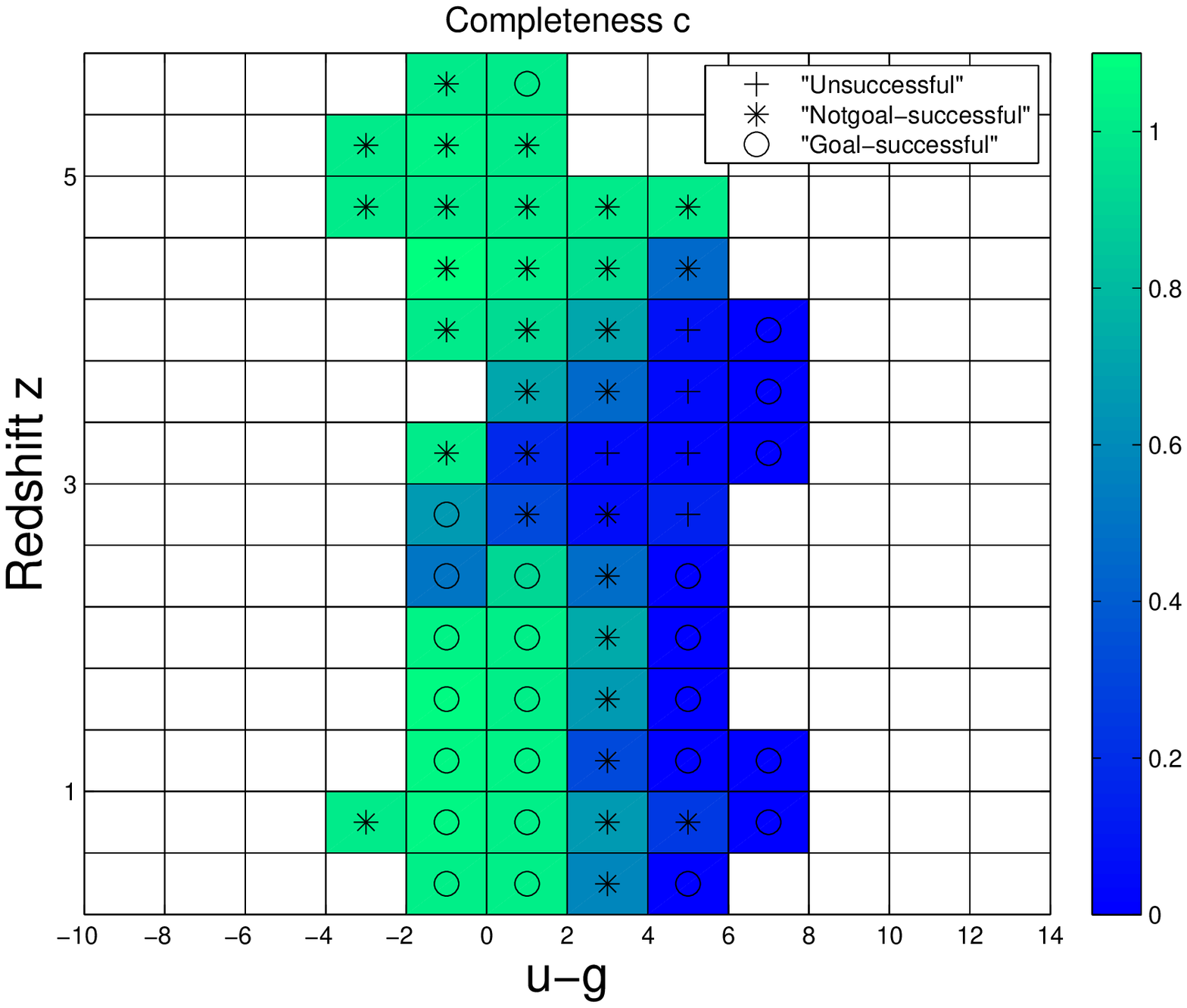}}\\
\caption{Efficiency (upper left-hand panel) and completeness (upper right-hand panel) in the redshift vs ($u-g$) plane for the labelling phase of the fourth experiment based on the S-S data set. Colour of the cell is associated to the efficiency and completeness of the selection process of the "goal-successful" clusters, while numbers contained in each cell represent the total number of candidates in that region of the $z$ vs $u-g$ plane. Lower left-hand and right hand panels contain the maximum efficiency and completeness for the selection process of the three types of clusters produced by the algorithm. The class of clusters contributing the maximum fraction to the efficiency and completeness represented by the symbol contained in each cell of the plane.}
\label{fig:paramparam_four_experiment}
\end{figure}

\begin{figure}
\centering
\includegraphics[width=12cm]{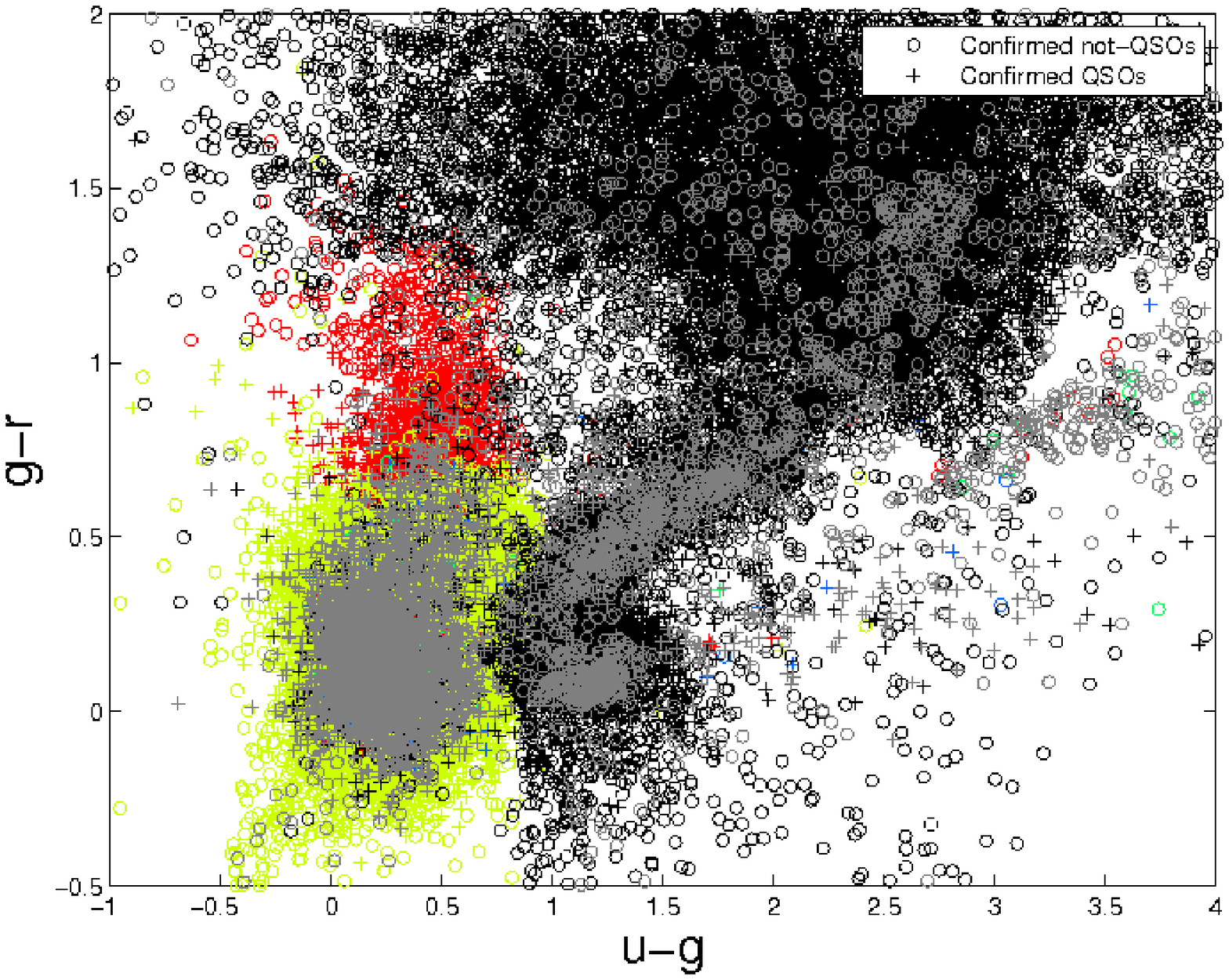}
\caption{Distribution of S-S sample points in the ($u-g$) vs ($g-r$) plane after the labelling 
phase of the fourth experiment. Black and grey symbols are associated respectively to members of
the "notgoal-successful" and "unsuccessful" clusters, while each single "goal-successful" cluster
is drawn using a different colour. All confirmed QSOs and not QSOs, regardless of their 
membership, are represented by crosses and circles respectively.}
\label{ugvsgr_fourth}
\end{figure}

The distribution of "goal-successful" candidates in the ceentral region of the optical colour-colour ($u-g$) vs ($g-r$) 
plane for the fourth experiment is shown in figure (\ref{ugvsgr_fourth}), while the local values of efficiency and completeness 
for the labelling phase of the same experiment are shown in figure (\ref{fig:paramparam_four_experiment}).

\subsection{Comparison of the results}

The estimates of the efficiency and completeness of the experiments described in the previous paragraphs show that the method
successfully recover a very high fraction of the spectroscopic QSOs when the BoK is based on the spectroscopic classification
provided by the SDSS. The differences in the performances of the selection process are due to the different  criteria used to select 
the samples of sources used for the training of the method in the different experiments.  
The results obtained in the four experiments can be summarized as it follows:

\begin{itemize}
\item First experiment (optical BoK and parameter space): the S-A sample composed by unresolved objects belonging 
to the "Star" table of the SDSS database and placed in the overlapping regions between SDSS-DR1 
and UKIDSS-DR1 LAS surveys, together with a BoK represented by the "specClass" spectroscopic 
classification index are used. Optical colours are employed as parameters. The highest performance 
is reached with a total efficiency $e_{tot} =  81.5\%$ and a completeness $c_{tot} = 89.3\% $, within 
$n_{gen} = 2 $ generations of clustering.
\item Second experiment (optical + near infrared BoK and optical parameter space):  the S-UK sample obtained selecting only the 
sources observed in both SDSS and UKIDSS-DR1 LAS surveys and classified as stars in both surveys, 
together with a BoK represented by the SDSS "specClass" spectroscopic classification index are used. 
Optical colours are employed as parameters. The best performance is reached with a total efficiency 
$e_{tot} = 92.3\% $ and a completeness $c_{tot} = 91.4\%$, within $n_{gen} = 1$ generation of clustering.
This experiment shows a significant improvement of the total efficiency and a slight improvement of the 
total completeness over the previous experiment.
\item Third experiment (optical + near infrared BoK and parameter space):  the same sample and BoK of the previous experiment 
are used. Optical and near infrared colours are employed as parameters. The best performance is reached 
with a total efficiency $e_{tot} = 97.2\% $ and a completeness $c_{tot} = 94.3\% $, within $n_{gen} = 1$ 
generation of clustering. The addition of  infrared photometric informations improves both efficiency 
and completeness of the candidate quasars selection process. 
\item Fourth experiment (optical BoK and parameter space): the S-S sample composed by all the candidate quasars  according 
to the native SDSS candidate QSOs selection algorithm in the whole DR5 database "Target" table, and 
the same BoK of the previous experiments are used. Only optical colours are employed as parameters. 
The best performance is reached with a total efficiency $e_{tot} = 95.4\%   $ and a completeness 
$c_{tot} = 94.7\%$, within $n_{gen} = 3 $ generations of clustering.
\end{itemize}

The most significant experiments, in terms of selection performances, appear to be the second and third ones, which 
take into account also part of the so-called ''optically dull''  quasars, i. e. those objects which show very little signature 
in the optical bands and are usually selected as QSOs by means of additional information of spectroscopic nature or, 
like in this case, of photometric nature (near infrared colours). The increase in efficiency between the first and second 
experiments, even if the parameter space has not been modified, is due to the fact that the BoK used for the second 
experiment is formed by sources that have been observed in both optical and near infrared bands and in both bands 
have been confirmed as QSOs. The different BoK represents a more specific family of QSOs (emitting strongly in NIR) 
whose location in the colour space is more easily traceable than the whole class of QSOs observed in the optical colours. 
The only slight increase in efficiency observed in the third experiment with respect to the second, caused by to the addition 
of near infrared colours to the photometric parameter (which allows to resolve more efficiently part of the degeneracies 
in the colour distribution between stars and QSOs), confirms that this method is much more sensitive to the features of the 
BoK that to the number of dimensions of the parameter space where the unsupervised clustering is carried out. The fourth 
experiment shows that this method can be applied to a large BoK composed by stellar sources which have received 
spectroscopic observation to perform the selection of photometric candidate QSOs. The same kind of experiment, even 
if with a different BoK, has been performed to extract the original catalogue of QSOs candidates described in section (\ref{Sec:catalogues}).

\section{The selection of the parameters}
\label{dependance_parameters}

In this section we shortly discuss how the performances of our candidate quasars
selection method depend upon the assumed parameters of the PPS and NEC algorithms 
respectively,  and on the set of features used for the characterization of the distribution of
objects inside the parameter space. More specifically, we have focused our attention  
on the dependance of the performance of the PPS algorithm on the number of latent variables (i.e. the 
number of preclusters produced by PPS), the dependance of the NEC algorithm on 
the critical value of the dissimilarity threshold $T_{cr}$, and the dependance
of the overall method from the particular set of colours used to define the parameter 
space inside which the clustering is performed.

\subsection{Dependence on PPS parameter}

As it has been discussed elsewhere, the PPS performances are rather independent on 
the choice of the parameters (clamping factor, width and orientation of principal surfaces, 
tolerance and number of iterations), with the exception of the number of 
latent variables. This number needs to be neither too large nor too small (within rather large 
boundaries)  An excessive number of latent variables produces clusters that are not agglomerated 
efficiently by the NEC algorithm, while setting the number of latent variable to a low number 
does not allow a proper separation of different groups of objects in distinct clusters. In absence 
of theoretical ways to estimate the correct number of latent variables, the only way is to guess 
it through a trial and error procedure. It needs to be stressed that even though rather demanding 
in terms of computing time, for a given problem and data set, this procedure needs to be run only once. 

\subsection{Dependence on colours choice}

In order to detect any bias introduced by the choice of a particular 
colours set in the results, all experiments have 
been repeated using colours derived by different combination of magnitudes 
(table (\ref{table_colours})) and keeping unchanged all the other parameters. 
These tests showed that the total efficiency $e_{tot}$ and completeness 
$c_{tot}$ are robust with respect to the particular parameter 
space where the distribution of objects is studied. Fluctuations affecting both
$e_{tot}$  and $c_{tot}$ for all possible sets of parameters are, in the worst cases, 
comparable with few percents of the optimal values obtained using the natural 
combination of colours.  This result can be understood from a theoretical 
point of view remembering that, given an initial set of colours $C_{0}$ derived from a certain 
photometric system, all other possible colours sets $C_{i}$ can be expressed 
as linear combinations of $C_{0}$, so that each parameter 
space $\Sigma_{i}$ generated by $C_{i}$ is the result of the application 
of a rigid rotation to the the parameter space $\Sigma_{0}$ associated to $C_{0}$. 
The transformation applied to the parameter space does not affect the relative positions and 
distances between points of the distribution and, as a consequence, the principal 
curves or surfaces generated by PPS algorithm in order to determine the best 
projection from the generic parameter space $\Sigma_{i}$ to the latent space 
remain unchanged.    

\begin{table}
\label{table_colours}     
\centering                 
\begin{tabular}{c c c c}        
\hline\hline                
Experiment &Colours & $e_{tot}$ & $c_{tot}$ \\    
\hline                      
   1 & natural  & 81.5\% & 89.3\% \\      
   1 &  ($u-r$, $g-i$, $r-z$, $i-u$)  & 81.7\%  & 89.5\% \\
   1 &  ($u-i$, $g-z$, $r-g$, $i-r$) &  80.8\%  & 89.3\% \\
   1 &  ($u-z$, $g-u$, $r-z$, $i-r$) & 82.0\%  & 89.0\%  \\
   1 &  ($u-g$, $g-z$, $r-i$, $i-z$) &  81.4\% & 89.7\%    \\
   2 & natural &  92.3\% & 91.4\%  \\
   2 &  ($u-r$, $g-i$, $r-z$, $i-u$, $Y-H$, $J-K$, $H-Y$) & 92.3\% &  91.5\% \\
   2 &  ($u-i$, $g-z$, $r-g$, $i-r$, $Y-K$, $J-Y$, $H-J$) &   92.7\%  &  91.8\% \\
   2 &  ($u-z$, $g-u$, $r-z$, $i-r$, $Y-H$, $J-K$, $H-Y$) &  91.9\% & 90.9\% \\
   2 &  ($u-Y$, $g-H$, $r-J$, $i-K$, $z-u$, $Y-g$, $H-r$)&  91.0\%  & 91.0\% \\
   2 &  ($u-H$, $g-J$, $r-K$, $i-u$, $z-g$, $Y-z$, $H-i$)& 90.9\% & 91.2\% \\
   2 &  ($u-J$, $g-K$, $r-u$, $i-g$, $z-r$, $Y-i$, $H-z$)&  92.2\%  & 91.5\% \\
   2 &  ($u-z$, $g-K$, $H-J$, $z-Y$, $r-u$, $z-i$, $i-H$)&  92.6\%  &  91.4\%\\
   3 &  natural & 97.3\% & 94.3 \%  \\
   3 & ($u-r$, $g-i$, $r-z$, $i-u$)  & 97.1\% &  94.8 \% \\
   3 & ($u-i$, $g-z$, $r-g$, $i-r$)  &  97.0\% &  93.9 \% \\
   3 & ($u-z$, $g-u$, $r-g$, $i-r$)  &   97.3\% &  94.0 \% \\		 
   3 &  ($u-g$, $g-z$, $r-i$, $i-z$) &   96.9\% &  94.9\% \\
   4 & natural &  94.5\%   & 93.0\% \\
   4 & ($u-r$, $g-i$, $r-z$, $i-u$)  & 95.2\%     &  93.9\% \\
   4 & ($u-i$, $g-z$, $r-g$, $i-r$)  & 95.0\%    &  94.0\%\\  
   4 & ($u-z$, $g-u$, $r-g$, $i-r$)  & 95.4\%    &  94.4\%\\
   4 &  ($u-g$, $g-z$, $r-i$, $i-z$) & 95.7\%    & 94.6\% \\
\hline                                   
\end{tabular}
\caption{Efficiency and completeness for the three experiments evaluated using sets of different colours. The natural 
combinations of colours corresponds, by definition, to ($u-g$, $g-r$, $r-i$, $i-z$) for the experiments making use of only 
optical colours, and ($u-g$, $g-r$, $r-i$, $i-z$, $Y-J$, $J-H$, $H-K$) for the experiment making use of both optical 
and infrared colours. Only a selection of all permutations is shown.}
\end{table}

\subsection{Dependence on clustering algorithm parameter}
\label{Sec:NECparameter}

The choice of the thresholds $\widetilde{sr}^{(g)}$ and $\widetilde{sr}^{(ng)}$ 
introduced in the definitions of "goal-succcessful"  and "notgoal-successful" 
clusters, respectively, only affects the total time required by algorithm in order 
to converge, i. e. the number of generations needed to select all possible 
candidate quasars in a given sample. The only requirement is that these 
thresholds need to be included in an interval of reasonable values (in this case, 
we have tested values of this parameter in the range [0.65, 0.90]).  Lower values 
of the thresholds would allow the algorithm to select many clusters 
with a very high level of contamination from "notgoal" objects ("goal" objects) 
at the cost of a reduction of the overall efficiency, while higher values would 
make practically impossible the selection of a major fraction of candidates, 
since only few clusters composed almost by only "goal" ("notgoal") 
sources would conform to the definition.

\subsection{Local estimates of efficiency and completeness}

The overall trends in the local estimates of the efficiency $e$ and completeness $c$ 
in the redshift vs colour planes for the various experiments described in the above paragraphs,
carry useful information on the performances of the algorithm and on the utility of the BoK. 
Even if the samples used for the experiments are slightly different, the main features of the 
efficiency and completeness plots can be traced back to the same properties of the general distribution 
of QSOs in the photometric parameter space.

\begin{enumerate}

\item In almost all plots showing the local values of efficiency calculated for 
the experiments examined in the present paper (for example, see upper-left panels 
of figures (\ref{fig:paramparam_first_experiment_ug}), (\ref{fig:paramparam_sec_experiment_ug}) and
(\ref{fig:paramparam_third_experiment})), few cells with lower values of $e$ are found 
for low values of the redshift. This effect does not seem to depend on the specific colour 
since a similar systematic is visible, for example, in the upper-left panels of figures (\ref{fig:paramparam_sec_experiment_ug}) and 
(\ref{fig:paramparam_sec_experiment_yj}),  where $e$ has been evaluated in both 
the z vs $(u-g)$ and z vs $(Y-J)$ planes respectively for the second experiment. Another interesting 
feature of these local estimates of the efficiency is the presence of two distinct clumps of cells
populated by few members of the BoK located at high redshift and approximately around $(u-g) \approx 2.2$
and  $(u-g) \approx 6.2$. These clumps can be immediately associated to two "goal-successful" clusters 
visible in the $u-g$ vs $g-r$ plot (see figure (\ref{ugvsgr_first})) around similar values of the $u-g$ colour. 
These two features of the local estimates of the efficiency, while clearly visible in the case of the first experiment (left-hand panel
of figure (\ref{fig:paramparam_first_experiment_ug})), do not show up in the same plots for 
the second and third experiments (left-hand panels of figures (\ref{fig:paramparam_sec_experiment_ug}) 
and (\ref{fig:paramparam_third_experiment})) where only the first and the second clump can be seen,
respectively. 
 
\item In almost all plots showing the local values of the completeness in the z vs $(u-g)$ plane
(see the upper right-hand panels of figures (\ref{fig:paramparam_first_experiment_ug}), (\ref{fig:paramparam_sec_experiment_ug}) and
(\ref{fig:paramparam_third_experiment})), a region with low values of $c$ is found for redshifts 
in the range $[2.4, 3.2]$ and $(u-g)$ colour from $\approx 1$ to $\approx 3$ for the first three experiments. 
These regions are more or less extended according to the experiment considered and to the widths of the 
bins used to determine the tessellation of the z vs $(u-g)$ plane. In terms of total completeness for each 
experiment, the number of confirmed QSOs lost in these underpopulated regions of the parameter space 
in nonetheless low, namely $\approx 5 \%$ of the total number of confirmed QSOs in the BOK for the first experiment, 
and $\approx 3\%$ and $\approx 7\%$ for the second and third experiments respectively.

\item All the efficiency and completeness plots show a lack of members of the BoK can be noticed in the 
upper left corners of the plots, roughly corresponding to a triangular region going from z $\approx 2$ to 
the upper limit of the redshift distribution, and on the other axis spanning approximately the interval 
$[-0.5, 1]$ for the $u-g$ colour and the interval $[-0.4, 0.4]$ along the $Y-J$ colour axis. This selection 
effect depends on the nature of the BoK (i.e. on the selection criteria for the spectroscopic
targets of the SDSS survey) used to perform the experiments (cf. \cite{stoughton_2002,richards_2002}) 
and is not caused in any regard by the selection method here discussed. 

\item The plots showing which type of clusters contributes most to the efficiency and completeness inside 
each cell of the redshift vs colour plane (lower panels of figures (\ref{fig:paramparam_first_experiment_ug}), 
(\ref{fig:paramparam_sec_experiment_ug}) and (\ref{fig:paramparam_third_experiment})) indicates that 
"goal-successful" clusters dominate the regions where the efficiency and completeness of the QSOs candidate
selection are almost unitary, as can be seen comparing these plots with the local efficiency plots and completeness 
(for examples, in the first experiment, compare the upper-left and lower-left side panels for the efficiency and the upper-right 
and lower-right panels for the completeness of figure (\ref{fig:paramparam_first_experiment_ug})). "Notgoal-successful" clusters, 
on the other hand, are mainly found in places of the redshift vs colour plane corresponding to the regions of the
parameter space occupied by the "stellar locus" which can be easily singled out in the colour-colour plots 
(figure (\ref{ugvsgr_first}), left-hand panel in figure (\ref{colourcolour_second}) and figure (\ref{ugvsgr_third})). "Unsuccessful"
clusters, i.e. clusters composed of similar fractions of confirmed QSOs and contaminant, in the first and third experiments, 
are usually dominant in the neighbourhood of the almost isolated and underpopulated regions of the plane 
where groups of few cells with unitary efficiency are located (high redshift and high value of the $(u-g)$ colour, 
see above for details). In these experiments few other cells in both efficiency and completeness plots are dominated by the contribution from unsuccessful clusters, while no cell of this type is determined for the second experiment.  

\end{enumerate}

It should be emphasised that all the systematic effects and biases here discussed and affecting the selection of the QSOs, though not hindering this candidate selection method from achieving performances significantly better than those produced by more classical algorithms found in the literature, are merely reflections and consequences of the same systematics and biases affecting the spectroscopic data set used to constitute the base of knowledge for the labelling phase of our method. We can therefore conclude that the improvement of this method is subordinate to the enhancement of the coverage of the photometric parameter space sampled by the spectroscopic data and to the improvement of the accuracy of the estimates of both spectroscopic and photometric observables making up the parameter space and the labels to be used during the unsupervised clustering phase of the algorithm. 

\section{The catalogue of candidate QSOs}
\label{Sec:catalogues}

The application of our algorithm to a base of knowledge composed of all stellar sources with spectroscopic observations (see (\ref{Subsec:baseknowledgecat}) for details) extracted from the SDSS DR7 Legacy has produced the clustering in the photometric parameter space used to select the QSOs candidates from a suitable catalogue of sources extracted from the SDSS DR7 photometric dataset. The next sections describe in detail the nature of the spectroscopic BoK used to determine the partition of the parameter space and of the photometric samples of objects from which the candidate QSOs have been extracted (\ref{Subsec:baseknowledgecat}), the process of the extraction of the catalogue and its characterization (\ref{Subsec:extractcat}), the expected completeness and efficiency of the catalogue (\ref{Subsec:efficcomplet}), the comparison with one the catalogues of photometric QSO candidates available in the literature (\ref{Subsec:comparison}) and a first astronomical application of the catalogue of candidates, namely the number counts derived from our candidate QSOs (\ref{Subsec:numbercounts}). 

\subsection{The base of knowledge}
\label{Subsec:baseknowledgecat}

\begin{figure}
\centering
\includegraphics[width=12cm]{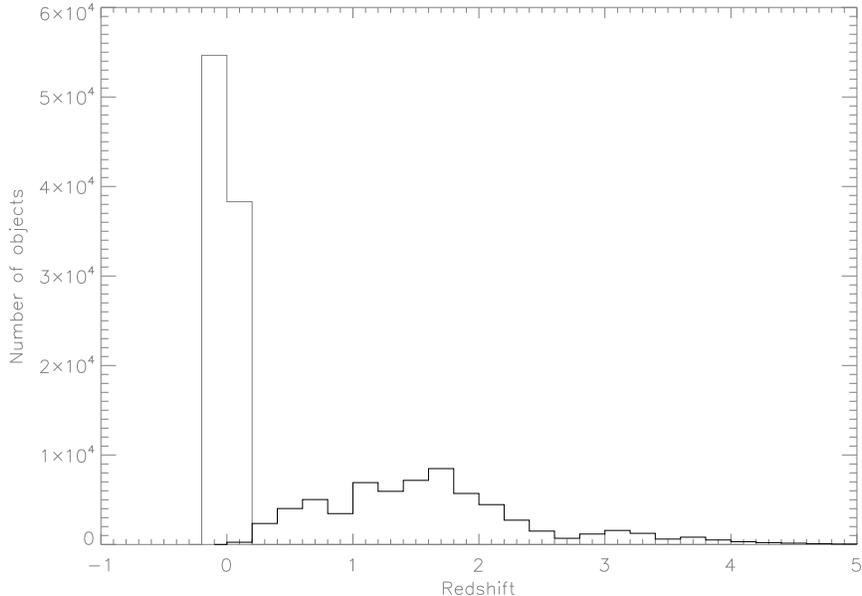}
\caption{Redshift distribution of the S-SCat sample. The grey line represents the distribution of stars and unknown sources according 
to the "specClass" ("specClass" $= \{1, 6, 0 \}$) classification while the black line is associated to QSOs ("specClass" $= \{3, 4 \}$).}
\label{S-SCat:redshifts}
\end{figure}

\begin{figure}
\centering
\includegraphics[width=12cm]{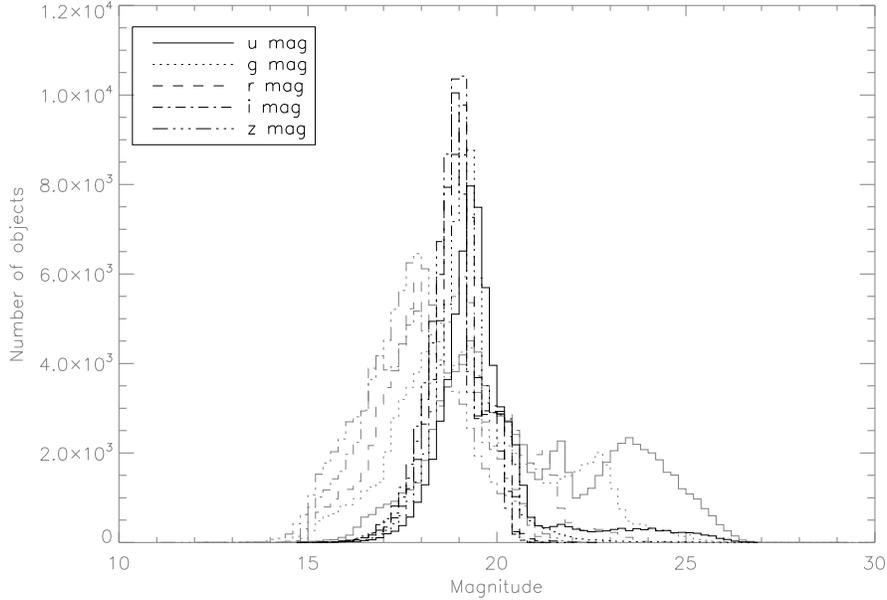}
\caption{Distribution in luminosity of the sources belonging to the S-SCat sample. The grey lines represent the distribution
of stars and unknown sources according to the "specClass" classification ("specClass" $= \{1, 6, 0 \}$), while the black lines are
associated to QSOs ("specClass" $= \{3, 4 \}$).}
\label{S-SCat:magnitudes}
\end{figure}

The dataset (hereafter S-SCat) used as BoK for the extraction of the QSO candidates from the SDSS photometric database is composed by all the primary ("mode" = 1) stellar sources belonging to the PhotoObjAll table of the DR7 data release (i.e. identified as point-like sources according to a purely photometric criterion, i.e. $"type" = 6$) with standard clean photometry (for all dereddened psf magnitudes) in all SDSS bands, reliable spectroscopic classification (with "specClass" = 1, 3, 4, 6, 0) and reliable spectroscopic redshift estimate ("zConf"$ > 0.95$). In order to avoid the inclusion in S-SCat of faint or saturated sources, following \cite{richards_2008}, a further constraint on the $i$ magnitude has been applied, requiring that $psfMag_i > 14.5 $ and $(psfMag_i - extinction_i) < 21.3$. The S-SCat dataset, regardless of the status of each member as QSO candidates according to the native SDSS candidate selection algorithm, has been used to reconstruct the distribution of spectroscopically confirmed QSOs and stars inside the SDSS colours space by the identification of "goal-successful" and "notgoal-successful" clusters using the unsupervised clustering method discussed in this paper. The total number of sources belonging to the S-SCat dataset and the fractions of different type of sources according to the spectral classification expressed by the "specClass" index are shown in table (\ref{table:2}).  The distribution of the spectroscopic redshift of the S-SCat and the distribution in magnitude are shown in figures (\ref{S-SCat:redshifts}) and (\ref{S-SCat:magnitudes}) respectively. 

\subsection{S-SCat experiment: efficiency and completeness}
\label{Subsec:efficcomplet}

\begin{table*}
\caption{Main features of the "goal-successful" clusters determined during the S-SCat experiment with the position of the centres of mass ("CoM") of the clusters in the colours space.}             
\label{table:cat3}      
\centering          
\begin{tabular}{l c c c c c c c c c c}    
\hline\hline       
Cluster Index & 1 & 2 & 3 & 4 & 5 & 6 & 7 & 8 & 9 & 10\\ 
\hline
Nm elements   & 1326 & 54368 & 3429  & 835 & 741 & 273  & 622 & 486  & 2491 & 595  \\  
Nm confirmed  & 1135 & 50012 & 3250 & 592 & 636 & 206 &546 & 380 &2057 &550 \\
Nm not-confirmed  & 191 & 4345& 179 & 242 & 105 & 67 & 75 & 106  & 434 & 45\\ 
confirmed/total & 0.86 &0.92 & 0.95  & 0.71 & 0.86 & 0.75 & 0.88 & 0.78  & 0.83 & 0.92 \\
not-confirmed/total & 0.14 & 0.08 & 0.05  & 0.29 & 0.14 & 0.25 & 0.12 & 0.22 & 0.17 & 0.08 \\
CoM $u-g$ position & 0.40 & 0.21 &  0.53 & 4.33 & 2.32 & 0.63  & 0.52 & 5.34  & 0.53 & 1.98 \\
CoM $g-r$ position & 0.42 & 0.12 &  0.04 & 0.82 & 0.34 & 0.06 & 0.46 & 0.90  & 0.40 & 0.23 \\
CoM $r-i$ position &  0.41 & 0.12 & 0.03 & 0.17 & 0.10 & -0.15 & 0.15 & 0.20  & 0.13 & 0.08 \\
CoM $i-z$ position & 0.17 & 0.04 & 0.24  & 0.11 & 0.06 & 0.31 & 0.50 & 0.13  & 0.14 & 0.06 \\
\hline                  
\end{tabular}
\end{table*}

\begin{figure}
\centering
\includegraphics[width=8cm]{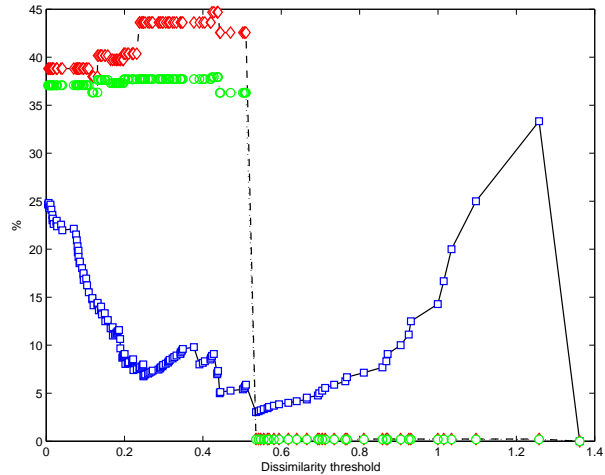} 
\caption{The fraction of 'goal-successful' clusters (squares), the total percentage 
of the sources of the BoK belonging to 'goal-successful' clusters (circles), the total 
percentage of objects belonging to 'goal-successful' clusters irrespective of their 
spectroscopic classification (diamonds) are plotted as a function of the dissimilarity 
threshold for the S-SCat experiment.}
\label{fracsucc_cat}
\end{figure}

\begin{figure}
\centering
\includegraphics[width=8cm]{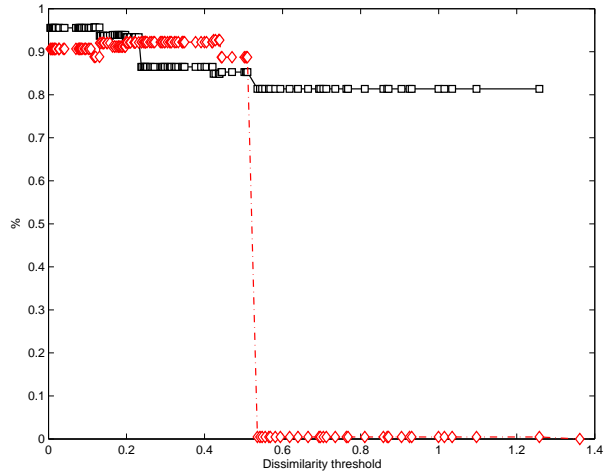}
\caption{Total estimated efficiency $e_{tot}$ (squares) and completeness $c_{tot}$ (diamonds) 
of the candidate QSOs selection process as functions of the dissimilarity threshold 
are plotted for the S-SCat experiment.}
\label{efficompl_cat}
\end{figure}

\begin{figure}
\centering
\includegraphics[width=11cm, angle=90]{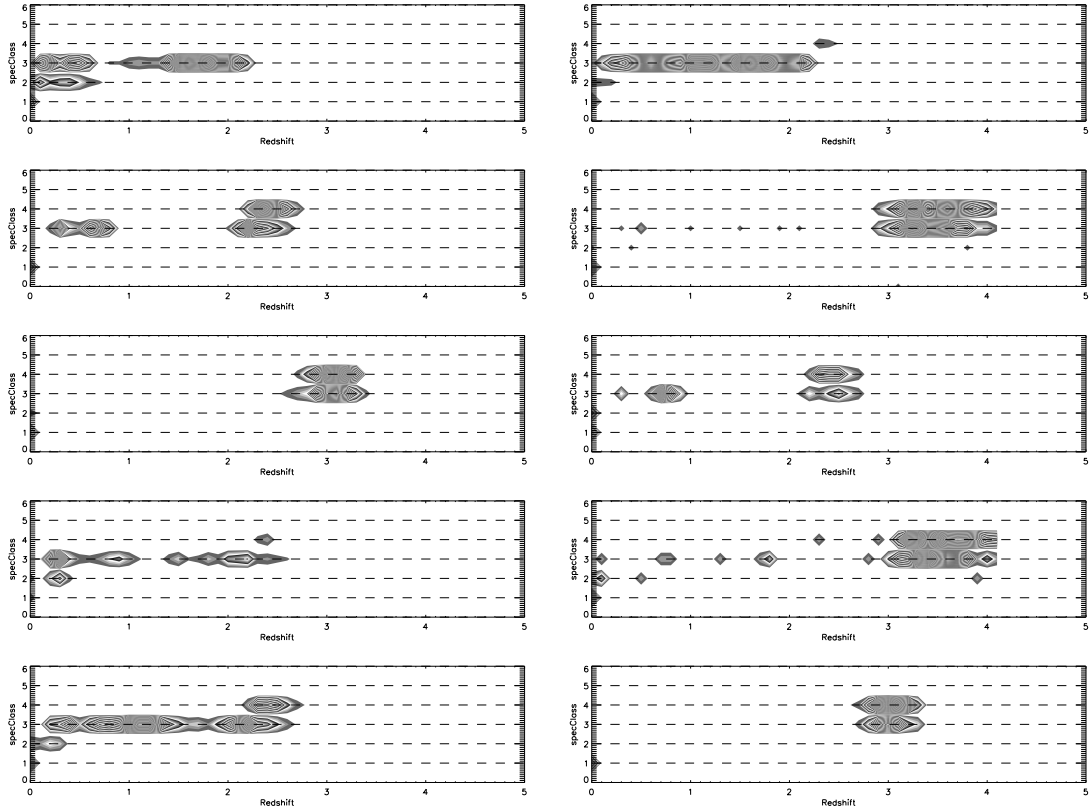} 
\caption{Distribution of members of the 10 "goal-successful" clusters determined during the S-SCat experiment in the 
"specClass"-redshift plane (cluster 1 to cluster 10 from left to right and from top to down).}
\label{zspec_cat}
\end{figure}

\begin{figure}
\centering
\includegraphics[width=12cm, angle=-90]{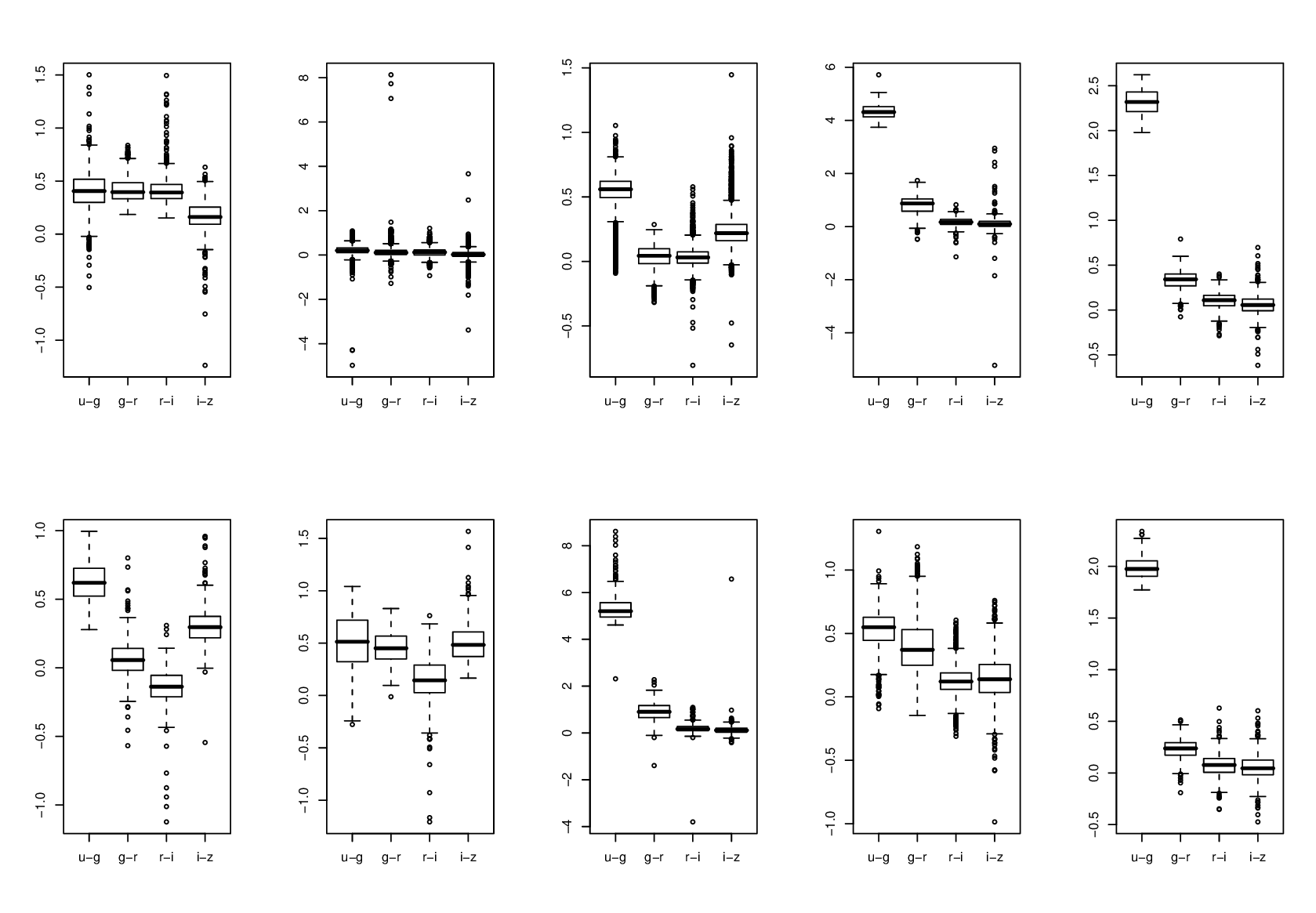} 
\caption{"Box and whisker" plots of the distribution of clusters produced in the experiment S-SCat (cluster 1 to cluster 10 from left to right and from top to down).}
\label{boxplots_cat}
\end{figure}

The number of latent variables (i.e. pre-clusters) produced by the PPS in the S-SCat experiment is 146 and the critical value of the dissimilarity threshold has been chosen according to the criteria explained in the previous paragraphs. The best clustering is produced for a value of the dissimilarity threshold $D_{th} = 0.20$ corresponding to a final number of "goal-successful" clusters equal to 10, whose main characteristics are summarized in the table (\ref{table:cat3}). The normalised success ratio and other statistical indicators of the clustering process as functions of the dissimilarity threshold $D_{th}$ are shown in figure (\ref{fracsucc_cat}), while the estimated efficiency and completeness of the selection process are shown in figure (\ref{efficompl_cat}) as functions of the dissimilarity threshold as well. The distribution of sources as a function of redshift and spectroscopic classification "specClass" in the "goal-successful" clusters produced in this experiment is shown in figure (\ref{zspec_cat}), while the main features of the distribution of sources associated to the "goal-successful" clusters of experiment S-SCat in the parameter space can be summerized and described through the "box and whisker" plots (see figure (\ref{boxplots_cat})).  

\noindent In what follows, a brief description of the characteristics of the distributions of sources associated to the 
10 "goal-successful" clusters selected in the S-SCat experiment, in terms of colours, redshift and classification is given:

\begin{itemize}
\item  Cluster 1: formed by sources having $u-g$, $g-r$ and $r-i$ colours centered at about 0.4 and 
$i-z$ with slightly smaller average ($\sim 0.2$). The members of this cluster are mainly QSOs and 
galaxies, according to the "specClass" classification, spanning in redshift from from 0 to 2.3. No 
"specClass" = 4 quasars (according to the SDSS classification) are found. A small fraction of contaminant galaxies
("specClass" = 2) is present at $z_{spec} < 0.6$
\item Cluster 2: composed by a very large fraction of "specClass" = 3 QSOs and few galaxies and "specClass" = 4
quasars, all located in the $[0, 2.5]$ redshift interval. The colours distribution of the members of this cluster is
characterized by very similar average for all four colours, approximately located at 0.2.   
\item Cluster 3: this cluster shows a widespread $u-g$ distribution centered around 0.5, with $g-r$ and $r-i$ colours
distributions peaked around 0 and $i-z$ with average $\sim 0.2$. The members of this cluster are almost totally
"specClass" = 3 and "specClass" = 4 QSOs located in two different regions in redshift, namely the intervals 
$[0, 1]$ and $[2, 2.8]$ respectively. Very few "specClass" = 1 stars can be found.       
\item Clusters 4 and 5: both these clusters are mostly populated by QSOs classified as "specClass" = 3 and 4 according to
the SDSS spectroscopic classification. The redshift distributions are different: both clusters have their lowest edge 
approximately at the $z_{spec}\sim 2.5$ but cluster 5 reaches redshifts slightly higher than 4 while cluster 4 is contained 
in the $[2.5, 3.3]$ interval. The colour distributions are similar, with $g-r$, $r-i$ and $i-z$ colours peaked between 0 and 0.6, 
while the $u-g$ colour has a remarkably high average in the fourth cluster (at $\sim 4$) while reaching $\sim 2.3$ for the cluster 5.
\item Cluster 6: this cluster shows $u-g$, $g-r$, $r-i$ and $i-z$ colours centered around 0.6, 0.0, -0.1 and 0.3 respectively, 
with a large dispersion for $g-r$ and $r-i$. It is mainly formed by "specClass" = 3 and 4 QSOs, with redshift distribution 
very similar to that of cluster 3. Few stars ("specClass" = 1) and galaxies ("specClass" = 2) can be found.
\item Cluster 7:  mainly formed by "specClass" = 3 QSOs spanning the redshift range from $\sim 0.1$ to 0.5, few "specClass" = 4 
quasars and galaxies are also present, located at $z_{spec}\sim 0.3$ and $z_{spec}\sim 2.2$ respectively. Colours 
distributions are centered between 0 and 0.5 with large dispersions around the averages. 
\item Cluster 8: the colours distribution is very similar to the distribution of cluster 4, with $u-g$ reaching even larger average 
($\sim 5.2$ instead of $\sim 4$.) Composition and redshift distribution are also similar to those of cluster 4, with a slightly 
higher fraction of contaminants represented by galaxies ("specClass" = 2) and stars ("specClass" = 1).     
\item Cluster 9: formed mainly by "specClass" = 3 and 4 QSOs with redshift to 2.6, there is a significant fraction of galaxies 
in the $z_{spec}$ interval from $\sim 0$ to 0.3. Colours distributions are peaked around 0.5, 0.4, and 0.2 for $u-g$ , $g-r$ 
and the remaining two colours respectively. 
\item Cluster 10: very similar to the cluster 4 and 5 in terms of colours distribution, it is almost exclusively composed by 
"specClass" = 3 and 4, with redshifts around 3 in the interval $[2.7, 3.3]$.
\end{itemize}

\begin{figure}
\centering
\subfigure[Efficiency in $z$ vs $u-g$ plane]{\includegraphics[width=8cm]{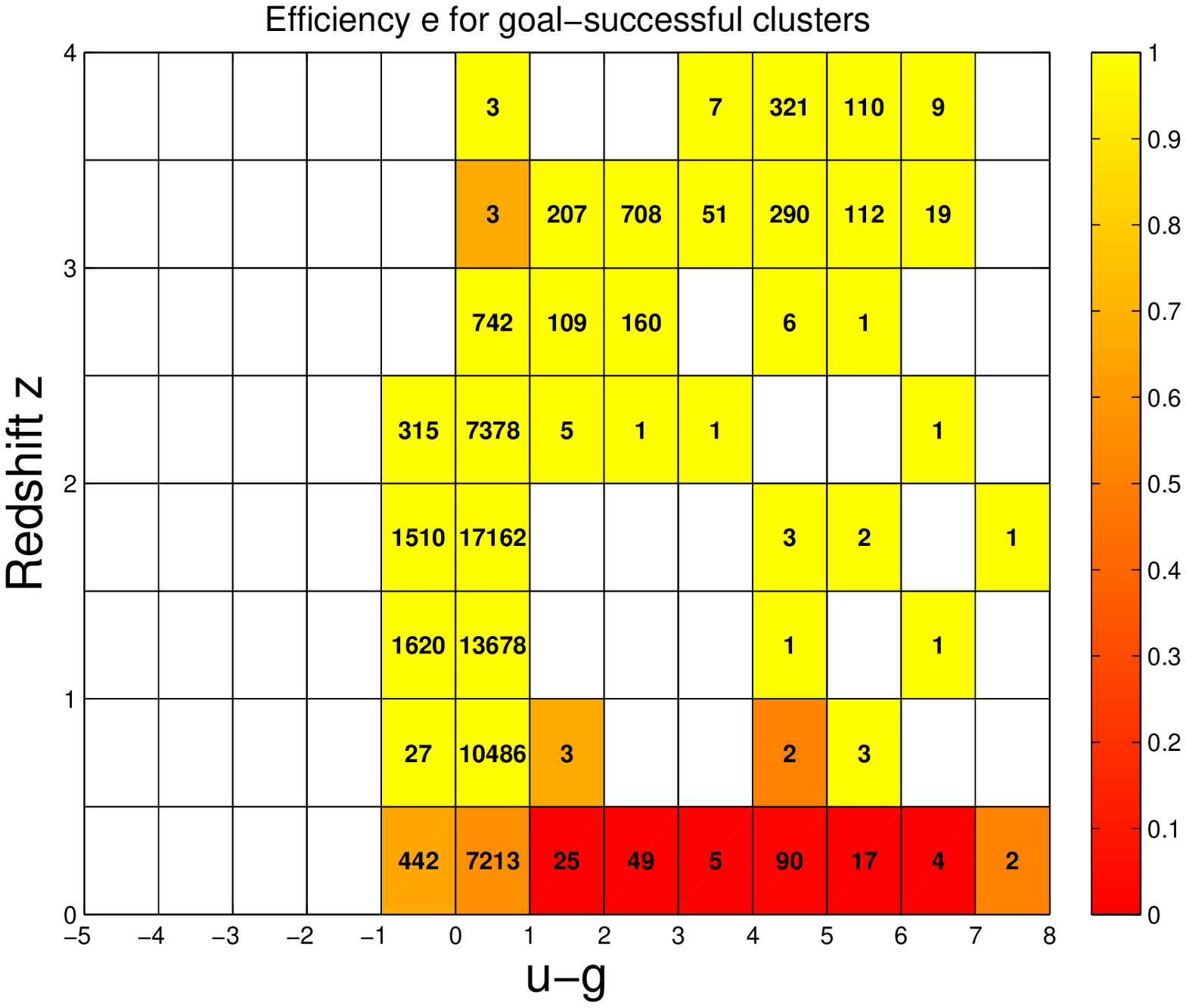}}
\subfigure[Efficiency in $z$ vs $g-r$ plane]{\includegraphics[width=8cm]{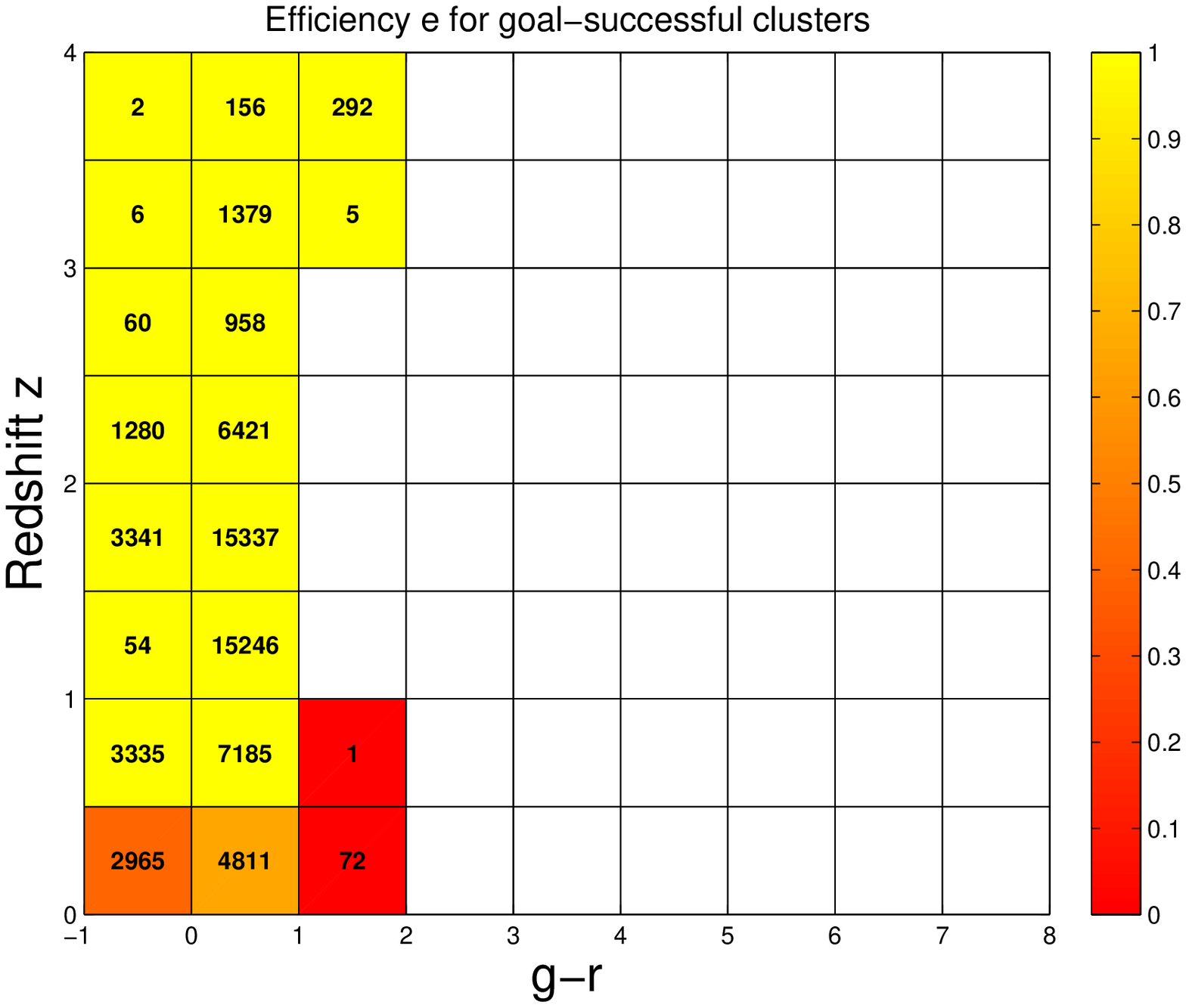}}\\
\subfigure[Efficiency in $z$ vs $r-i$ plane]{\includegraphics[width=8cm]{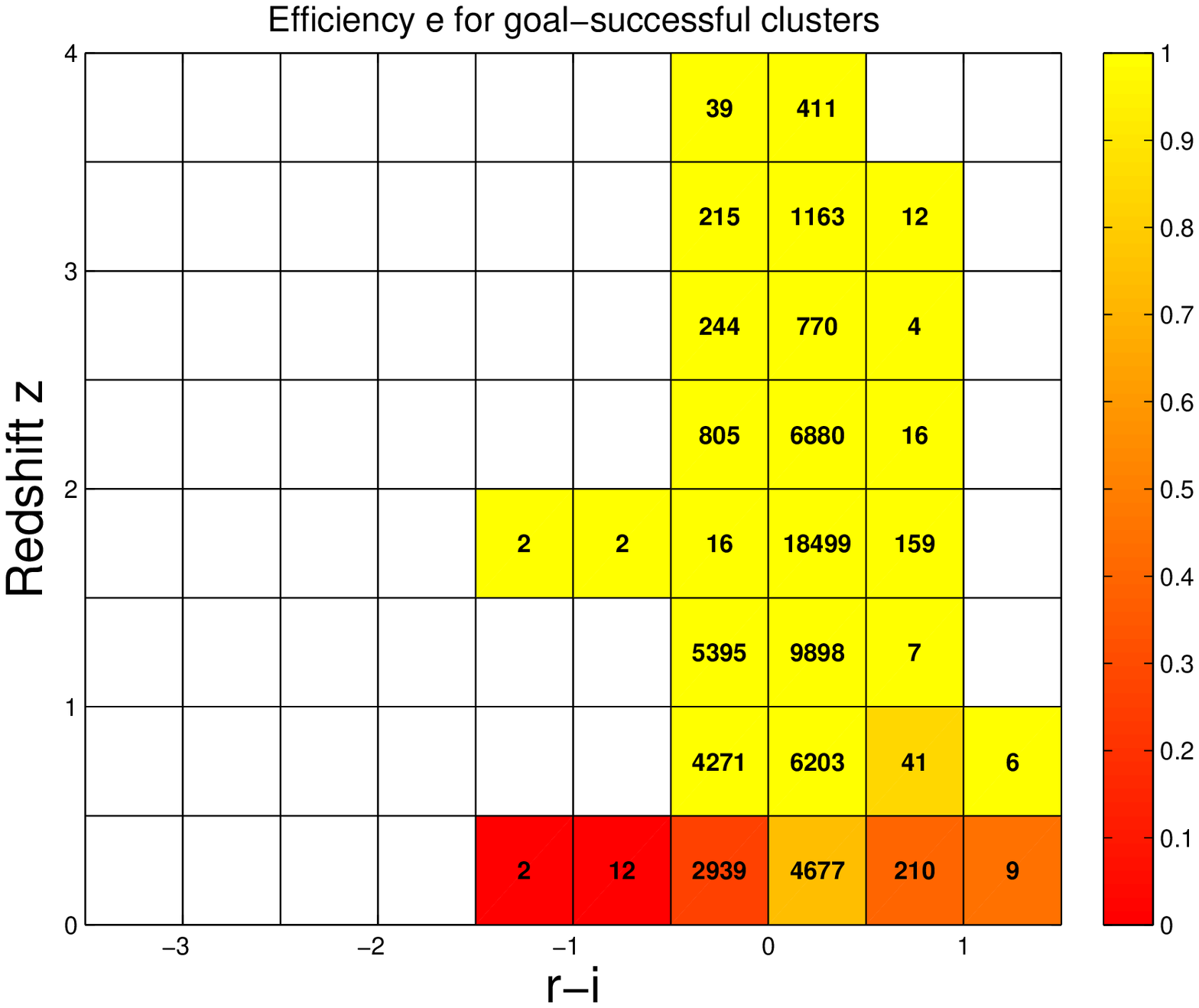}}
\subfigure[Efficiency in $z$ vs $i-z$ plane]{\includegraphics[width=8cm]{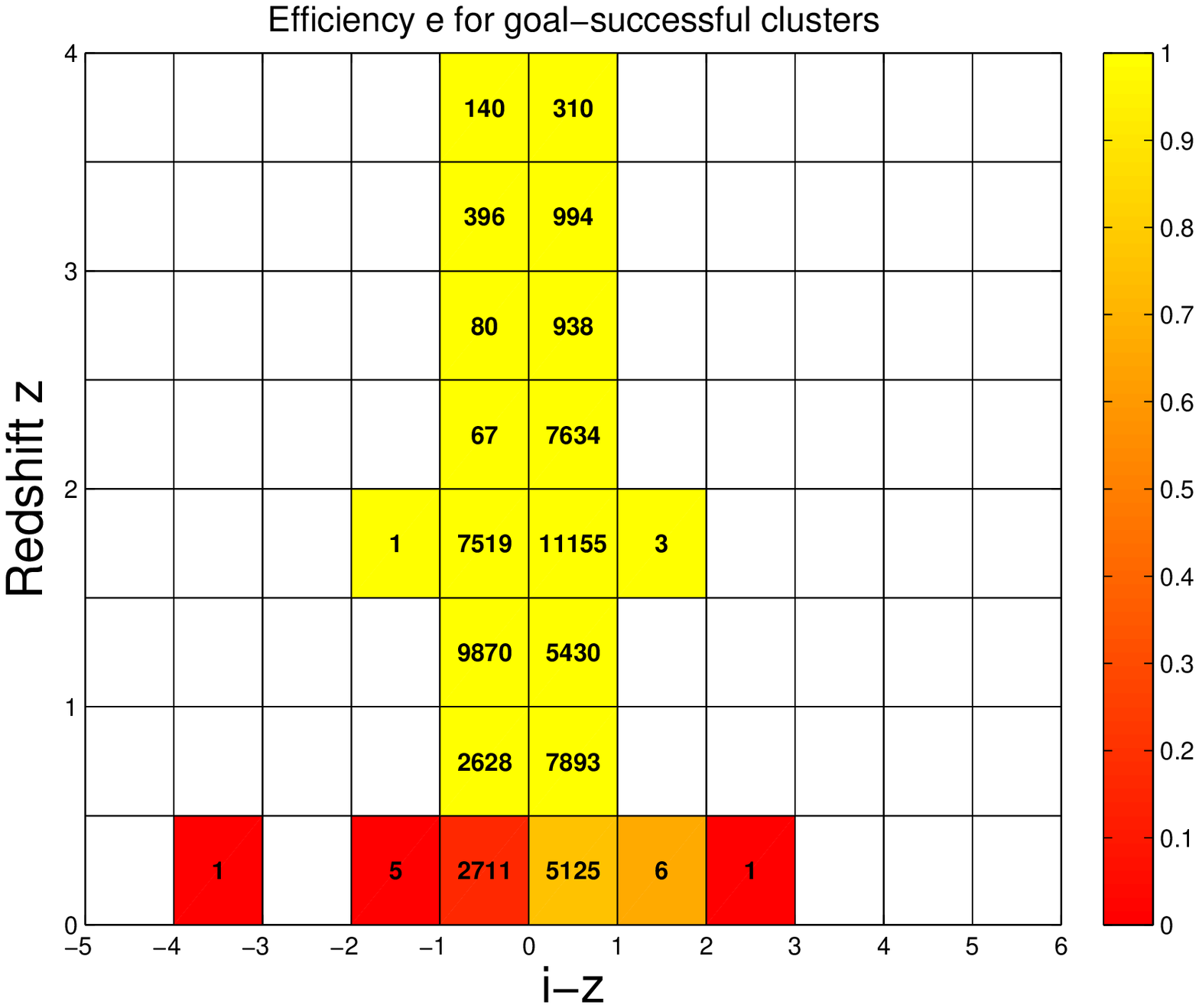}}\\
\caption{Local efficiency estimates in the redshift vs colour planes for the "goal-successful" clusters of experiment S-SCat. Numbers represent
the total number of sources of the BoK located in each cell.}
\label{zcolourseffic_cat}
\end{figure}

\begin{figure}
\centering
\subfigure[Completeness in $z$ vs $u-g$ plane]{\includegraphics[width=8cm]{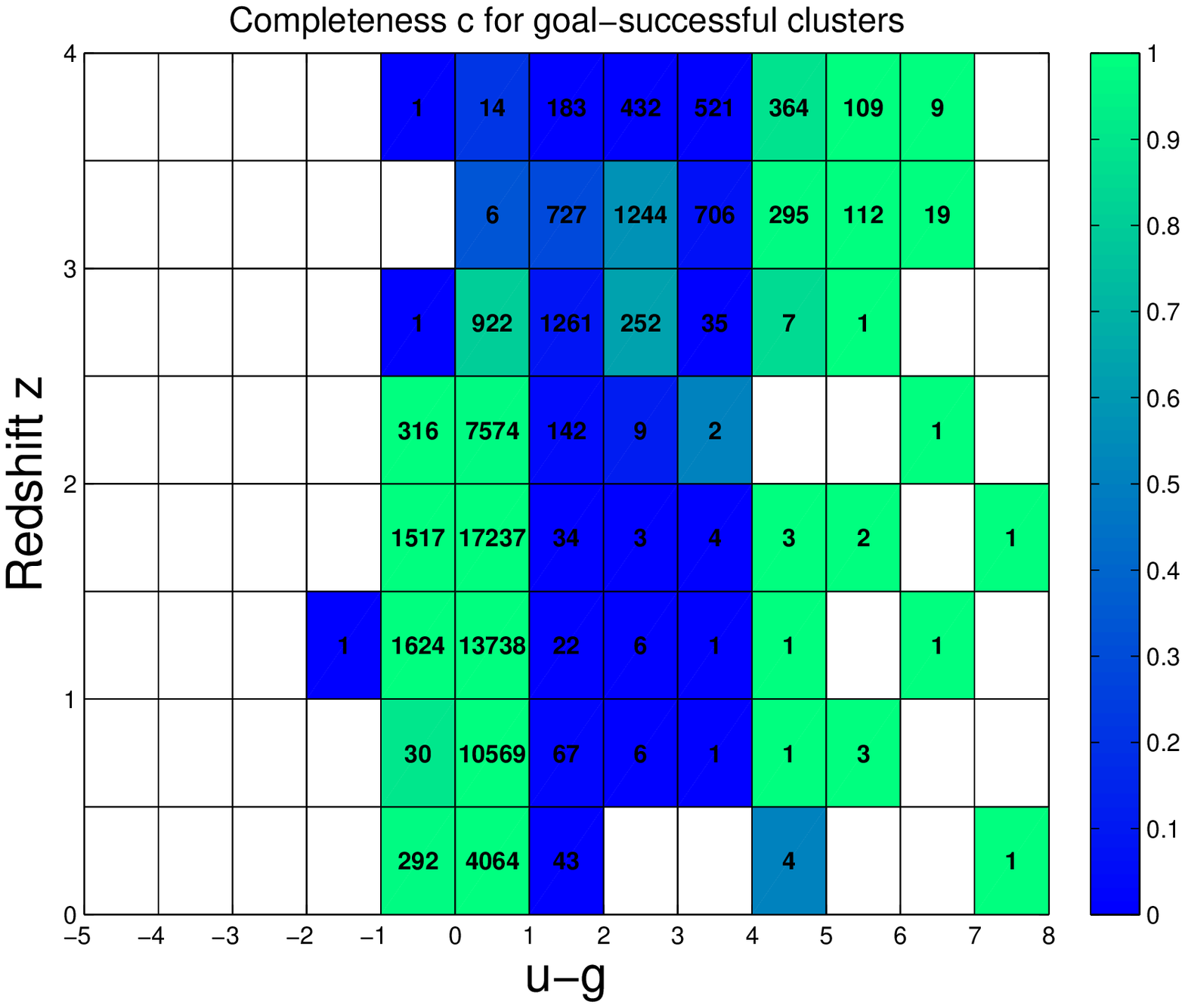}}
\subfigure[Completeness in $z$ vs $g-r$ plane]{\includegraphics[width=8cm]{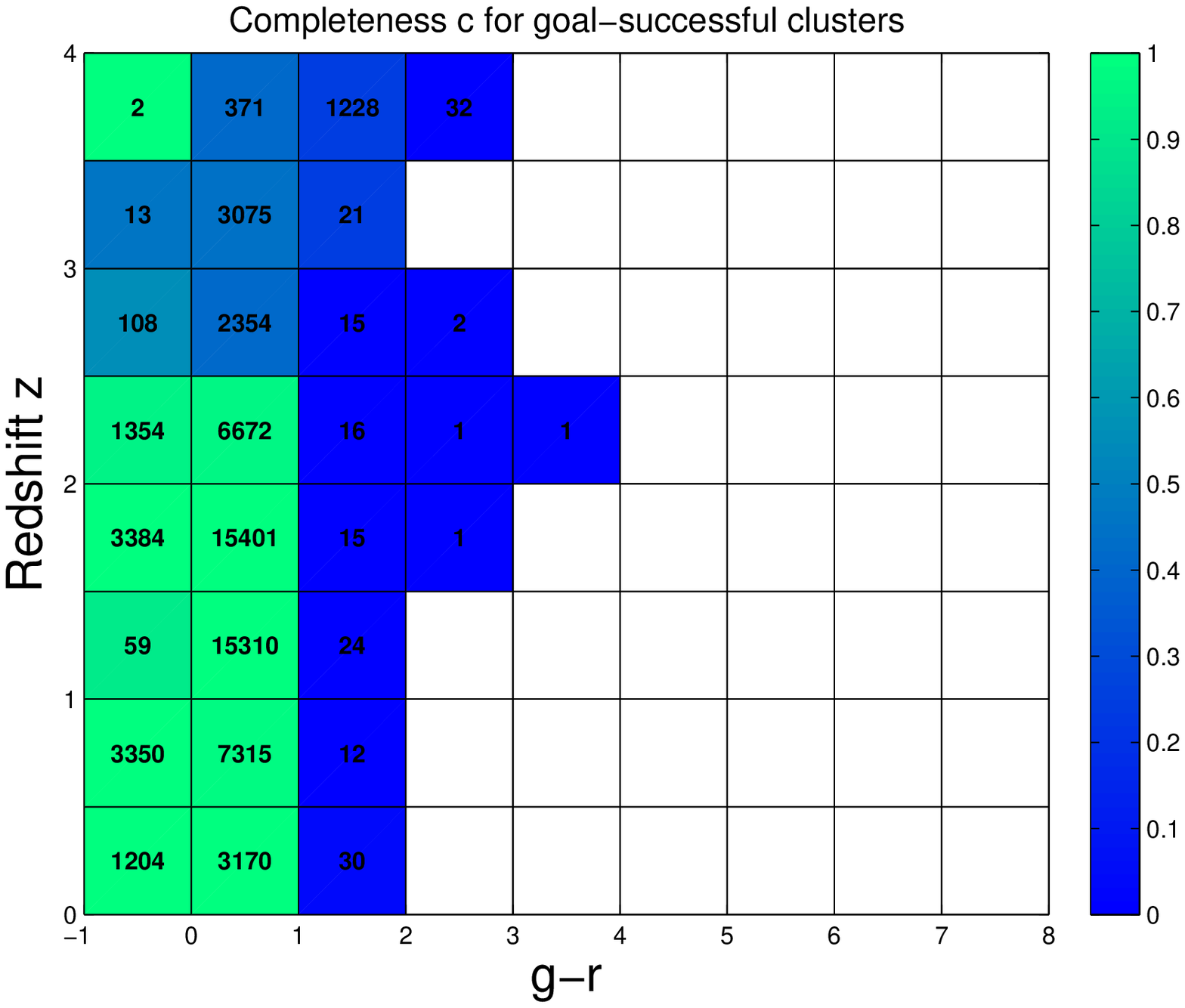}}\\
\subfigure[Completeness in $z$ vs $r-i$ plane]{\includegraphics[width=8cm]{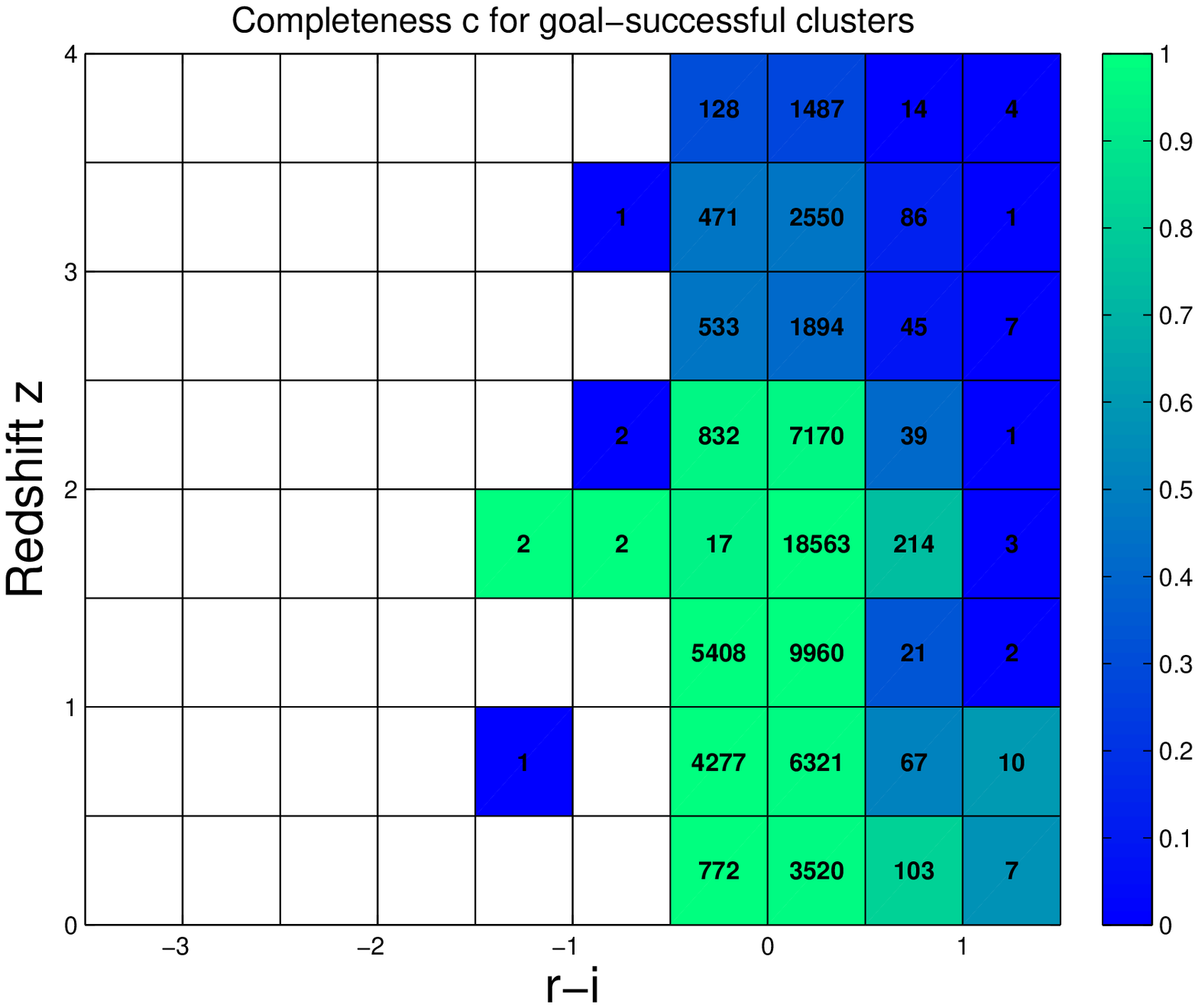}}
\subfigure[Completeness in $z$ vs $u-g$ plane]{\includegraphics[width=8cm]{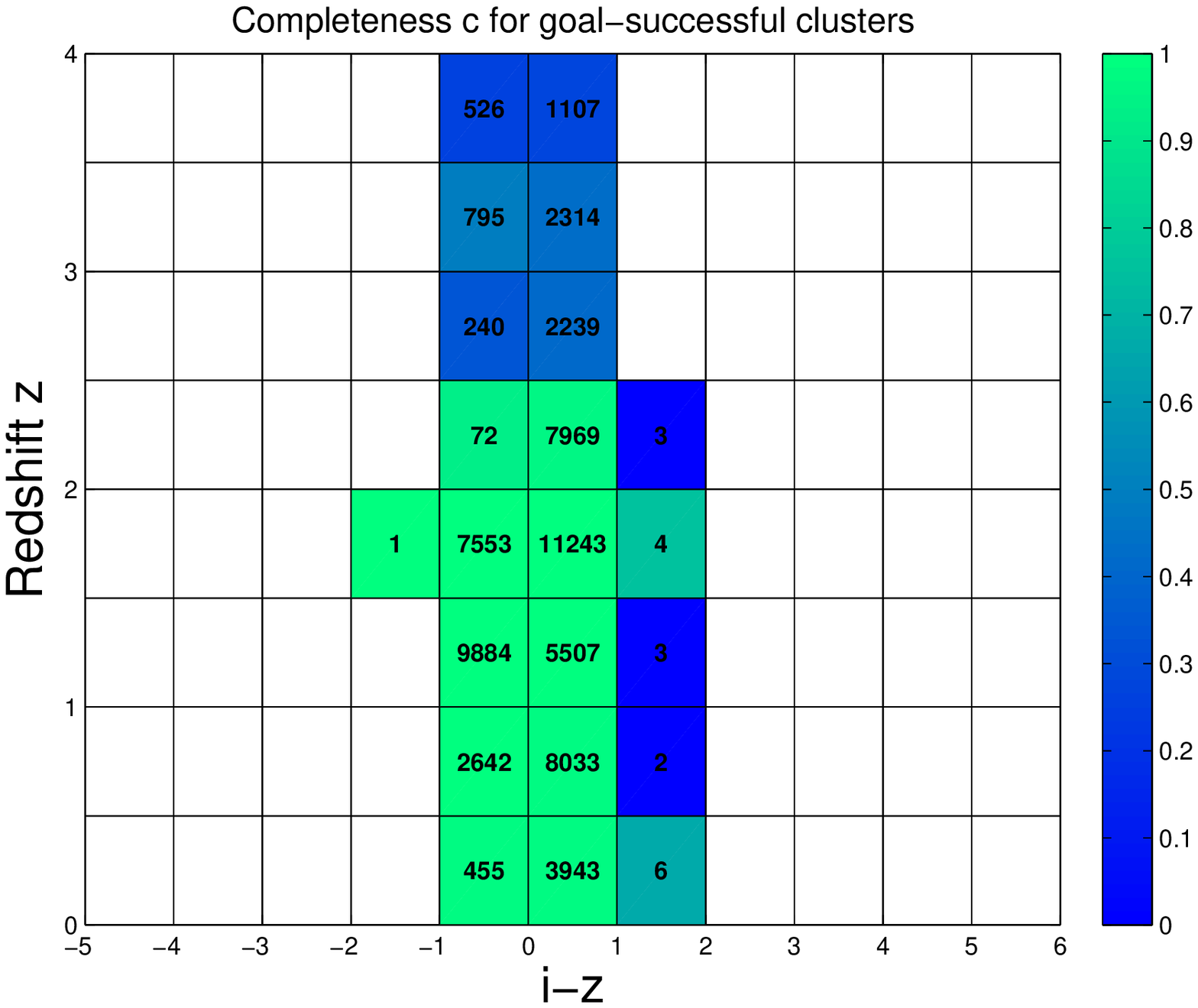}}\\
\caption{Local completeness estimates in the redshift vs colour planes for the "goal-successful" clusters of experiment S-SCat. Numbers represent
the total number of sources of the BoK located in each cell.}
\label{zcolourscompl_cat}
\end{figure}

\begin{figure}
\centering
\subfigure[Efficiency in $u-g$ vs $g-r$ plane]{\includegraphics[width=8cm]{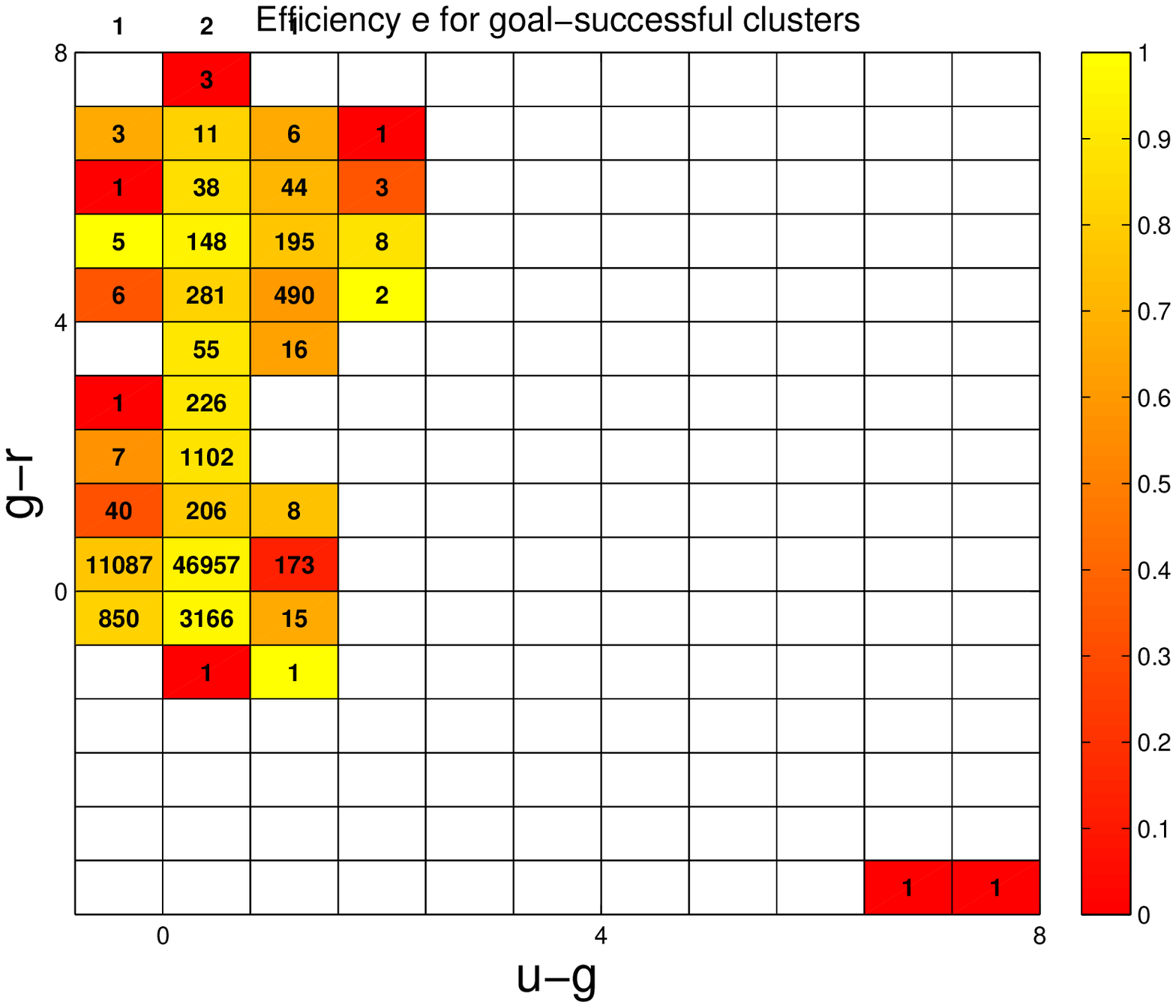}}
\subfigure[Efficiency in $u-g$ vs $r-i$ plane]{\includegraphics[width=8cm]{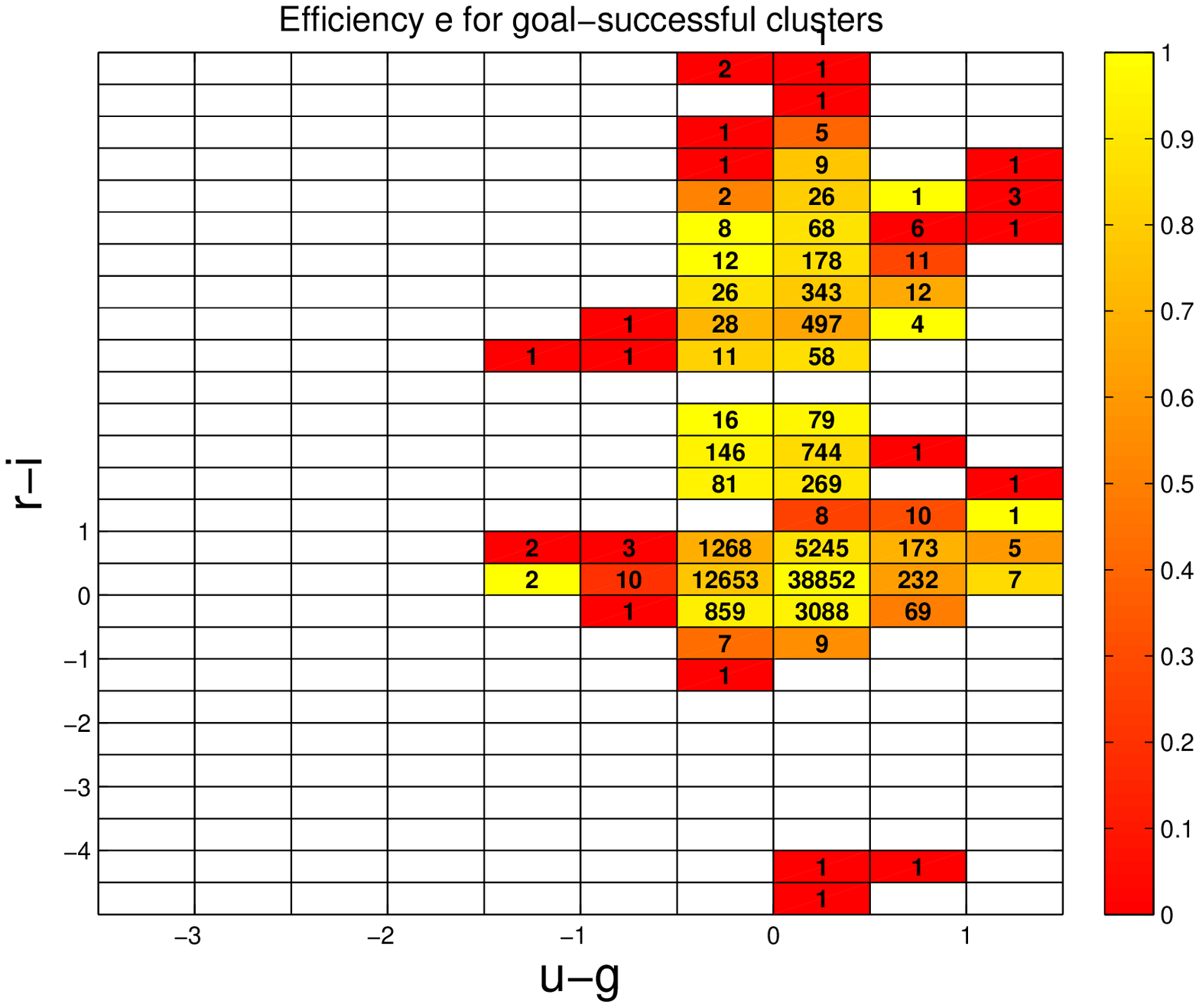}}\\
\subfigure[Efficiency in $u-g$ vs $i-z$ plane]{\includegraphics[width=8cm]{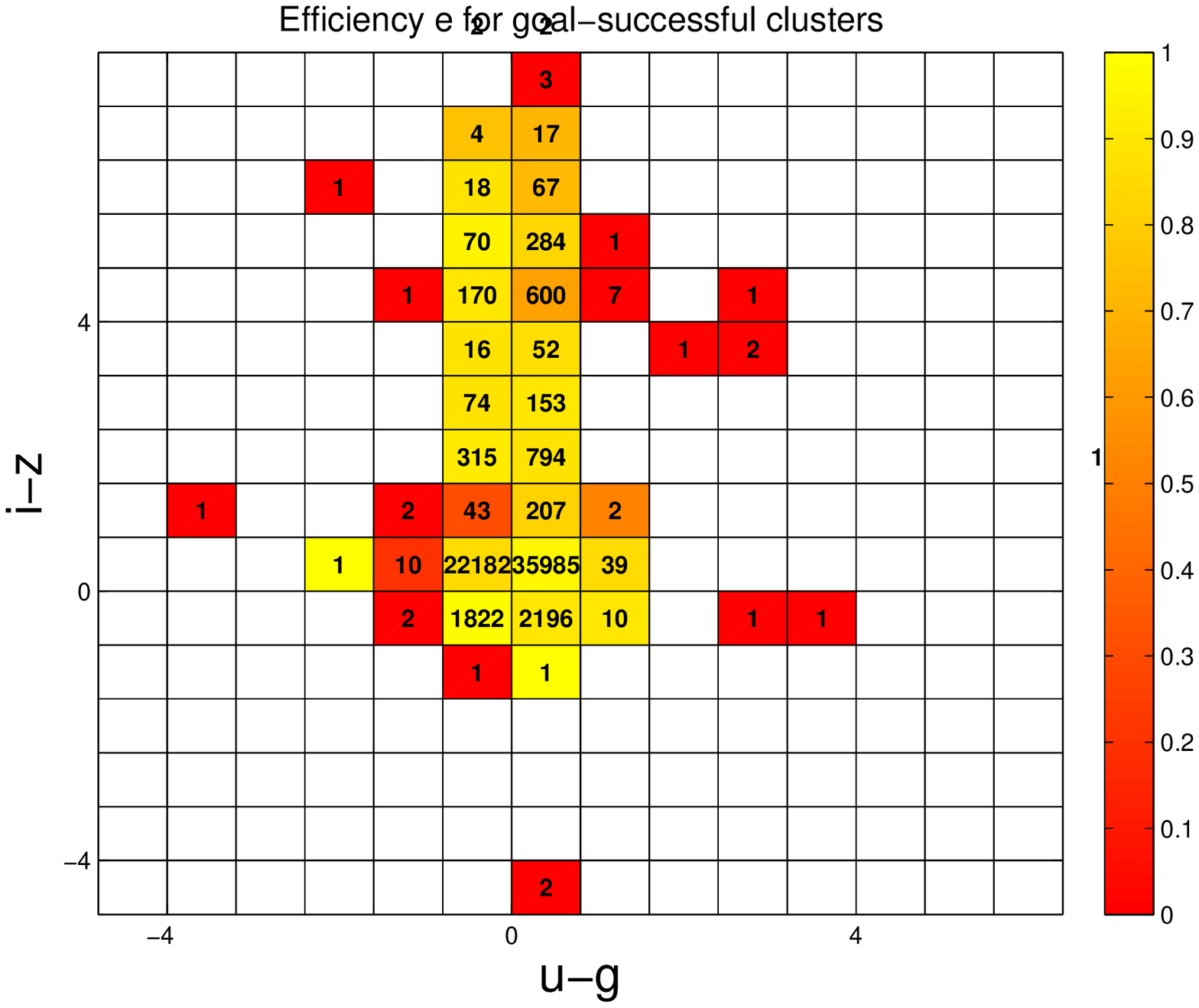}}
\subfigure[Efficiency in $g-r$ vs $r-i$ plane]{\includegraphics[width=8cm]{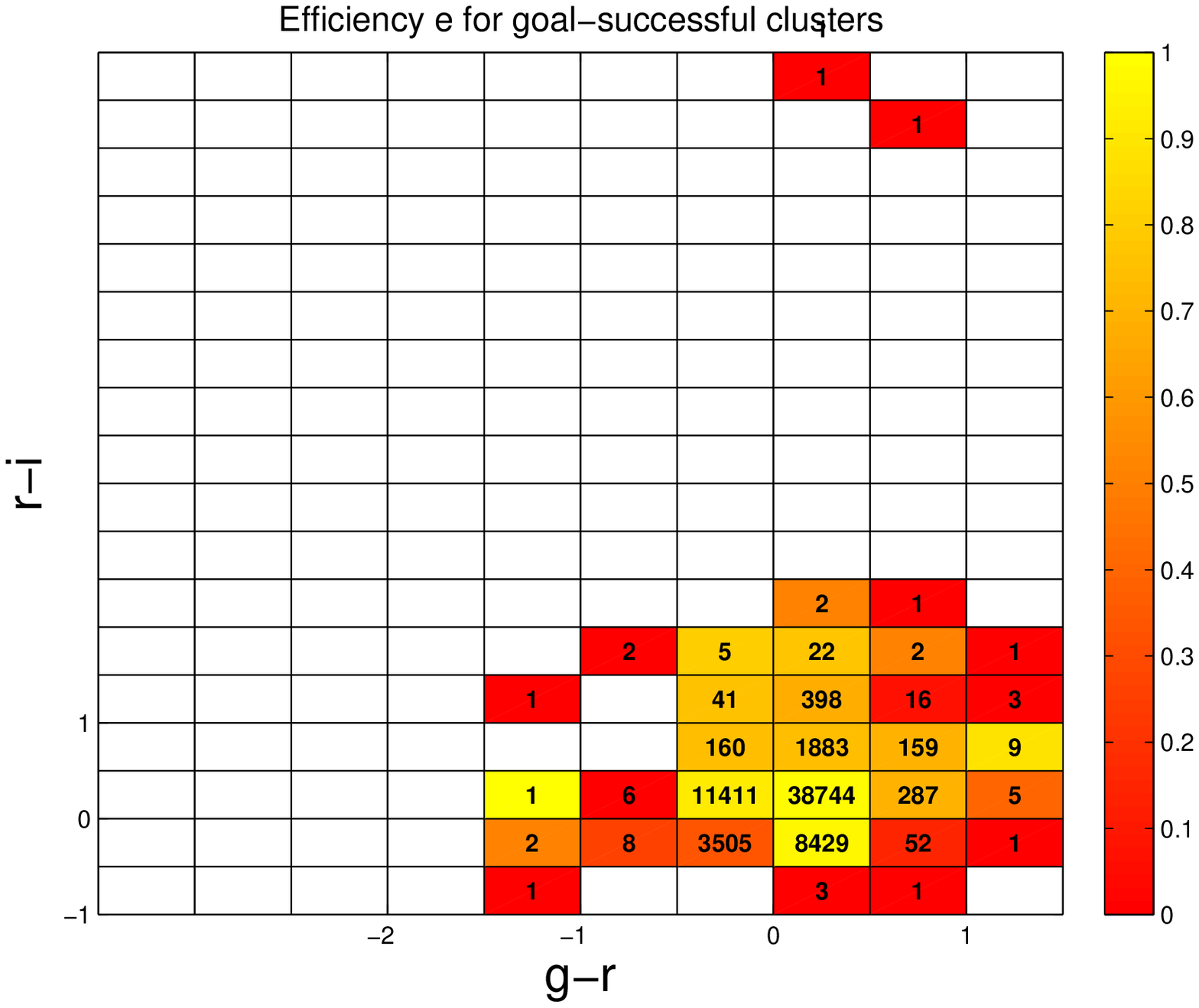}}\\
\subfigure[Efficiency in $g-r$ vs $i-z$ plane]{\includegraphics[width=8cm]{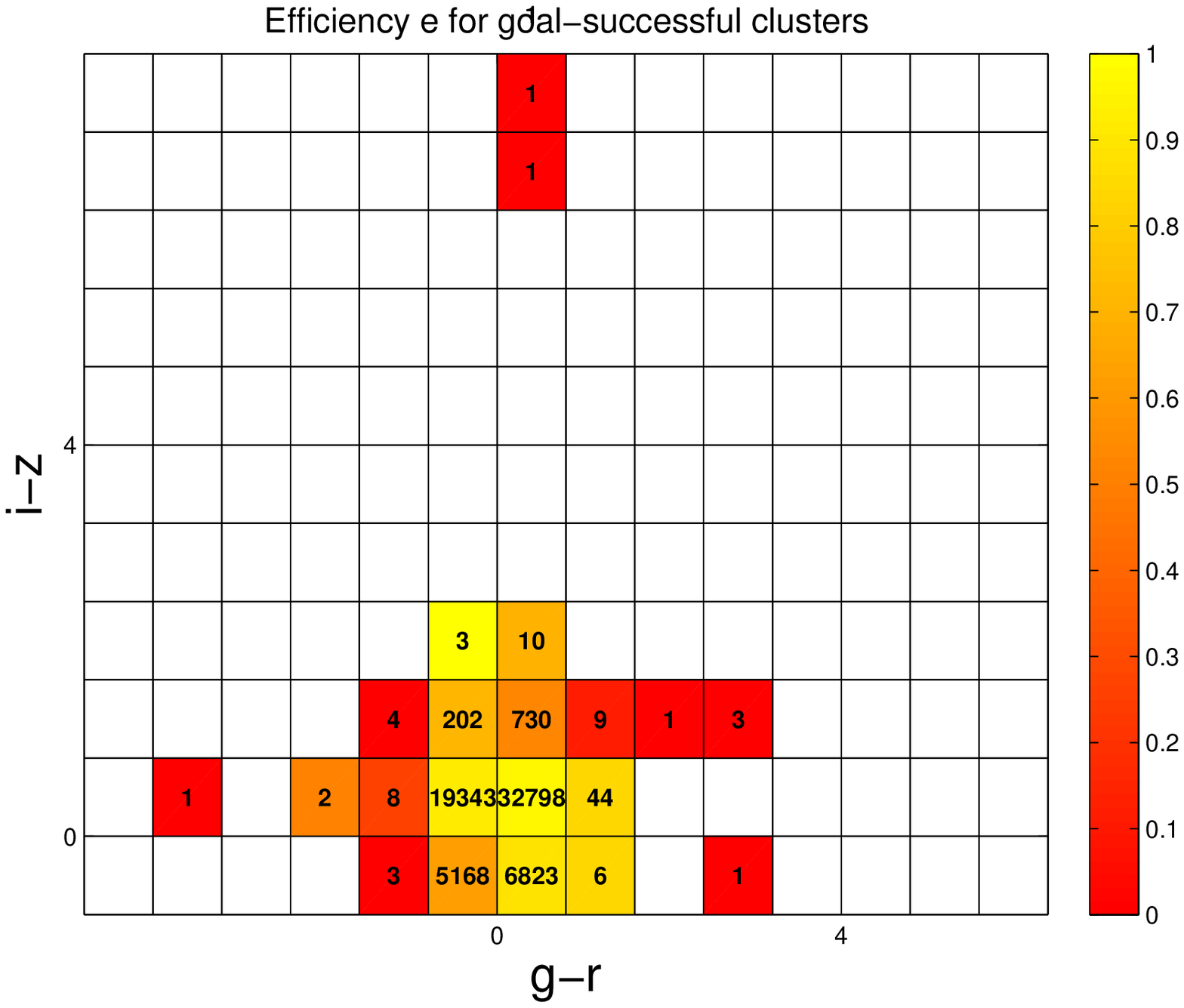}}
\subfigure[Efficiency in $r-i$ vs $i-z$ plane]{\includegraphics[width=8cm]{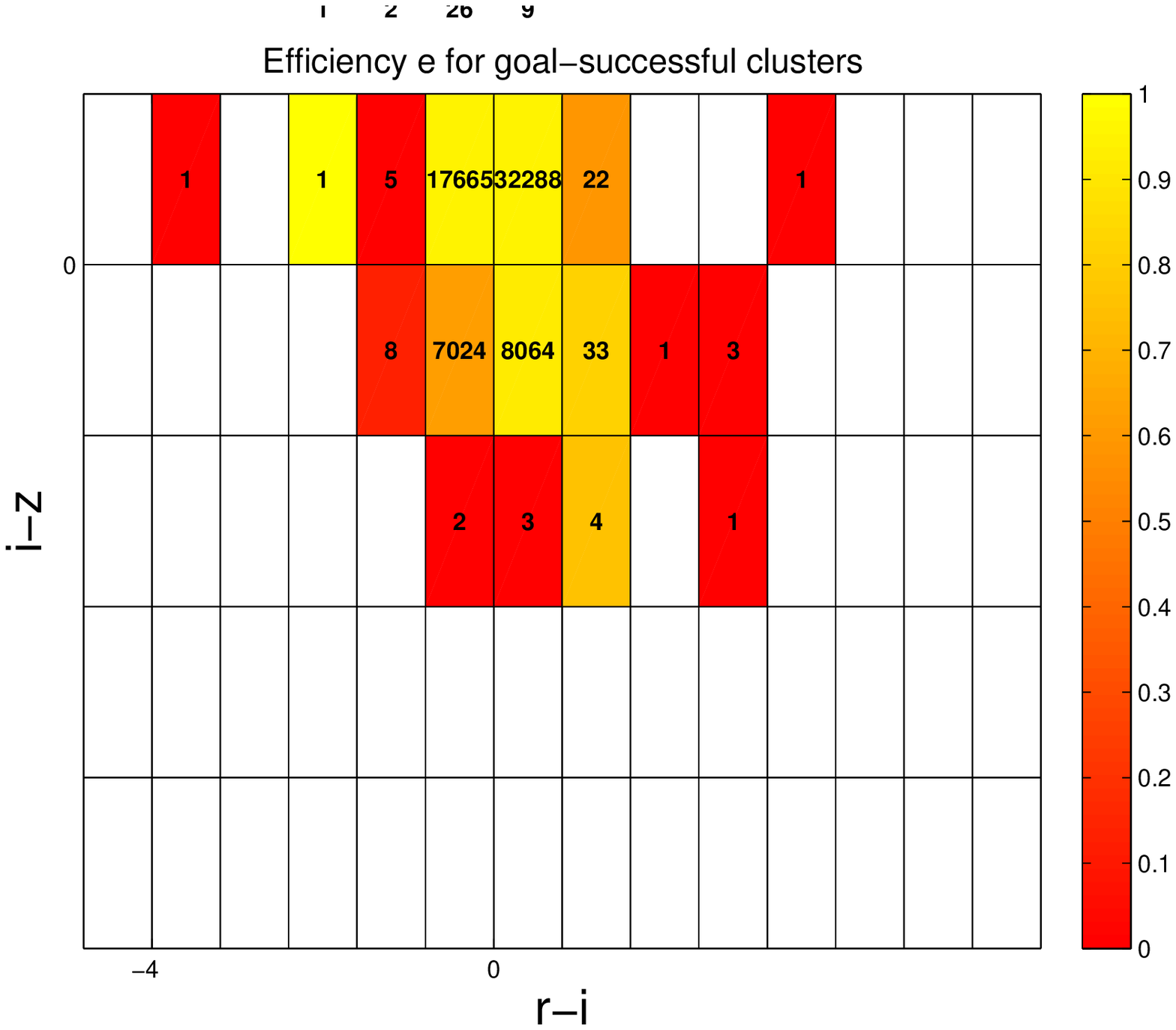}}\\
\caption{Local efficiency estimates in the colour vs colour planes for the "goal-successful" clusters of experiment S-SCat. Numbers represent
the total number of sources of the BoK located in each cell.}
\label{colourcoloureffic_cat}
\end{figure}

\begin{figure}
\centering
\subfigure[Completeness in $u-g$ vs $g-r$ plane]{\includegraphics[width=8cm]{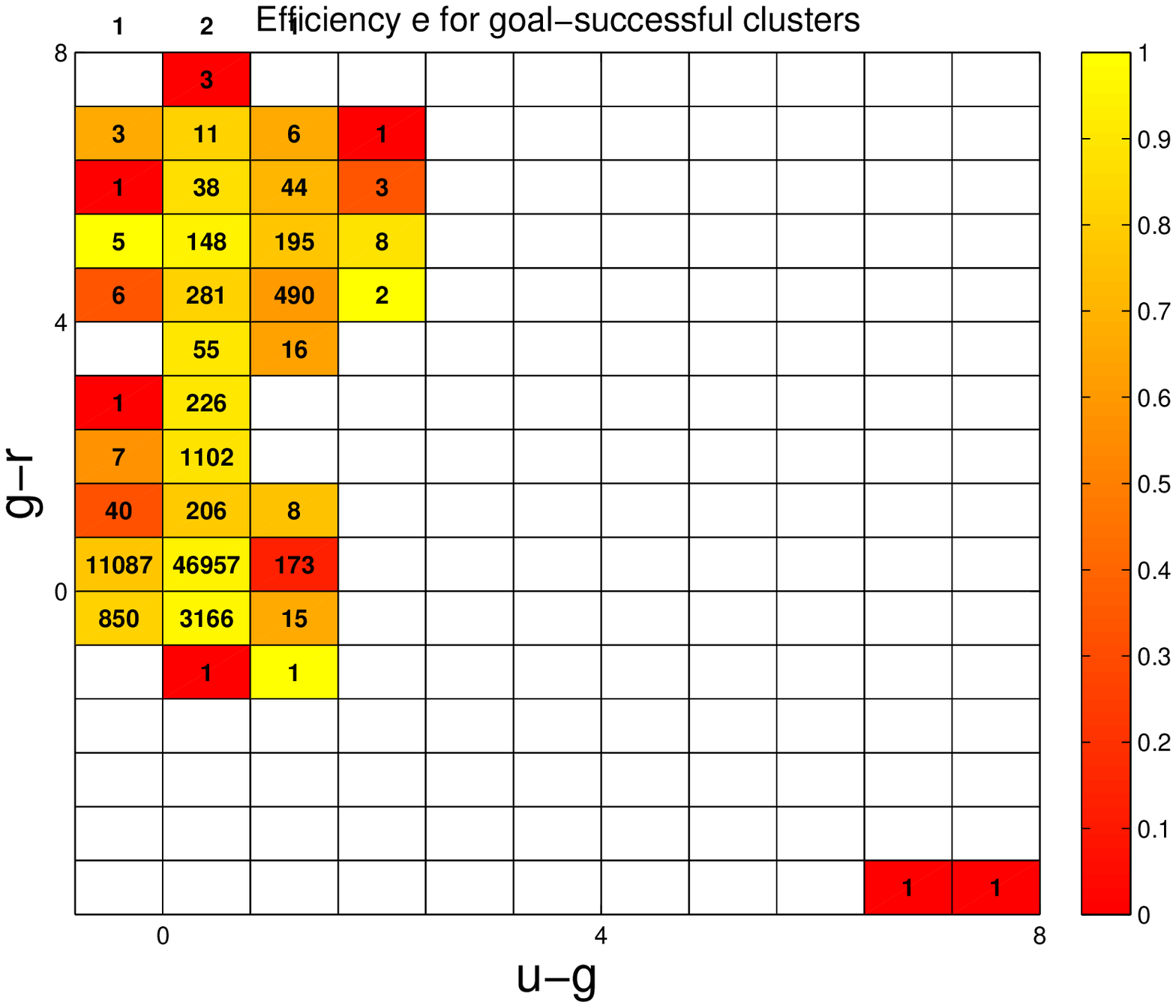}}
\subfigure[Completeness in $u-g$ vs $r-i$ plane]{\includegraphics[width=8cm]{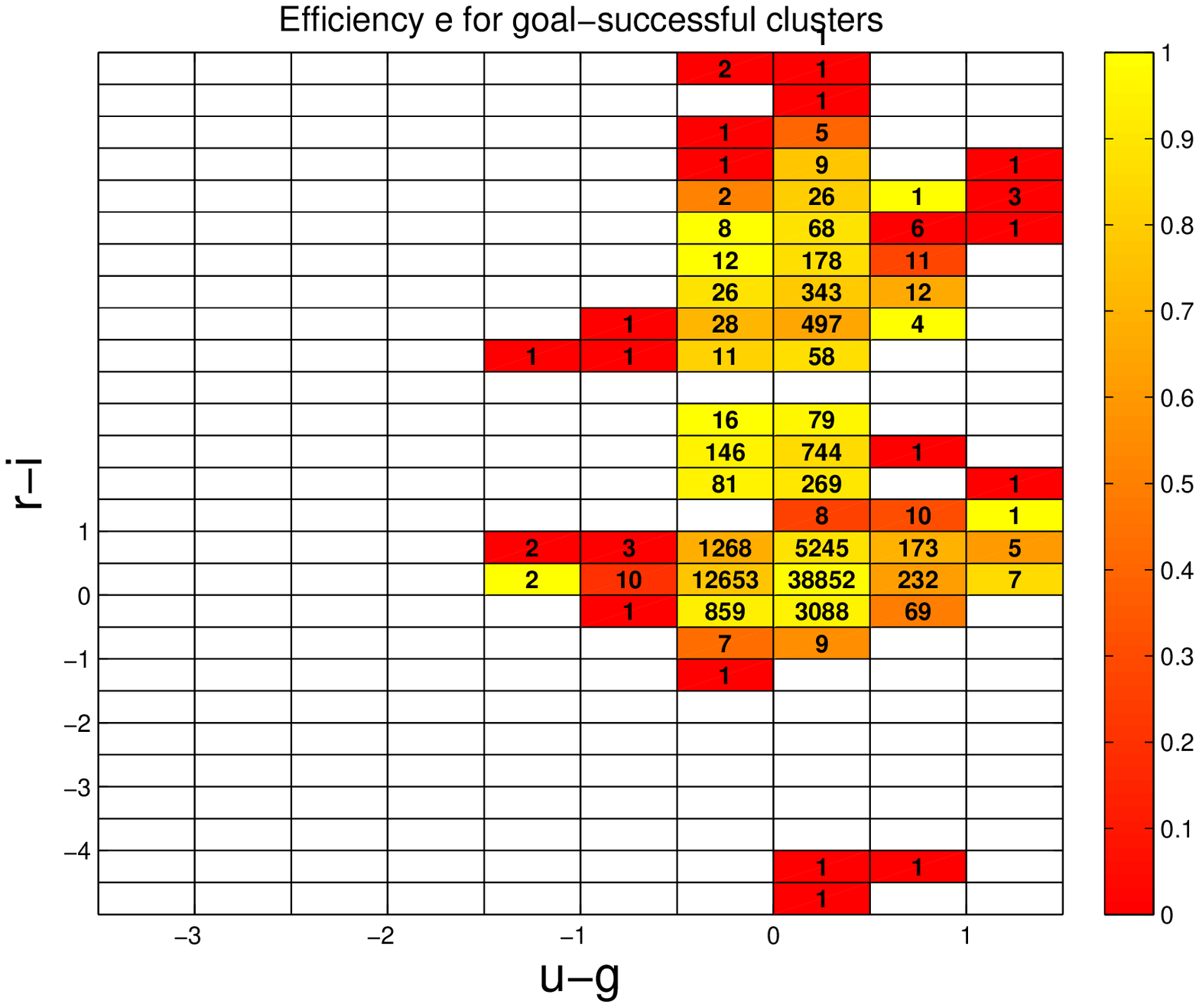}}\\
\subfigure[Completeness in $u-g$ vs $i-z$ plane]{\includegraphics[width=8cm]{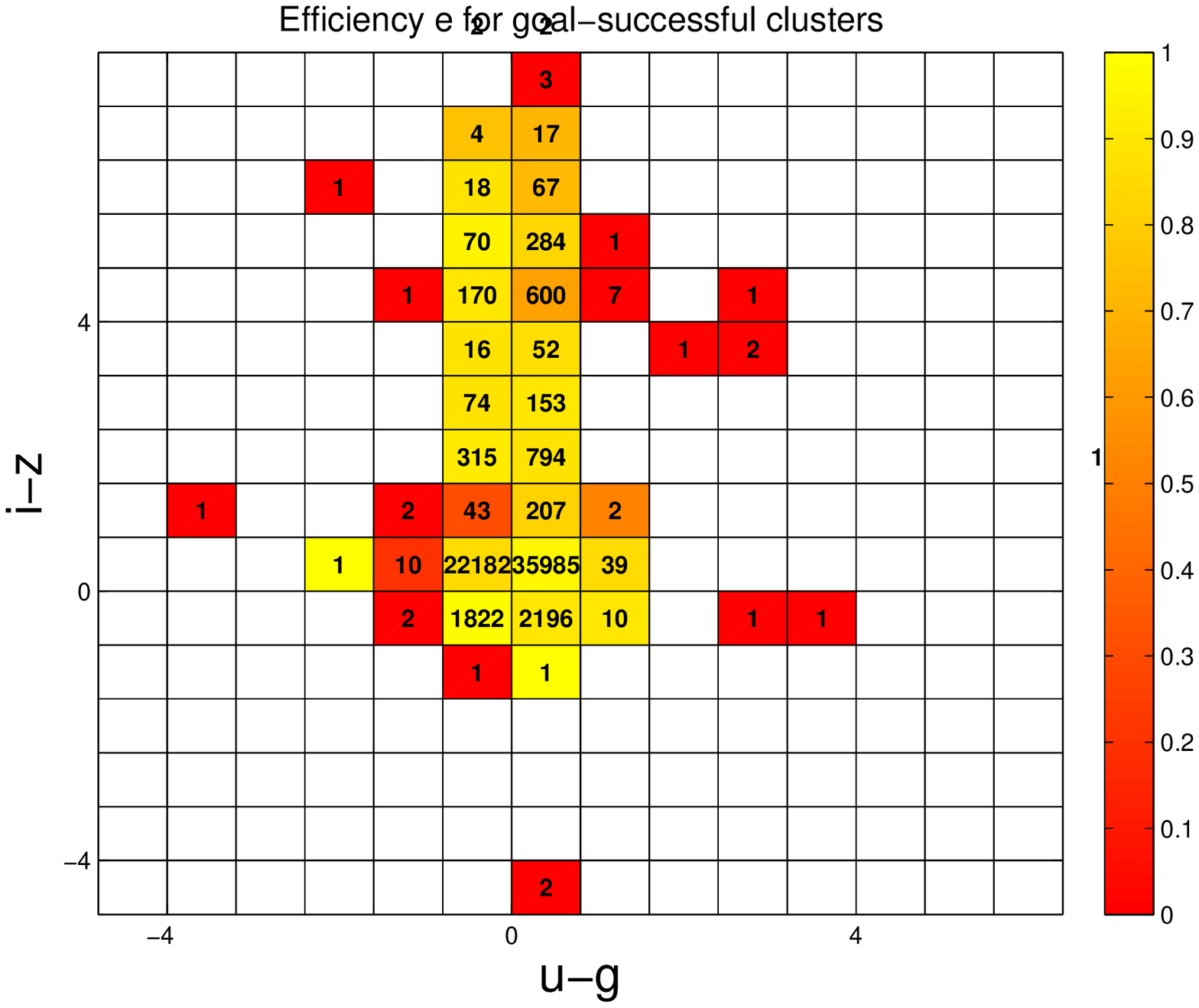}}
\subfigure[Completeness in $g-r$ vs $r-i$ plane]{\includegraphics[width=8cm]{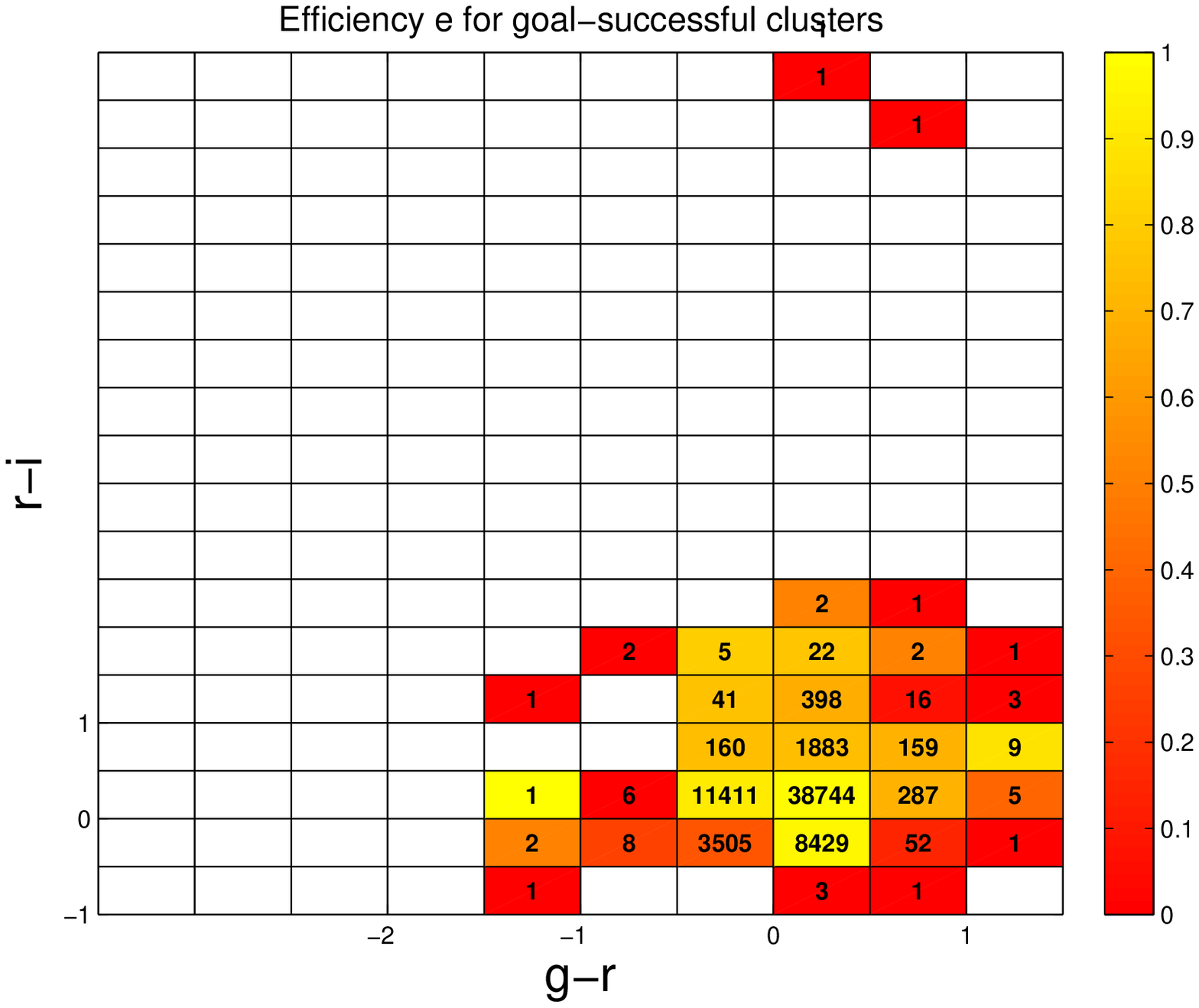}}\\
\subfigure[Completeness in $g-r$ vs $i-z$ plane]{\includegraphics[width=8cm]{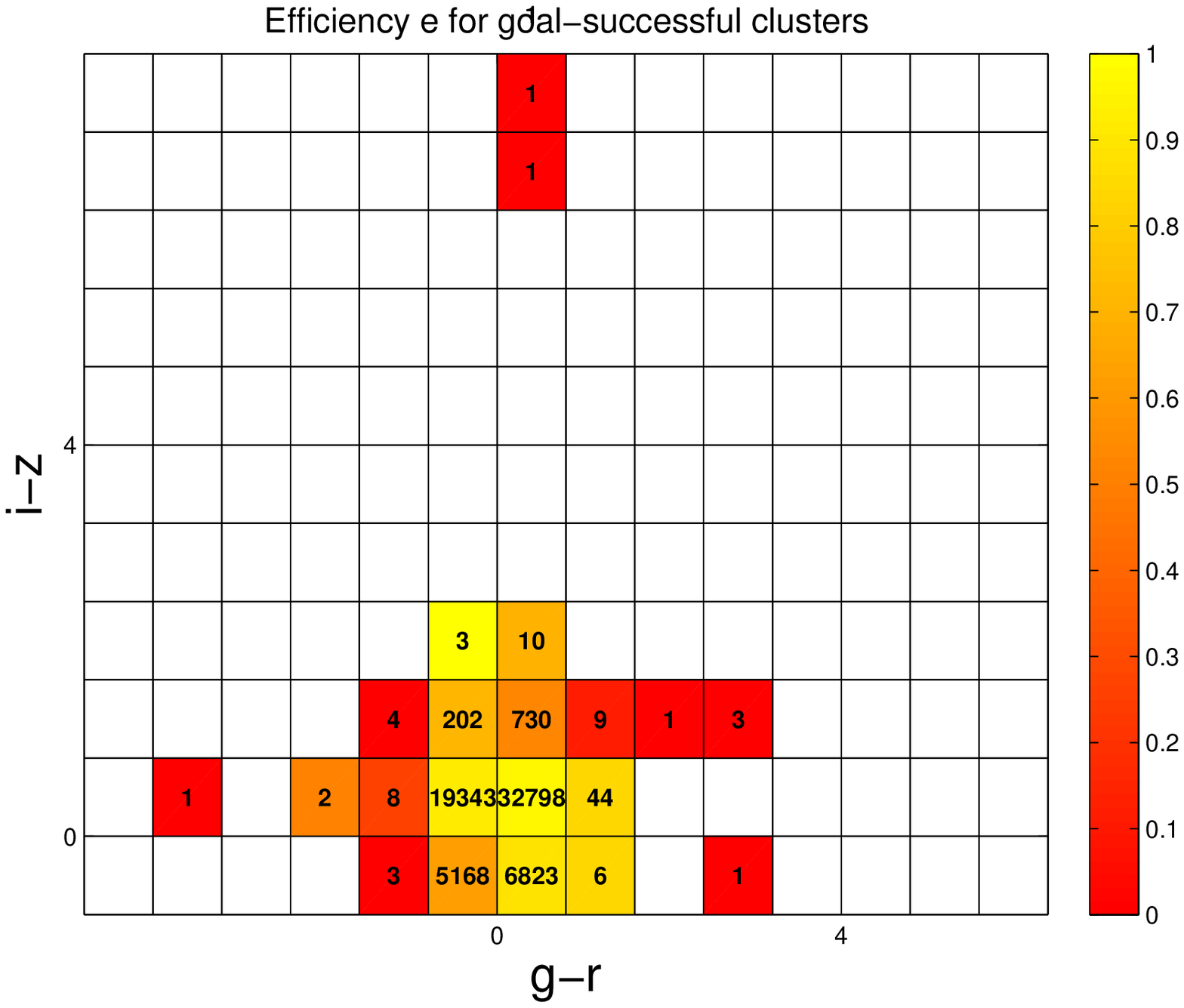}}
\subfigure[Completeness in $r-i$ vs $i-z$ plane]{\includegraphics[width=8cm]{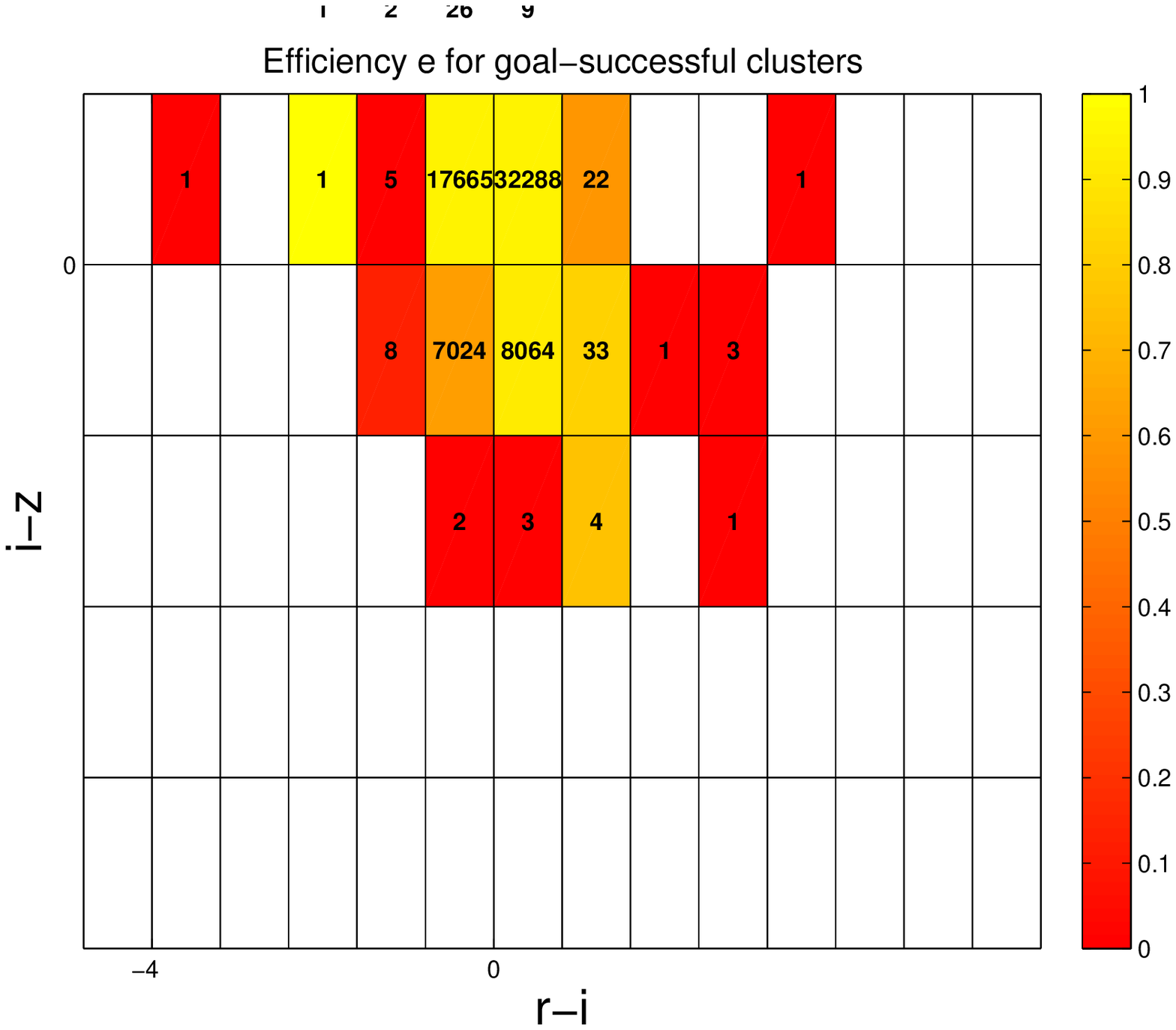}}\\
\caption{Local completeness estimates in the colour vs colour planes for the "goal-successful" clusters of experiment S-SCat. Numbers represent
the total number of sources of the BoK located in each cell.}
\label{colourcolourcompl_cat}
\end{figure}

\begin{figure}
\centering
\subfigure[Efficiency in $u-g$ vs $g-r$ plane for all clusters]{\includegraphics[width=8cm]{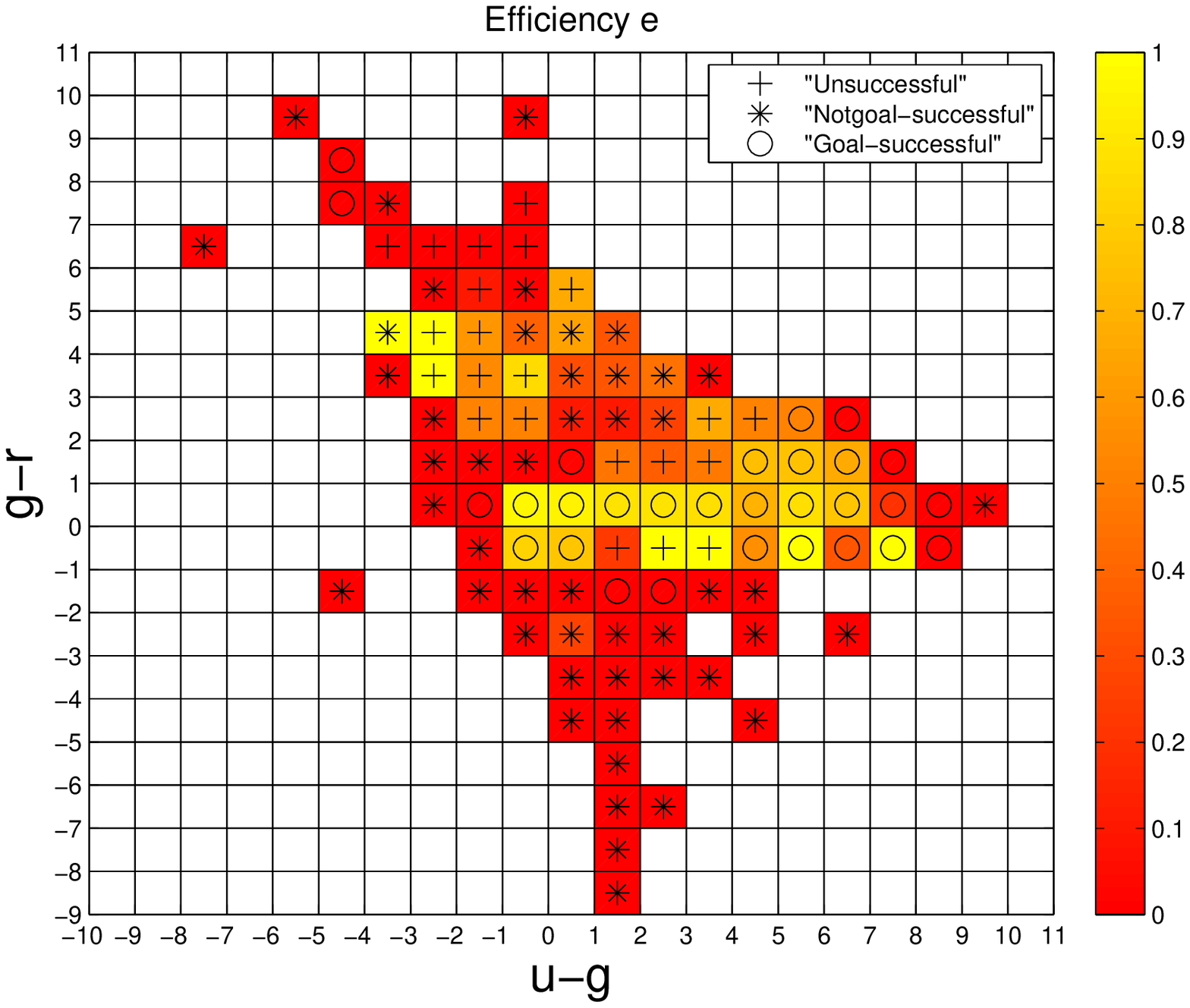}}
\subfigure[Efficiency in $u-g$ vs $r-i$ plane for all clusters]{\includegraphics[width=8cm]{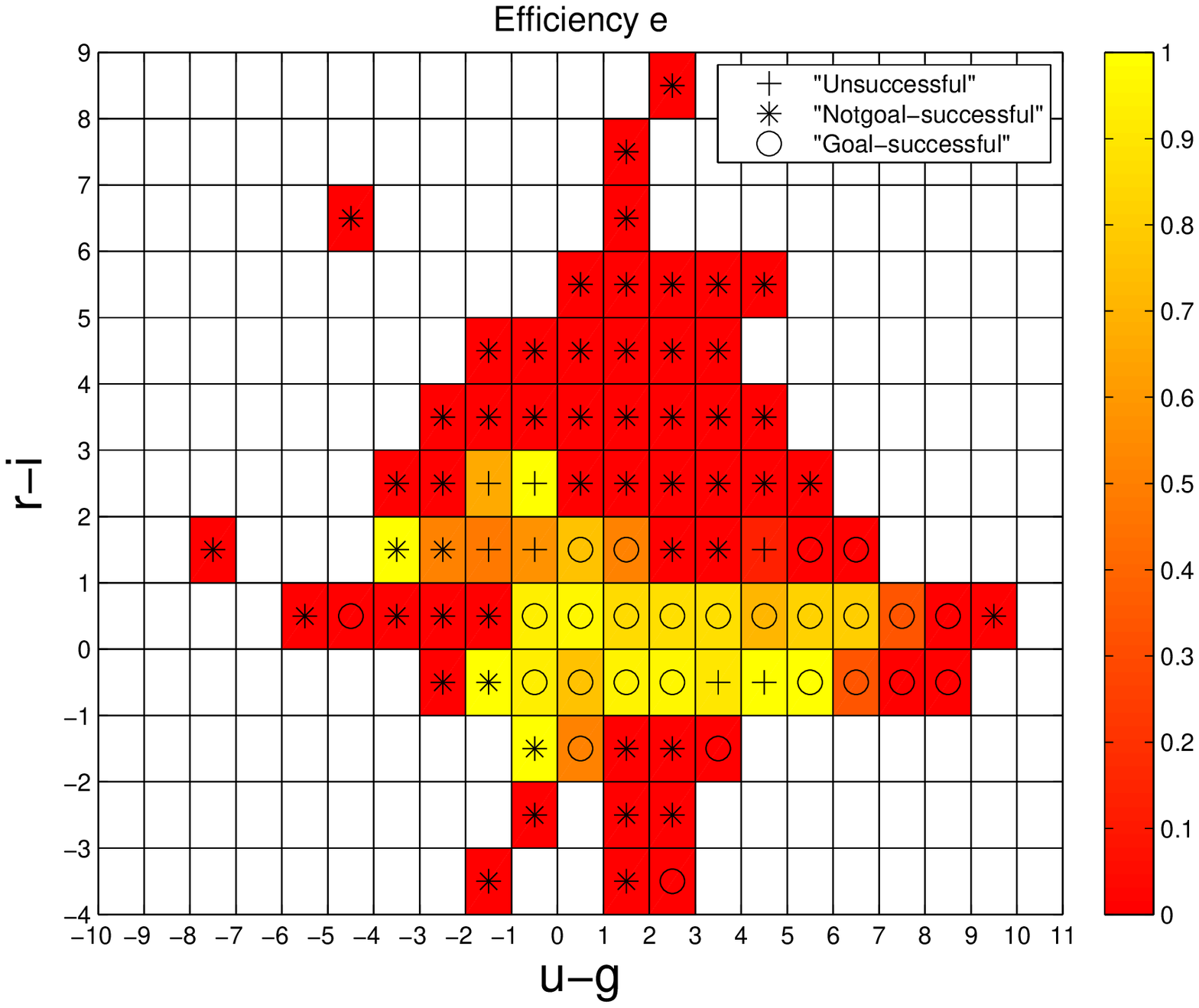}}\\
\subfigure[Efficiency in $u-g$ vs $i-z$ plane for all clusters]{\includegraphics[width=8cm]{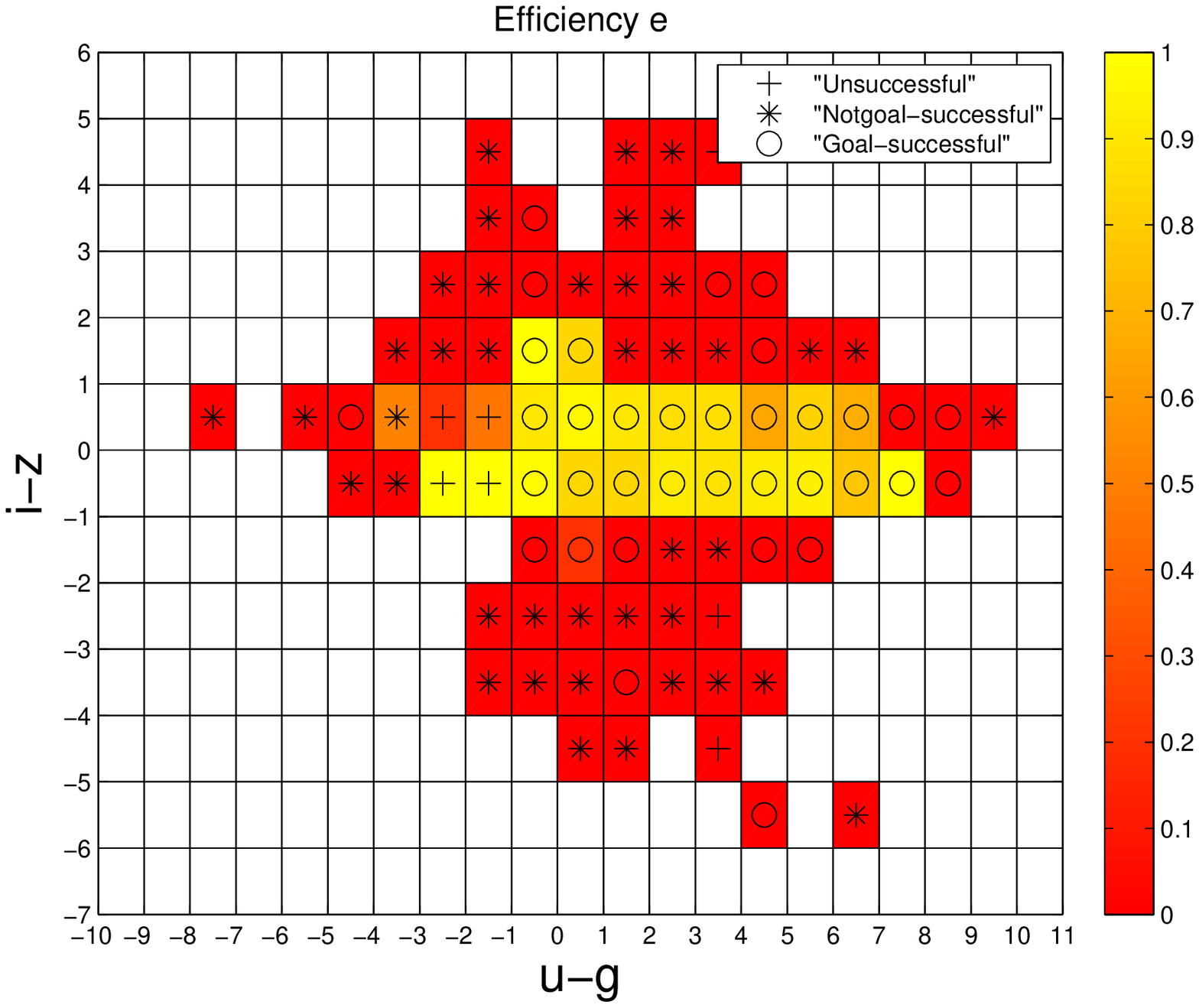}}
\subfigure[Efficiency in $g-r$ vs $r-i$ plane for all clusters]{\includegraphics[width=8cm]{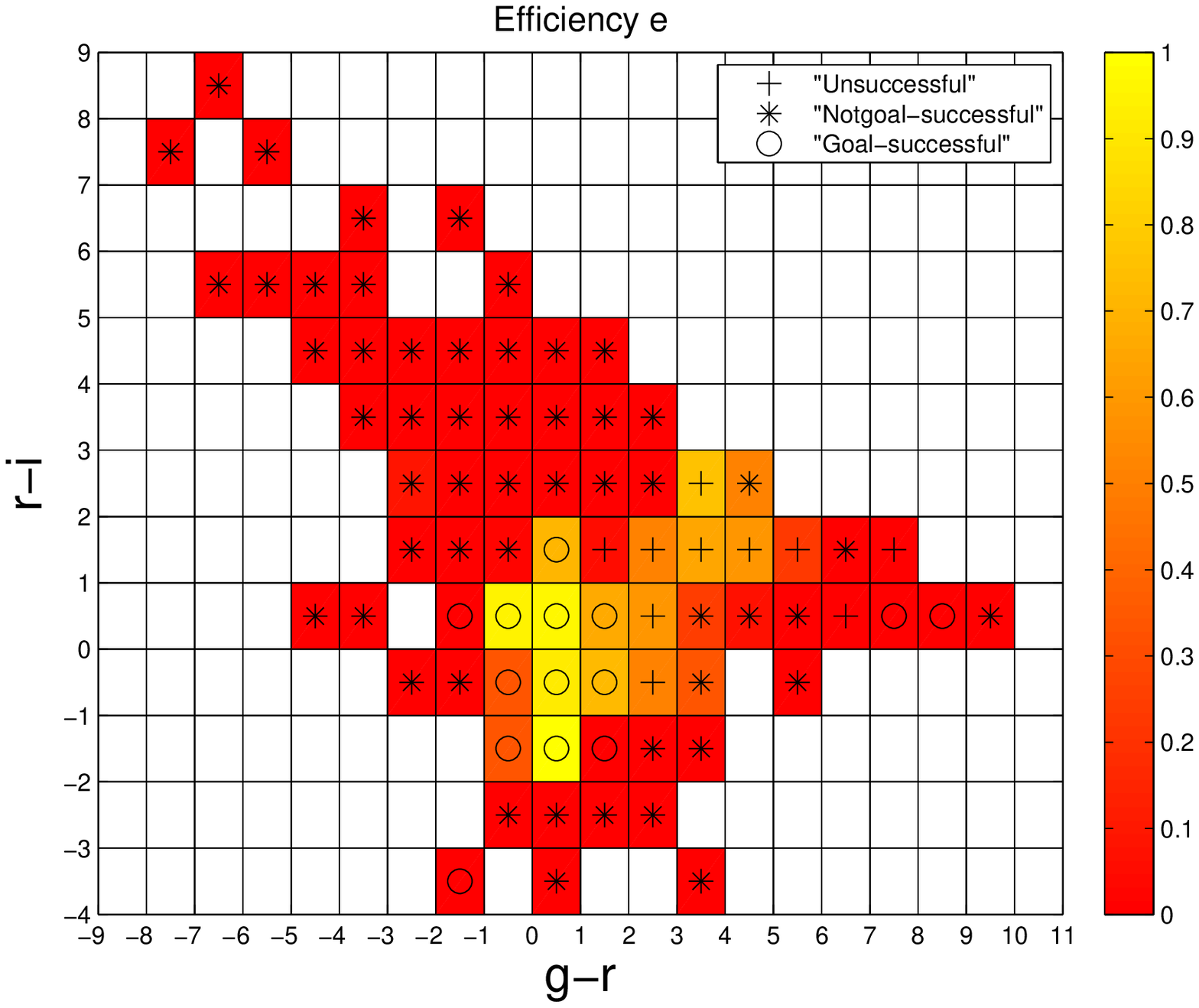}}\\
\subfigure[Efficiency in $g-r$ vs $i-z$ plane for all clusters]{\includegraphics[width=8cm]{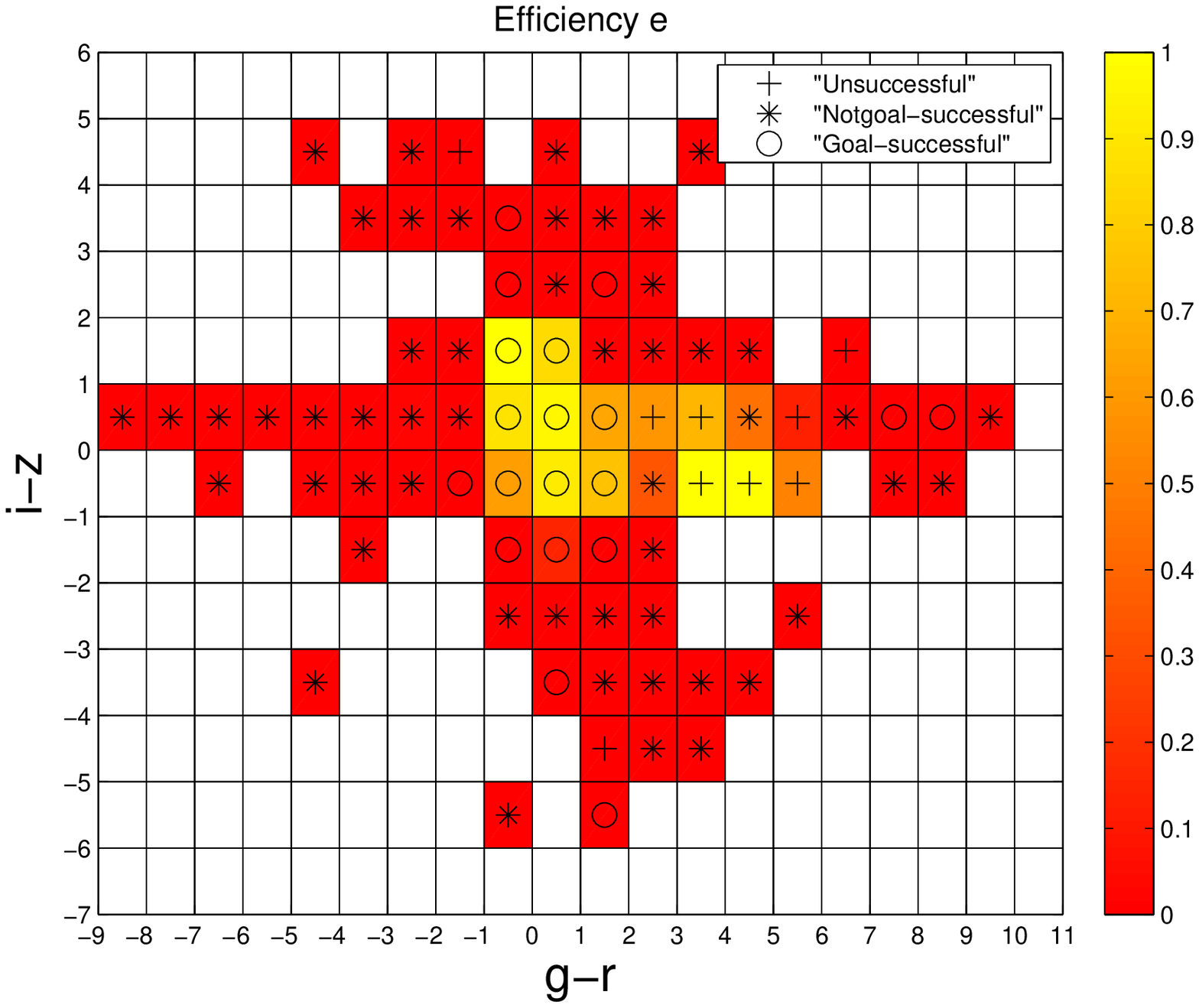}}
\subfigure[Efficiency in $r-i$ vs $i-z$ plane for all clusters]{\includegraphics[width=8cm]{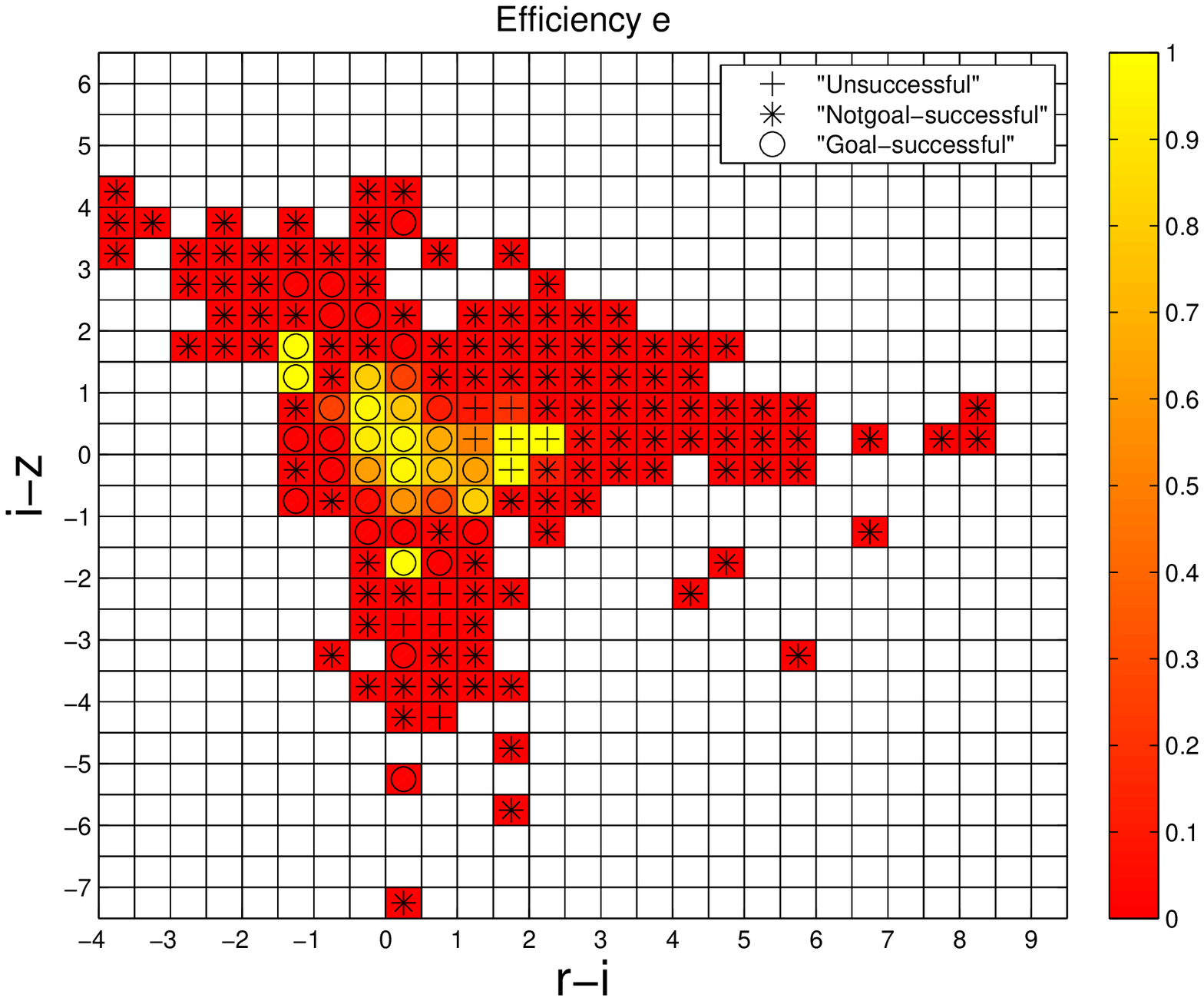}}\\
\caption{Local efficiency estimates in the colour vs colour planes for the all types of clusters of experiment S-SCat.}
\label{colourcoloureffictot_cat}
\end{figure}

\begin{figure}
\centering
\subfigure[Completeness in $u-g$ vs $g-r$ plane for all clusters]{\includegraphics[width=8cm]{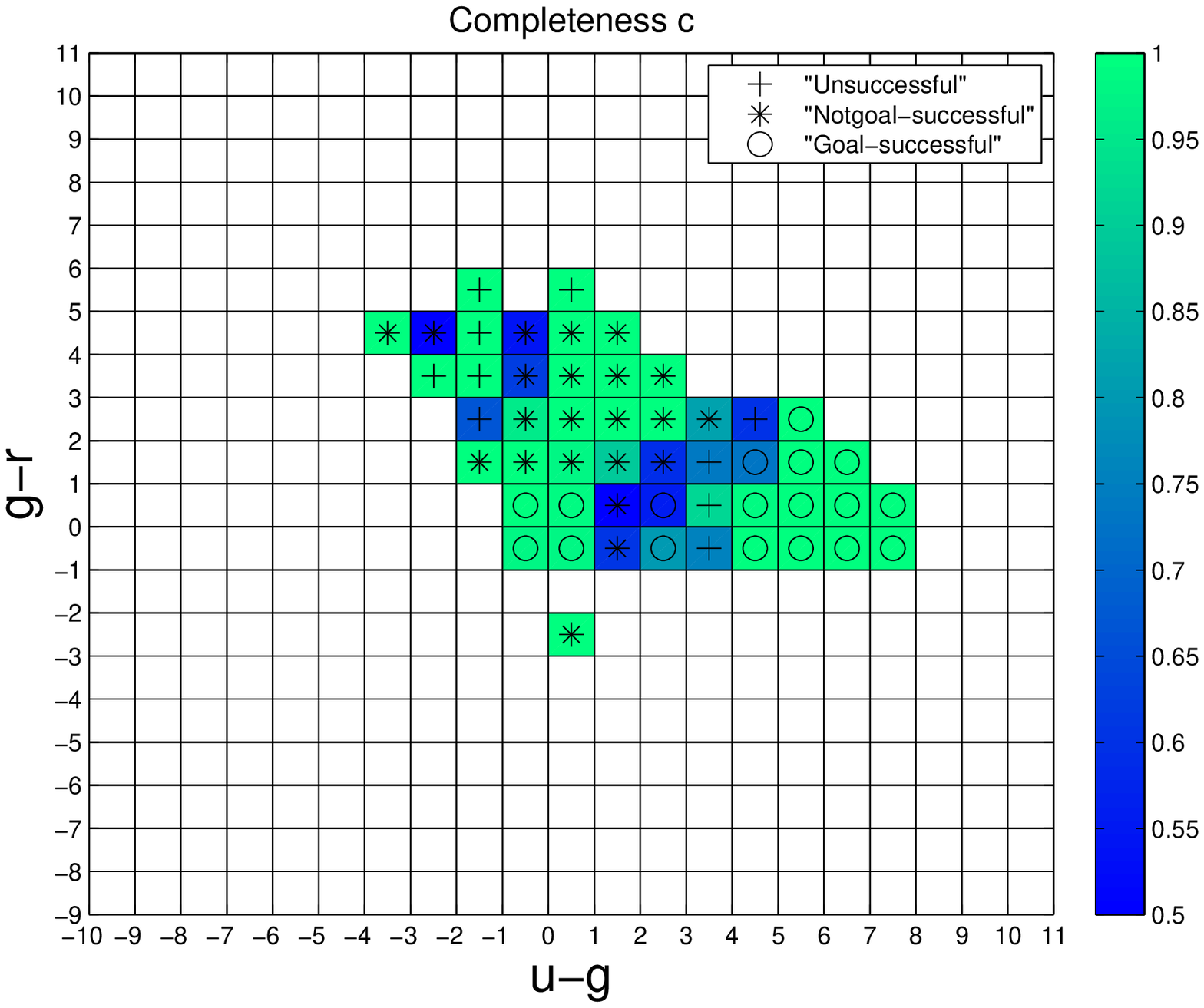}}
\subfigure[Completeness in $u-g$ vs $r-i$ plane for all clusters]{\includegraphics[width=8cm]{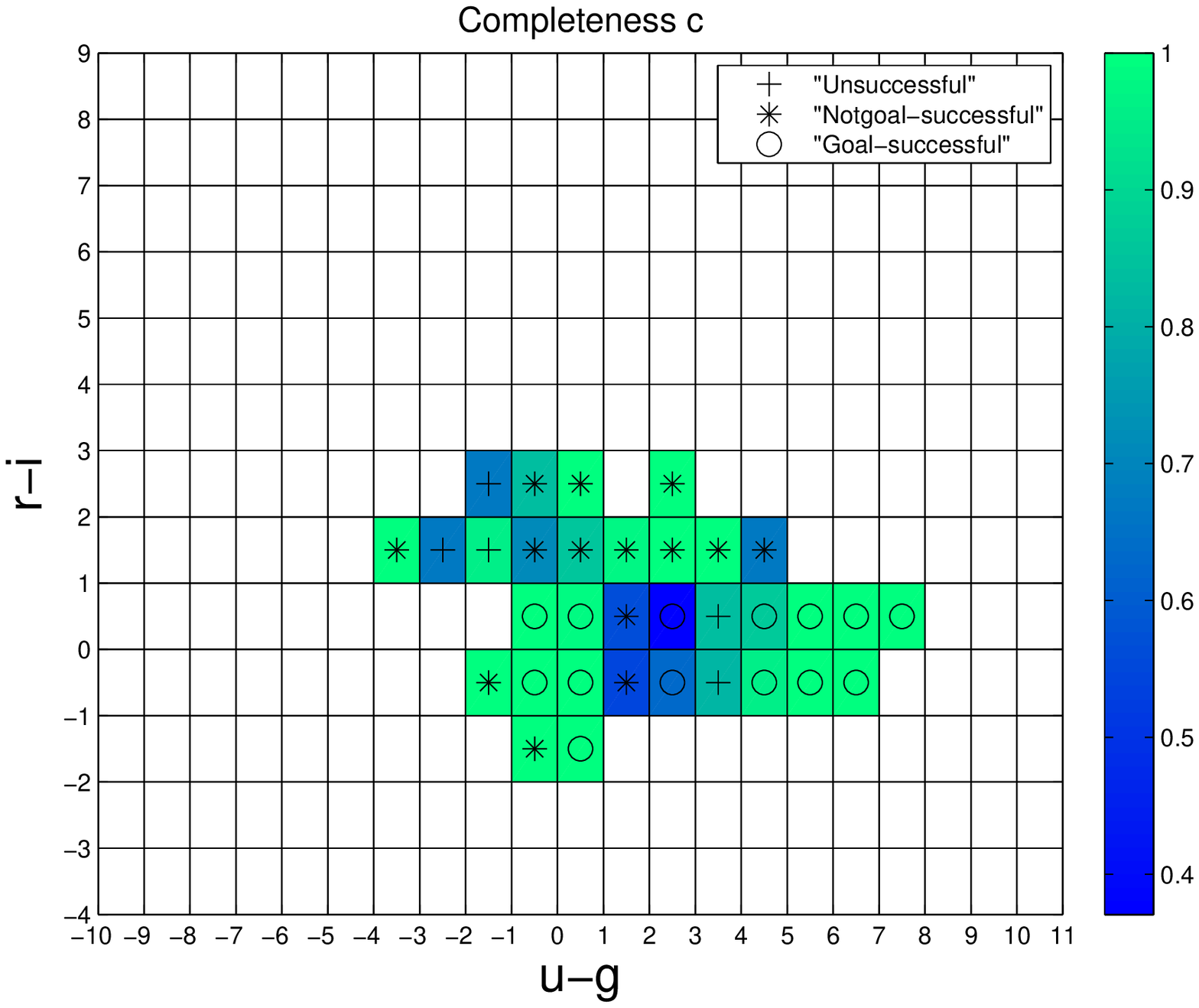}}\\
\subfigure[Completeness in $u-g$ vs $i-z$ plane for all clusters]{\includegraphics[width=8cm]{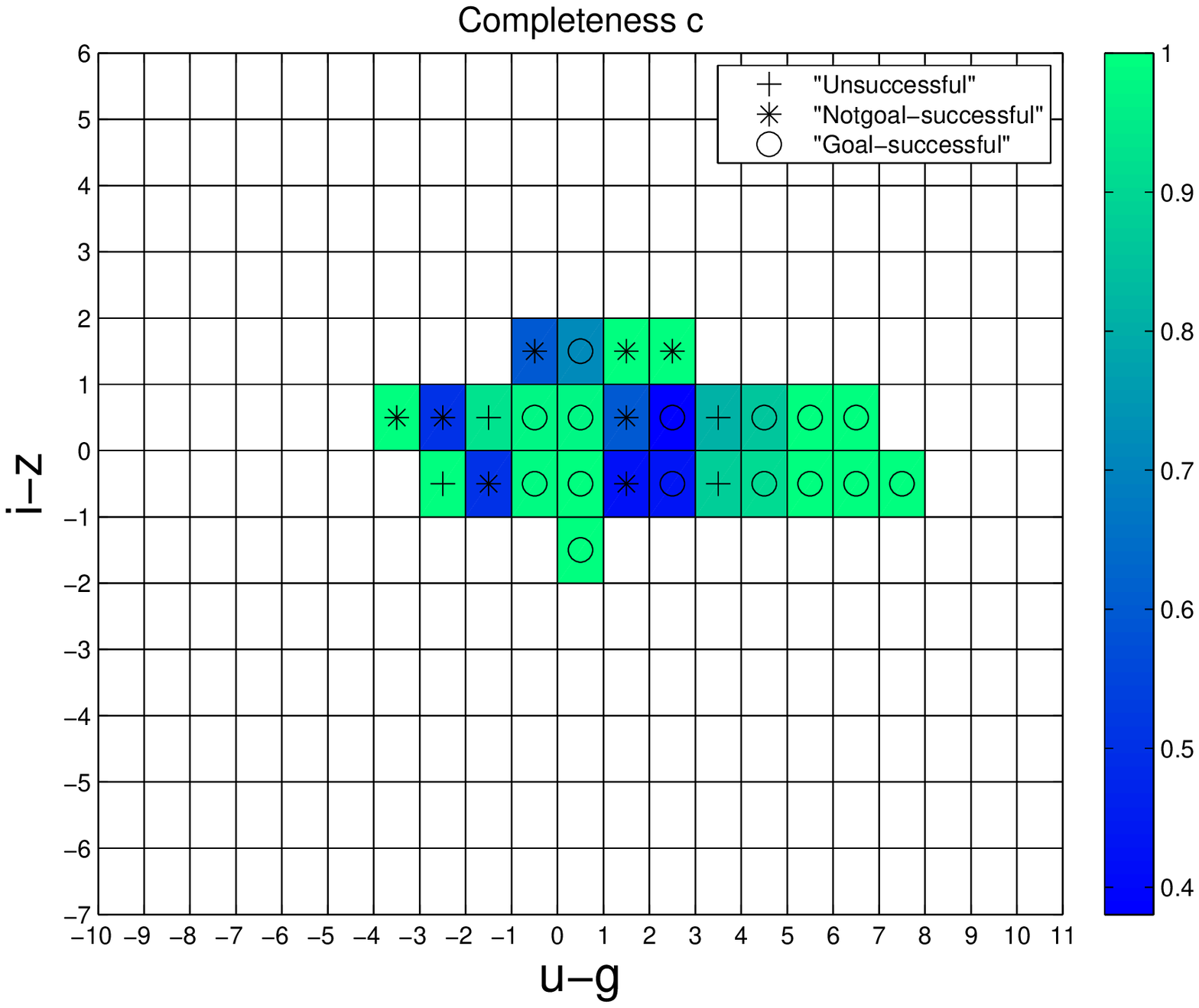}}
\subfigure[Completeness in $g-r$ vs $r-i$ plane for all clusters]{\includegraphics[width=8cm]{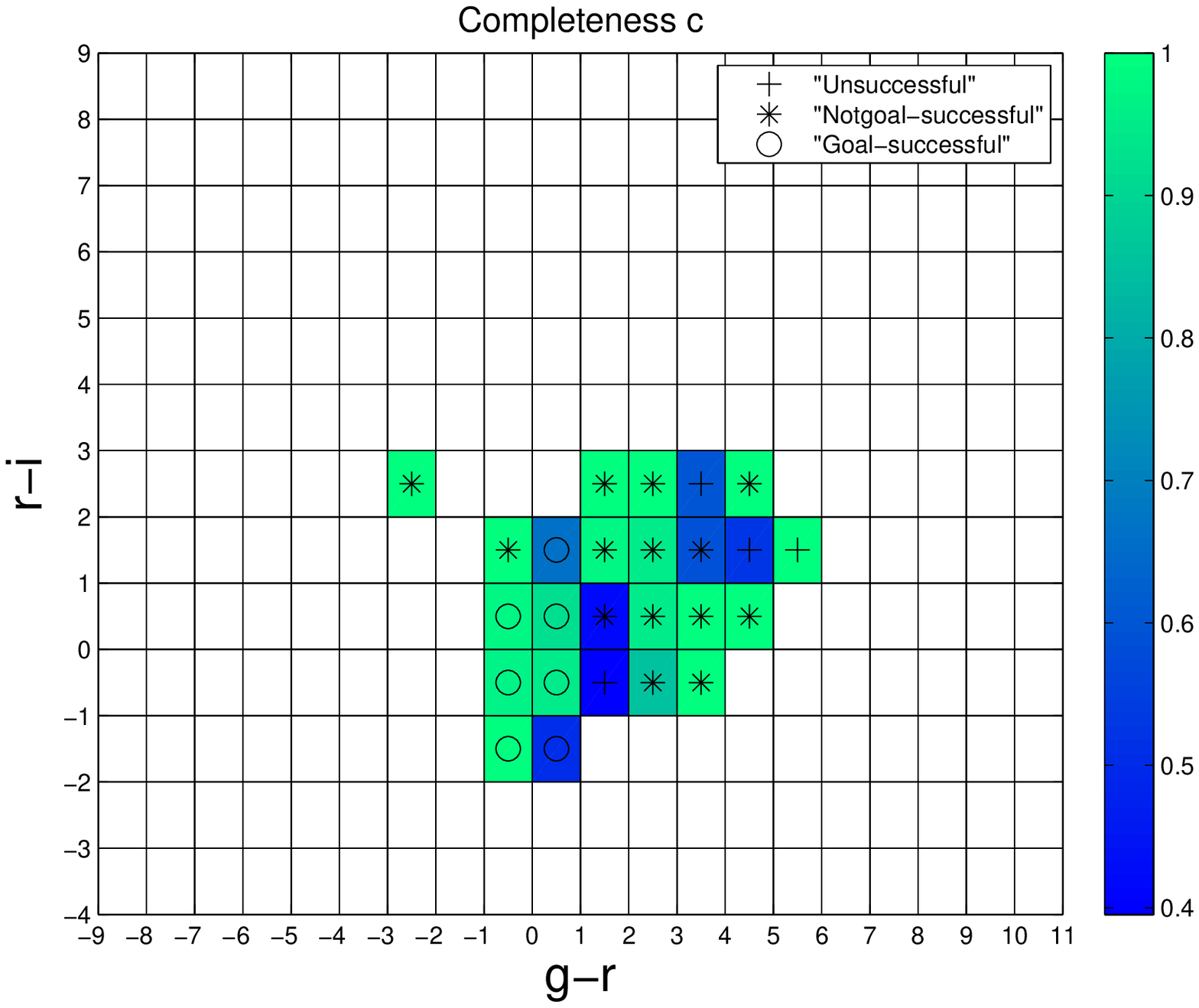}}\\
\subfigure[Completeness in $g-r$ vs $i-z$ plane for all clusters]{\includegraphics[width=8cm]{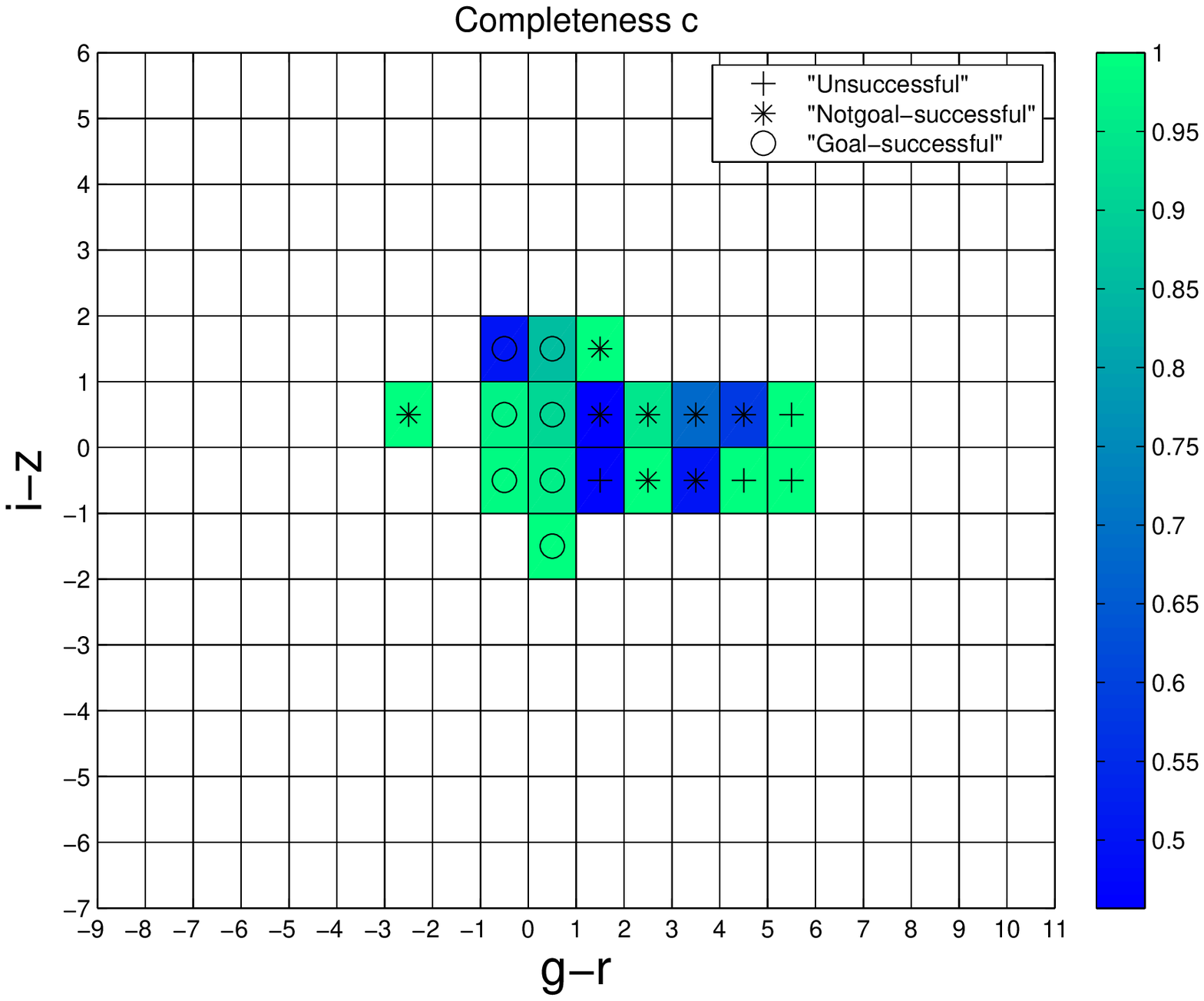}}
\subfigure[Completeness in $r-i$ vs $i-z$ plane for all clusters]{\includegraphics[width=8cm]{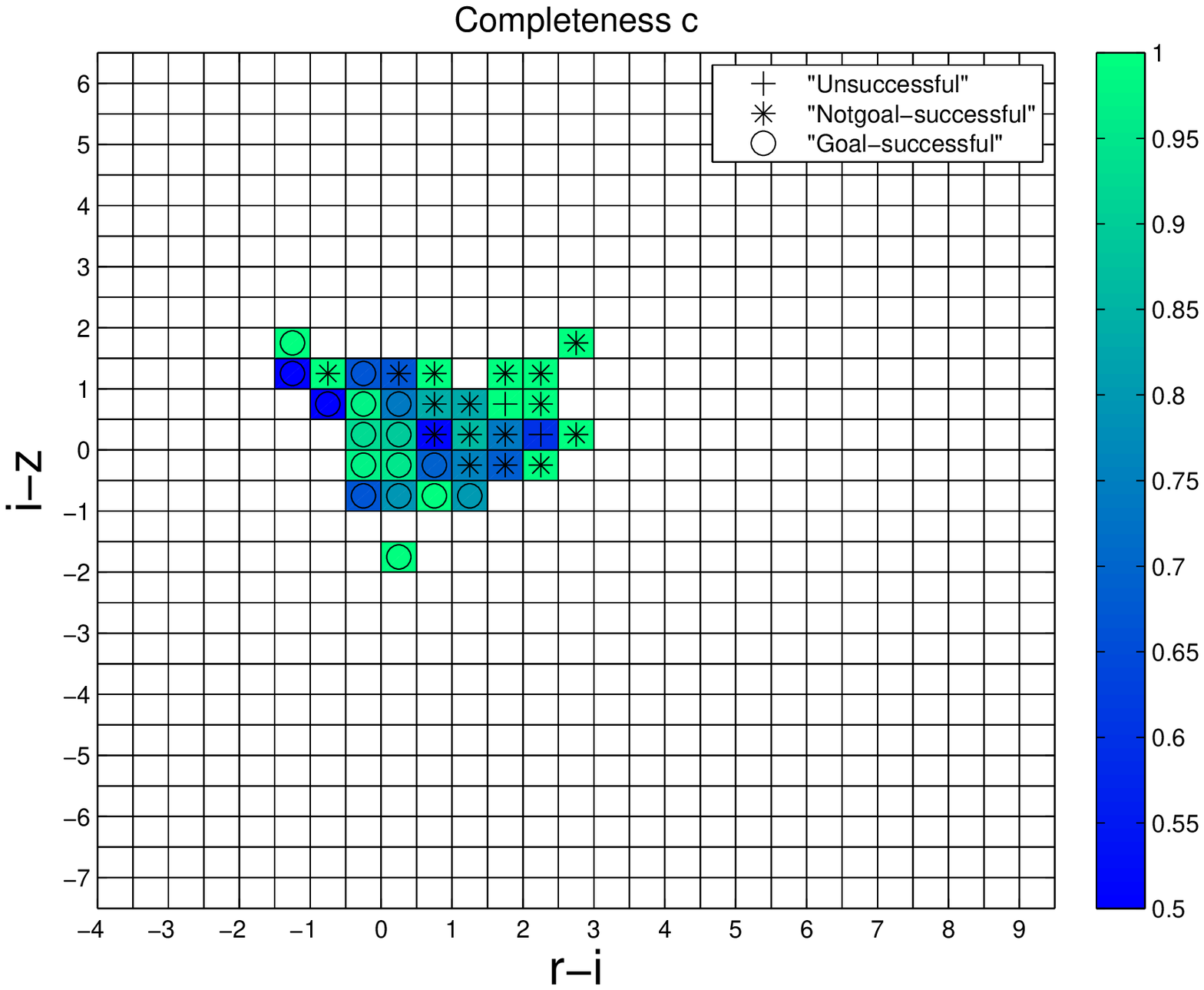}}\\
\caption{Local completeness estimates in the colour vs colour planes for the all types of clusters of experiment S-SCat.}
\label{colourcolourcompltot_cat}
\end{figure}

The local values of the efficiency and completeness of the selection process in the redshift vs colour and colour vs colour planes for the "goal-successful" clusters, evaluated using the BoK of the S-SCat experiment, are shown in figures (\ref{zcolourseffic_cat}, \ref{zcolourscompl_cat}) and (\ref{colourcoloureffic_cat}, \ref{colourcolourcompl_cat}) respectively. The meaning of these plots is described in section (\ref{Sec:experiments}). These plots can be used to inspect on a local scale the efficiency and completeness of the selection process in termes of colours and redshifts of the members of the BoK used for the experiment S-SCat. A more general view of the partition of the parameter space produced during the S-SCat experiment can be seen in figures (\ref{colourcoloureffictot_cat}) and (\ref{colourcolourcompltot_cat}), where the local efficiency and completeness of the selection process are shown in the colour vs colour planes for all types of clusters (i. e., not only "goal-successful", but also "notgoal-successful" and"unsuccessful" clusters). The meaning of the symbols is also explained in section (\ref{Sec:experiments}).

\subsection{The catalogue}
\label{Subsec:extractcat}

A catalogue of QSO candidates has been extracted from the SDSS DR7 photometric catalogue using "Method III", 
described in section (\ref{subsubsec:method3}) and based on the application of the Mahalanobis' distance to the 
distribution of photometric sources in the colour space and on the results of the experiment S-SCat. The total number 
of members of the catalogue of QSO candidates is $\sim 1,120,000$, which is in good agreement with the number 
of candidate quasars found in \cite{richards_2008}. The general efficiency and completeness of the candidate QSOs selection process are 92$\%$ and 93$\%$ respectively, evaluated according to the method described in paragraph (\ref{Sec:label}). The catalogue of candidate QSOs, for consistency with the SDSS footprint organization, has been splitted in several files, each corresponding to a different SDSS \emph{stripe} of the observed sky. In the SDSS jargon, a \emph{stripe} is defined by a line of constant survey latitude 
$\eta$, bounded on the north and south by the edges of the two \emph{strips} (scans along a constant $\eta$ value), 
and bounded on the east and west by lines of constant lambda. Because both \emph{strips} and \emph{stripes} are 
defined in "observed" space, they are rectangular areas which overlap as one approaches the poles (for more details see 
the web-page {\it http://www.sdss.org}). These files can be downloaded at the URL {\it http://voneural.na.infn.it/qso.html} 
as FITS files. Each file contains the fundamental SDSS parameters of the sources selected as candidates in the format listed in table (\ref{table_format}).

\begin{table*}
\label{table_format}      
\caption{Format of the catalogues of QSOs candidates.}             
\centering          
\begin{tabular}{rrr}    
\hline\hline       
Column & Format  & Description\\ 
\hline                    
1 & I7 & Unique catalogue number\\  
2 & A19 & SDSS Object ID\\
3 & F12.7 & Right Ascension in deg (J2000)\\   
4 & F11.7 & Declination in deg (J2000)\\
5 & F7.3 & $u$ psf dereddened magnitude\\
6 & F6.3 & $g$ psf dereddened magnitude\\
7 & F6.3 & $r$ psf dereddened magnitude\\
8 & F6.3 & $i$ psf dereddened magnitude\\
9 & F6.3 & $z$ psf dereddened magnitude\\  
10 & I7 & "Goal-successful" cluster index\\ 
\hline                  
\end{tabular}
\end{table*}

\subsection{Comparison with the literature}
\label{Subsec:comparison}

The catalogue of QSO candidates here discussed is composed by stellar sources which have been selected 
because associated, in the parameter space, to one the "goal-successful" clusters produced during the S-SCat 
experiment. In order to assess the validity of the extraction process against other catalogues of candidate quasars 
already available in the literature, we have crossmatched our catalogue with the "NBC Quasar Candidate Catalog", 
consisting of 100,563 unresolved, UV-excess (UVX) quasar candidates with magnitudes as faint as 21 in the $g$ band 
from 2099 square degrees of the SDSS DR1 imaging data (see \cite{richards_2004}). This catalogue, extracted using 
the KDE method described in some details in paragraph (\ref{Subsec:kdeselecalgorithm}) (the same method applied 
recently applied to the SDSS DR6 imaging data for the extraction of a larger catalogue of candidates not yet publicly 
available, see \cite{richards_2008}), has been cross-matched with our catalogue of candidate quasars. The comparison 
of these two catalogues, even if the limiting magnitudes and covered areas of the photometric observations from which 
the candidates have been extractred are different, shows that less than $1\%$ of the candidates located in the overlapping 
regions of the sky and belonging to the KDE catalogue are not found inside our catalogue, while there is a $\sim 3\%$ of 
our candidates (once the same photometric cuts have been applied to the catalogues) which do not match any source 
listed in the KDE catalogue.

\subsection{Number counts}
\label{Subsec:numbercounts}

Even though the efficiency and completeness of our photometrically selected sample of candidate QSOs are not unitary, 
the number count distribution of QSO candidates can be determined and used to characterize the candidates distribution 
in luminosity and for further comparison with other catalogues of QSOs candidate available in the literature. 

\begin{figure}
\centering
\includegraphics[width=12cm]{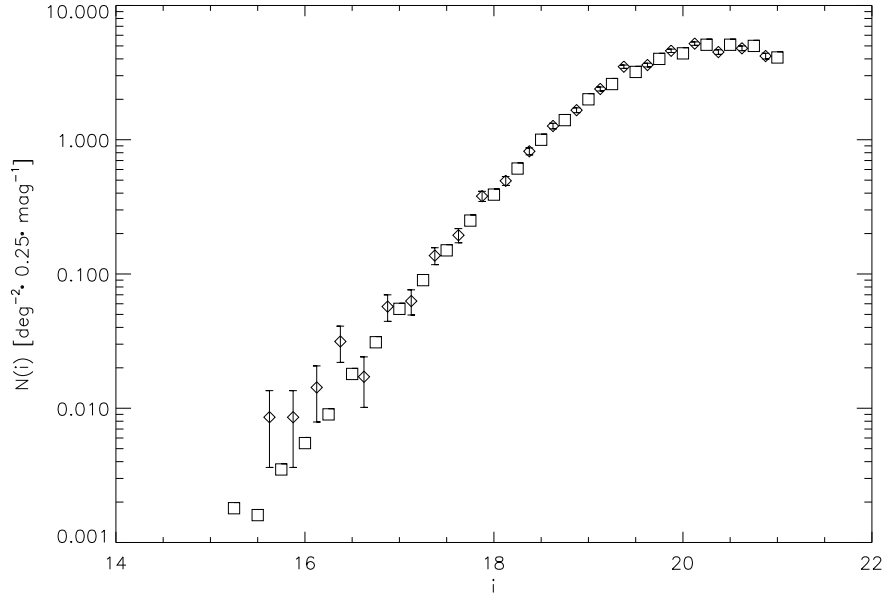}
\caption{Number counts of quasars candidates in the SDSS $i$ band. Diamonds indicate the number counts of our candidates while the squares represent the number counts obtained in the same SDSS $i$ band using the quasar candidates extracted from a reference catalogue of candidate QSOs.}
\label{number_counts}
\end{figure}

The number counts (NC) of our candidates have been calculated in the redshift interval $0.3 < z < 2.2$. The sample of candidates used for the NC is formed only by those sources belonging to the "goal-successful" clusters 1, 2 and 9 since, as can be noticed in the figure (\ref{zspec_cat}) and is pointed out in the description of the clusters in paragraph (\ref{Subsec:efficcomplet}), such clusters are formed by groups of sources located at redshifts lower than 2.2. Selecting QSOs candidates in redshift by distinct clusters in the parameter space is reasonable since such clusters, formed by sources belonging to the BoK used for the esperiment, are clearly separated in redshift and the photometric QSOs candidates are expected to reflect the same partition along the redshift axis. In order to obtain an estimate of the NC in the $i$ band (figure \ref{number_counts}), the efficiency e($i$) and completeness c($i$) have been used to correct for the performances of the selection process according to the following recipe. $N_{cat}(i)$, the density of candidates extracted from the catalogue in the generic magnitude interval $[i, i + di]$, is multiplied and divided by the efficiency e($i$) and completeness c($i$) respectively, measured in the same magnitude interval and yielding the corrected NC density $N(i) = N_{cat}(i) \cdot e(i)/c(i)$. Our NC estimates (diamonds in \ref{number_counts}) are in fair agreement with the reference NC derived from the catalogue of candidate quasars produced using the KDE method in \cite{richards_2008} (open squares). The error bars represent poissonian errors on the data points. 

\section{Discussion and conclusions}
\label{Sec:conclusions}
    
We have presented a new unsupervised method to perform quasar candidate selection, based on the clustering of astronomical sources in the parametric space defined by the photometric colours. The method requires a suitably large base of knowledge (BoK) which is used only for labelling purposes. The method consists of three steps: 
i) an unsupervised clustering performed by Probabilistic Principal Surfaces algorithm; ii) an agglomerative process driven 
by a measure of negative entropy, and iii) a fine tuning of the clustering performances through the exploitation of 
the information contained in the BoK. In the experiments described here, the BoK consists of the spectroscopic classification 
provided by the "specClass" SDSS flag which is available for a relatively small fraction of the photometric sources observed by 
SDSS. Extensive testing on different data sets and parameter combinations and the comparison of the performance obtained in this
experiments with the performance of the methods found in the literature and based on other techniques have shown 
that this method, which is based on a completely data-mining ground and employs tools never applied before to 
astronomical problems, can achieve better performances than most methods published in the literature so far (see section
(\ref{Sec:experiments}). The method has also been applied to the SDSS-DR7 data sets using the results of a specific experiment 
(S-SCat) and a catalogue of QSO candidates has been produced. This catalogue is publicly available at the web page {\it http://voneural.na.infn.it/qso.html}. Even though the method of unsupervised clustering in astronomical  parameter spaces described in this 
paper has been adjusted to the case of SDSS QSOs selection, such fine tuning depends only on the information 
contained in the BoK and it can be applied to any similar case provided that a suitably large and complete base of 
knowledge is available. In this context, the role played by the Virtual Observatory for the future evolution 
of the data mining approach to classification/selection problems in astronomy, will be extremely important 
because the VO will allow the construction of BoKs of unprecedented accuracy and completeness obtained by 
federating and standardizing the information contained in most (if not all) astronomical databases worldwide, 
thus providing the most natural environment for the further development and exploitation of similar techniques.\\

\noindent {\it Acknowledgements:} 
This work was funded through the Ph.D. program of the University Federico II in Napoli, 
the European VOTECH consortium and the PON-SCOPE (financed by the Italian MIUR) and the
italian Ministry of Foreign Affairs in the framework of the bilateral Italy-Usa agreement. R.D'A. acknowledges the Marie Curie EARA-EST action for financial support during his visit at the Institute of Astronomy, Cambridge (UK). The authors are grateful to the anonymous referee for useful suggestions that greatly helped to improve the paper. The authors also wish to thank M. Brescia, 
S. Cavuoti, E. De Filippis, O. Laurino, M. Paolillo and D. Capozzi for many useful discussions. A special thank goes to 
A. Staiano and A. Ciaramella for making available to us their implementations of the PPS and NEC algorithms.

\appendix

\section{Local estimates of the efficiency and completeness}
\label{appendix}

\begin{figure}
\centering
\subfigure[Efficiency in $z$ vs $u-g$ plane]{\includegraphics[width=6cm]{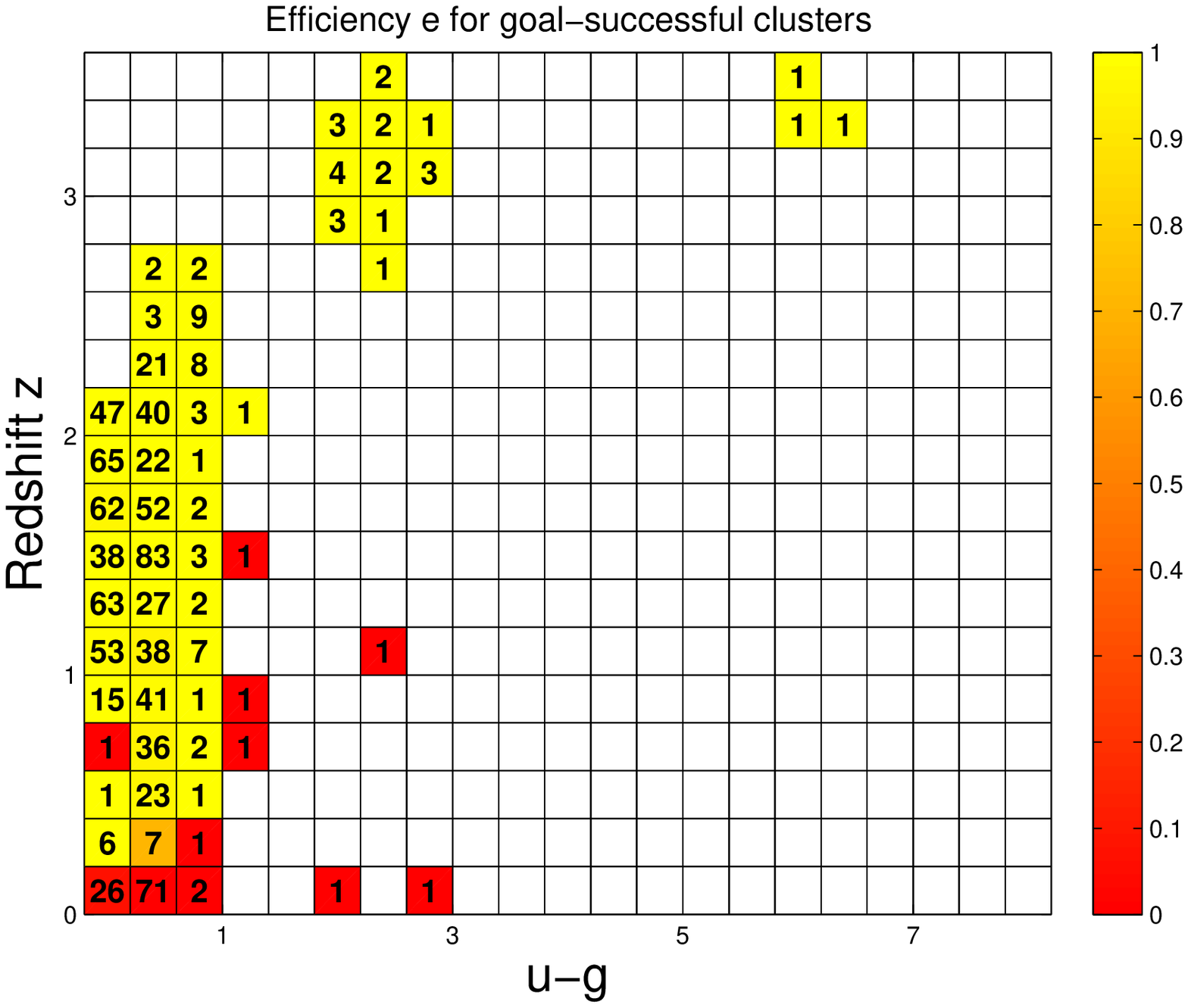}}
\subfigure[Completeness in $z$ vs $u-g$ plane]{\includegraphics[width=6cm]{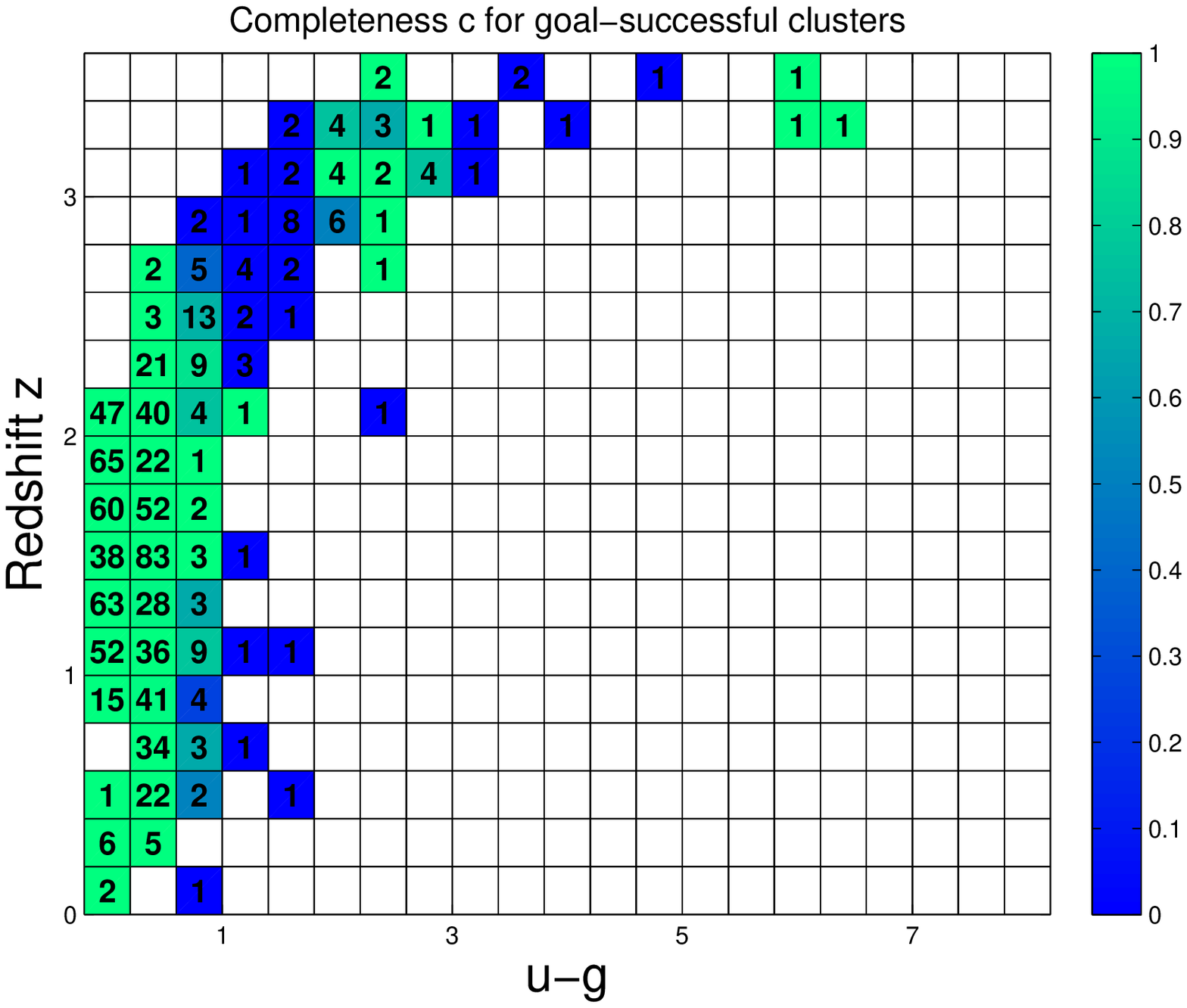}}\\
\subfigure[Efficiency in $z$ vs $g-r$ plane]{\includegraphics[width=6cm]{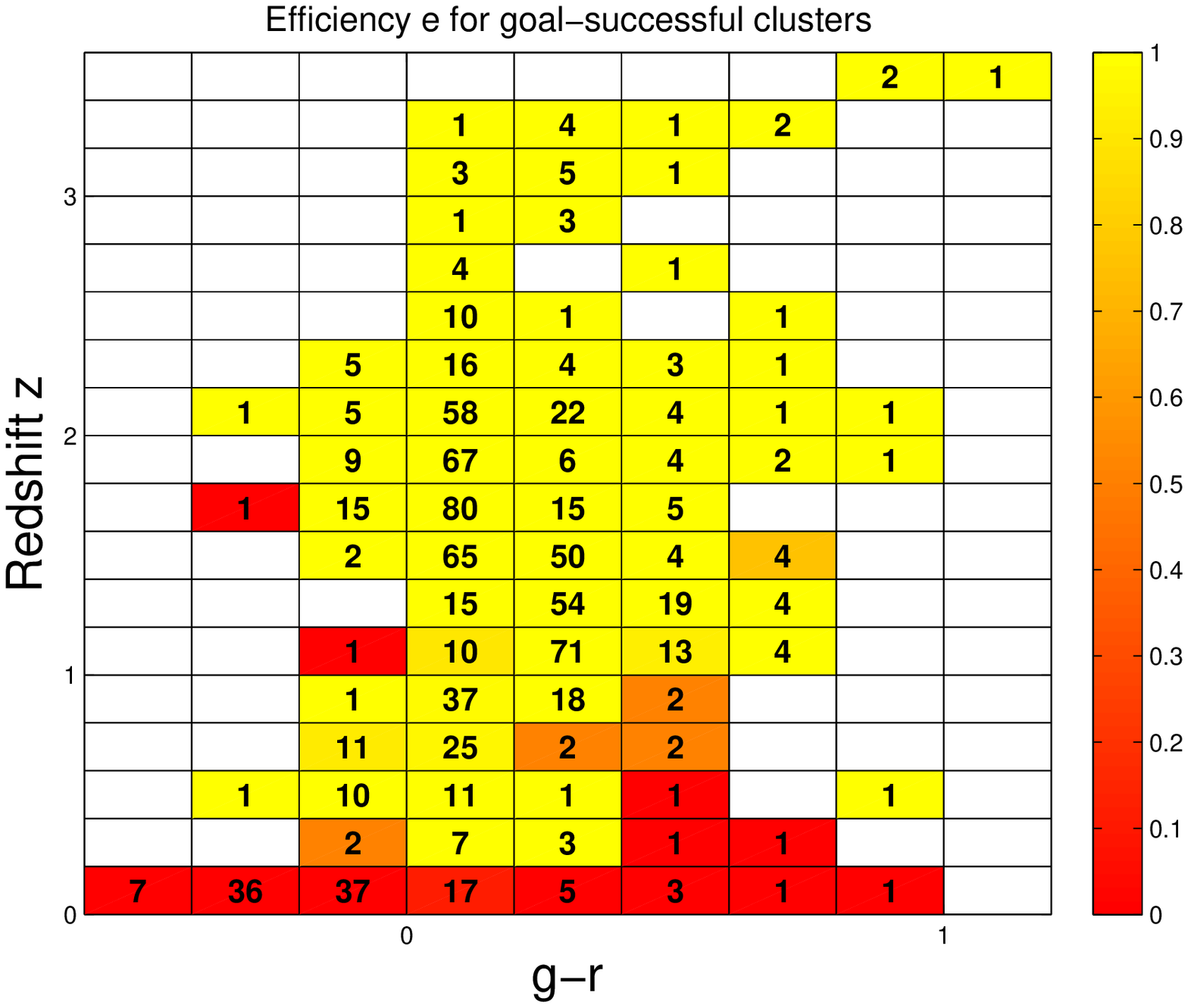}}
\subfigure[Completeness in $z$ vs $g-r$ plane]{\includegraphics[width=6cm]{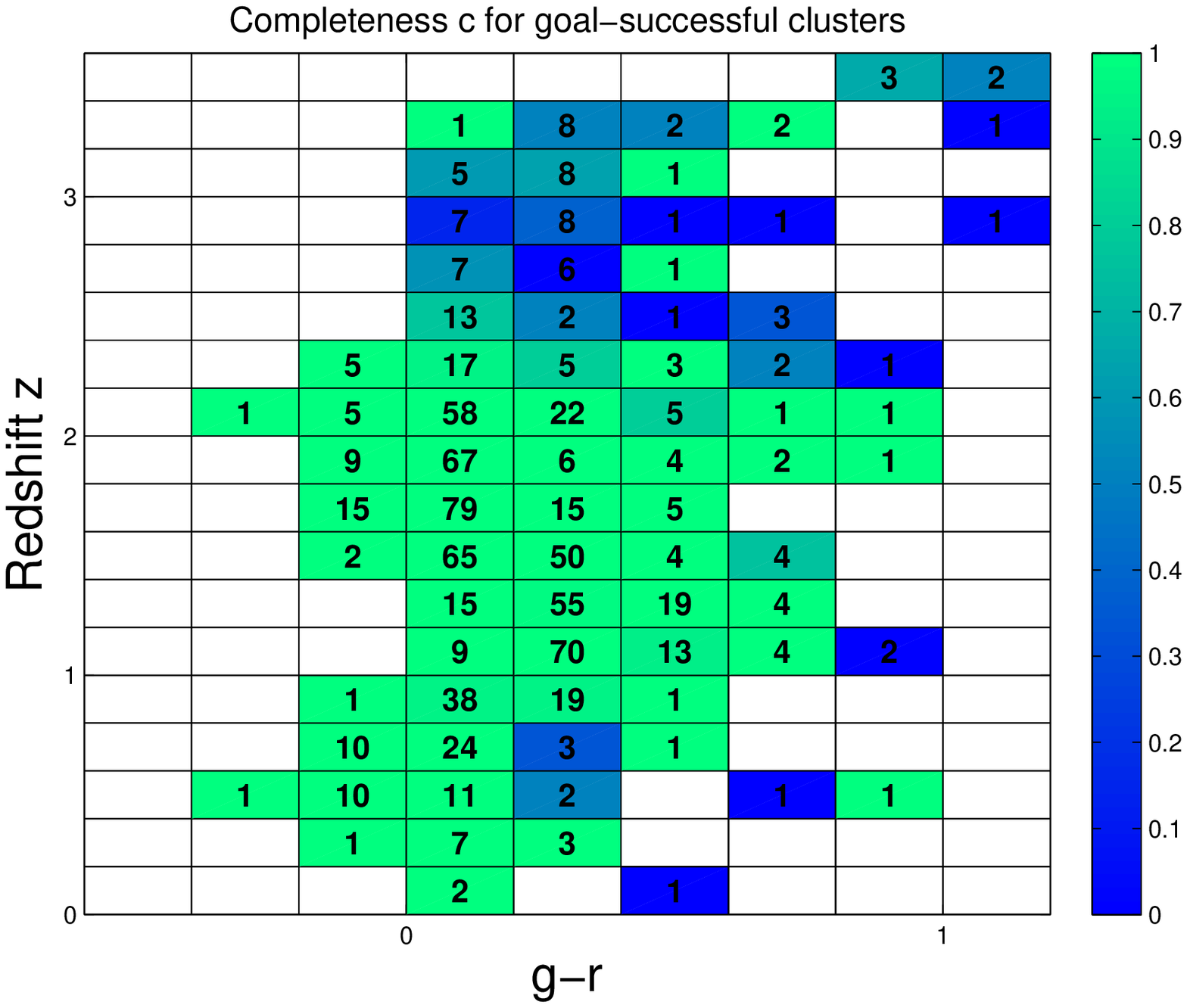}}\\
\subfigure[Efficiency in $z$ vs $r-i$ plane]{\includegraphics[width=6cm]{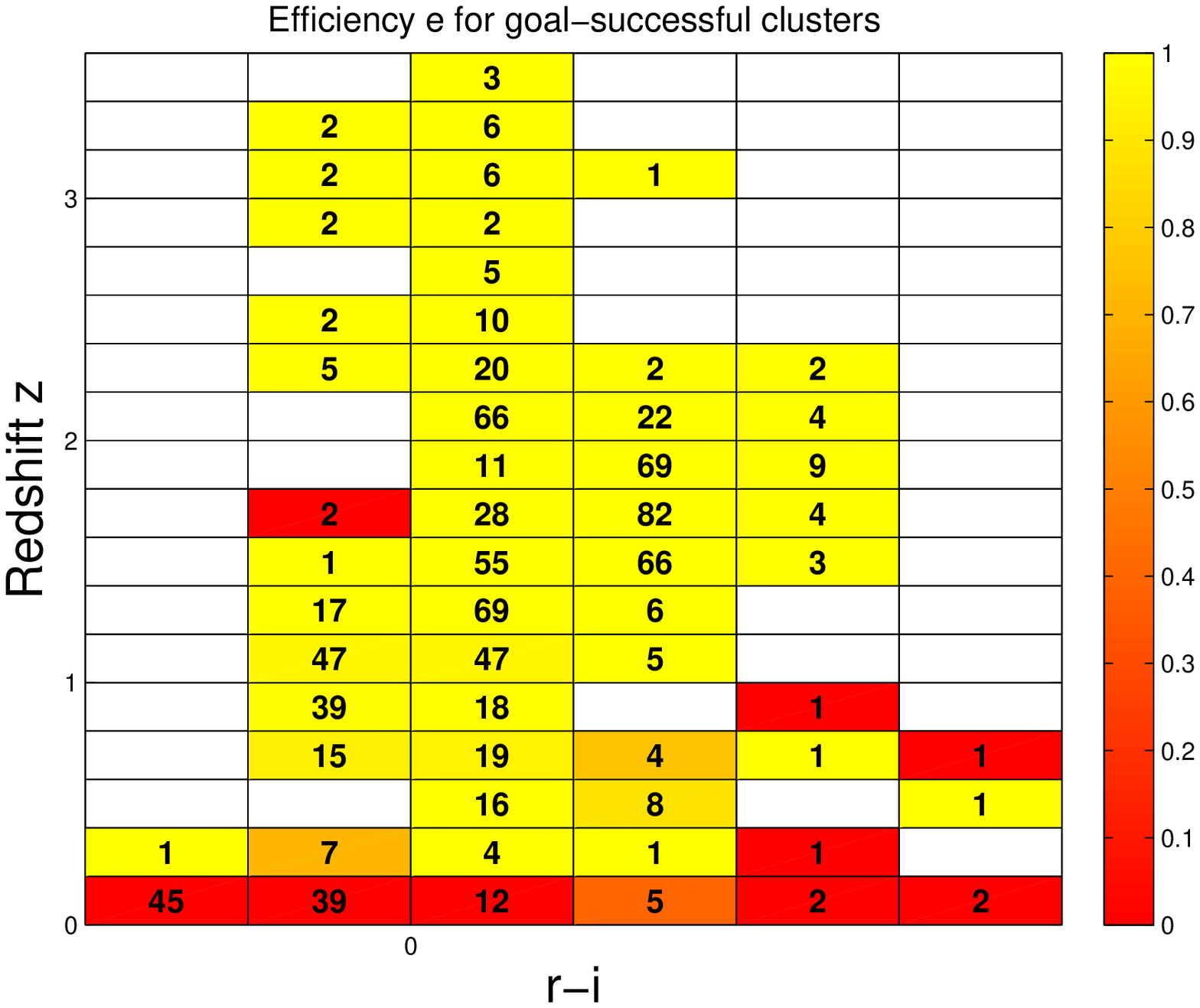}}
\subfigure[Completeness in $z$ vs $r-i$ plane]{\includegraphics[width=6cm]{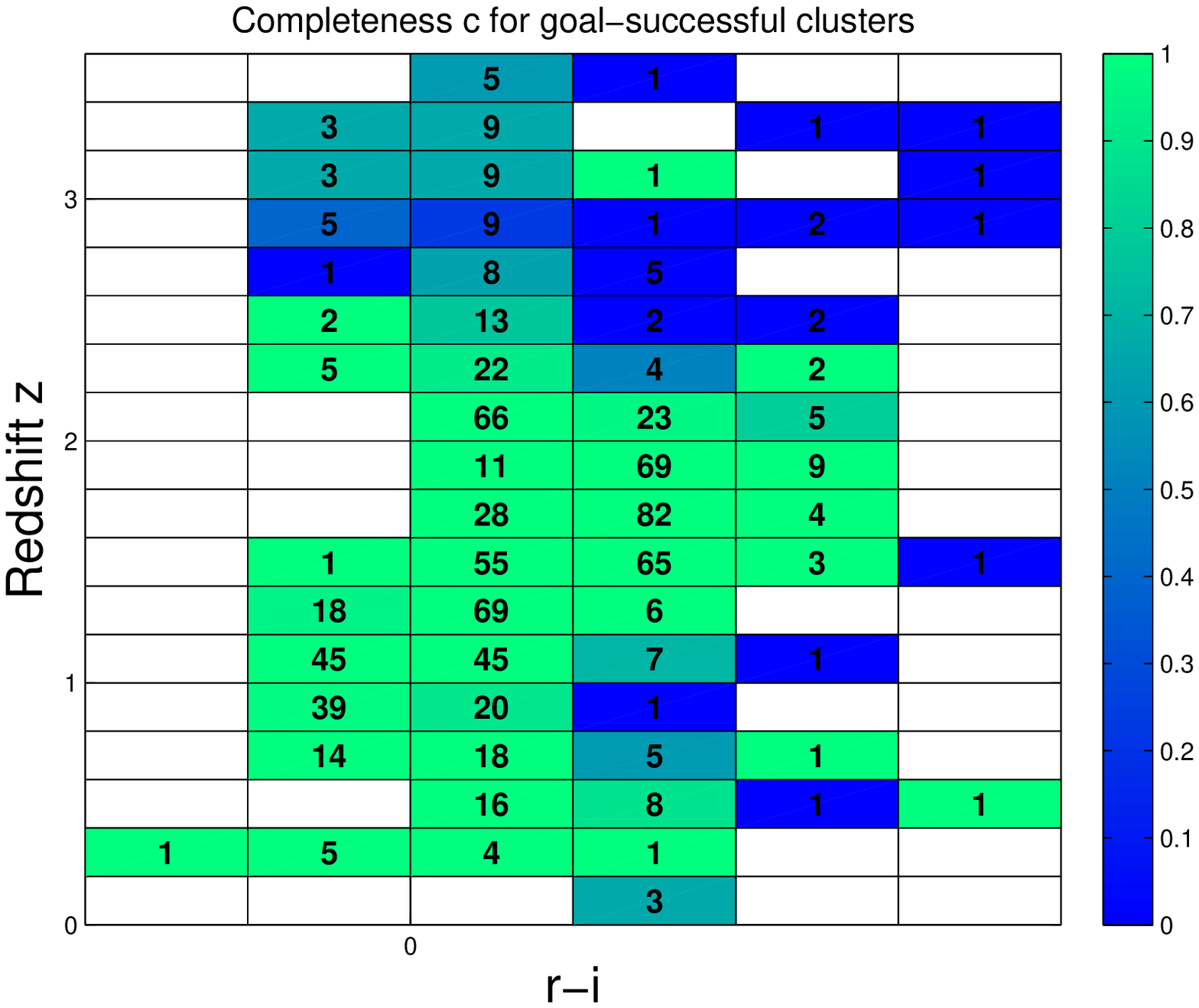}}\\
\subfigure[Efficiency in $z$ vs $i-z$ plane]{\includegraphics[width=6cm]{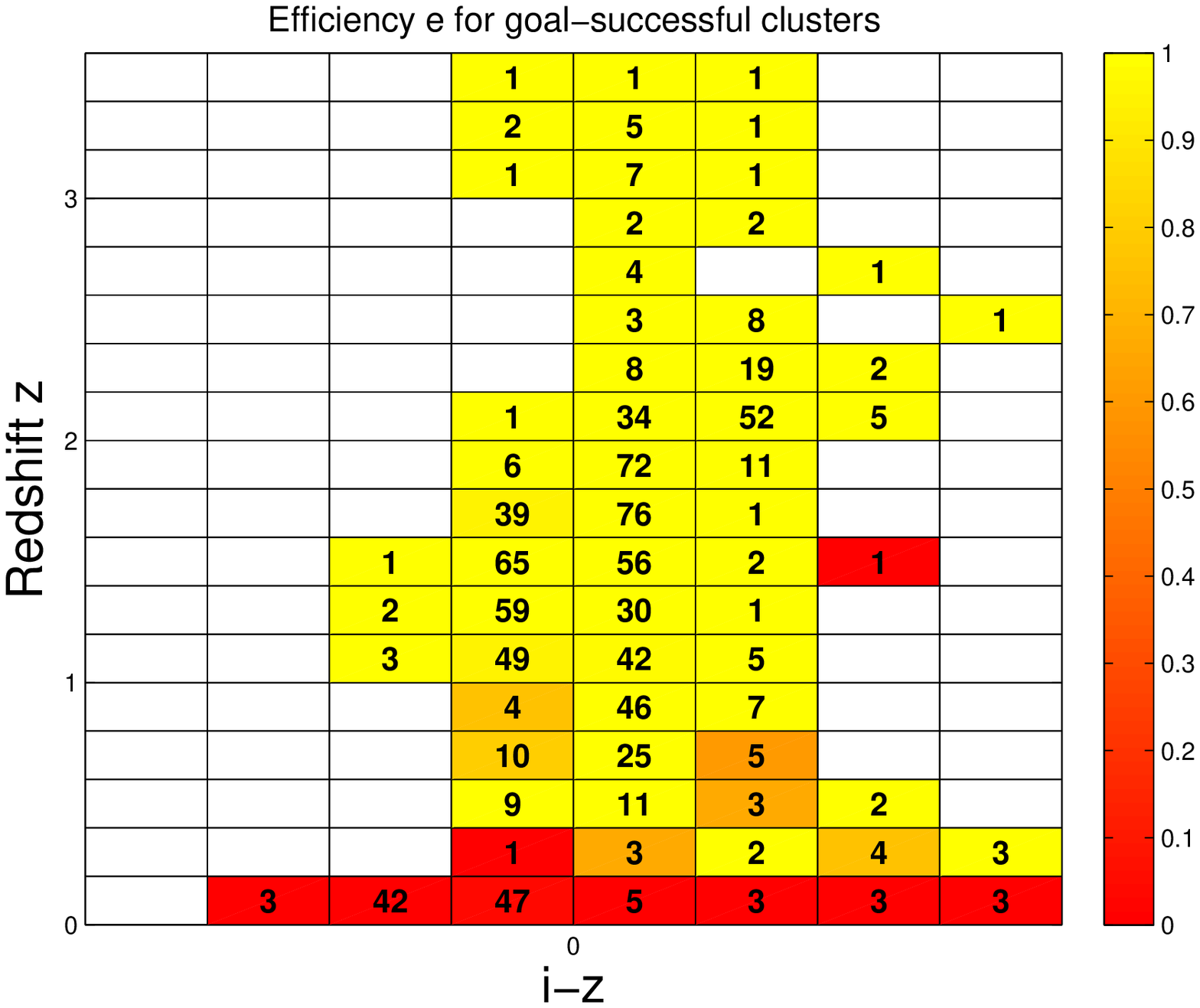}}
\subfigure[Completeness in $z$ vs $i-z$ plane]{\includegraphics[width=6cm]{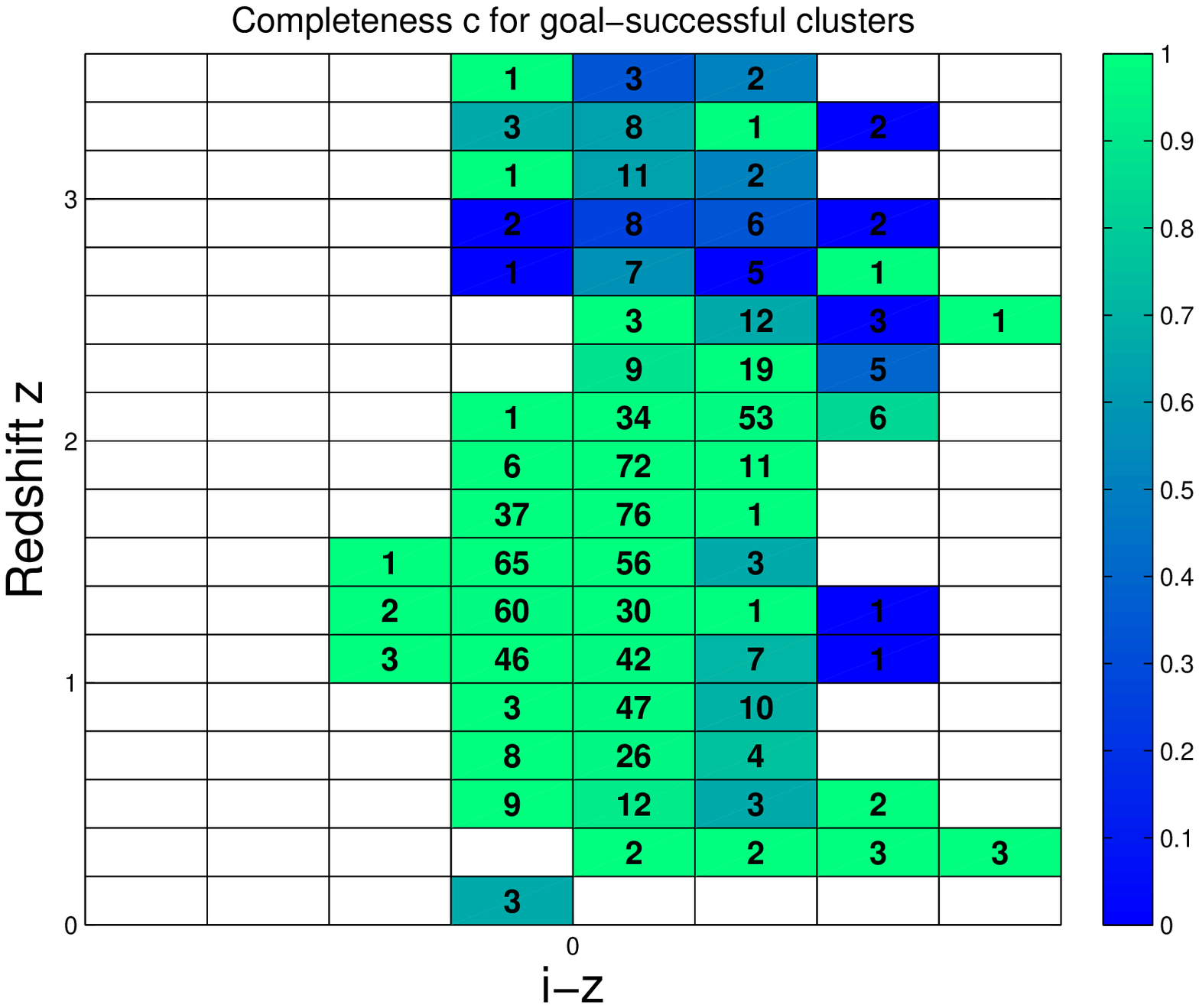}}\\
\caption{Local efficiency (left-hand panels) and completeness (right-hand panels) estimates in the redshift vs optical colours planes for the labelling
phase of the first experiment. Colour of the cell is associated to the efficiency and completeness of the selection process
of the "goal-successful" clusters, while numbers contained in each cell represent the total number of candidates in that 
region of the plane. Lower left-hand and right hand panels contain the maximum efficiency and completeness for the selection 
process of the three types of clusters produced by the algorithm. The class of clusters contributing the maximum fraction to the efficiency
and completeness represented by the symbol contained in each cell of the plane.}
\label{fig:paramparam_1app_1exp}
\end{figure}

\begin{figure}
\centering
\subfigure[Efficiency in $z$ vs $u-g$ plane]{\includegraphics[width=6cm]{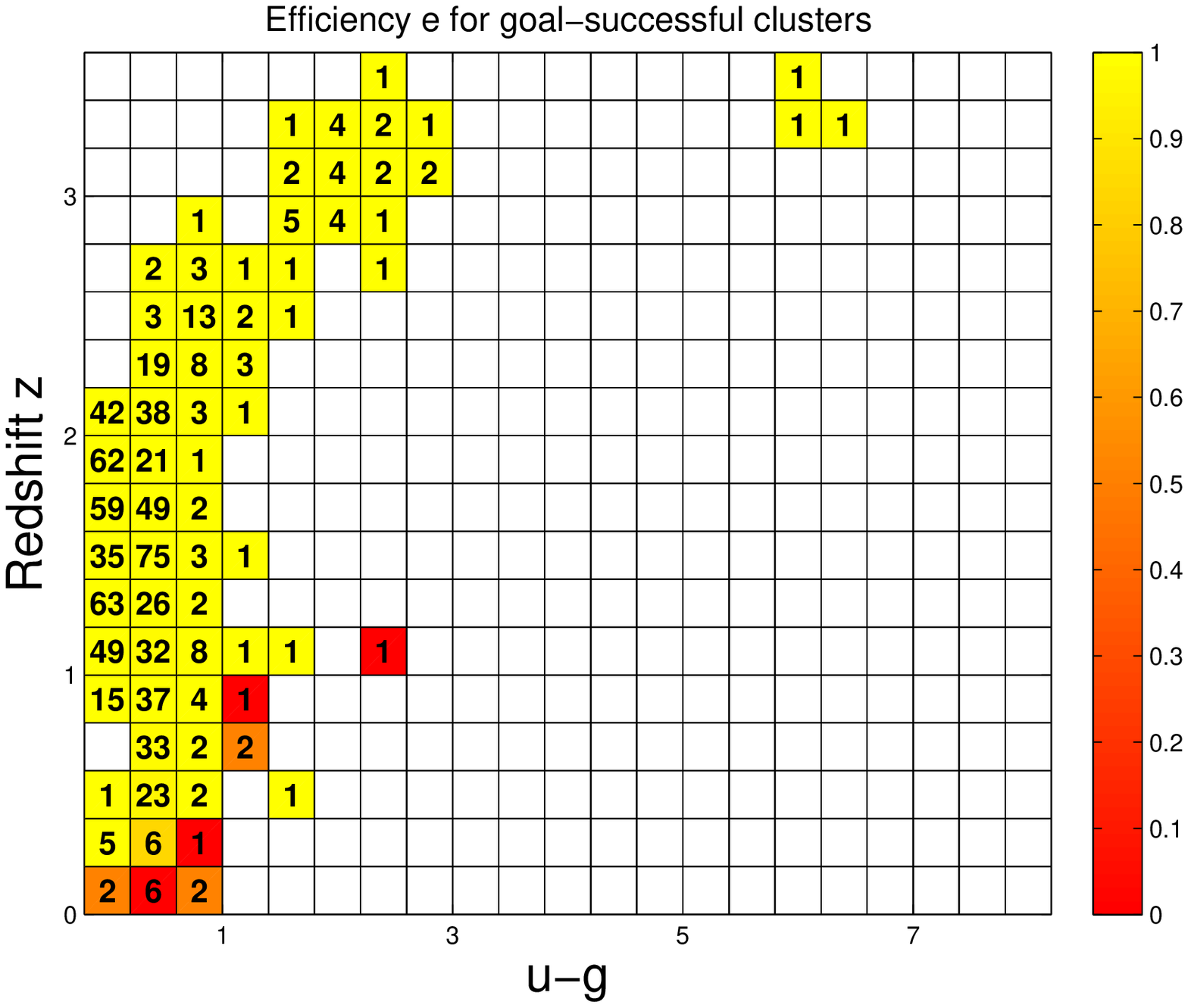}}
\subfigure[Completeness in $z$ vs $u-g$ plane]{\includegraphics[width=6cm]{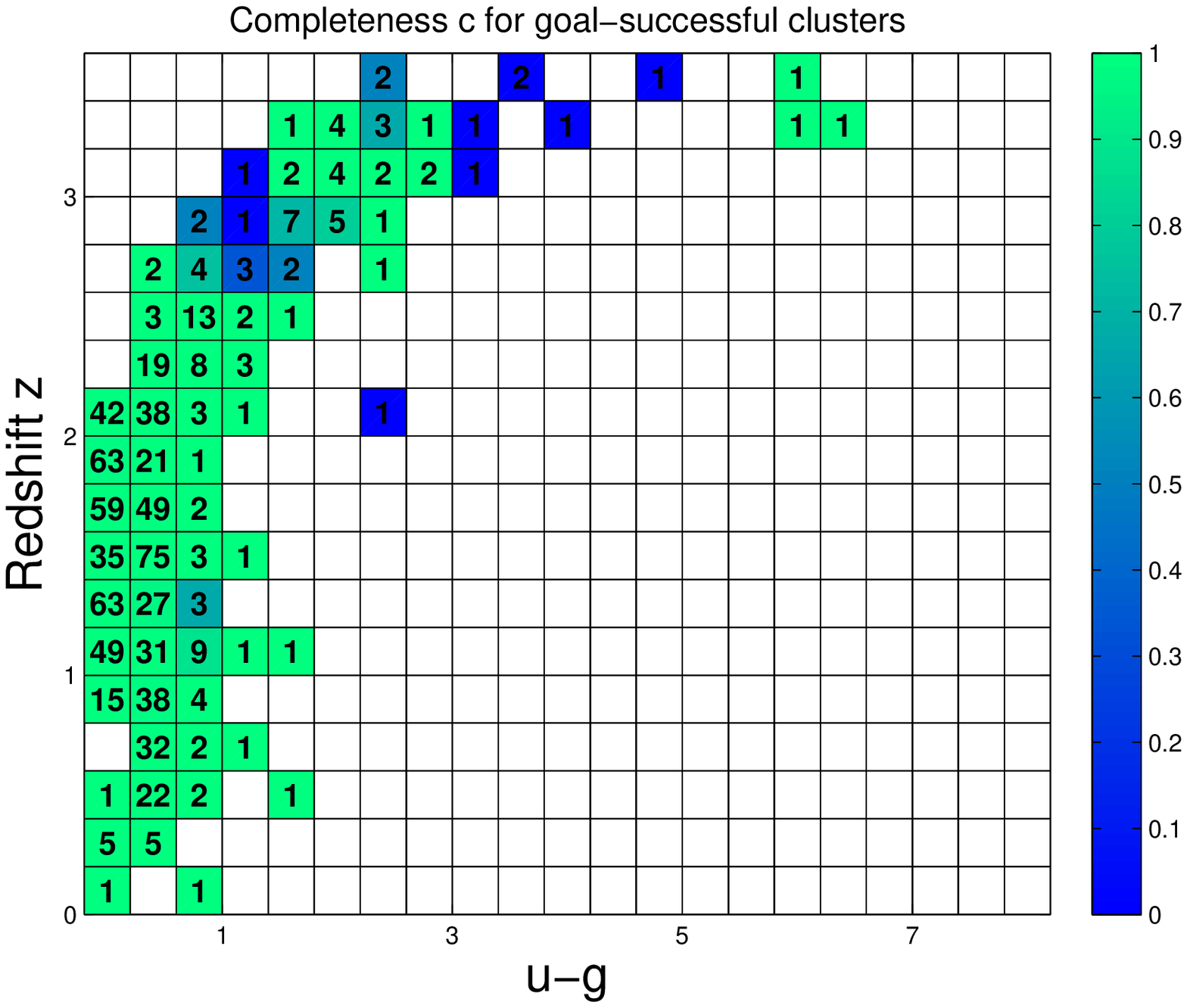}}\\
\subfigure[Efficiency in $z$ vs $g-r$ plane]{\includegraphics[width=6cm]{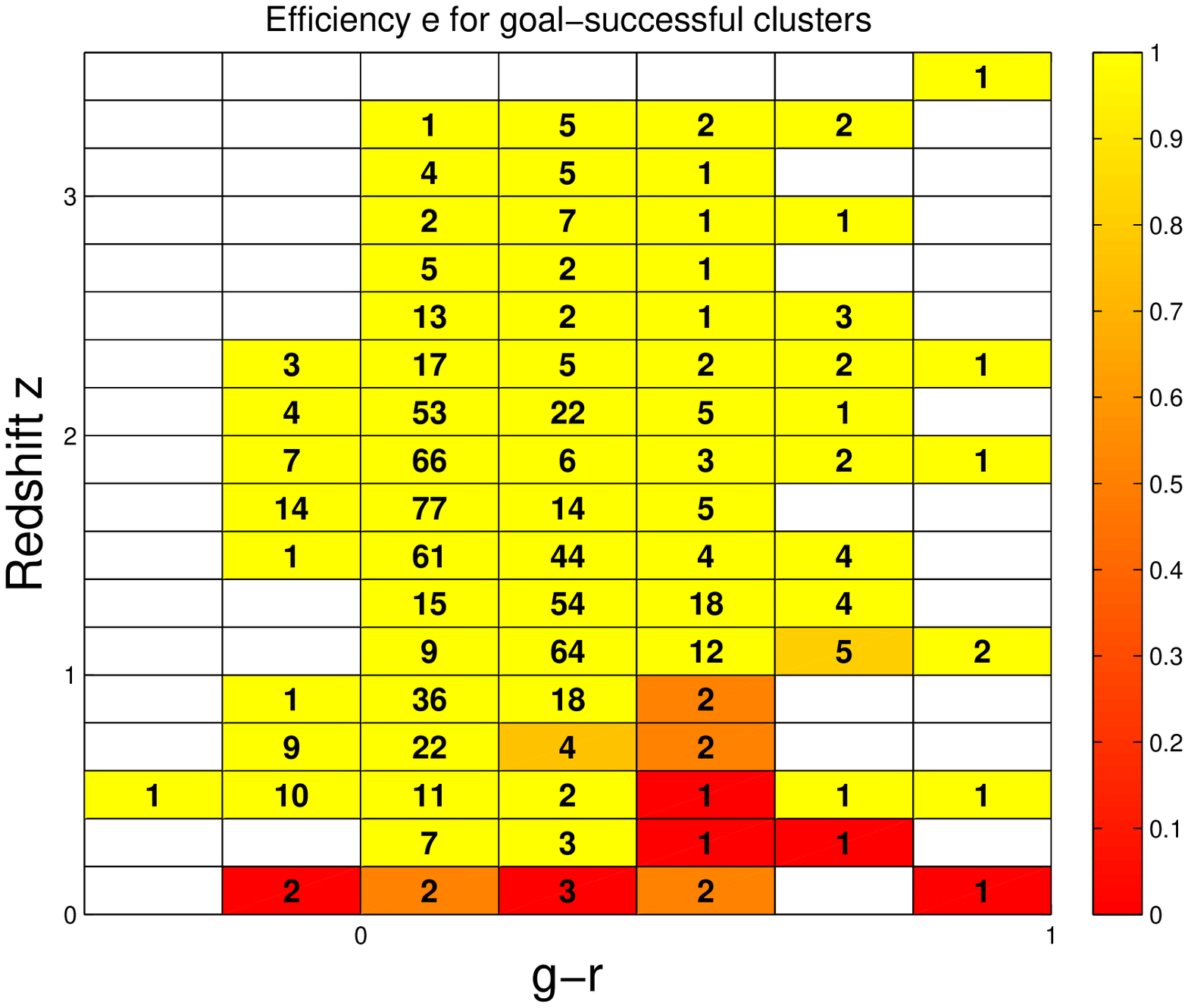}}
\subfigure[Completeness in $z$ vs $g-r$ plane]{\includegraphics[width=6cm]{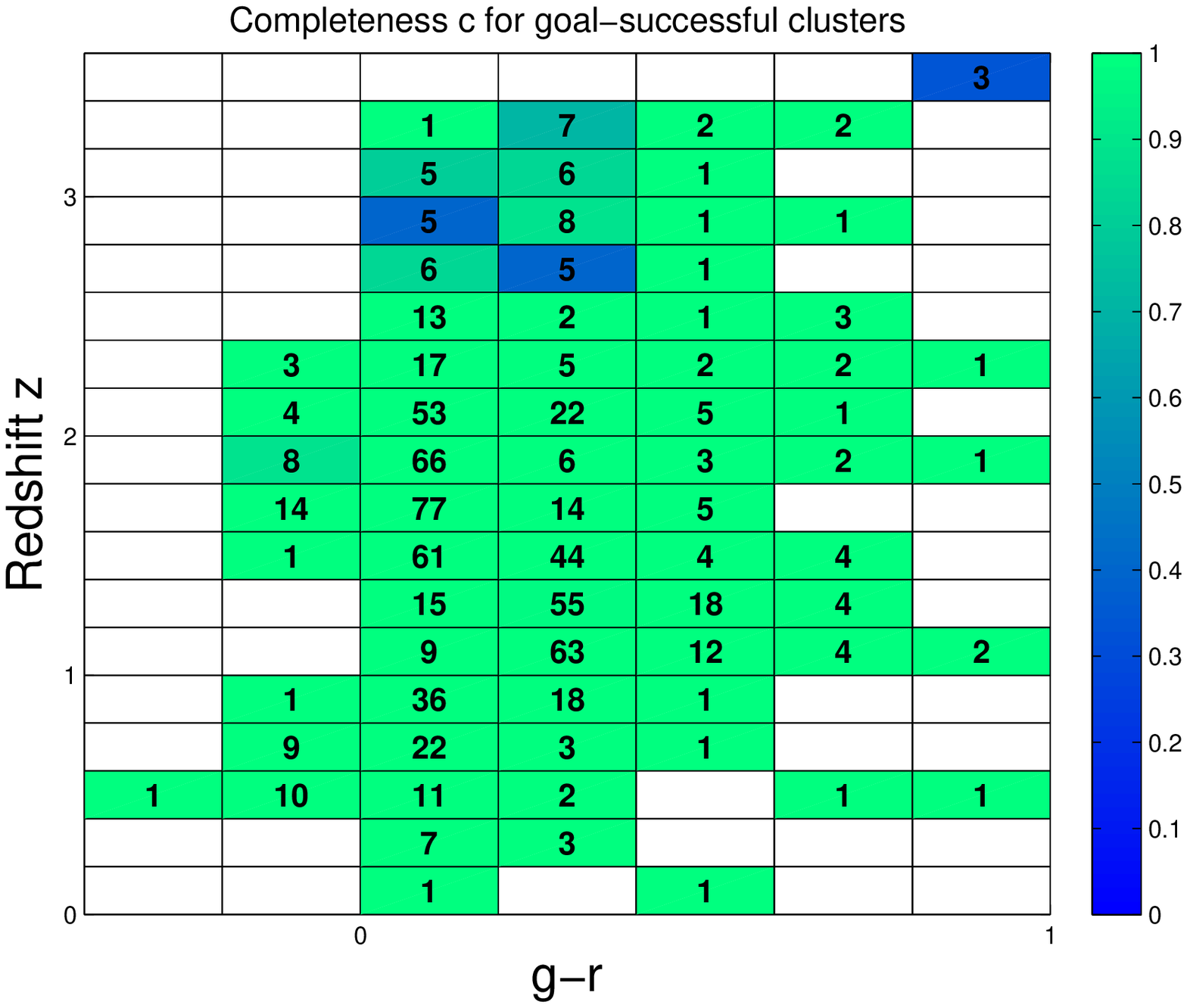}}\\
\subfigure[Efficiency in $z$ vs $r-i$ plane]{\includegraphics[width=6cm]{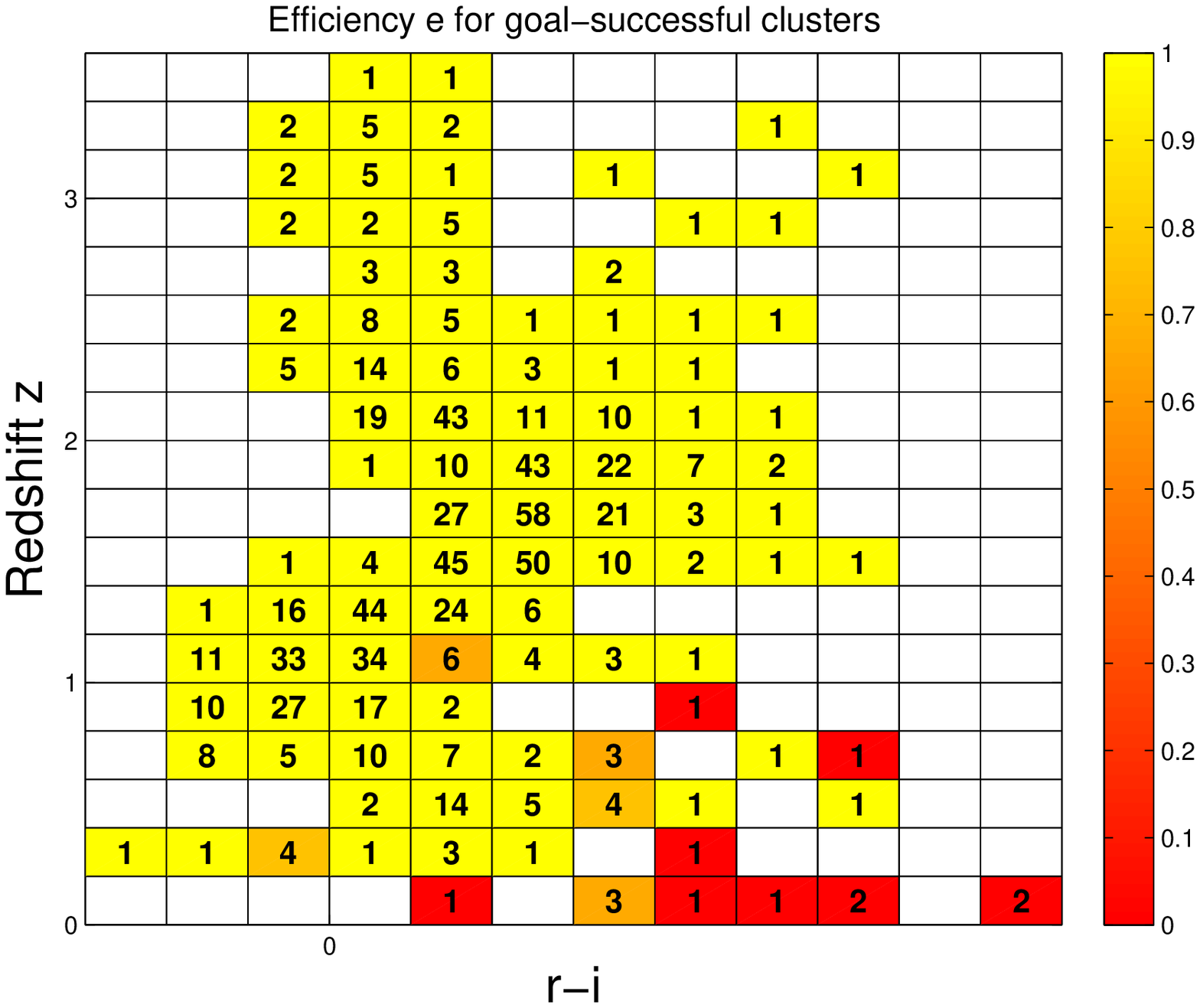}}
\subfigure[Completeness in $z$ vs $r-i$ plane]{\includegraphics[width=6cm]{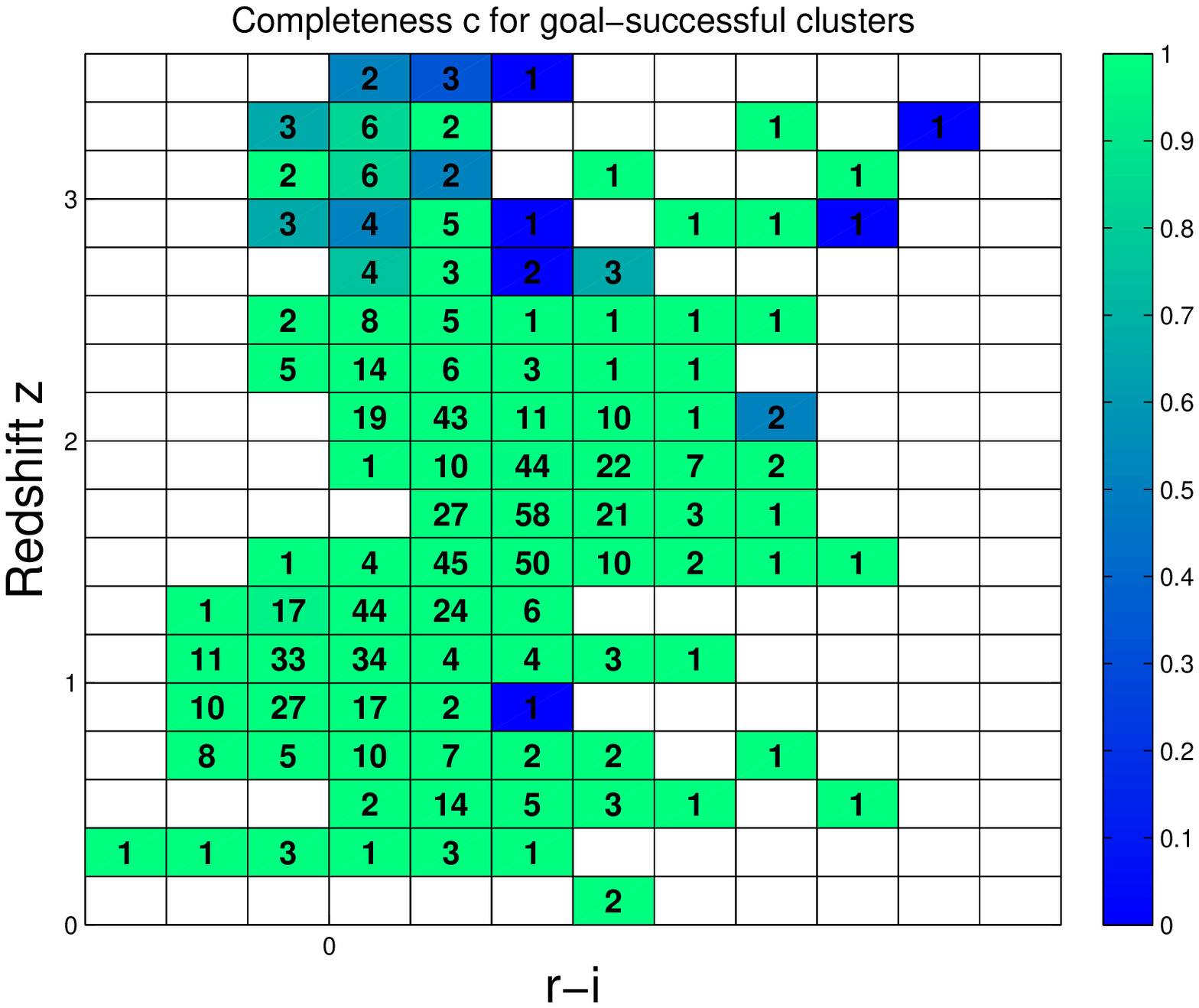}}\\
\subfigure[Efficiency in $z$ vs $i-z$ plane]{\includegraphics[width=6cm]{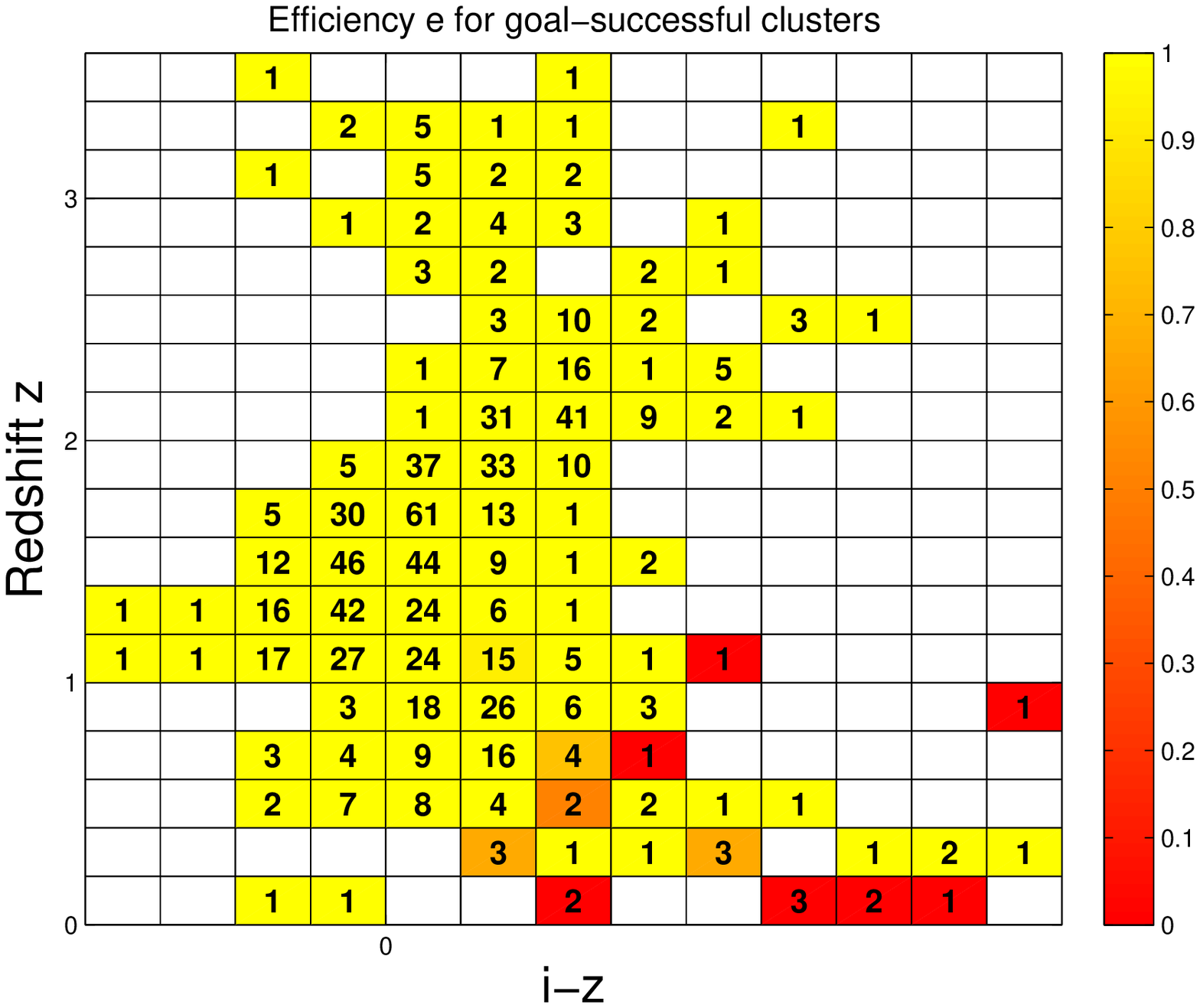}}
\subfigure[Completeness in $z$ vs $i-z$ plane]{\includegraphics[width=6cm]{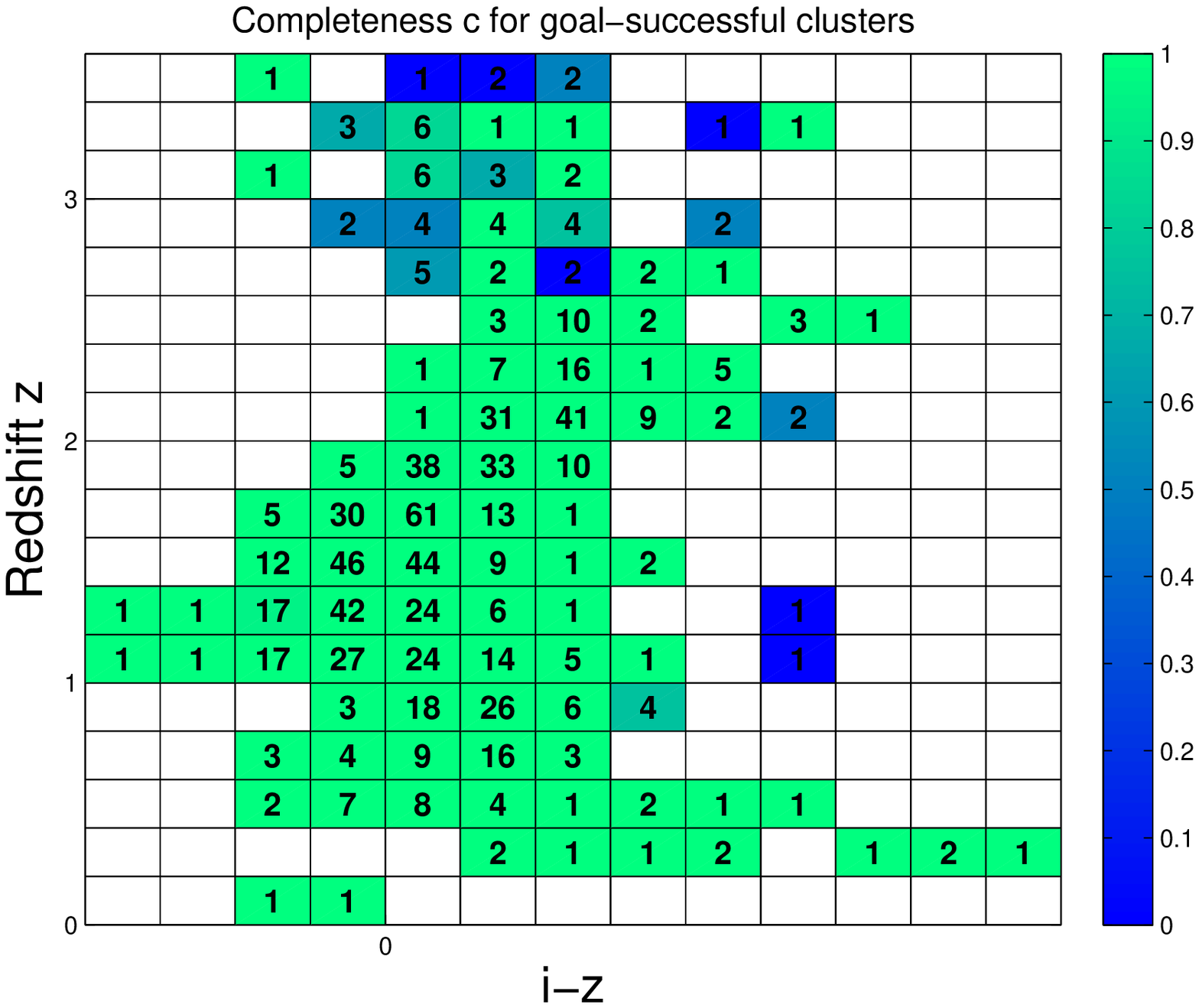}}\\
\caption{Local efficiency and completeness for the second experiment (optical colours). See previous figure for details.}
\label{fig:paramparam_1app_2exp_opt}
\end{figure}

\begin{figure}
\centering
\subfigure[Efficiency in $z$ vs $Y-J$ plane]{\includegraphics[width=6cm]{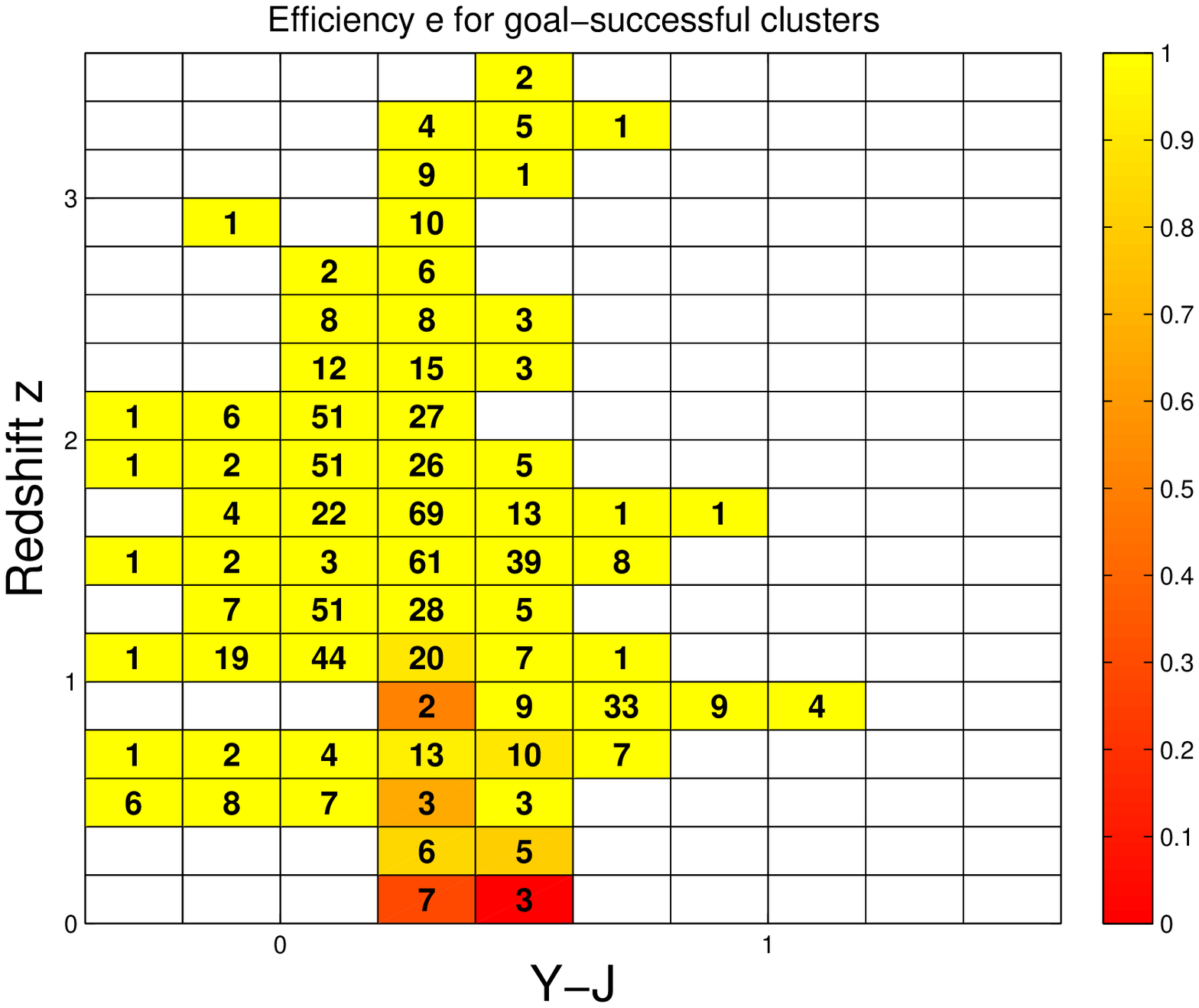}}
\subfigure[Completeness in $z$ vs $Y-J$ plane]{\includegraphics[width=6cm]{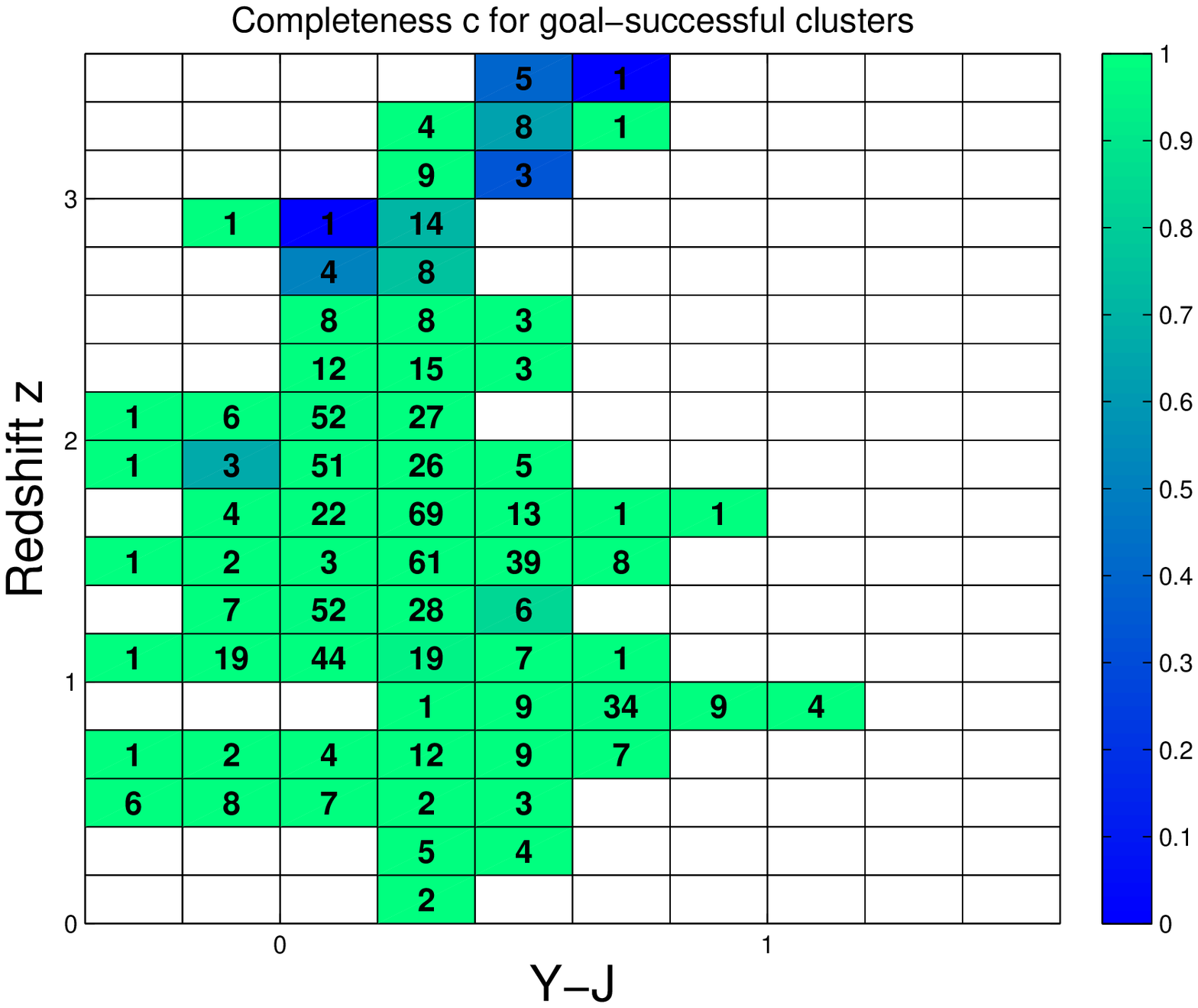}}\\
\subfigure[Efficiency in $z$ vs $J-H$ plane]{\includegraphics[width=6cm]{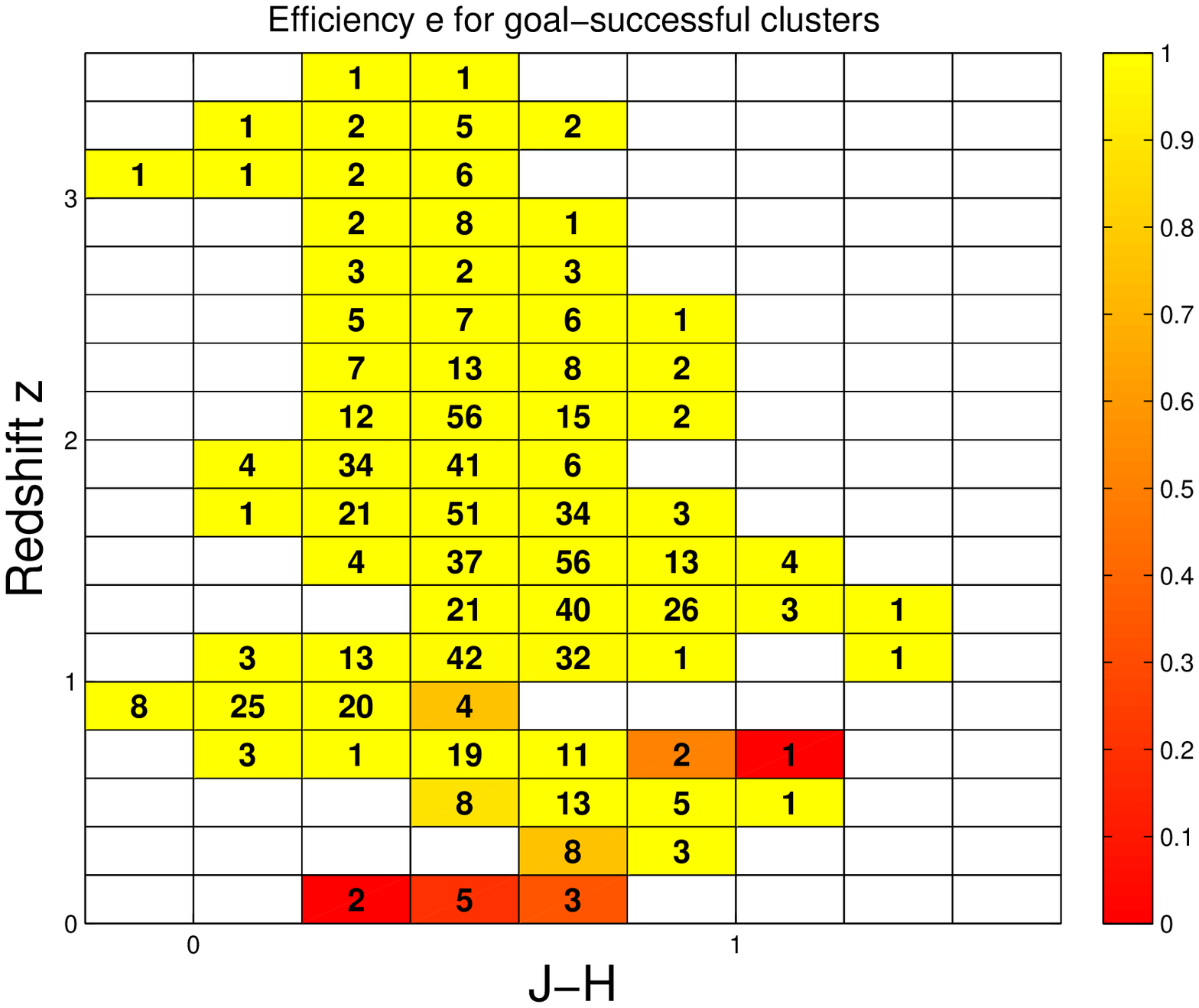}}
\subfigure[Completeness in $z$ vs $J-H$ plane]{\includegraphics[width=6cm]{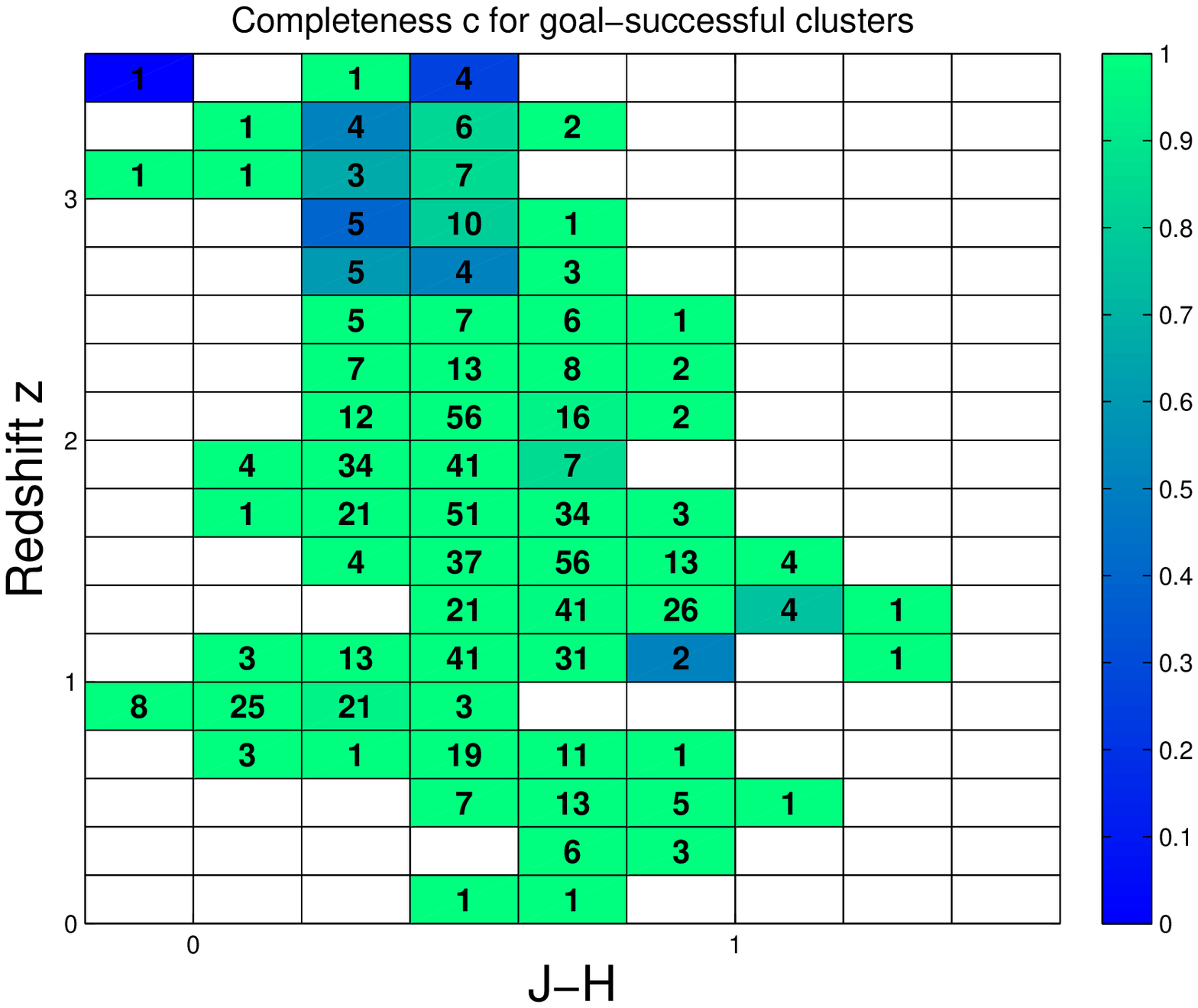}}\\
\subfigure[Efficiency in $z$ vs $H-K$ plane]{\includegraphics[width=6cm]{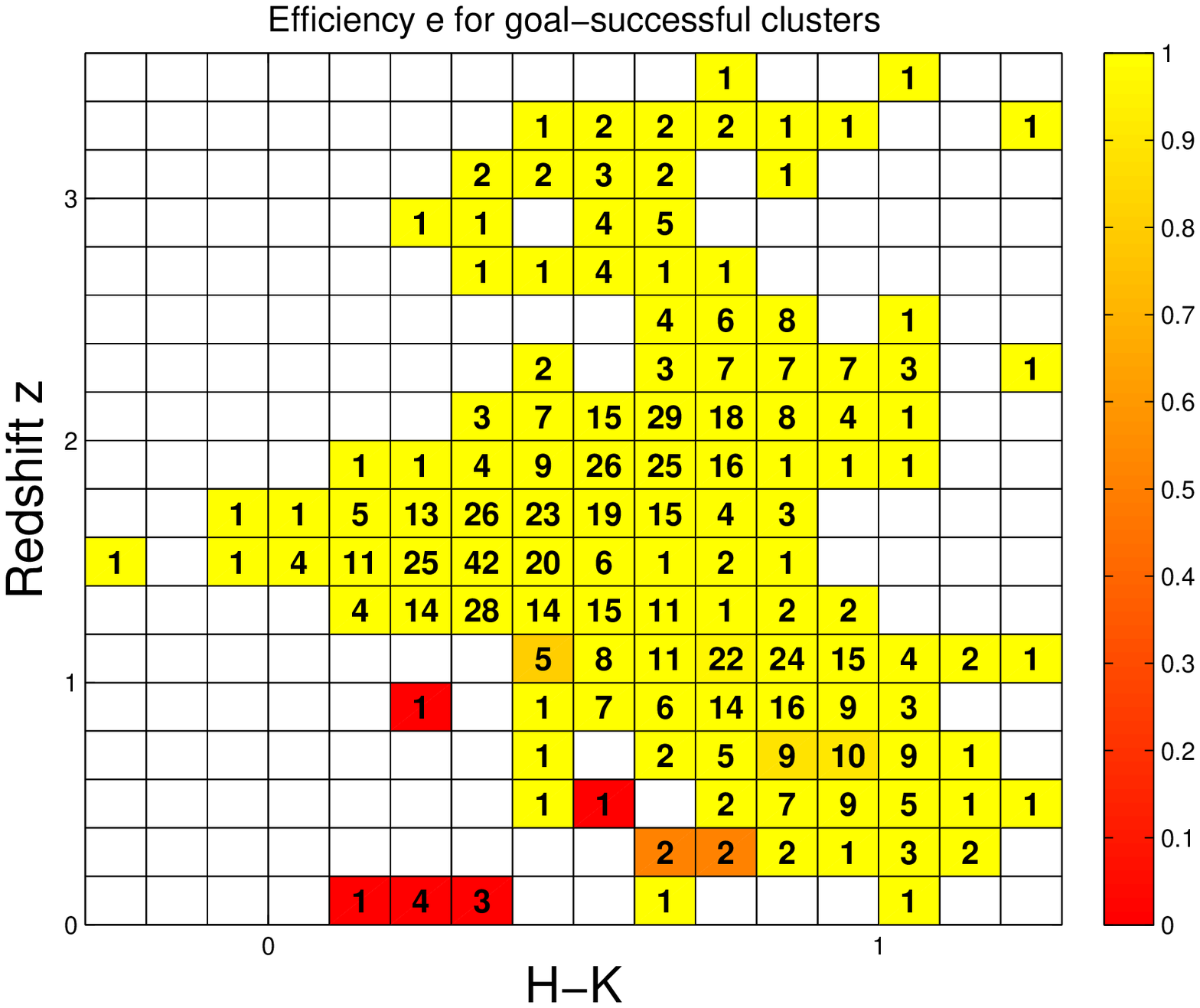}}
\subfigure[Completeness in $z$ vs $H-K$ plane]{\includegraphics[width=6cm]{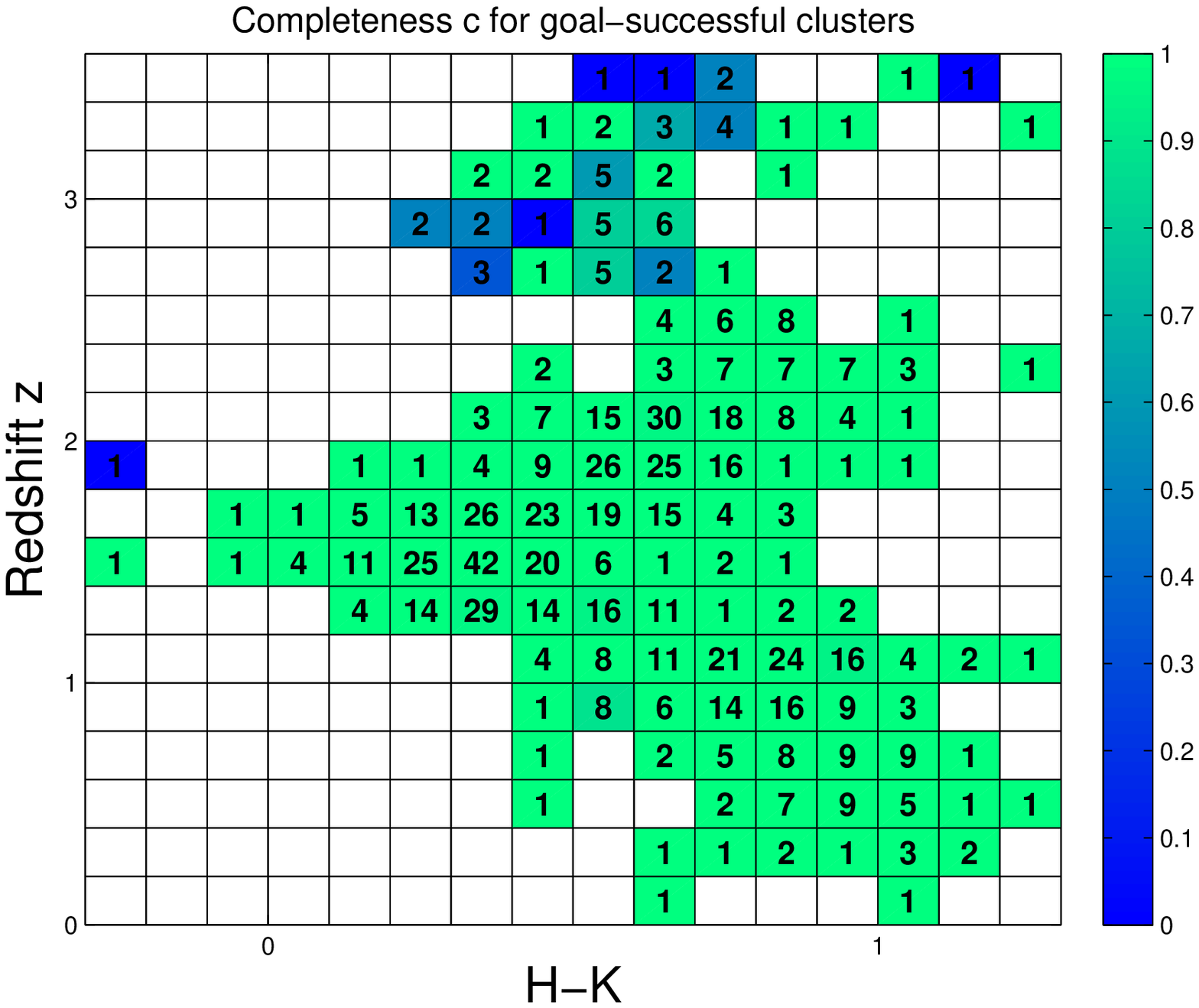}}\\
\caption{Local efficiency and completeness for the second experiment (infrared colours).}
\label{fig:paramparam_1app_2exp_infrared}
\end{figure}

\begin{figure}
\centering
\subfigure[Efficiency in $z$ vs $u-g$ plane]{\includegraphics[width=6cm]{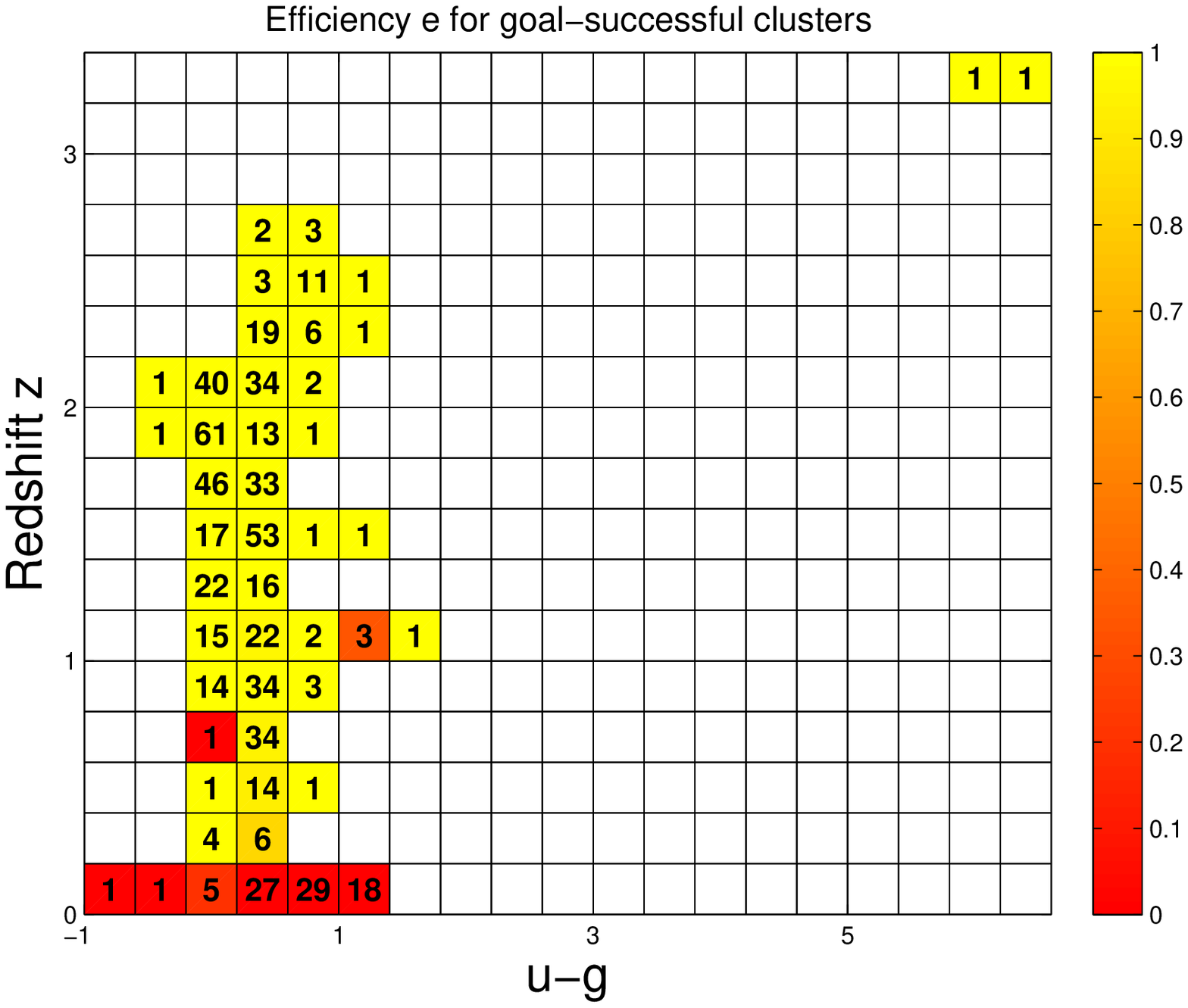}}
\subfigure[Completeness in $z$ vs $u-g$ plane]{\includegraphics[width=6cm]{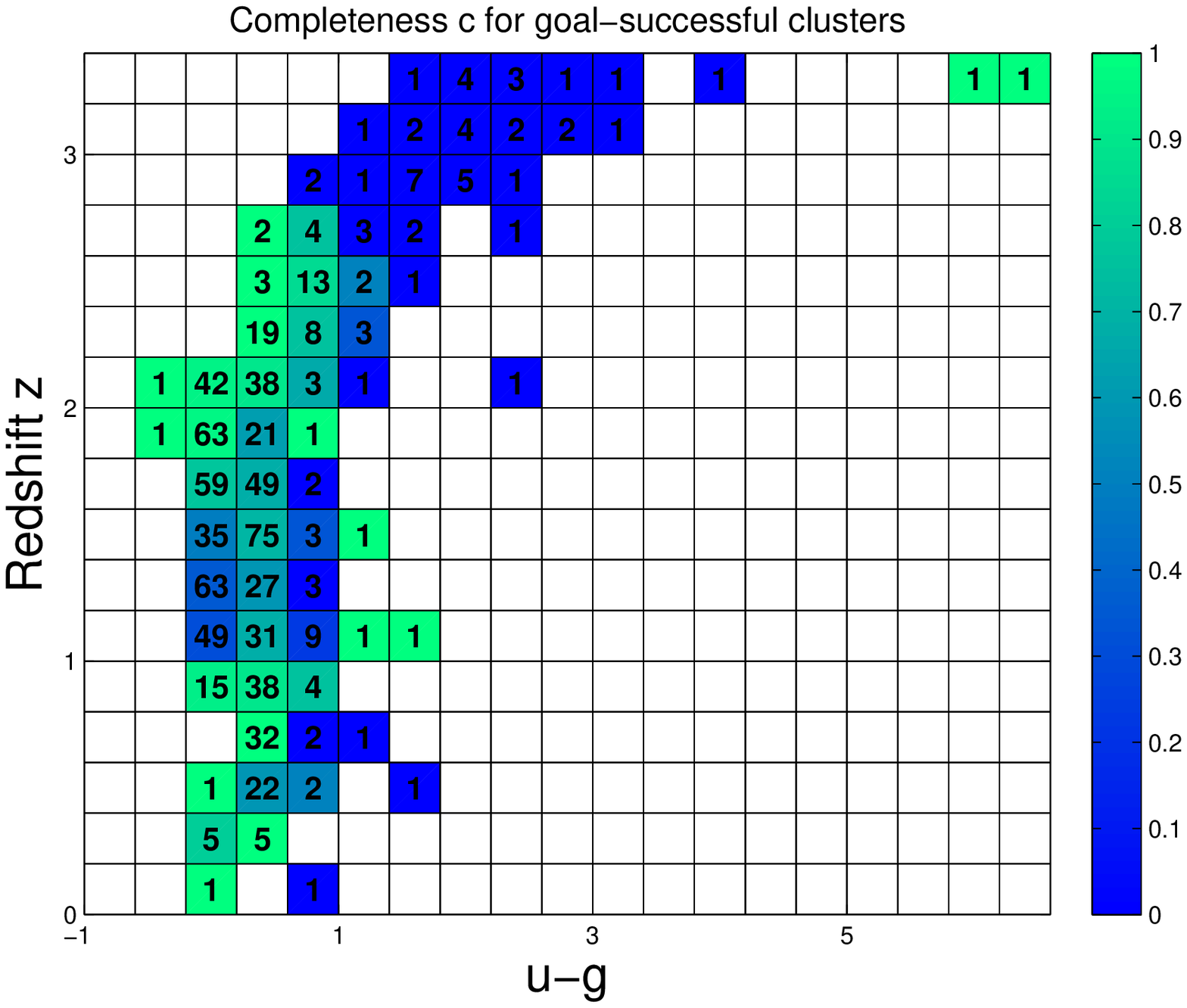}}\\
\subfigure[Efficiency in $z$ vs $g-r$ plane]{\includegraphics[width=6cm]{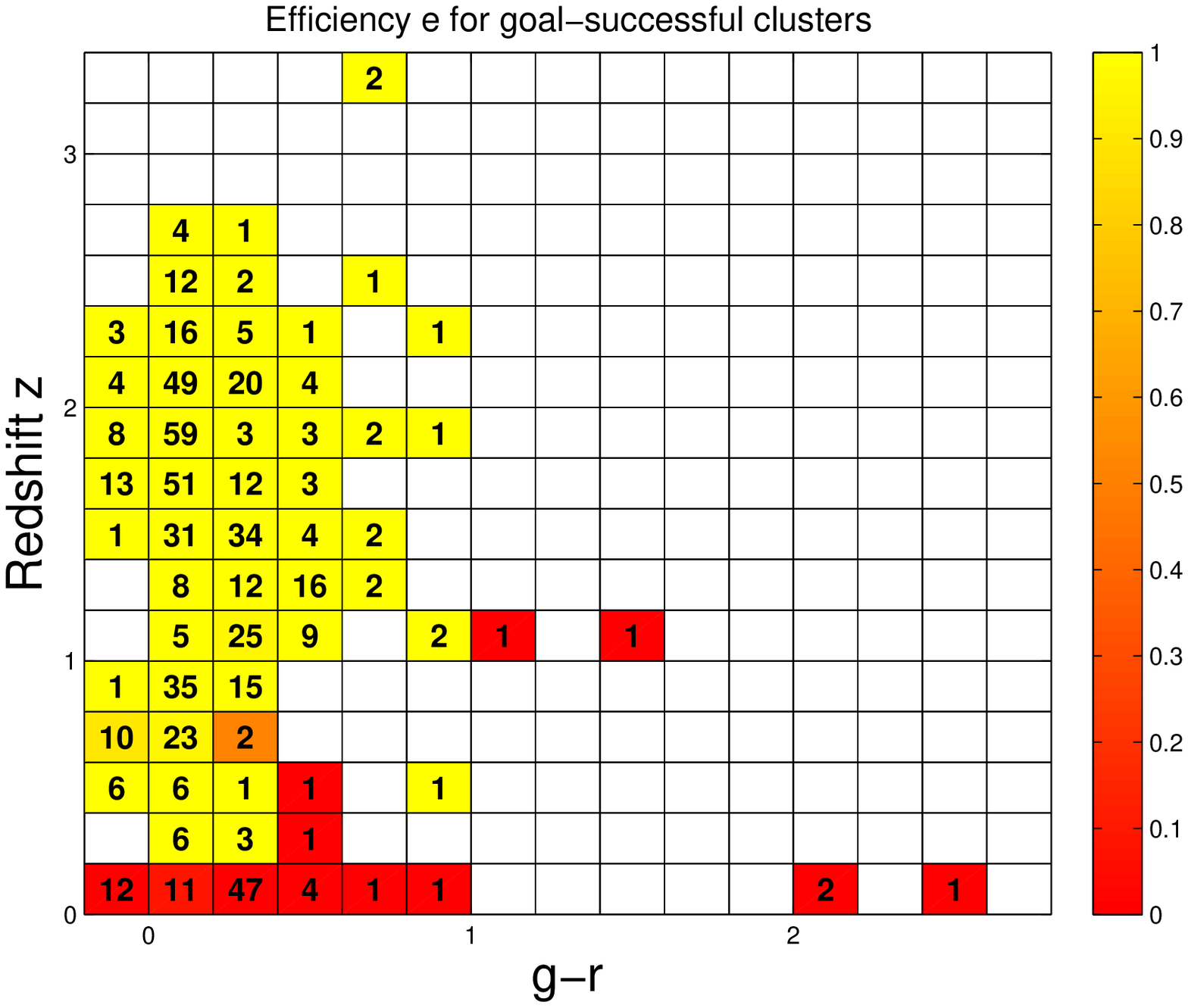}}
\subfigure[Completeness in $z$ vs $g-r$ plane]{\includegraphics[width=6cm]{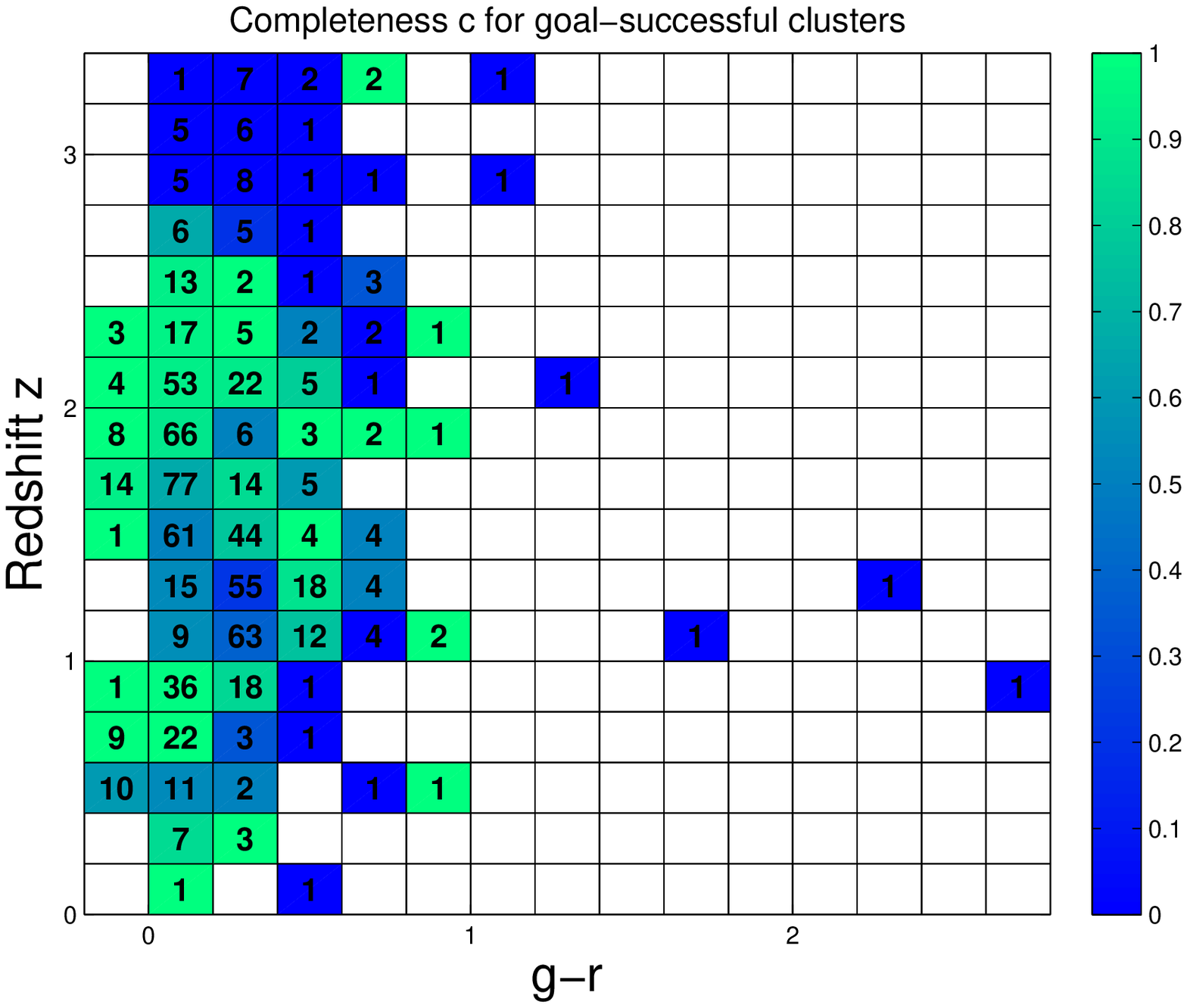}}\\
\subfigure[Efficiency in $z$ vs $r-i$ plane]{\includegraphics[width=6cm]{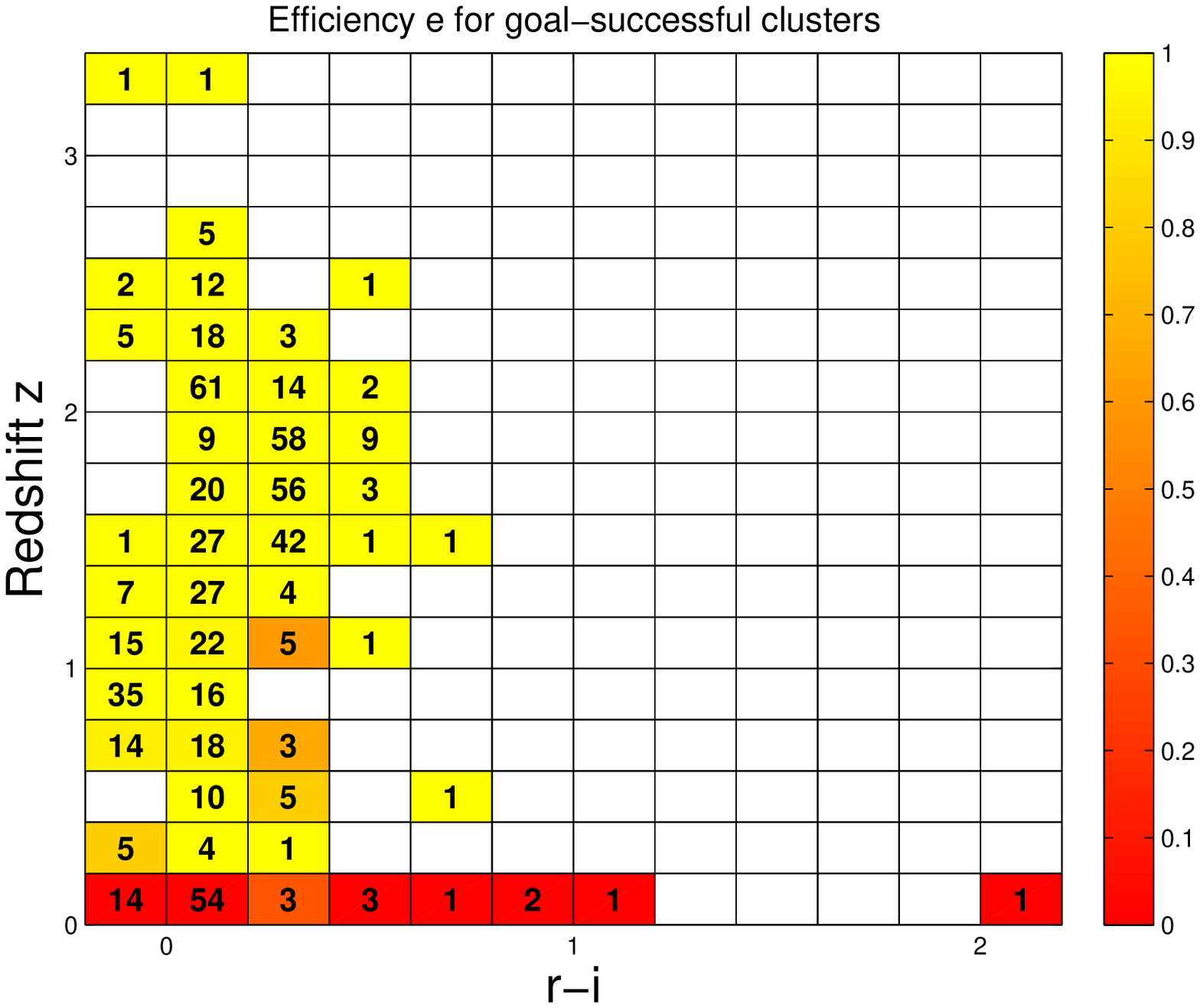}}
\subfigure[Completeness in $z$ vs $r-i$ plane]{\includegraphics[width=6cm]{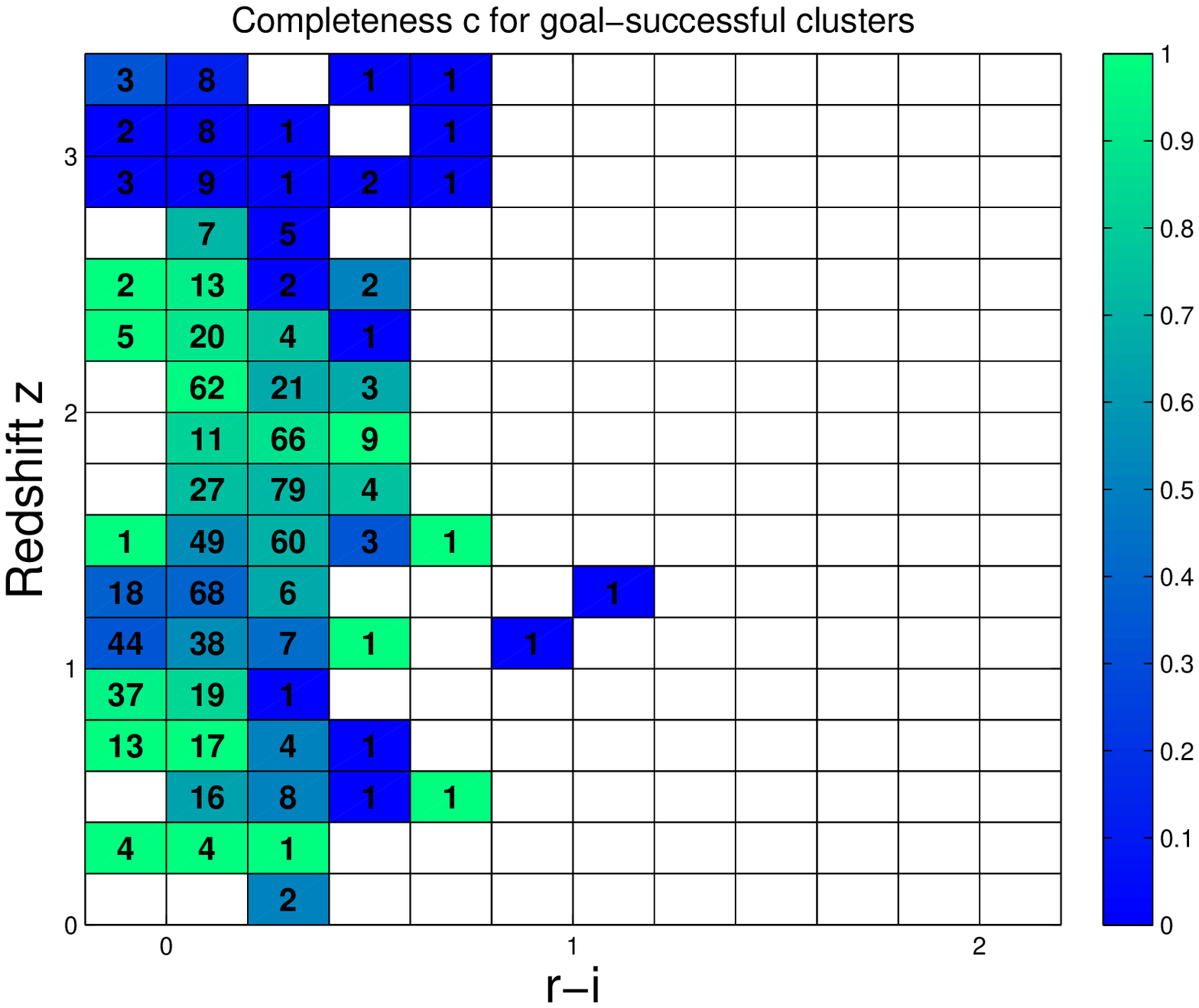}}\\
\subfigure[Efficiency in $z$ vs $i-z$ plane]{\includegraphics[width=6cm]{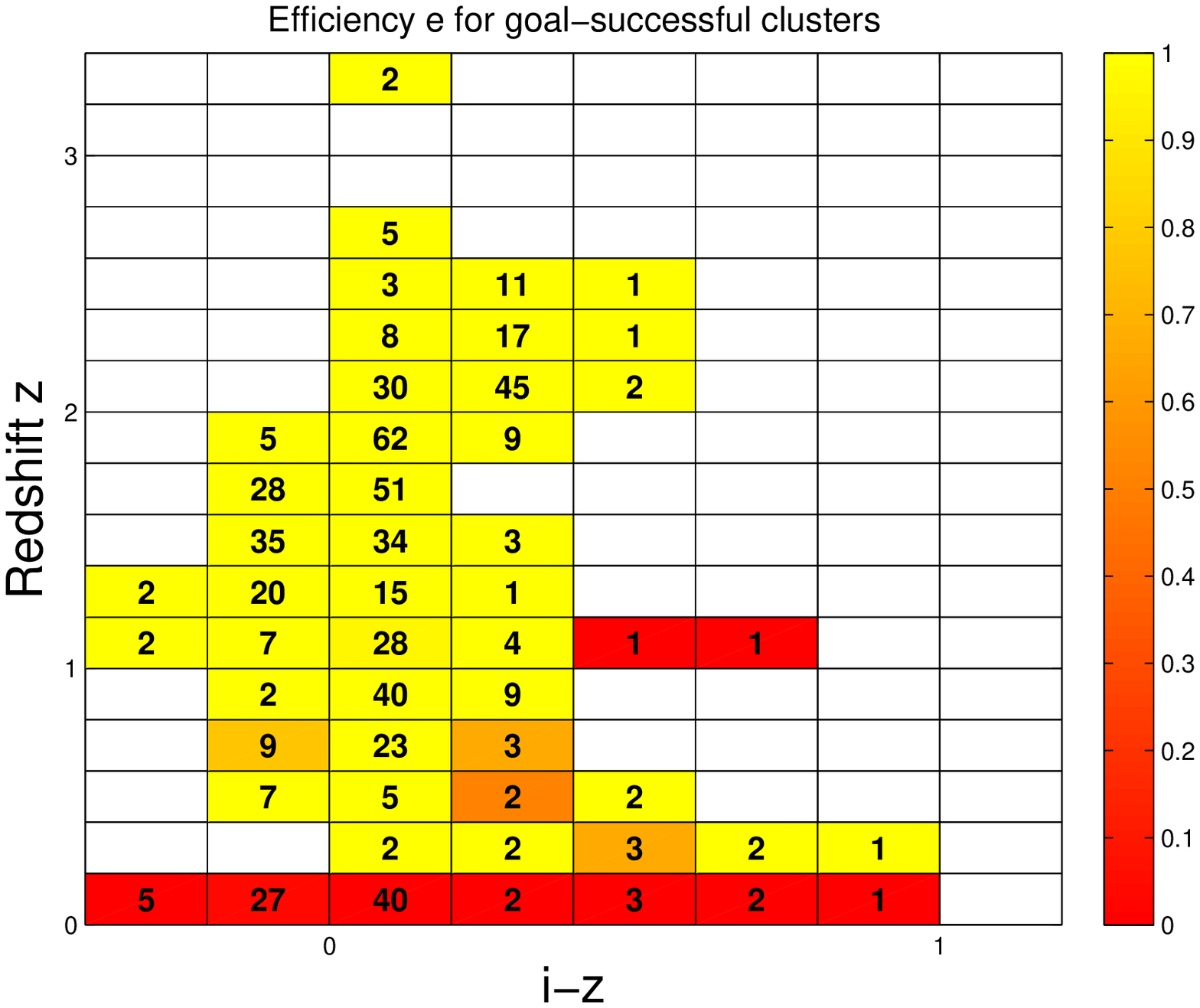}}
\subfigure[Completeness in $z$ vs $i-z$ plane]{\includegraphics[width=6cm]{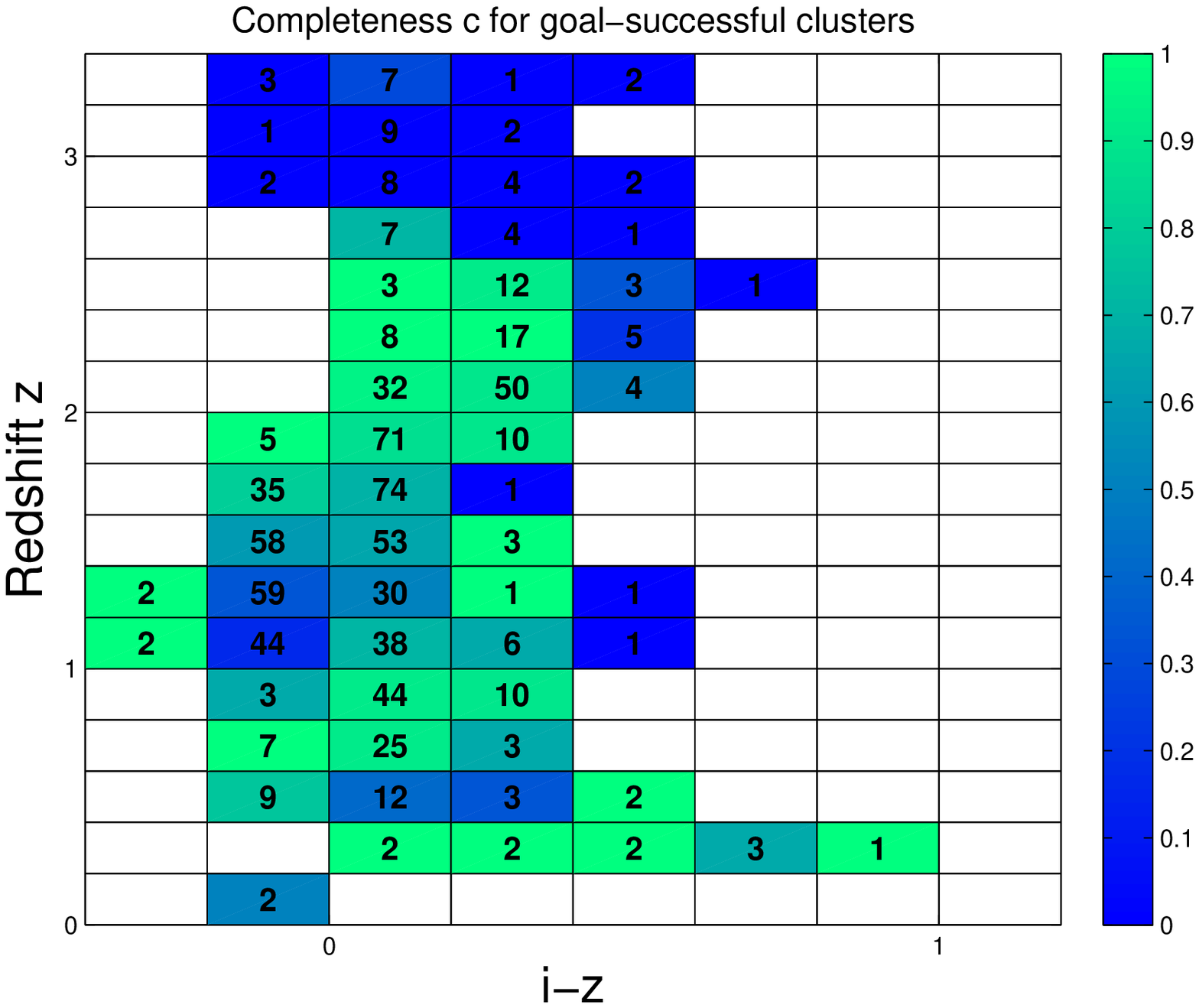}}\\
\caption{Local efficiency and completeness for the third experiment.}
\label{fig:paramparam_1app_3exp}
\end{figure}

\begin{figure}
\centering
\subfigure[Efficiency in $z$ vs $u-g$ plane]{\includegraphics[width=6cm]{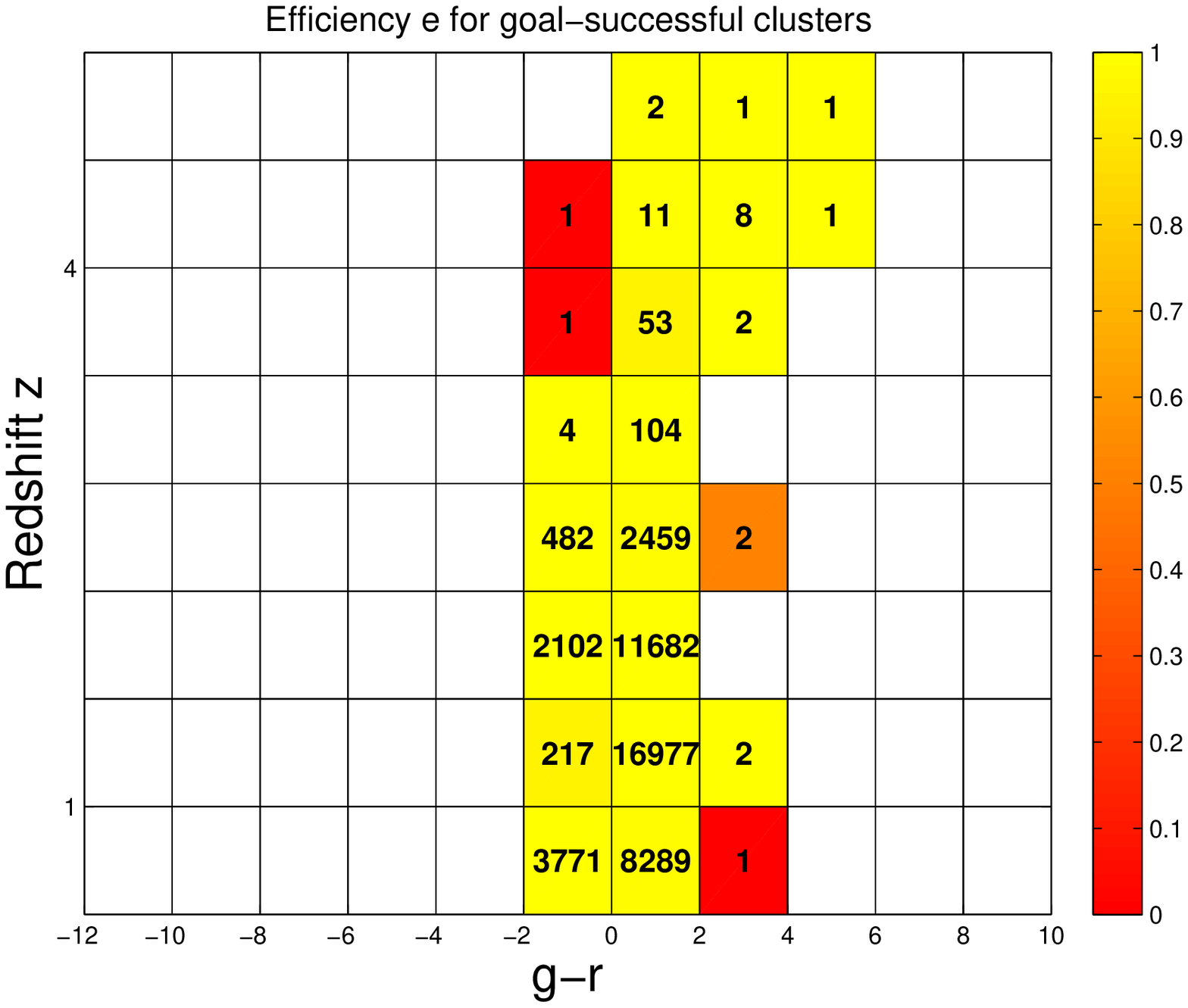}}
\subfigure[Completeness in $z$ vs $u-g$ plane]{\includegraphics[width=6cm]{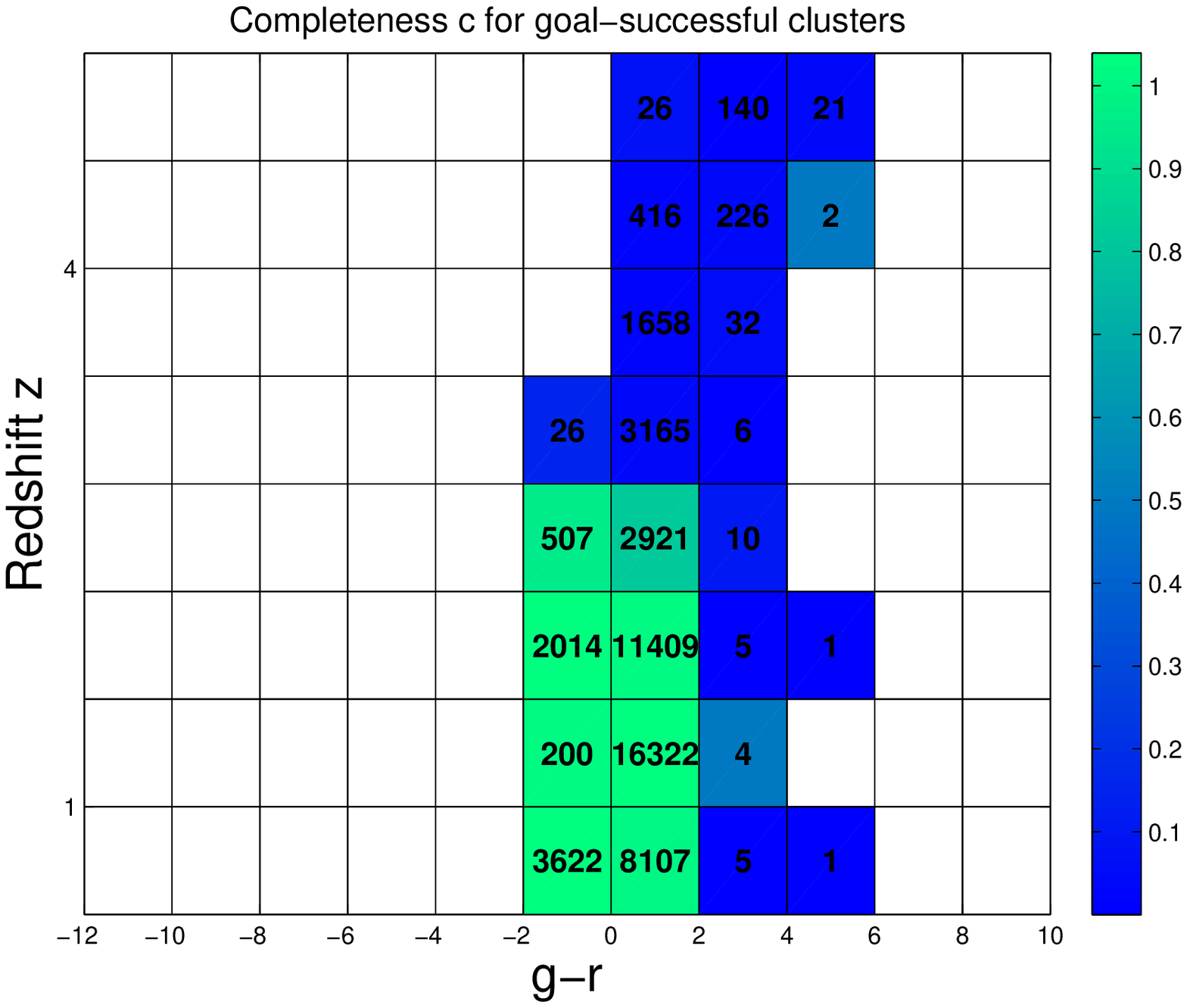}}\\
\subfigure[Efficiency in $z$ vs $g-r$ plane]{\includegraphics[width=6cm]{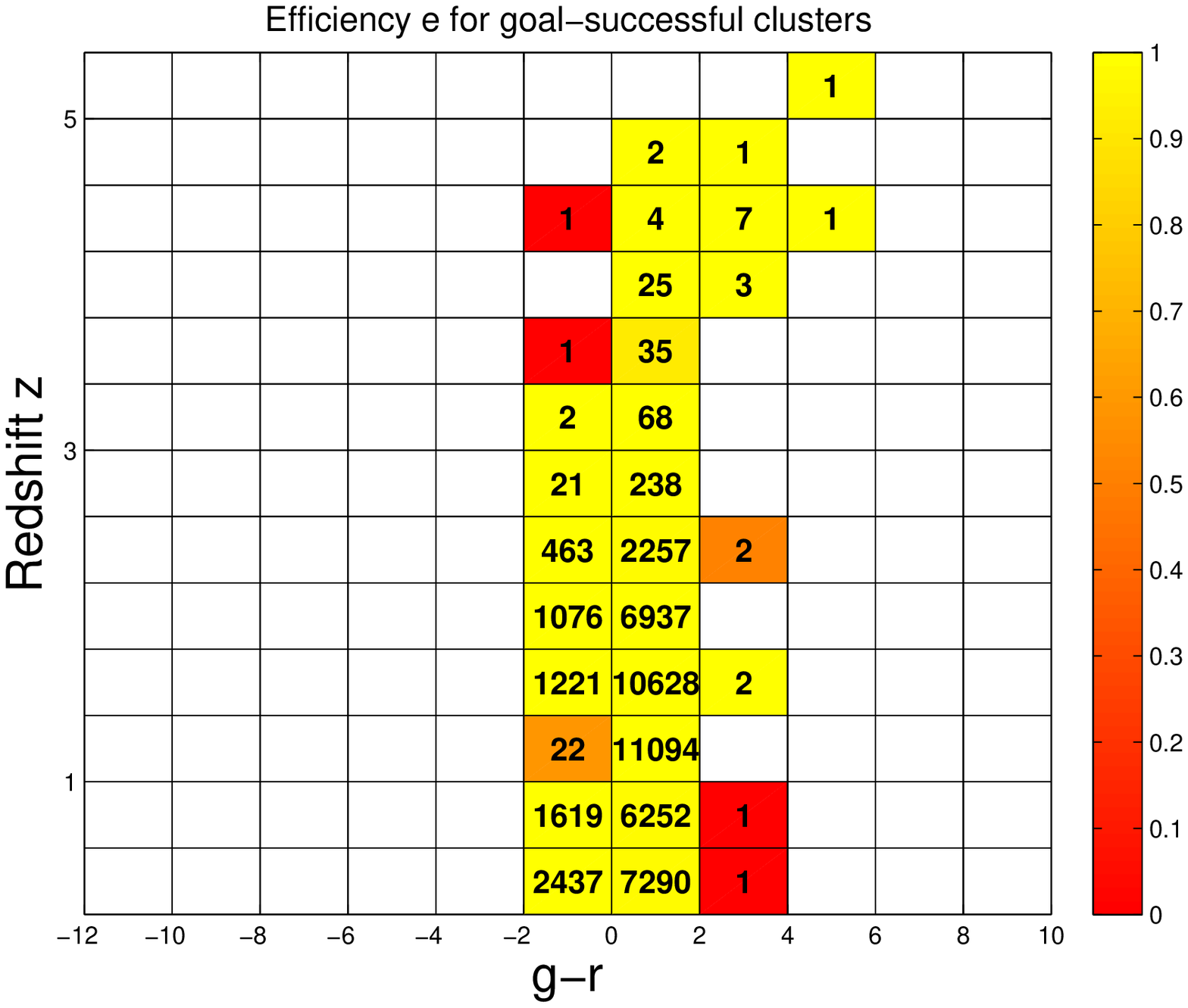}}
\subfigure[Completeness in $z$ vs $g-r$ plane]{\includegraphics[width=6cm]{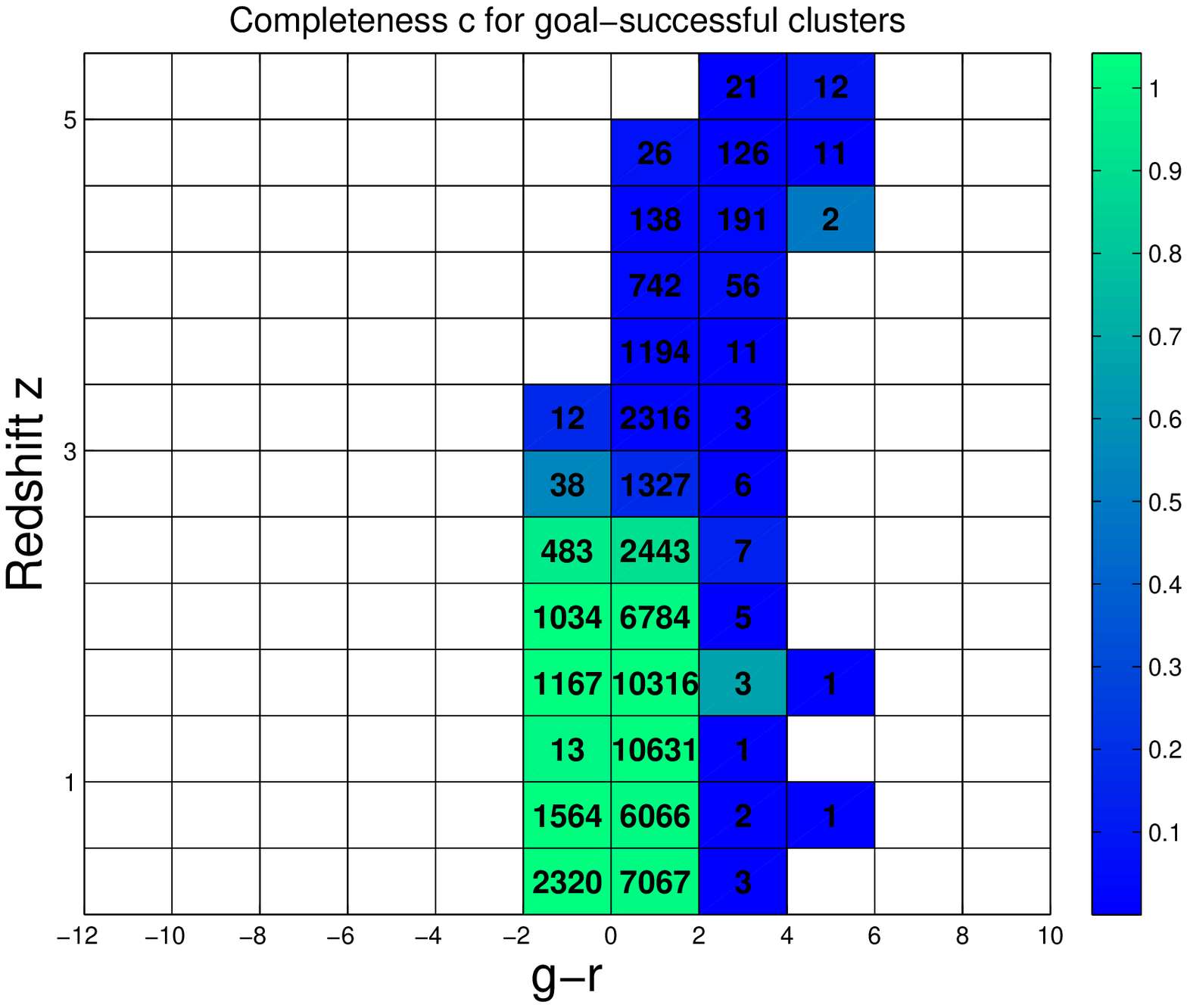}}\\
\subfigure[Efficiency in $z$ vs $r-i$ plane]{\includegraphics[width=6cm]{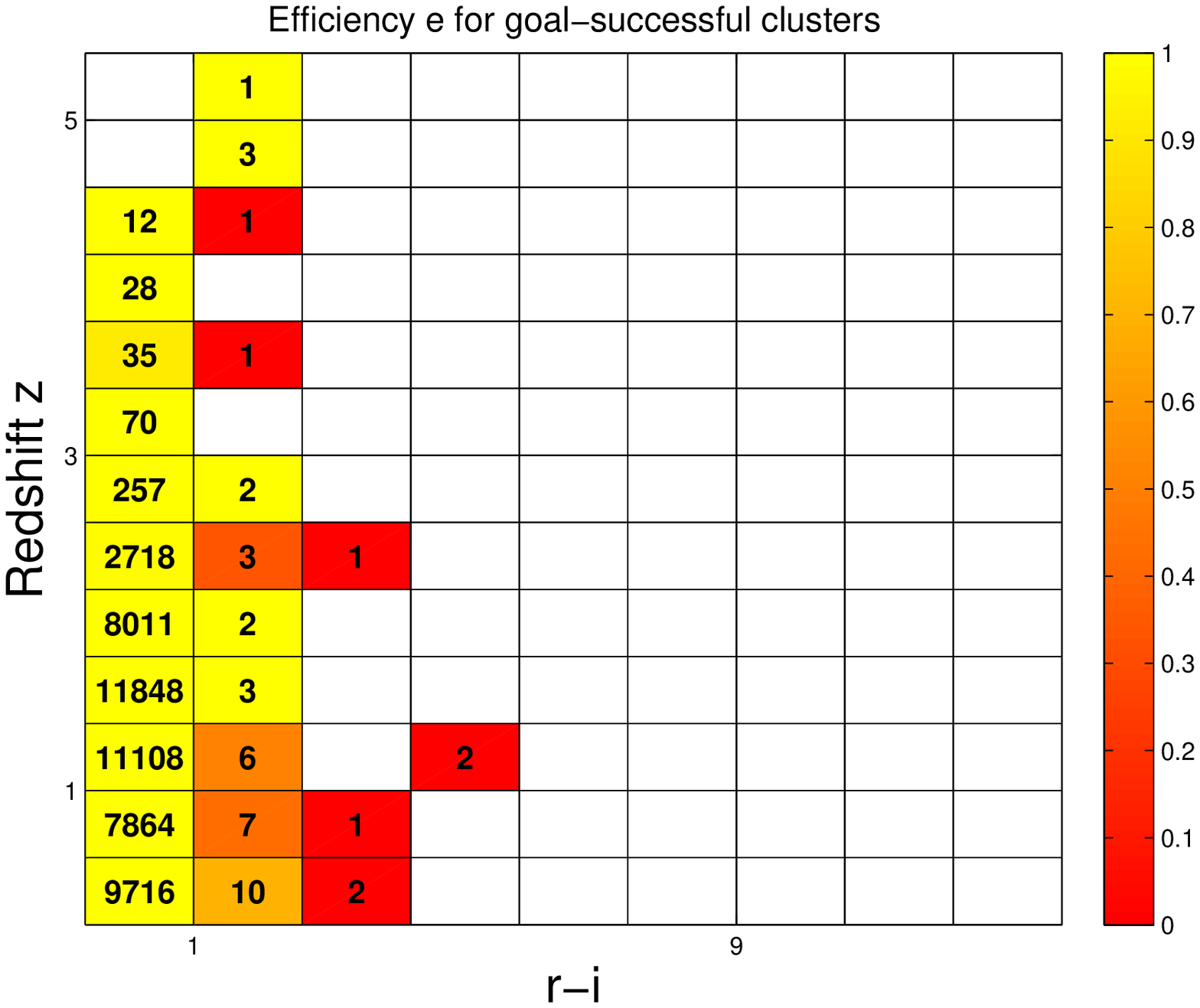}}
\subfigure[Completeness in $z$ vs $r-i$ plane]{\includegraphics[width=6cm]{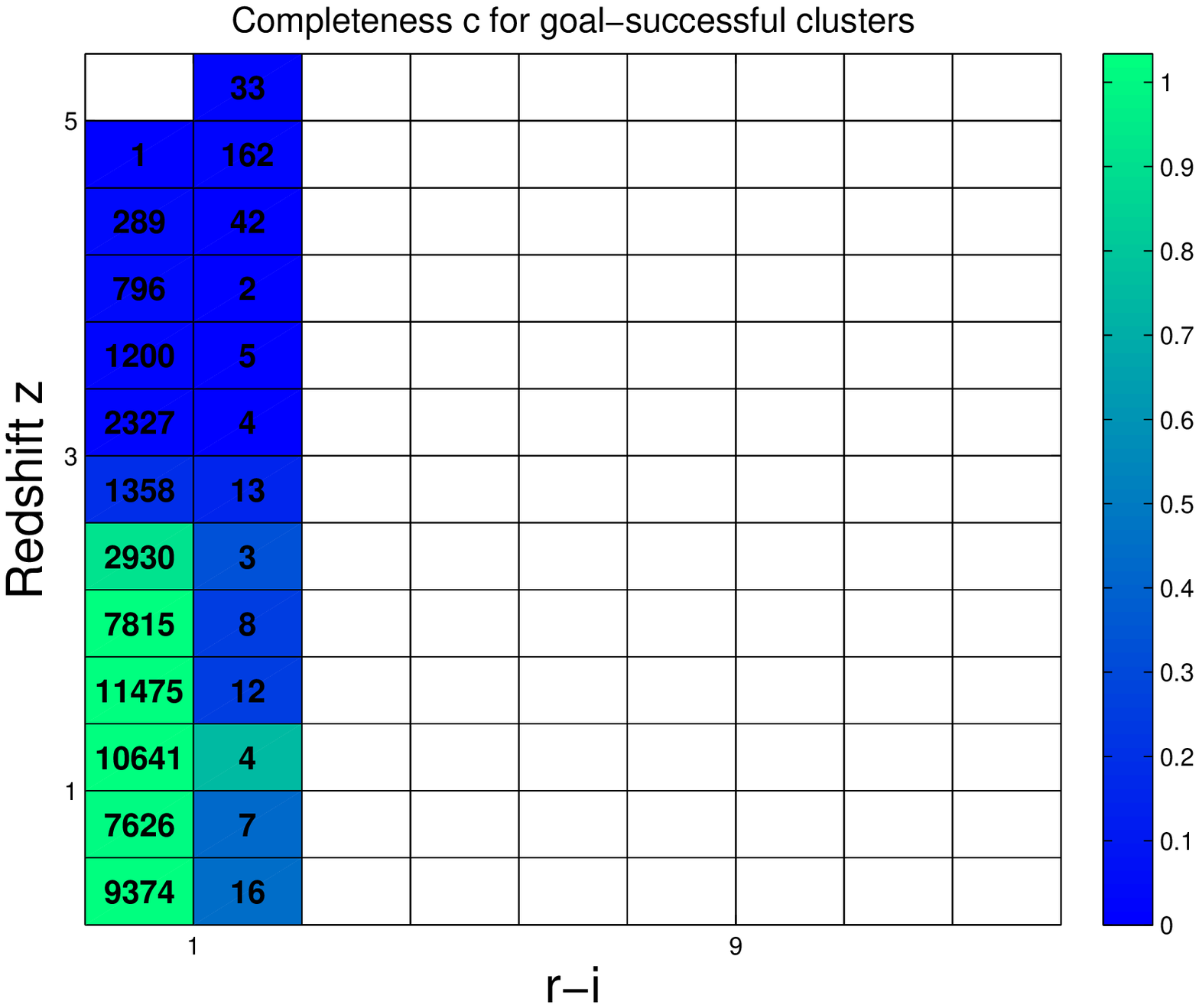}}\\
\subfigure[Efficiency in $z$ vs $i-z$ plane]{\includegraphics[width=6cm]{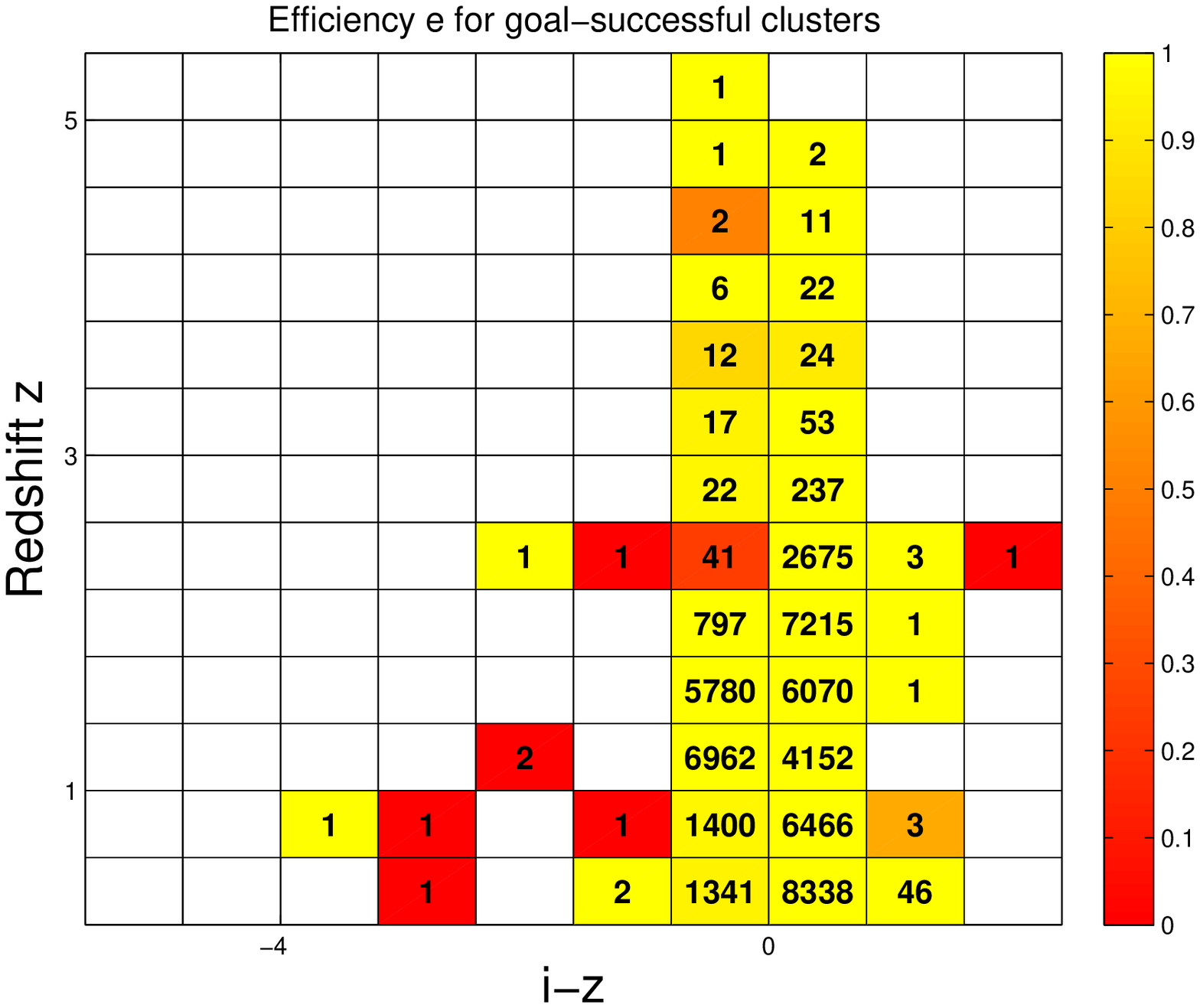}}
\subfigure[Completeness in $z$ vs $i-z$ plane]{\includegraphics[width=6cm]{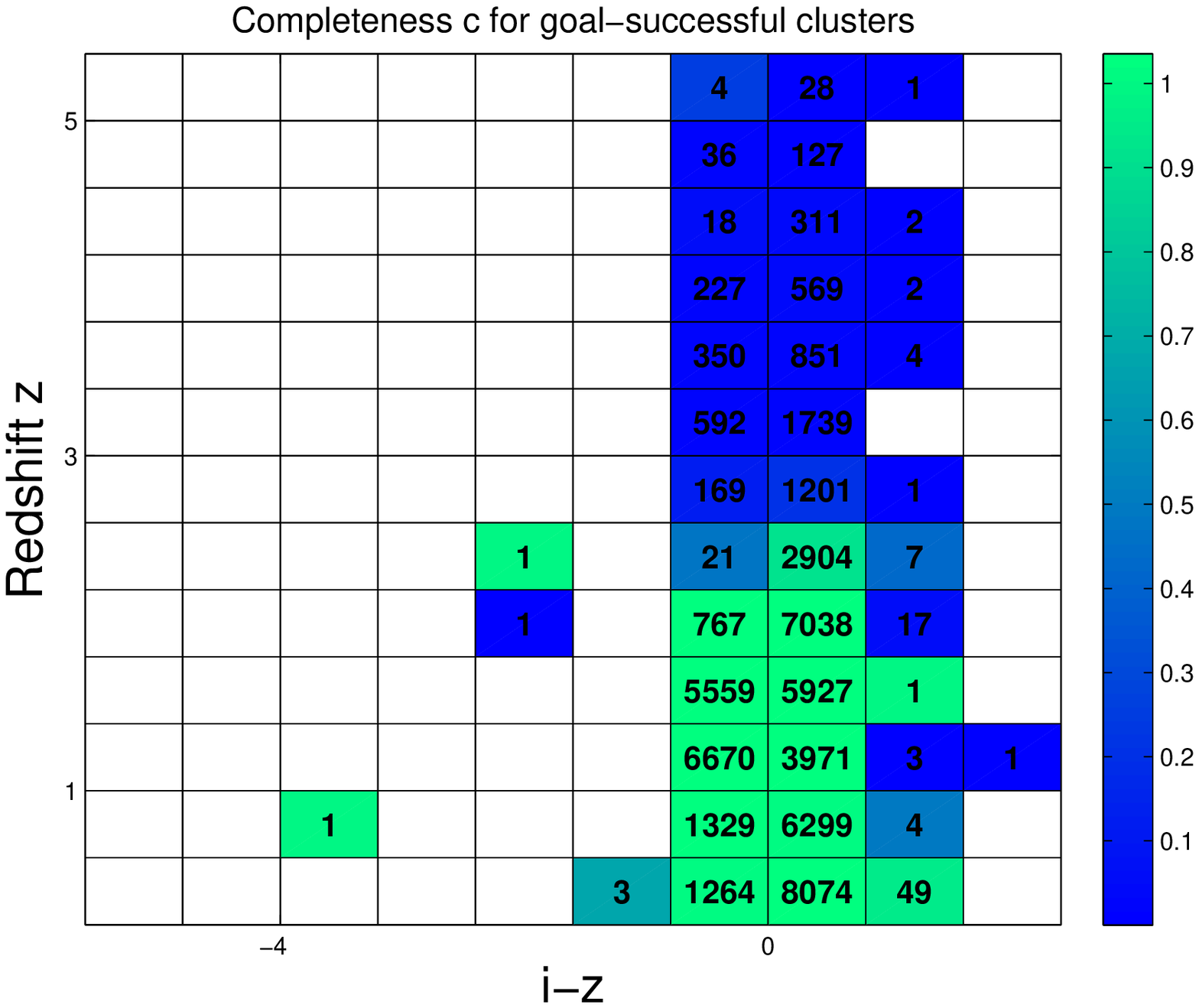}}\\
\caption{Local efficiency and completeness for the fourth experiment.}
\label{fig:paramparam_1app_4exp}
\end{figure}

In this appendix, the plots of the local estimates of the efficiency $e$ and the completeness $c$
in the planes determined by the redshift and each photometric parameter (i.e., each colour) are shown for all the experiments described in the
paper. Figure (\ref{fig:paramparam_1app_1exp}) represents the first experiment, while figures (\ref{fig:paramparam_1app_2exp_opt})
and (\ref{fig:paramparam_1app_2exp_infrared}) describe the second experiment, and the remaining figures (\ref{fig:paramparam_1app_3exp})
and (\ref{fig:paramparam_1app_4exp}) are associated to the third and fourth experiments, respectively. 

\end{document}